\documentclass{tADP2e}
\pdfoutput=1

\usepackage[latin9]{inputenc}
\usepackage{graphics,graphicx}         %you probably want to use figures
\usepackage{amsmath,amssymb,mathrsfs}  %you probably will use some math symbols
\usepackage{verbatim}          % for multiline comments
\usepackage{color,enumerate,esint,epstopdf,dcolumn}
\usepackage{breakurl}
\usepackage[normalem]{ulem} % to cross out text

\newcommand{\be}{\begin{equation}}
\newcommand{\ee}{\end{equation}}
\newcommand{\beq}{\begin{eqnarray}}
\newcommand{\eeq}{\end{eqnarray}}
\newcommand{\bem}{\begin{multline}}
\newcommand{\eem}{\end{multline}}

\def\la{\langle}
\def\ra{\rangle}

\begin{document}

\markboth{From Quantum Chaos and Eigenstate Thermalization to Statistical Mechanics and Thermodynamics}{Advances in Physics}

\articletype{Review}

\title{From Quantum Chaos and Eigenstate Thermalization to \\Statistical Mechanics and Thermodynamics}

\author{Luca D'Alessio$^{\rm a,b}$, Yariv Kafri$^{\rm c}$, Anatoli Polkovnikov$^{\rm b}$,
and Marcos Rigol$^{\rm a}$
\\\vspace{6pt}
$^{a}${\em{Department of Physics, The Pennsylvania State University, University Park, PA 16802, USA}};\\
$^{b}${\em{Department of Physics, Boston University, Boston, MA 02215, USA}};\\
$^{c}${\em{Department of Physics, Technion, Haifa 32000, Israel}}
\\\received{v0.0 released September 2015} 
}

\maketitle

\begin{abstract}

This review gives a pedagogical introduction to the eigenstate thermalization hypothesis (ETH), its basis, and its implications to statistical mechanics and thermodynamics. In the first part, ETH is introduced as a natural extension of ideas from quantum chaos and random matrix theory. To this end, we present a brief overview of classical and quantum chaos, as well as random matrix theory and some of its most important predictions. The latter include the statistics of energy levels, eigenstate components, and matrix elements of observables. Building on these, we introduce the ETH and show that it allows one to describe thermalization in isolated chaotic systems without invoking the notion of an external bath. We examine numerical evidence of eigenstate thermalization from studies of many-body lattice systems. We also introduce the concept of a quench as a means of taking isolated systems out of equilibrium, and discuss results of numerical experiments on quantum quenches. The second part of the review explores the implications of quantum chaos and ETH to thermodynamics. Basic thermodynamic relations are derived, including the second law of thermodynamics, the fundamental thermodynamic relation, fluctuation theorems, the fluctuation-dissipation relation, and the Einstein and Onsager relations. In particular, it is shown that quantum chaos allows one to prove these relations for individual Hamiltonian eigenstates and thus extend them to arbitrary stationary statistical ensembles. In some cases, it is possible to extend their regimes of applicability beyond the standard thermal equilibrium domain. We then show how one can use these relations to obtain nontrivial universal energy distributions in continuously driven systems. At the end of the review, we briefly discuss the relaxation dynamics and description after relaxation of integrable quantum systems, for which ETH is violated. We present results from numerical experiments and analytical studies of quantum quenches at integrability. We introduce the concept of the generalized Gibbs ensemble, and discuss its connection with ideas of prethermalization in weakly interacting systems. 


\vfill

\begin{flushright}
 {\it Verily at the first Chaos came to be, \\ but next wide-bosomed Earth, \\the ever-sure foundations of all\ldots}\\
 Hesiod, {\it Theogony}
\end{flushright}

\newpage

\centerline{\bfseries Contents}\medskip
\hbox to \textwidth{\hsize\textwidth\vbox{\hsize48pc

\hspace*{-12pt} {1.  Introduction}\\
{2. Chaos and Random Matrix Theory (RMT)}\\
\hspace*{8pt} {2.1. Classical Chaos}\\
\hspace*{8pt} {2.2. Random Matrix Theory}\\
\hspace*{25pt} {2.2.1. Chaotic Eigenfunctions}\\
\hspace*{25pt} {2.2.2. The Structure of the Matrix Elements of Operators}\\ 
\hspace*{8pt} {2.3. Berry-Tabor Conjecture}\\
\hspace*{8pt} {2.4. The Semi-Classical Limit and Berry's Conjecture}\\
{3. Quantum Chaos in Physical Systems}\\
\hspace*{8pt} {3.1. Examples of Wigner-Dyson and Poisson Statistics}\\
\hspace*{8pt} {3.2. The Structure of Many-Body Eigenstates}\\
\hspace*{8pt} {3.3. Quantum Chaos and Entanglement}\\
\hspace*{8pt} {3.4. Quantum Chaos and Delocalization in Energy Space}\\
{4. Eigenstate Thermalization}\\
\hspace*{8pt} {4.1. Thermalization in Quantum Systems}\\
\hspace*{8pt} {4.2. The Eigenstate Thermalization Hypothesis (ETH)}\\
\hspace*{25pt} {4.2.1. ETH and Thermalization}\\
\hspace*{25pt} {4.2.2. ETH and the Quantum Ergodic Theorem}\\
\hspace*{8pt} {4.3. Numerical Experiments in Lattice Systems}\\
\hspace*{25pt} {4.3.1. Eigenstate Thermalization}\\
\hspace*{25pt} {4.3.2. Quantum Quenches and Thermalization in Lattice Systems}\\
{5. Quantum Chaos and the Laws of Thermodynamics}\\
\hspace*{8pt} {5.1. General Setup and Doubly Stochastic Evolution}\\
\hspace*{25pt} {5.1.1. Properties of Master Equations and Doubly Stochastic Evolution}\\
\hspace*{8pt} {5.2. General Implications of Doubly Stochastic Evolution}\\
\hspace*{25pt} {5.2.1. The Infinite Temperature State as an Attractor of Doubly-Stochastic Evolution}\\
\hspace*{25pt} {5.2.2. Increase of the Diagonal Entropy Under Doubly Stochastic Evolution}\\
\hspace*{25pt} {5.2.3. The Second Law in the Kelvin Formulation for Passive Density Matrices}\\
\hspace*{8pt} {5.3. Implications of Doubly-Stochastic Evolution for Chaotic Systems}\\
\hspace*{25pt} {5.3.1. The Diagonal Entropy and the Fundamental Thermodynamic Relation}\\
\hspace*{25pt} {5.3.2. The Fundamental Relation vs the First Law of Thermodynamics}\\
{6. Quantum Chaos, Fluctuation Theorems, and Linear Response Relations}\\
\hspace*{8pt} {6.1. Fluctuation Theorems}\\
\hspace*{25pt} {6.1.1. Fluctuation Theorems for Systems Starting from a Gibbs State}\\
\hspace*{25pt} {6.1.2. Fluctuation Theorems for Quantum Chaotic Systems}\\
\hspace*{8pt} {6.2. Detailed Balance for Open Systems}\\
\hspace*{8pt} {6.3. Einstein's Energy Drift-Diffusion Relations for Isolated and Open Systems}\\
\hspace*{8pt} {6.4. Fokker-Planck Equation for the Heating of a Driven Isolated System}\\
\hspace*{8pt} {6.5. Fluctuation Theorems for Two (or More) Conserved Quantities}\\
\hspace*{8pt} {6.6. Linear Response and Onsager Relations}\\
\hspace*{8pt} {6.7. Non-linear Response Coefficients}\\
\hspace*{8pt} {6.8. ETH and the Fluctuation-Dissipation Relation for a Single Eigenstate}\\
\hspace*{8pt} {6.9. ETH and Two-Observable Correlation Functions}\\
{7. Application of Einstein's Relation to Continuously Driven Systems}\\
\hspace*{8pt} {7.1. Heating a Particle in a Fluctuating Chaotic Cavity}\\
\hspace*{8pt} {7.2. Driven Harmonic System and a Phase Transition in the Distribution Function}\\
\hspace*{8pt} {7.3. Two Equilibrating Systems}\\
{8. Integrable Models and the Generalized Gibbs Ensemble (GGE)}\\
\hspace*{8pt} {8.1. Constrained equilibrium: the GGE}\\
\hspace*{25pt} {8.1.1. Noninteracting Spinless Fermions}\\
\hspace*{25pt} {8.1.2. Hard-core Bosons}\\
\hspace*{8pt} {8.2. Generalized Eigenstate Thermalization}\\
\hspace*{25pt} {8.2.1. Truncated GGE for the Transverse Field Ising Model}\\
\hspace*{8pt} {8.3. Quenches in the XXZ model} \\
\hspace*{8pt} {8.4. Relaxation of Weakly Non-Integrable Systems: Prethermalization and Quantum Kinetic Equations}\\
{Appendix A. The Kicked Rotor} \\
{Appendix B. Zeros of the Riemann Zeta Function}  \\
{Appendix C. The Infinite Temperature State as an Attractor}  \\
{Appendix D. Birkhoff's Theorem and Doubly Stochastic Evolution}  \\
{Appendix E. Proof of $\langle W\rangle\ge 0$ for Passive Density Matrices and Doubly Stochastic Evolution} \\
{Appendix F. Derivation of the Drift Diffusion Relation for Continuous Processes} \\
{Appendix G. Derivation of Onsager Relations}
}}

\end{abstract}

\newpage

%%%%%%%%%%%%%%%%%%%%%%%%%%%%%%%%%%%%%%%%%%%%%%%%%%%%%%%%%%%%%%%%%%%%%%%%%%%%%%%%%%%%%%%%%%
%%%%%%%%%%%%%%%%%%%%%%%%%%%%%%%%%%%%%%%%%%%%%%%%%%%%%%%%%%%%%%%%%%%%%%%%%%%%%%%%%%%%%%%%%%
%%\chapter{PART 1: Eigenstate Thermalization}
%%%%%%%%%%%%%%%%%%%%%%%%%%%%%%%%%%%%%%%%%%%%%%%%%%%%%%%%%%%%%%%%%%%%%%%%%%%%%%%%%%%%%%%%%%
%%%%%%%%%%%%%%%%%%%%%%%%%%%%%%%%%%%%%%%%%%%%%%%%%%%%%%%%%%%%%%%%%%%%%%%%%%%%%%%%%%%%%%%%%%

%%%%%%%%%%%%%%%%%%%%%%%%%%%%%%%%%%%%%%%%%%%%%%%%%%%%%%%%%%%%%%%%%%%%%%%%%%%%%%%%%%%%%%%%%%
\section{Introduction\label{sec:sec1}}
%%%%%%%%%%%%%%%%%%%%%%%%%%%%%%%%%%%%%%%%%%%%%%%%%%%%%%%%%%%%%%%%%%%%%%%%%%%%%%%%%%%%%%%%%%

Despite the huge success of statistical mechanics in describing the macroscopic behavior of physical systems~\cite{ma_book_85,feynman_book_98}, its relation to the underlying microscopic dynamics has remained a subject of debate since the foundations were laid~\cite{boltzmann_96,boltzmann_97}. One of the most controversial topics has been the reconciliation of the time reversibility of most microscopic laws of nature and the apparent irreversibility of the laws of thermodynamics.

Let us first consider an isolated {\it classical} system subject to some macroscopic constraints (such as conservation of the total energy and confinement to a container). To derive its equilibrium properties, within statistical mechanics, one takes a fictitious ensemble of systems evolving under the same Hamiltonian and subject to the same macroscopic constraints. Then, a probability is assigned to each member of the ensemble, and the macroscopic behavior of the system is computed by averaging over the fictitious ensemble~\cite{gibbs_book_65}. For an isolated system, the ensemble is typically chosen to be the microcanonical one. To ensure that the probability of each configuration in phase space does not change in time under the Hamiltonian dynamics, as required by equilibrium, the ensemble includes, with equal probability, all configurations compatible with the macroscopic constraints. The correctness of the procedure used in statistical mechanics to describe real systems is, however, far from obvious. In actual experiments, there is generally no ensemble of systems -- there is one system -- and the relation between the calculation just outlined and the measurable outcome of the underlying microscopic dynamics is often unclear. To address this issue, two major lines of thought have been offered.

In the first line of thought, which is found in most textbooks, one invokes the ergodic hypothesis~\cite{penrose_book_65} (refinements such as mixing are also invoked~\cite{lebowitz_penrose_73}). This hypothesis states that during its time evolution an ergodic system visits every region in phase space (subjected to the macroscopic constraints) and that, in the long-time limit, the time spent in each region is proportional to its volume. Time averages can then be said to be equal to ensemble averages, and the latter are the ones that are ultimately computed \cite{penrose_book_65}. The ergodic hypothesis essentially implies that the ``equal probability" assumption used to build the microcanonical ensemble is the necessary ingredient to capture the long-time average of observables. This hypothesis has been proved for a few systems, such as the Sinai billiard~\cite{sinai_63,sinai_70}, the Bunimovich stadium~\cite{bunimovich_79}, and systems with more than two hard spheres on a $d$-dimensional torus ($d\geq 2$)~\cite{simanyi_04}.

Proving that there are systems that are ergodic is an important step towards having a mathematical foundation of statistical mechanics. However, a few words of caution are necessary. First, the time scales needed for a system to explore phase space are exponentially large in the number of degrees of freedom, that is, they are irrelevant to what one observes in macroscopic systems. Second, the ergodic hypothesis implies thermalization only in a \textit{weak sense}. Weak refers to the fact that the ergodic hypothesis deals with \textit{long-time averages of observables} and not with the \textit{values of the observables at long times}. These two can be very different. Ideally, one would like to prove thermalization in a \textit{strong sense}, namely, that instantaneous values of observables approach the equilibrium value predicted by the microcanonical ensemble and remain close to it at almost all subsequent times. This is what is seen in most experiments involving macroscopic systems. Within the strong thermalization scenario, the instantaneous values of observables are nevertheless expected to deviate, at some rare times, from their typical value. For a system that starts its dynamics with a non-typical value of an observable, this is, in fact, guaranteed by the Poincar\'e recurrence theorem \cite{poincare_90}. This theorem states that during its time evolution any finite system eventually returns arbitrarily close to the initial state. However, the Poincar\'e recurrence time is exponentially long in the number of degrees of freedom and is not relevant to observations in macroscopic systems. Moreover, such recurrences are not at odds with statistical mechanics, which allows for atypical configurations to occur with exponentially small probabilities. We should stress that, while the ergodic hypothesis is expected to hold for most interacting systems, there are notable exceptions, particularly in low dimensions. For example, in one dimension, there are many known examples of (integrable or near integrable\footnote{We briefly discuss classical integrability and chaos in Sec.~\ref{sec:sec2} and quantum integrability in Sec.~\ref{sec:sec8}}) systems that do not thermalize, not even in the weak sense~\cite{faddeev_tachtajan_07}. A famous example is the Fermi-Pasta-Ulam numerical experiment in a chain of {\em anharmonic} oscillators, for which the most recent results show no (or extremely slow) thermalization~\cite{fermi_pasta_55, dauxois_08}. This problem had a major impact in the field of nonlinear physics and classical chaos (see, e.g., Ref.~\cite{berman_izrailev_2005}). Relaxation towards equilibrium can also be extremely slow in turbulent systems~\cite{kozik_svistunov_09} and in glassy systems~\cite{bouchad_92}.

In the second, perhaps more appealing, line of thought one notes that macroscopic observables essentially exhibit the same values in almost all configurations in phase space that are compatible with a given set of macroscopic constraints. In other words, almost all the configurations are equivalent from the point of view of macroscopic observables. For example, the number of configurations in which the particles are divided equally (up to non-extensive corrections) between two halves of a container is exponentially larger than configurations in which this is not the case.  Noting that ``typical" configurations vastly outnumber ``atypical" ones and that, under chaotic dynamics each configuration is reached with equal probability, it follows that ``atypical" configurations quickly evolve into ``typical" ones, which almost never evolve back into ``atypical" configurations. Within this line of thought, thermalization boils down to reaching a ``typical" configuration. This happens much faster than any relevant exploration of phase space required by ergodicity. Note that this approach only applies when the measured quantity is macroscopic (such as the particle number mentioned in the example considered above). If one asks for the probability of being in a specific microscopic configuration, there is no meaning in separating ``typical" from ``atypical" configurations and the predictive power of this line of reasoning is lost. As appealing as this line of reasoning is, it lacks rigorous support.

Taking the second point of view, it is worth noting that while most configurations in phase space are ``typical", such configurations are difficult to create using external perturbations. For example, imagine a piston is moved to compress air in a container. If the piston is not moved slowly enough, the gas inside the container will not have time to equilibrate and, as a result, during the piston's motion (and right after the piston stops) its density will not be uniform, that is, the system is not in a ``typical" state. In a similar fashion, by applying a radiation pulse to a system, one will typically excite some degrees of freedom resonantly, for example, phonons directly coupled to the radiation. As a result, right after the pulse ends, the system is in an atypical state. Considering a wide range of experimental protocols, one can actually convince oneself that it is generally difficult to create ``typical" configurations if one does not follow a very slow protocol or without letting the system evolve by itself.

At this point, a comment about time-reversal symmetry is in order. While the microscopic laws of physics usually exhibit time-reversal symmetry, notable exceptions include systems with external magnetic fields, the resulting macroscopic equations used to describe thermodynamic systems do not exhibit such a symmetry. This can be justified using the second line of reasoning -- it is exponentially rare for a system to evolve into an ``atypical" state by itself. Numerical experiments have been done (using integer arithmetic) in which a system was started in an atypical configuration, was left to evolve, and, after some time, the velocities of all particles were reversed. In those experiments, the system was seen to return to the initial (atypical) configuration~\cite{levesque_verlet_93}. Two essential points to be highlighted from these numerical simulations are: (i) after reaching the initial configuration, the system continued its evolution towards typical configurations (as expected, in a time-symmetric fashion) and (ii) the time-reversal transformation needed to be carried out with exquisite accuracy to observe a return to the initial (atypical) configuration (hence, the need of integer arithmetic). The difficulty in achieving the return increases dramatically with increasing system size and with the time one waits before applying the time-reversal transformation. It is now well understood, in the context of fluctuation theorems \cite{jarzynski_97,crooks_99}, that violations of the second law (i.e., evolution from typical to atypical configurations) can occur with a probability that decreases exponentially with the number of degrees of freedom in the system. These have been confirmed experimentally (see, e.g., Ref.~\cite{collin_05}). As part of this review, we derive fluctuation theorems in the context of quantum mechanics~\cite{kurchan_00, tasaki_00, campisi_hanggi_11}. 

Remarkably, a recent breakthrough \cite{deutsch_91,srednicki_94,rigol_dunjko_08} has put the understanding of thermalization in quantum systems on more solid foundations than the one discussed so far for classical systems. This breakthrough falls under the title of the eigenstate thermalization hypothesis (ETH). This hypothesis can be formulated as a mathematical ansatz with strong predictive powers \cite{srednicki_99}. ETH and its implications for statistical mechanics and thermodynamics are the subject of the review. As we discuss, ETH combines ideas of chaos and typical configurations in a clear mathematical form that is unparalleled in classical systems. This is remarkable considering that, in some sense, the relation between microscopic dynamics and statistical mechanics is more subtle in quantum mechanics than in classical mechanics. In fact, in quantum mechanics one usually does not use the notion of phase space as one cannot measure the positions and momenta of particles simultaneously. The equation dictating the dynamics (Schr\"odinger's equation) is linear which implies that the key ingredient leading to chaos in classical systems, that is, nonlinear equations of motion, is absent in quantum systems.

As already noted by von Neumann in 1929, when discussing thermalization in isolated quantum systems one should focus on physical observables as opposed to wave functions or density matrices describing the entire system \cite{vonneumann_29}. This approach is similar to the one described above for classical systems, in which the focus is put on macroscopic observables and ``typical" configurations. In this spirit, ETH states that the eigenstates of generic quantum Hamiltonians are ``typical" in the sense that the statistical properties of physical observables\footnote{We will explain what we mean by physical observables when discussing the ETH ansatz.} are the same as those predicted by the microcanonical ensemble. As we will discuss, ETH implies that the expectation values of such observables as well as their fluctuations in isolated quantum systems far from equilibrium relax to (nearly) time-independent results that can be described using traditional statistical mechanics ensembles \cite{deutsch_91,srednicki_94,rigol_dunjko_08}. This has been verified in several quantum lattice systems and, according to ETH, should occur in generic many-body quantum systems. We also discuss how ETH is related to quantum chaos in many-body systems, a subject pioneered by Wigner in the context of Nuclear Physics \cite{wigner_55}. Furthermore, we argue that one can build on ETH not only to understand the emergence of a statistical mechanics description in isolated quantum systems, but also to derive basic thermodynamic relations, linear response relations, and fluctuation theorems.

When thinking about the topics discussed in this review some may complain about the fact that, unless the entire universe is considered, there is no such thing as an isolated system, i.e., that any description of a system of interest should involve a bath of some sort. While this observation is, strictly speaking, correct, it is sometimes experimentally irrelevant. The time scales dictating internal equilibration in ``well-isolated'' systems can be much faster than the time scales introduced by the coupling to the ``outside world''. It then makes sense to question whether, in experiments with well-isolated systems, observables can be described using statistical mechanics on time scales much shorter than those introduced by the coupling to the outside world.  This question is of relevance to current experiments with a wide variety of systems. For example, in ultracold quantum gases that are trapped in ultrahigh vacuum by means of (up to a good approximation) conservative potentials \cite{bloch_dalibard_review_08,cazalilla_citro_11}. The near unitary dynamics of such systems has been observed in beautiful experiments on collapse and revival phenomena of bosonic \cite{greiner_mandel_02,will_best_10,will_best_11} and fermionic \cite{will_iyer_15} fields, lack of relaxation to the predictions of traditional ensembles of statistical mechanics \cite{kinoshita_wenger_06,gring_kuhnert_12,langen_erne_15}, and dynamics in optical lattices that were found to be in very good agreement with numerical predictions for unitary dynamics \cite{trotzky_chen_12}. In optical lattice experiments, the energy conservation constraint, imposed by the fact that the system are ``isolated'', has also allowed the observation of counterintuitive phenomena such as the formation of stable repulsive bound atom pairs \cite{winkler_thalhammer_06} and quantum distillation \cite{xia_zundel_15} in ultracold bosonic systems. Other examples of nearly isolated systems include nuclear spins in diamond~\cite{childress_dutt_06}, pump-probe experiments in correlated materials in which dynamics of electrons and holes are probed on time scales much faster than the relaxation time associated with electron-phonon interactions~\cite{wall_brida_11, basov_averitt_11}, and ensembles of ultra-relativistic particles generated in high-energy collisions~\cite{berges_boguslavski_14}.

This review can naturally be separated in two parts, and an addendum. In the first part, Secs.~\ref{sec:sec2}--\ref{sec:sec4}, we briefly introduce the concept of quantum chaos, discuss its relation to random matrix theory (RMT), and calculate its implications to observables. We then introduce ETH, which is a natural extension of RMT, and discuss its implications to thermalization in isolated systems, that is, relaxation of observables to the thermal equilibrium predictions. We illustrate these ideas with multiple numerical examples. In the second part, Secs.~\ref{sec:sec5}--\ref{sec:sec7}, we extend our discussion of the implications of quantum chaos and ETH to dynamical processes. We show how one can use quantum chaos and ETH to derive various thermodynamic relations (such as fluctuation theorems, fluctuation-dissipation relations, Onsager relations, and Einstein relations), determine leading finite-size corrections to those relations, and, in some cases, generalize them (e.g., the Onsager relation) beyond equilibrium. Finally, in the addendum (Sec.~\ref{sec:sec8}), we discuss the relaxation dynamics and description after relaxation of integrable systems after a quench. We introduce the generalized Gibbs ensemble (GGE), and, using time-dependent perturbation theory, show how it can be used to derive kinetic equations. We note that some of these topics have been discussed in other recent reviews \cite{dziarmaga_10,polkovnikov_sengupta_11,yukalov_11,nandkishore_huse_14,eisert_friesdorf_15,gogolin_eisert_15} and special journal issues \cite{cazalilla_rigol_10,daley_rigol_14}.

%%%%%%%%%%%%%%%%%%%%%%%%%%%%%%%%%%%%%%%%%%%%%%%%%%%%%%%%%%%%%%%%%%%%%%%%%%%%%%%%%%%%%%%%%%
\section{Chaos and Random Matrix Theory (RMT)}\label{sec:sec2}
%%%%%%%%%%%%%%%%%%%%%%%%%%%%%%%%%%%%%%%%%%%%%%%%%%%%%%%%%%%%%%%%%%%%%%%%%%%%%%%%%%%%%%%%%%

\subsection{Classical Chaos}
In this section, we very briefly discuss chaotic dynamics in classical systems. We refer the readers to Refs.~\cite{lichtenberg_liebermann_book_85,chaosbook}, and the literature therein, for further information about this topic. As the focus of this review is on quantum chaos and the eigenstate thermalization hypothesis, we will not attempt to bridge classical chaos and thermalization. This has been a subject of continuous controversies.  

While there is no universally accepted rigorous definition of chaos, a system is usually considered chaotic if it exhibits a strong (exponential) sensitivity of phase-space trajectories to small perturbations. Although chaotic dynamics are generic, there is a class of systems for which dynamics are not chaotic. They are known as integrable systems~\cite{arnold_89}. Specifically, a classical system whose Hamiltonian is $H({\bf p},{\bf q})$, with canonical coordinates ${\bf q}=(q_1,\cdots,q_N)$ and momenta ${\bf p}=(p_1,\cdots,p_N)$, is said to be integrable if it has as many functionally independent conserved quantities ${\bf I}=(I_1,\cdots,I_N)$ in involution as degrees of freedom $N$:
\be
\left\lbrace I_j,H\right\rbrace =0,
\quad \left\lbrace I_j,I_k \right\rbrace =0, \quad
\text{where} \quad\left\lbrace f,g \right\rbrace = \sum_{j=1,N} 
\frac{\partial f}{\partial q_j}\frac{\partial g}{\partial p_j} -
\frac{\partial f}{\partial p_j}\frac{\partial g}{\partial q_j}.
\ee
From Liouville's integrability theorem \cite{jose_saletan_book_98}, it follows that there is a canonical transformation $(p,q)\rightarrow (I,\Theta)$ (where $I,\Theta$ are called action-angle variables) such that $H(p,q)=H(I)$ \cite{arnold_89}. As a result, the solutions of the equations of motion for the action-angle variables are trivial: $I_j(t)=I_j^0=\mathrm{constant}$, and $\Theta_j(t)=\Omega_j t+\Theta_j(0)$. For obvious reasons, the motion is referred to as taking place on an $N$-dimensional torus, and it is not chaotic.

\begin{figure}[!t]
\includegraphics[width=14cm]{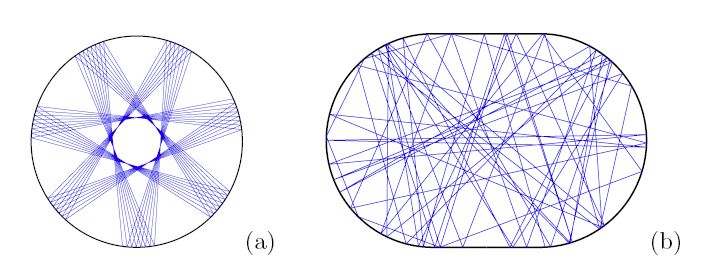}
\caption{Examples of trajectories of a particle bouncing in a cavity: (a) non-chaotic circular and (b) chaotic Bunimovich stadium. The images were taken from scholarpedia~\cite{stockmann_10}.}
\label{fig:billiards}
\end{figure}

To get a feeling for the differences between integrable and chaotic systems, in Fig.~\ref{fig:billiards}, we illustrate the motion of a particle in both an integrable and a chaotic two-dimensional cavity~\cite{stockmann_10}. Figure~\ref{fig:billiards}(a) illustrates the trajectory of a particle in an integrable circular cavity. It is visually apparent that the trajectory is a superposition of two periodic motions along the radial and angular directions. This is a result of the system having two conserved quantities, energy and angular momentum~\cite{landau_lifshitz_1_76}. Clearly, the long-time average of the particle density does not correspond to a uniform probability which covers phase space. Figure~\ref{fig:billiards}(b), on the other hand, shows a trajectory of a particle in a chaotic Bunimovich stadium~\cite{bunimovich_79}, which looks completely random. If one compares two trajectories that are initially very close to each other in phase space one finds that, after a few bounces against the walls, they become uncorrelated both in terms of positions and directions of motion. This is a consequence of chaotic dynamics. 

There are many examples of dynamical systems that exhibit chaotic behavior. A necessary, and often sufficient, condition for chaotic motion to occur is that the number of functionally independent conserved quantities (integrals of motion), which are in involution, is smaller than the number of degrees of freedom. Otherwise, as mentioned before, the system is integrable and the dynamics is ``simple". This criterion immediately tells us that the motion of one particle, without internal degrees of freedom, in a one-dimensional system, described by a static Hamiltonian, is integrable. The energy provides a unique (up to a sign) relation between the coordinate and the momentum of the particle. In two dimensions, energy conservation is not sufficient to constrain the two components of the momentum at a given position in space, and chaos is possible. However, if an additional conservation law is present, e.g., angular momentum in the example of Fig.~\ref{fig:billiards}(a), then the motion is regular. As a generalization of the above, a many-particle system is usually considered chaotic if it does not have an extensive number of conserved quantities. For example, an ensemble of noninteracting particles in high-dimensional systems is not chaotic in this sense, even if each particle exhibits chaotic motion in the part of phase space associated with its own degrees of freedom. This due to the fact that the energy of each particle is separately conserved. However, one expects that interactions between the particles will lead to chaotic motion. 

It is natural to ask what happens to an integrable system in the presence of a small integrability breaking perturbation. The KAM theorem (after Kolmogorov, Arnold, and Moser \cite{kolmogorov_54,arnold_63,moser_62}) states that, under quite general conditions and for systems with a finite number of degrees of freedom, most of the tori that foliate phase space in the integrable limit persist under small perturbations \cite{jose_saletan_book_98}. This means that, in finite systems, there is a crossover between regular and chaotic dynamics.

It is instructive to see how chaos emerges in simple system. The easiest way to do this is to study one particle in one dimension and remove the energy conservation by applying a time-dependent protocol. Very well-studied examples of such driven systems (usually exhibiting a coexistence of chaotic and regular motion in different parts of phase space) include the Fermi-Ulam model~\cite{lichtenberg_lieberman_92}, the Kapitza pendulum~\cite{broer_hoveijn_04}, and the kicked rotor~\cite{chirikov_79a,chirikov_79b}. The latter example provides, perhaps, the simplest realization of a chaotic system. As an illustration, we discuss it in detail in Appendix~\ref{app:kicked}.

\subsection{Random Matrix Theory}

A focus of this review is on eigenstate thermalization which, as we argue in the following, is closely related to quantum chaos (see, e.g., Refs.\cite{santos_rigol_10a,santos_rigol_10b}, for numerical studies that discuss it). In this section, we review results from quantum chaos that will be needed later. We refer the readers to more complete reviews on quantum chaos and RMT for further details~\cite{guhr_muller_98, alhassid_00, mehta_04, kravtsov_09}.

From the early days of quantum mechanics, it was clear that the classical notion of chaos does not directly apply to quantum-mechanical systems. The main reason is that Schr\"odinger's equation is linear and therefore cannot have exponentially departing trajectories for the wave functions. As a matter of fact, the overlap between two different quantum states, evolved with the same Hamiltonian, is constant in time. Also, while quantum mechanics can be formulated in a phase-space language, for example, using the Wigner-Weyl quantization~\cite{hillery_oconnell_84,polkovnikov_10}, one still does not have the notion of a trajectory (and thus its sensitivity to small perturbations) since coordinates and momenta of particles cannot be defined simultaneously due to the uncertainty principle. It is then natural to ask what is the analogue of chaotic motion in quantum systems.

To better understand this question, let us first consider the single-particle classical limit. For integrable systems, the physics was understood in the early days of quantum mechanics, based on Bohr's initial insight. Along allowed trajectories, the classical reduced action satisfies the quantization condition:
\be
\oint pdq\approx 2\pi \hbar n \,.
\ee
Namely, the classical action is quantized in units of $\hbar$. In 1926, this conjecture was formalized by what is now known as the WKB (after Wentzel, Kramers, and Brillouin) approximation~\cite{landau_lifshitz_3_81}. Essentially, the WKB quantization implies that, in the semi-classical limit, one has to discretize (quantize) classical trajectories. In chaotic systems, the situation remained unclear for a very long time. In particular, it was not clear how to quantize classical chaotic trajectories, which are not closed (in phase space). Initial attempts to resolve these issues go back to Einstein who wrote a paper about them already in 1917 (see Ref.~\cite{stone_05} for details). However, the question was largely ignored until the 1970s when, after a pioneering work by Gutzwiller~\cite{gutzwiller_71}, it became the focus of much research broadly falling under the title of quantum chaos. To this day many questions remain unresolved, including the precise definition of quantum chaos~\cite{rudnik_08}. 

A set of crucial results on which quantum chaos builds came from works of Wigner~\cite{wigner_55, wigner_57, wigner_58} who, followed by Dyson~\cite{dyson_62} and others, developed a theory for understanding the spectra of complex atomic nuclei. This theory is now known as RMT~\cite{mehta_04}. RMT became one of the cornerstones of modern physics and, as we explain later, underlies our understanding of eigenstate thermalization. Wigner's original idea was that it is hopeless to try to predict the exact energy levels and corresponding eigenstates of complex quantum-mechanical systems such as large nuclei. Instead, one should focus on their statistical properties. His second insight was that, if one looks into a small energy window where the density of states is constant, then the Hamiltonian, in a non {\it fine-tuned} basis, will look essentially like a random matrix. Therefore, by studying statistical properties of random matrices (subject to the symmetries of the Hamiltonian of interest, such as time-reversal symmetry), one can gain insights on the statistical properties of energy levels and eigenstates of complex systems. This latter insight was very revolutionary and counterintuitive. It should be noted that whenever we attempt to diagonalize many-body physical Hamiltonians, we usually write them in special bases in which the resulting matrices are very sparse and the nonzero matrix elements are anything but random. This, however, does not contradict Wigner's idea which deals with ``generic" bases.

The main ideas of RMT and the statistics of the energy levels (known as Wigner-Dyson statistics) can be understood using $2\times 2$ Hamiltonians whose entries are random numbers taken from a Gaussian distribution \cite{guhr_muller_98, alhassid_00, mehta_04, kravtsov_09}:
\be \label{eq:2x2}
\hat{H}\stackrel{.}{=}
\left[\begin{array}{cc}
\varepsilon_1\, & \frac{V}{\sqrt{2}}\\
\frac{V^\ast}{\sqrt{2}}\, & \varepsilon_2\;
\end{array}
\right].
\ee
Here the factor $1/\sqrt{2}$ in the off-diagonal matrix elements is introduced since, as it will become clear soon, this choice leaves the form of the Hamiltonian invariant under basis rotations. The Hamiltonian in Eq.~\eqref{eq:2x2} can be easily diagonalized and the eigenvalues are
\be\label{eq:energlevel2x2}
E_{1,2}={\varepsilon_1+\varepsilon_2\over 2}\pm {1\over 2}\sqrt{(\varepsilon_1-\varepsilon_2)^2+ 2|V|^2}.
\ee
If the system is invariant under time reversal (e.g., there is no external magnetic field) then the Hamiltonian can be written as a real matrix, so $V=V^\ast$. For simplicity, we draw $\varepsilon_1, \, \varepsilon_2,$ and $V$ from a Gaussian distribution with zero mean and variance $\sigma$. Using Eq.~\eqref{eq:energlevel2x2} one can compute the statistics of the level separations $P(E_1-E_2=\omega)\equiv P(\omega)$ (here and in what follows, unless otherwise specified, we set $\hbar$ to unity):
\be
P(\omega)={1\over (2\pi)^{3/2}\sigma^3}\int d\varepsilon_1\int d\varepsilon_2\int dV 
\,\delta\left(\sqrt{(\varepsilon_1-\varepsilon_2)^2+2V^2}-\omega\right) 
\exp\left(-{\varepsilon_1^2+\varepsilon_2^2+V^2\over 2\sigma^2}\right).
\ee
Before evaluating the integral over $\varepsilon_1$, we make a change of variables $\varepsilon_2=\varepsilon_1+\sqrt{2}\xi$. Then, integrating over $\varepsilon_1$, which is a Gaussian integral, we are left with
\be
P(\omega)={1\over 2\pi\sigma^2}\int\int d\xi d V 
\,\delta\left(\sqrt{2\xi^2+2V^2}-\omega\right)\exp\left(-{\xi^2+V^2\over 2\sigma^2}\right).
\ee
The latter integrals can be evaluated using cylindrical coordinates, $V=r\cos(x)$, $\xi=r\sin(x)$, and one finds:
\be
\label{eq:pomega2x2}
P(\omega)={\omega\over 2\sigma^2} \exp\left[-{\omega^2\over 4\sigma^2}\right].
\ee
In the absence of time-reversal symmetry, $\Re [V]$ and $\Im [V]$ can be treated as independent random variables and, carrying out a similar calculation using spherical coordinates, leads to:
\be
\label{eq:pomega2x2_beta2}
P(\omega)={\omega^2 \over 2 \sqrt{\pi} \left(\sigma^2\right)^{3/2}} \exp\left[-{\omega^2\over 4\sigma^2}\right].
\ee

These distributions exhibit some remarkable (generic) properties: (i) there is level repulsion since the probability $P(\omega)$ of having energy separation $\omega$ vanishes as $\omega\rightarrow0$ and (ii) the probability decays as a Gaussian at large energy separation. The two distributions~\eqref{eq:pomega2x2} and \eqref{eq:pomega2x2_beta2} can be written as
\be
P(\omega)=A_\beta\, \omega^{\beta}\, \exp[-B_\beta\omega^2],
\label{eq:Wigner_Surmise}
\ee
where $\beta=1$ in systems with time-reversal symmetry and $\beta=2$ in systems that do not have time-reversal symmetry. The coefficients $A_\beta$ and $B_\beta$ are found by normalizing $P(\omega)$ and fixing the mean level spacing. The normalized distributions, with an average level spacing set to one, are given by
\be
P_1(\omega)={\pi\over 2}\omega \exp\left[-{\pi\over 4}\omega^2\right],\qquad
P_2(\omega)={32\over \pi^2}\omega^2\exp\left[-{4\over \pi}\omega^2\right].
\label{wigner}
\ee

It turns out that the features described above are not unique to the $2\times 2$ Hamiltonian~\eqref{eq:2x2}. In fact, this simple example can be generalized to larger matrices. In particular, one can define an ensemble of matrices drawn from a random Gaussian distribution~\cite{alhassid_00}:
\be \label{eq:lgauss}
P(\hat H)\propto \exp\left[-{\beta \over 2 a^2}{\rm Tr}(\hat{H}^2)\right]\equiv  
\exp\left[-{\beta \over 2 a^2}\sum_{ij} H_{ij} H_{ji}\right],
\ee
where $a$ sets the overall energy scale and, as before, $\beta=1$ refers to systems with time-reversal symmetry where all entries in the Hamiltonian are real and satisfy $H_{ij}= H_{ji}$, that is, the so-called Gaussian orthogonal ensemble (GOE), and $\beta=2$ refers to systems without time-reversal symmetry, where the entities are complex and satisfy $ H_{ij}= H_{ji}^\ast$, that is, the so-called Gaussian unitary ensemble (GUE).\footnote{There is a third ensemble, corresponding to $\beta=4$, known as the Gaussian simplectic ensemble (GSE). We will not discuss here.} Note that the factor of $\sqrt{2}$ in Eq.~\eqref{eq:2x2} ensures that the Hamiltonian is described by the distribution~\eqref{eq:lgauss}. 

The choice of the ensemble in Eq.~\eqref{eq:lgauss} is a natural one. The ensemble must be invariant under any orthogonal (GOE) or unitary (GUE) transformation, so the probability distribution can only depend on the invariant ${\rm Tr}(\hat{H}^2)$. It is Gaussian because ${\rm Tr}(\hat{H}^2)$ is a sum of many independent contributions and should therefore satisfy the central limit theorem. We will not discuss the details of the derivations of the level statistics for such random ensembles, which can be found in Refs.~\cite{kravtsov_09,guhr_muller_98, mehta_04, alhassid_00,reichl_04}. We only point out that the exact level spacing distributions (known as Wigner-Dyson distributions) do not have a closed analytic form. However, they are qualitatively (and quantitatively) close to the Wigner Surmise~\eqref{eq:Wigner_Surmise}. 

Following Wigner's ideas, it was possible to explain the statistical properties of the spectra of complex nuclei. However, for a long time it was not clear which are the ``complex systems'' for which RMT is generally applicable. In 1984, Bohigas, Giannoni, and Schmit, studying a single particle placed in an infinite potential well with the shape of a Sinai billiard, found that at high energies (i.e., in the semi-classical limit), and provided that one looks at a sufficiently narrow energy window, the level statistics is described by the Wigner-Dyson distribution~\cite{bohigas_giannoni_84}. Based on this discovery, they conjecture that the level statistics of quantum systems that have a classically chaotic counterpart are described by RMT (this is known as the BGS conjecture). This conjecture has been tested and confirmed in many different setups (we will show some of them in the next section). To date, only non-generic counterexamples, such as arithmetic billiards, are known to violate this conjecture~\cite{bogomolny_georgeot_92}. Therefore, the emergence of Wigner-Dyson statistics for the level spacings is often considered as a defining property of quantum chaotic systems, whether such systems have a classical counterpart or not.

\subsubsection{Chaotic Eigenfunctions}
RMT allows one to make an important statement about the eigenvectors of random matrices. The joint probability distribution of components of eigenvectors can be written as \cite{brody_flores_81,alhassid_00}
\be \label{eq:joprob}
P_{\rm GOE}(\psi_1,\psi_2,\dots,\psi_N)\propto \delta\left(\sum_j \psi_j^2-1\right),\quad 
P_{\rm GUE}(\psi_1,\psi_2,\dots,\psi_N)\propto \delta\left(\sum_j |\psi_j|^2-1\right),
\ee
where $\psi_j$ are the components of the wave functions in some fixed basis. This form follows from the fact that, because of the orthogonal (unitary) invariance of the random matrix ensemble, the distribution can depend only on the norm $\sqrt{\sum_j \psi_j^2}$ \big($\sqrt{\sum_j |\psi_j^2|}$\big) of the eigenvector, and must be proportional to the $\delta$-functions in Eq.~\eqref{eq:joprob} because of the normalization \cite{brody_flores_81}. Essentially, Eq.~\eqref{eq:joprob} states that the eigenvectors of random matrices are random unit vectors, which are either real (in the GOE) or complex (in the GUE). Of course, different eigenvectors are not completely independent since they need to be orthogonal to each other. However, because two uncorrelated random vectors in a large-dimensional space are, in any case, nearly orthogonal, in many instances the correlations due to this orthogonality condition can be ignored.

One may wonder about the classical limit of quantum eigenvectors. The latter are stationary states of the system and should therefore correspond to stationary (time-averaged) trajectories in the classical limit. In integrable systems with a classical limit, the quantum eigenstates factorize into a product of WKB-like states describing the stationary phase-space probability distribution of a particle corresponding to one of the trajectories \cite{landau_lifshitz_3_81}. However, if the system is chaotic, the classical limit of the quantum eigenstates is ill-defined. In particular, in the classical limit, there is no smooth (differentiable) analytic function that can describe the eigenstates of chaotic systems. This conclusion follows from the BGS conjecture, which implies that the eigenstates of a chaotic Hamiltonian in non-fine-tuned bases, including the real space basis, are essentially random vectors with no structure. 

Let us address a point that often generates confusion. Any given (Hermitian) Hamiltonian, whether it is drawn from a random matrix ensemble or not, can be diagonalized and its eigenvectors form a basis. In this basis, the Hamiltonian is diagonal and RMT specifies the statistics of the eigenvalues. The statistical properties of the eigenstates are specified for an ensemble of random Hamiltonians {\it in a fixed basis}. If we fix the basis to be that of the eigenkets of the first random Hamiltonian we diagonalize, that basis will not be special for other randomly drawn Hamiltonians. Therefore, all statements made will hold for the ensemble even if they fail for one of the Hamiltonians. The issue of the basis becomes more subtle when one deals with physical Hamiltonians. Here, one can ask what happens if we diagonalize a physical Hamiltonian and use the eigenvectors obtained as a basis to write a slightly modified version of the same Hamiltonian (which is obtained, say, by slightly changing the strength of the interactions between particles). As we discuss below (see also Ref.~\cite{alhassid_00}), especially in the context of many-body systems, the eigenstates of chaotic quantum Hamiltonians [which are away from the edge(s) of the spectrum]\footnote{That one needs to be away from the edges of the spectrum can already be inferred from the fact that BGS found that the Wigner-Dyson distribution occurs only at sufficiently high energies \cite{bohigas_giannoni_84}. We will discuss this point in detail when presenting results for many-body quantum systems.} are very sensitive to small perturbations. Hence, one expects that the perturbed Hamiltonian will look like a random matrix when written in the unperturbed basis. In that sense, writing a Hamiltonian in its own basis can be considered to be a fine-tuning of the basis. It is in this spirit that one should take Wigner's insight. The sensitivity just mentioned in chaotic quantum systems is very similar to the sensitivity of classical chaotic trajectories to either initial conditions or the details of the Hamiltonian. 

\subsubsection{The Structure of the Matrix Elements of Operators}

Let us now analyze the structure of matrix elements of Hermitian operators
\be
 \hat{O}=\sum_i O_{i}|i\rangle \langle i|, \quad \text{where}\quad \hat{O}|i\rangle=O_{i}|i\rangle,
\ee
within RMT. For any given random Hamiltonian, for which the eigenkets are denoted by 
$|m\rangle$ and $|n \rangle$,
\be
O_{mn}\equiv\langle m|\hat{O}|n\rangle=
\sum_{i}O_i\langle m|i\rangle\langle i|n\rangle=
\sum_i O_{i}(\psi^m_i)^\ast\psi^n_i.
\label{eq:ramdom_vec}
\ee
Here, $\psi_i^m \equiv \langle i | m \rangle$ and similarly for $\psi_i^n$. Recall that the eigenstates of random matrices in any basis are essentially random orthogonal unit vectors. Therefore, to leading order in $1/\mathcal D$, where $\mathcal D$ is the dimension of the Hilbert space, we have
\be
\overline{(\psi_i^m)^\ast (\psi_j^n)}={1\over \mathcal D}\delta_{mn}\delta_{ij}\,,
\label{eq:ramdom_vec_2}
\ee
where the average $\overline{(\psi_i^m)^\ast (\psi_j^n)}$ is over random eigenkets $|m\rangle$ and $|n\rangle$. This implies that one has very different expectation values for the diagonal and off-diagonal matrix elements of $\hat{O}$. Indeed, using Eqs.~\eqref{eq:ramdom_vec} and \eqref{eq:ramdom_vec_2}, we find
\be
\overline{O_{mm}}={1\over \mathcal D}\sum_i O_{i}\equiv\bar O,
\label{eq:mean1}
\ee
and 
\be
\overline{O_{mn}}=0,\quad \text{for} \quad m\ne n.
\label{eq:mean2}
\ee
Moreover, the fluctuations of the diagonal and off-diagonal matrix elements are suppressed by the size of the Hilbert space. For the diagonal matrix elements
\beq
\overline{O_{mm}^2}-\overline{O_{mm}}^2&=&
\sum_{i,j} O_{i} O_{j} \overline{(\psi^m_i)^\ast \psi^m_i\,(\psi^m_j)^\ast \psi^m_j}-
\sum_{i,j} O_{i} O_{j} \overline{(\psi^m_i)^\ast \psi^m_i}\,
\overline{(\psi^m_j)^\ast \psi^m_j}\nonumber\\&=&
\sum_i O_{i}^2 \left(\overline{|\psi^m_i|^4}-(\overline{|\psi_i^m|^2})^2\right)
={3-\beta\over \mathcal D^2}\sum_i O_{i}^2\equiv{3-\beta\over \mathcal D}\,\overline {O^2},
\label{eq:var1}
\eeq
where, as before, $\beta=1$ for the GOE and $\beta=2$ for the GUE. For the GOE ($\psi_i^m$ are real numbers), we used the relation $\overline{(\psi^m_i)^4}=3[\overline{(\psi_i^m)^2}]^2$, while for the GUE ($\psi_i^m$ are complex numbers), we used the relation $\overline{|\psi^m_i|^4}=2(\overline{|\psi_i^m|^2})^2$. These results are a direct consequence of the Gaussian distribution of the components of the random vector $\psi_i^m$. Assuming that none of the eigenvalues $O_{i}$ scales with the size of the Hilbert space, as is the case for physical observables, we see that the fluctuations of the diagonal matrix elements of $\hat{O}$ are inversely proportional to the square root of the size of the Hilbert space.

Likewise, for the absolute value of the off-diagonal matrix elements, we have
\be
\overline{|O_{mn}|^2}-\left|\overline{O_{mn}}\right|^2=
\sum_{i} O_{i}^2 \overline {|\psi_i^m|^2|\psi_i^n|^2}={1\over \mathcal D}\overline {O^2}.
\label{eq:var2}
\ee
Combining these expressions, we see that, to leading order in $1/D$, the matrix elements of any operator can be written as
\be
O_{mn}\approx\bar O\delta_{mn}+\sqrt{ \frac{\overline{O^2}}{\mathcal D} }R_{mn},
\label{eth_rmt}
\ee
where $R_{mn}$ is a random variable (which is real for the GOE and complex for the GUE) with zero mean and unit variance (for the GOE, the variance of the diagonal components $R_{mm}$ is 2). It is straightforward to check that the ansatz~(\ref{eth_rmt}) indeed correctly reproduces the mean and the variance of the matrix elements of $\hat O$ given by Eqs.~(\ref{eq:mean1})--(\ref{eq:var2}).

In deriving Eqs.~(\ref{eq:mean1})--(\ref{eq:var2}), we averaged over a fictitious ensemble of random Hamiltonians. However, from Eq.~\eqref{eth_rmt}, it is clear that for large $\mathcal D$ the fluctuations of operators are small and thus one can use the ansatz~\eqref{eth_rmt} for a given fixed Hamiltonian.

\subsection{Berry-Tabor Conjecture}

In classical systems, an indicator of whether they are integrable or chaotic is the temporal behavior of nearby trajectories. In quantum systems, the role of such an indicator is played by the energy level statistics. In particular, for chaotic systems, as we discussed before, the energy levels follow the Wigner-Dyson distribution. For quantum integrable systems, the question of level statistics was first addressed by Berry and Tabor in 1977~\cite{berry_tabor_77}. For a particle in one dimension, which we already said exhibits non-chaotic classical dynamics if the Hamiltonian is time independent, we know that if we place it in a harmonic potential all levels are equidistant, while if we place it in an infinite well the spacing between levels increases as the energy of the levels increases. Hence, the statistics of the level spacings strongly depends on the details of the potential considered. This is unique to one particle in one dimension. It changes if one considers systems whose classical counterparts have more than one degree of freedom, for example, one particle in higher dimensional potentials, or many particles in one dimension. A very simple example of a non-ergodic system, with many degrees of freedom, would be an array of independent harmonic oscillators with incommensurate frequencies. These can be, for example, the normal modes in a harmonic chain. Because these oscillators can be diagonalized independently, the many-body energy levels of such a system can be computed as
\be\label{eq:manyosc}
E=\sum_j n_j \omega_j,
\ee
where $n_j$ are the occupation numbers and $\omega_j$ are the mode frequencies. If we look into high energies, $E$, when the occupation numbers are large, nearby energy levels (which are very closely spaced) can come from very different sets of $\{n_j\}$. This means that the energy levels $E$ are effectively uncorrelated with each other and can be treated as random numbers. Their distribution then should be described by Poisson statistics, that is, the probability of having $n$ energy levels in a particular interval $[E,E+\delta E]$ will be 
\be
P_n={\lambda^n\over n!}\exp[-\lambda],
\ee
where $\lambda$ is the average number of levels in that interval. Poisson and Wigner-Dyson statistics are very different in that, in the former there is no level repulsion. For systems with Poisson statistics, the distribution of energy level separations $\omega$ (with mean separation set to one) is 
\be\label{eq:berrytabor}
P_0(\omega)=\exp[-\omega],
\ee
which is very different from the Wigner Surmise~\eqref{wigner}. The statement that, for quantum systems whose corresponding classical counterpart is integrable, the energy eigenvalues generically behave like a sequence of independent random variables, that is, exhibit Poisson statistics, is now known as the Berry-Tabor conjecture \cite{berry_tabor_77}. While this conjecture describes what is seen in many quantum systems whose classical counterpart is integrable, and integrable quantum systems without a classical counterpart, there are examples for which it fails (such as the single particle in the harmonic potential described above and other harmonic systems~\cite{pandey_ramaswamy_91}). Deviations from Poisson statistics are usually the result of having symmetries in the Hamiltonian that lead to extra degeneracies resulting in commensurability of the spectra. 

The ideas discussed above are now regularly used when dealing with many-particle systems. The statistics of the energy levels of many-body Hamiltonians serves as one of the main indicators of quantum chaos or, conversely, of quantum integrability. As the energy levels become denser, the level statistics asymptotically approaches either the Wigner-Dyson or the Poisson distribution. It is interesting to note that in few-particle systems, like a particle in a billiard, the spectra become denser by going to the semi-classical limit by either increasing the energy or decreasing Planck's constant, while in many-particle systems one can achieve this by going to the thermodynamic limit. This means that the level statistics indicators can be used to characterize whether a quantum system is chaotic or not even when it does not have a classical limit. This is the case, for example, for lattice systems consisting of spins 1/2 or interacting fermions described within the one-band approximation.

Finally, as we discuss later, the applicability of RMT requires that the energy levels analyzed are far from the edges of the spectrum and that the density of states as a function of energy is accounted for. The first implies that one needs to exclude, say, the ground state and low-lying excited states and the states with the highest energies (if the spectrum is bounded from above). It is plausible that in generic systems only states within subextensive energy windows near the edges of the spectrum are not described by RMT. 

\subsection{The Semi-Classical Limit and Berry's Conjecture}

One of the most remarkable connections between the structure of the eigenstates of chaotic systems in the semi-classical limit and classical chaos was formulated by Berry~\cite{berry_77}, and is currently known as Berry's conjecture. (In this section, we return $\hbar$ to our equations since it will be important for taking the classical limit $\hbar \to 0$.)

In order to discuss Berry's conjecture, we need to introduce the Wigner function  $W({\bf x},{\bf p})$, which is defined as the Wigner-Weyl transform of the density matrix $\hat \rho$ \cite{hillery_oconnell_84,polkovnikov_10}.  For a pure state, $\hat \rho\equiv|\psi\ra\la \psi|$, one has
\be
W({\bf x},{\bf p})=\frac{1}{(2\pi\hbar)^{3N}}\int d^{3N}\xi\,\psi^\ast\left({\bf x}+{{\bm \xi}\over 2}\right) 
\psi\left({\bf x} -{{\bm \xi}\over 2}\right)\exp\left[-i {{\bf p}{\bm \xi}\over \hbar}\right],
\ee
where ${\bf x},\,{\bf p}$ are the coordinates and momenta of $N$-particles spanning a $6N$-dimensional phase space. For a mixed state one replaces the product
\be
\psi^\ast\left({\bf x}+{{\bm \xi}\over 2}\right) \psi\left({\bf x} -{{\bm \xi}\over 2}\right)\to 
\left<{\bf x} -{{\bm \xi}\over 2}\right| \hat{\rho}\left| {\bf x} +{{\bm \xi}\over 2}\right>\equiv
\rho\left({\bf x}-{{\bm \xi}\over 2},{\bf x} +{{\bm \xi}\over 2}\right),
\ee
where $\hat \rho$ is the density matrix. One can check that, similarly, the Wigner function can be defined by integrating over momentum
\be
W({\bf x},{\bf p})=\frac{1}{(2\pi\hbar)^{3N}}\int d^{3N}\eta\,\phi^\ast\left({\bf p}+{{\bm \eta}\over 2}\right) 
\phi\left({\bf p} -{{\bm \eta}\over 2}\right)\exp\left[i {{\bf x}{\bm \eta}\over \hbar}\right],
\label{eq:Wig_momentum}
\ee
where $\phi({\bf p})$ is the Fourier transform of $\psi({\bf x})$. From either of the two representations it immediately follows that
\be
\int{d^{3N}p\, W({\bf x},{\bf p})} = |\psi({\bf x})|^2\quad\text{and}\quad
\int{d^{3N}x\, W({\bf x},{\bf p})} = |\phi({\bf p})|^2\;.
\ee

The Wigner function is uniquely defined for any wave function (or density matrix) and plays the role of a quasi-probability distribution in phase space. In particular, it allows one to compute an expectation value of any observable $\hat{O}$ as a standard average~\cite{hillery_oconnell_84,polkovnikov_10}:
\be\label{eq:wignerobs}
\la \hat{O}\ra=\int d^{3N}x\, d^{3N}p\,O_W({\bf x},{\bf p})W({\bf x}, {\bf p}),
\ee
where $O_W({\bf x},{\bf p})$ is the Weyl symbol of the operator $\hat{O}$
\be
O_W({\bf x},{\bf p})=\frac{1}{(2\pi\hbar)^{3N}}\int d^{3N}\xi \left<{\bf x}-{ {\bm\xi}\over 2}\right| 
\hat{O} \left| {\bf x}
+{{\bm\xi}\over 2} \right>\exp\left[-i {{\bf p}{\bm \xi}\over \hbar}\right].
\ee
We note that instead of the coordinate and momentum phase-space variables one can, for example, use coherent state variables to represent electromagnetic or matter waves, angular momentum variables to represent spin systems, or any other set of canonically conjugate variables~\cite{polkovnikov_10}.

Berry's conjecture postulates that, in the semi-classical limit of a quantum system whose classical counterpart is chaotic, the Wigner function of energy eigenstates (averaged over a vanishingly small phase space) reduces to the microcanonical distribution. More precisely, define
\be
\overline{W({\bf X},{\bf P})}=\int_{\Delta \Omega_1} {dx_1 dp_1\over (2\pi\hbar)} 
\dots \int_{\Delta \Omega_N} {dx_N dp_N\over (2\pi\hbar)} W({\bf x},{\bf p}),
\label{eq:barwigner}
\ee
where $\Delta \Omega_j$ is a small phase-space volume centered around the point $X_j, P_j$. This volume is chosen such that, as $\hbar\to 0$, $\Delta\Omega_j\to 0$ and at the same time $\hbar/\Delta\Omega_j\to0$. Berry's conjecture then states that, as $\hbar\to 0$,
\be\label{eq:berry}
\overline {W({\bf X},{\bf P})}= {1\over \int{d^{3N}X d^{3N}P\,\delta [E-H({\bf X},{\bf P})] }} 
\delta [E-H({\bf X},{\bf P})],
\ee
where $H({\bf X}, {\bf P})$ is the classical Hamiltonian describing the system, and $\delta[\dots]$ is a one-dimensional Dirac delta function. In Berry's words, ``$\psi({\bf x})$ is a Gaussian random function of ${\bf x}$ whose spectrum at ${\bf x}$ is simply the local average of the Wigner function $\overline {W({\bf x},{\bf p})}$''. Berry also considered quantum systems whose classical counterpart is integrable. He conjectured that the structure of the energy eigenstates of such systems is very different \cite{berry_77}.

It follows from Berry's conjecture, see Eqs.~\eqref{eq:wignerobs} and~\eqref{eq:berry}, that the energy eigenstate expectation value of any observable in the semi-classical limit (of a quantum system whose classical counterpart is chaotic) is the same as a microcanonical average. 

For a  dilute gas of hard spheres, this was studied by Srednicki in 1994~\cite{srednicki_94}. Let us analyze the latter example in detail. Srednicki argued that the eigenstate corresponding to a high-energy eigenvalue $E_n$ can chosen to be real and written as
\be \label{eq:mark1}
\psi_n({\bf x})=\mathcal N_n \int d^{3N}p \,A_n({\bf p}) \delta({\bf p}^2-2m E_n) 
\exp[i {\bf p x}/\hbar],
\ee
where $\mathcal N_n$ is a normalization constant, and $A_n^\ast({\bf p})=A_n(-{\bf p})$. In other words the energy eigenstates with energy $E_n$ are given by a superposition of plane-waves with momentum $\bf p$ such that $E_n={\bf p}^2/(2m)$.  Assuming Berry's conjecture applies, $A_n({\bf p})$ was taken to be a Gaussian random variable satisfying 
\be \label{eq:mark2}
\la A_m({\bf p}) A_n({\bf p'})\ra_{EE}=\delta_{mn}{\delta^{3N}({\bf p}+{\bf p'})\over 
\delta (|{\bf p}|^2-|{\bf p'}|^2)}.
\ee
Here the average should be understood as over a fictitious ``eigenstate ensemble'' of energy eigenstates of the system indicated by ``EE". This replaces the average over a small phase-space volume used by Berry. The denominator in this expression is needed for proper normalization.  

From these assumptions, it follows that
\be\label{eq:mark3}
\la \phi^*_m({\bf p})\phi_n({\bf p}') \ra_{EE}= \delta_{mn}\, N^2_n\, (2\pi\hbar)^{3N} 
\delta({\bf p}^2-2m E_n) \delta_V^{3N} ({\bf p}-{\bf p'}),
\ee
where $\phi_n({\bf p})$ is the $3N$-dimensional Fourier transform of $\psi_n({\bf x})$, and $\delta_V({\bf p})\equiv (2\pi\hbar)^{-3N}\int_\text{V}d^{3N}x \exp[i{\bf p} {\bf x}/\hbar]$ and $V=L^3$ is the volume of the system. It is straightforward to check, using Eqs.~\eqref{eq:mark3} and (\ref{eq:Wig_momentum}), that the Wigner function averaged over the eigenstate ensemble is indeed equivalent to the microcanonical distribution [Eq.~\eqref{eq:berry}].  

Using Eqs.~\eqref{eq:mark1}--\eqref{eq:mark3} one can compute observables of interest in the eigenstates of the Hamiltonian. For example, substituting $\delta_V ({\bf 0})\rightarrow [L/(2\pi\hbar)]^{3N}$, one can calculate the momentum distribution function of particles in the eigenstate ensemble
\be
\la \phi_{nn}({\bf p}_1) \ra_{EE}\equiv \int d^3p_2 \ldots d^3p_N 
\la \phi^*_n({\bf p})\phi_n({\bf p}) \ra_{EE}
=N^2_n\,L^{3N} \int d^3p_2 \ldots d^3p_N \delta({\bf p}^2-2m E_n).
\ee
Finally, using the fact that $I_N(A)\equiv\int d^N\! p \delta({{\bf{p}}^2}-A)= (\pi A)^{N/2}/[\Gamma(N/2)A]$, which through the normalization of $\phi_n({\bf p})$ allows one to determine $N^{-2}_n=L^{3N}I_{3N}(2mE_n)$, one obtains
\be
\la \phi_{nn}({\bf p}_1) \ra_{EE}=\frac{I_{3N-3}(2mE_n-{\bf p}^2_1)}{I_{3N}(2mE_n)}
=\frac{\Gamma(3N/2)}{\Gamma[3(N-1)/2]} \left(\frac{1}{2\pi mE_n}\right)^{\frac32} 
\left(1-\frac{{\bf p}^2_1}{2mE_n}\right)^{\frac{(3N-5)}{2}}.
\ee
Defining a microcanonical temperature using $E_n=3Nk_BT_n/2$, where $k_B$ is the Boltzmann constant, and taking the limit $N\rightarrow\infty$ one gets
\be\label{eq:mxboltsrd}
\la \phi_{nn}({\bf p}_1) \ra_{EE}=\left(\frac{1}{2\pi mk_BT_n}\right)^{\frac32}
\exp\left(-\frac{{\bf p}^2_1}{2mk_BT_n}\right),
\ee
where we used that $\lim_{N\rightarrow\infty}\Gamma(N+B)/[\Gamma(N)N^B]=1$, which is valid for $B\in{\bf R}$. We immediately see that the result obtained for $\la \phi_{nn}({\bf p}_1) \ra_{EE}$ is the Maxwell-Boltzmann distribution of momenta in a thermal gas. 

One can go a step further and show that the fluctuations of $\phi_{nn}({\bf p}_1)$ about $\la \phi_{nn}({\bf p}_1) \ra_{EE}$ are exponentially small in the number of particles \cite{srednicki_94}. Furthermore, it can be shown that if one requires the wave function $\psi_n({\bf x})$ to be completely symmetric or completely antisymmetric one obtains, instead of the Maxwell-Boltzmann distribution in Eq.~\eqref{eq:mxboltsrd}, the (canonical) Bose-Einstein or Fermi-Dirac distributions, respectively \cite{srednicki_94}. An approach rooted in RMT was also used by Flambaum and Izrailev to obtain, starting from statistical properties of the structure of chaotic eigenstates, the Fermi-Dirac distribution function in interacting fermionic systems \cite{flambaum_izrailev_97a,flambaum_izrailev_97b}. These ideas underlie the eigenstate thermalization hypothesis, which is the focus of this review.

In closing this section, let us note that formulating a slightly modified conjecture one can also recover the classical limit from the eigenstates. Namely, one can define a different coarse-graining procedure for the Wigner function:
\be
\la W({\bf x}, {\bf p})\ra={1\over N_{\delta E}}\sum_{m\in E_m\pm \delta E} W_m({\bf x},{\bf p}),
\ee
where the sum is taken over all, $N_{\delta E}$, eigenstates in a window $\delta E$, which vanishes in the limit $\hbar\to 0$ but contains an exponential (in the number of degrees of freedom) number of levels. We anticipate that in the limit $\hbar\to 0$ the function $\la W({\bf x}, {\bf p})\ra$ also reduces to the right-hand-side (RHS) of Eq.~\eqref{eq:barwigner}, that is,
\be
\la W({\bf x}, {\bf p})\ra = {1\over \int{d^{3N}X d^{3N}P\,\delta [E-H({\bf X},{\bf P})] }} 
\delta [E-H({\bf X},{\bf P})].
\ee
However, this result does not require the system to be ergodic. While this conjecture has little to do with chaos and ergodicity, it suggests a rigorous way of defining classical microcanonical ensembles from quantum eigenstates. This is opposed to individual quantum states, which as we discussed do not have a well-defined classical counterpart.

%%%%%%%%%%%%%%%%%%%%%%%%%%%%%%%%%%%%%%%%%%%%%%%%%%%%%%%%%%%%%%%%%%%%%%%%%%%%%%%%%%%%%%%%%%
\section{Quantum Chaos in Physical Systems}\label{sec:sec3}
%%%%%%%%%%%%%%%%%%%%%%%%%%%%%%%%%%%%%%%%%%%%%%%%%%%%%%%%%%%%%%%%%%%%%%%%%%%%%%%%%%%%%%%%%%

\subsection{Examples of Wigner-Dyson and Poisson Statistics}\label{sssec:levstat}

\begin{figure}[!b]
\begin{center}
\includegraphics[width=12cm]{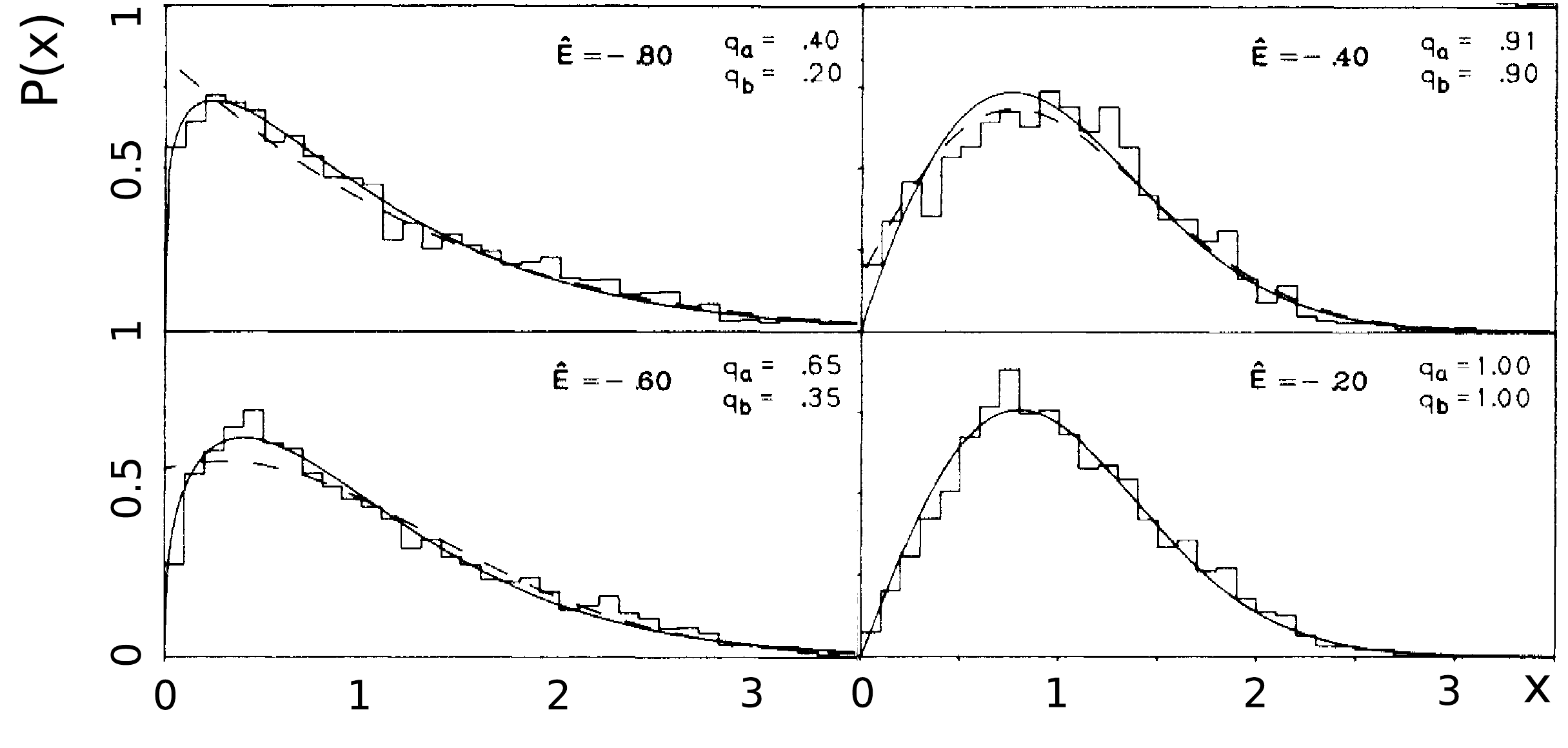}
\end{center}
\caption{The level spacing distribution of a hydrogen atom in a magnetic field. Different plots correspond to different mean dimensionless energies $\hat E$, measured in natural energy units  proportional to $B^{2/3}$, where $B$ is the magnetic field. As the energy increases one observes a crossover between Poisson and Wigner-Dyson statistics. The numerical results are fitted to a Brody distribution (solid lines)~\cite{brody_flores_81}, and to a semi-classical formula due to Berry and Robnik (dashed lines) \cite{berry_robnik_84}. From Ref.~\cite{wintgen_friedrich_87}.}
\label{fig:level_hydrogen}
\end{figure}

Random matrix statistics has found many applications since its introduction by Wigner. They extend far beyond the framework of the original motivation, and have been intensively explored in many fields (for a recent comprehensive review, see Ref.~\cite{kota_14}). Examples of quantum systems whose spectra exhibit Wigner-Dyson statistics are: (i) heavy nuclei \cite{bochhoff_83}, (ii) Sinai billiards (square or rectangular cavities with circular potential barriers in the center) \cite{bohigas_giannoni_84}, which are classically chaotic as the Bunimovich stadium in Fig.~\ref{fig:billiards}, (iii) highly excited levels\footnote{The low-energy spectra of this system exhibits Poissonian level statistics. This is understandable as, at low energies, the motion of the equivalent classical system is regular \cite{wintgen_friedrich_87}. See also Fig.~\ref{fig:level_hydrogen}.} of the hydrogen atom in a strong magnetic field \cite{wintgen_friedrich_87}, (iv) Spin-1/2 systems and spin-polarized fermions in one-dimensional lattices \cite{santos_rigol_10a,santos_rigol_10b}. Interestingly, the Wigner-Dyson statistics is also the distribution of spacings between zeros of the Riemann zeta function, which is directly related to prime numbers. In turn, these zeros can be interpreted as Fisher zeros of the partition function of a particular system of free bosons (see Appendix~\ref{app:riemann}). In this section, we discuss in more detail some examples originating from over 30 years of research.

\begin{figure}[ht]
\begin{center}
 \includegraphics[width=8cm]{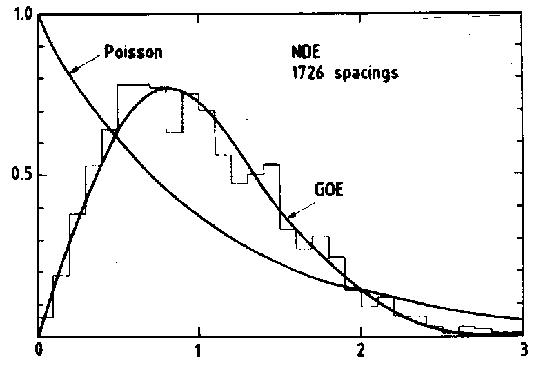}
\end{center}
\caption{Nearest neighbor spacing distribution for the ``Nuclear Data Ensemble'' comprising 1726 spacings (histogram) versus normalized (to the mean) level spacing. The two lines represent predictions of the random matrix GOE ensemble and the Poisson distribution. Taken from Ref.~\cite{bohigas_haq_83}. See also Ref.~\cite{guhr_muller_98}.}
\label{fig:heavy_nuclei}
\end{figure}

\noindent {\bf Heavy nuclei -} Perhaps the most famous example demonstrating the Wigner-Dyson statistics is shown in Fig.~\ref{fig:heavy_nuclei}. That figure depicts the cumulative data of the level spacing distribution obtained from slow neutron resonance data and proton resonance data of around 30 different heavy nuclei~\cite{bohigas_haq_83,guhr_muller_98}. All spacings are normalized by the mean level spacing. The data are shown as a histogram and the two solid lines depict the (GOE) Wigner-Dyson distribution and the Poisson distribution. One can see that the Wigner-Dyson distribution works very well, confirming Wigner's original idea.

\begin{figure}[!t]
\begin{center}
\includegraphics[width=7cm]{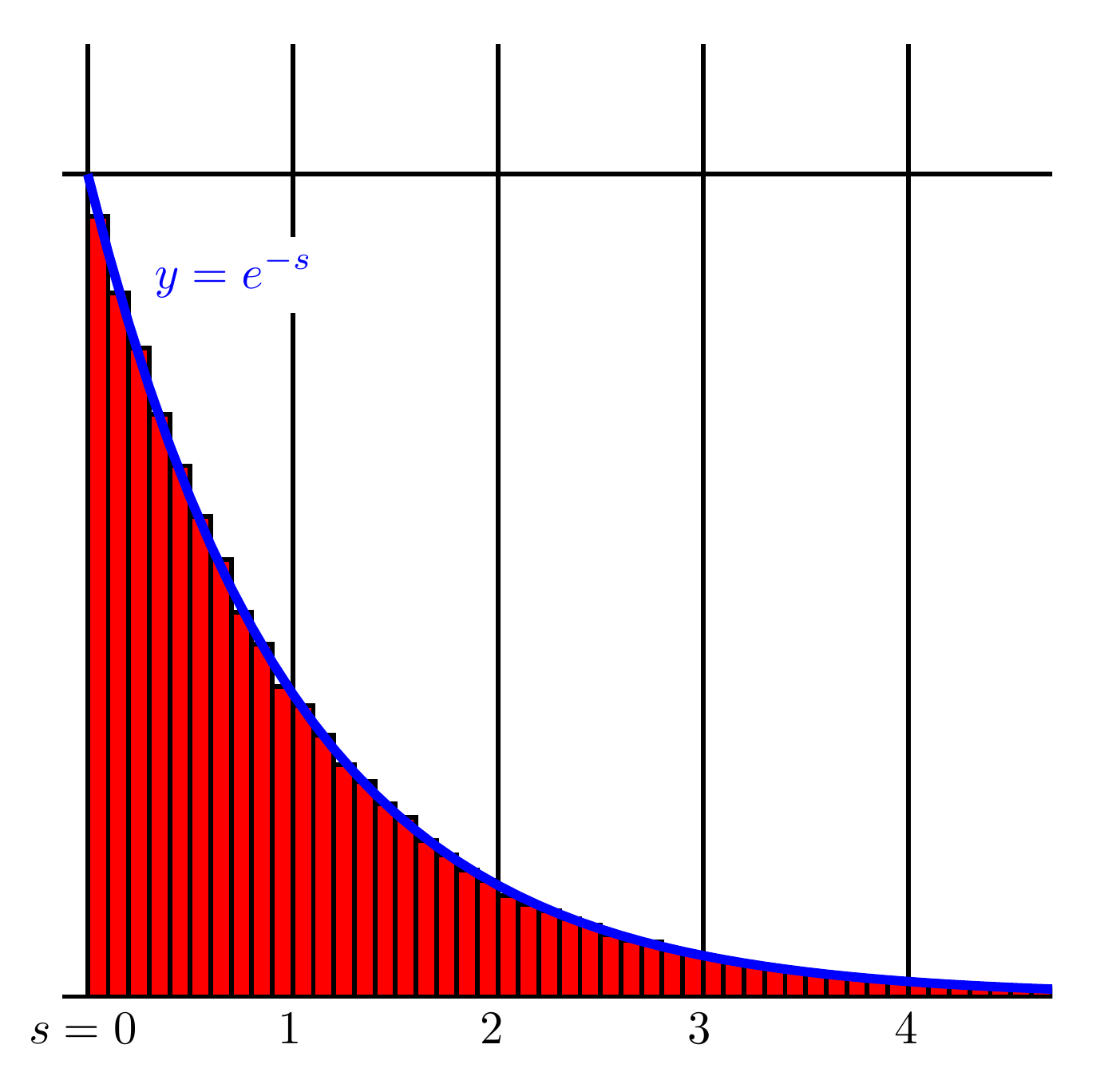}
\includegraphics[width=7cm]{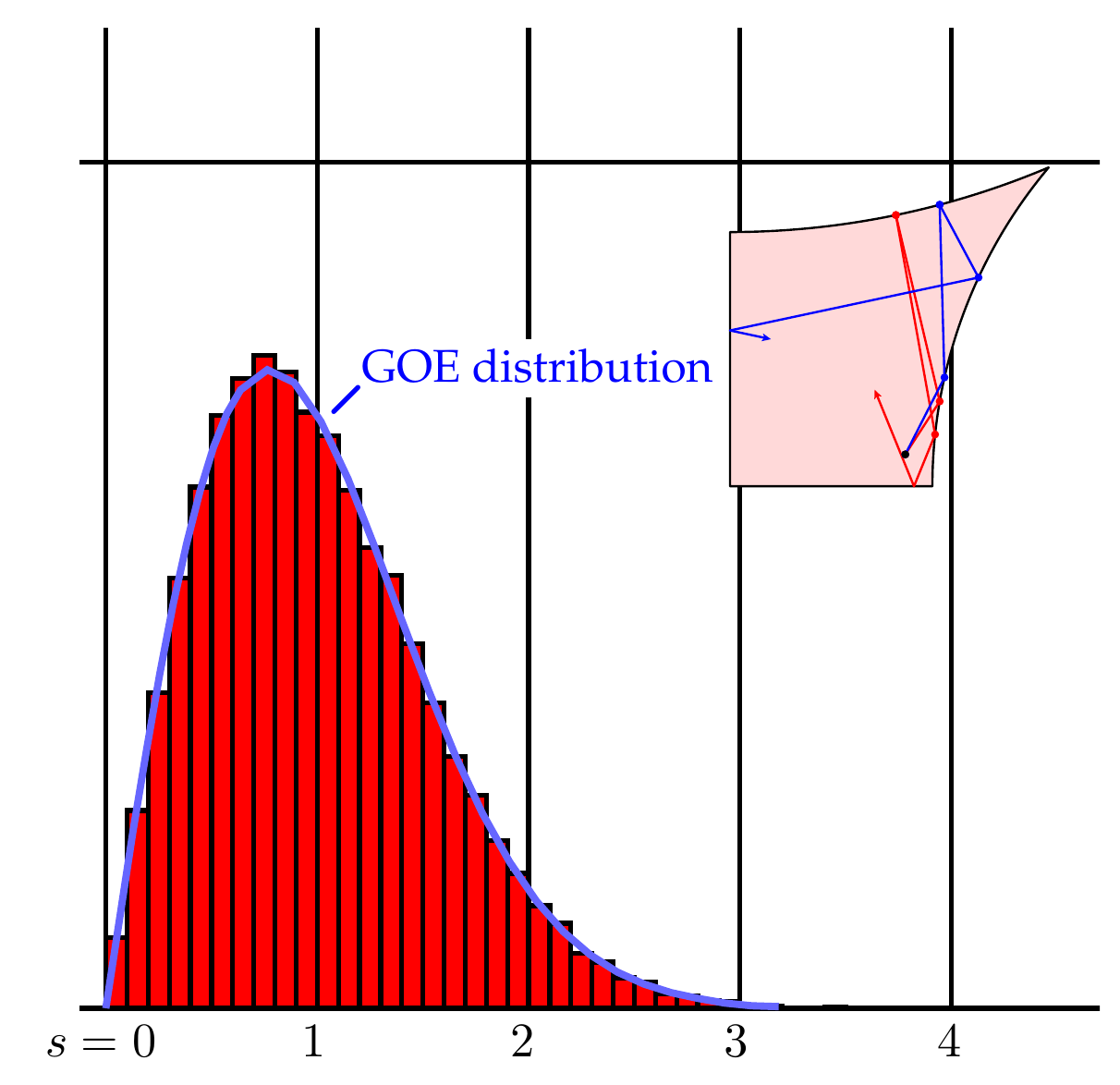}
\end{center}
\caption{(Left panel) Distribution of 250,000 single-particle energy level spacings in a rectangular two-dimensional box with sides $a$ and $b$ such that $a/b=\sqrt[4]{5}$ and $ab=4\pi$. (Right panel) Distribution of 50,000 single-particle energy level spacings in a chaotic cavity consisting of two arcs and two line segments (see inset). The solid lines show the Poisson (left panel) and the GOE (right panel) distributions. From Ref.~\cite{rudnik_08}.}
\label{fig:level_cavities}
\end{figure}

\noindent {\bf Single particle in a cavity -} Next, let us consider a much simpler setup, namely, the energy spectrum of a single particle in a cavity. Here, we can contrast the Berry-Tabor and BGS conjectures. To this end, in Fig.~\ref{fig:level_cavities}, we show the distribution of level spacings for two cavities: (left panel) an integrable rectangular cavity with sides $a$ and $b$ such that $a/b=\sqrt[4]{5}$ and $ab=4\pi$ and (right panel) a chaotic cavity constructed from two circular arcs and two line segments (see inset)~\cite{rudnik_08}. These two plots beautifully confirm the two conjectures. The distribution on the left panel, as expected from the Berry-Tabor conjecture, is very well described by the Poisson distribution. This occurs despite the fact that the corresponding classical system has only two degrees of freedoms [recall that in the argument used to justify the Berry-Tabor conjecture, Eqs.~\eqref{eq:manyosc}--\eqref{eq:berrytabor}, we relied on having many degrees of freedom]. The right panel depicts a level distribution that is in perfect agreement with the GOE, in accordance with the BGS conjecture.

\noindent {\bf Hydrogen atom in a magnetic field -} A demonstration of a crossover between Poisson statistics and Wigner-Dyson statistics can be seen in another single-particle system -- a hydrogen atom in a magnetic field. The latter breaks the rotational symmetry of the Coulomb potential and hence there is no conservation of the total angular momentum. As a result, the classical system has coexistence of regions with both regular (occurring at lower energies) and chaotic (occurring at higher energies) motion~\cite{rudnik_81}. Results of numerical simulations (see Fig.~\ref{fig:level_hydrogen}) show a clear interpolation between Poisson and Wigner-Dyson level statistics as the dimensionless energy (denoted by $\hat{E}$) increases~\cite{wintgen_friedrich_87}. Note that at intermediate energies the statistics is neither Poissonian nor Wigner-Dyson, suggesting that the structure of the energy levels in this range is richer. In the plots shown, the numerical results are fitted to a Brody distribution (solid lines) \cite{brody_flores_81}, which interpolates between the Poisson distribution and the GOE Wigner surmise, and to a semi-classical formula due to Berry and Robnik (dashed lines) \cite{berry_robnik_84}. We are not aware of a universal description of Hamiltonian ensembles corresponding to the intermediate distribution.

\noindent {\bf Lattice models -} As we briefly mentioned earlier, RMT theory and the Berry-Tabor conjecture also apply to interacting many-particle systems that do not have a classical counterpart. There are several models in one-dimensional lattices that fall in this category. They allow one to study the crossover between integrable and nonintegrable regimes by tuning parameters of the Hamiltonian~\cite{cazalilla_citro_11}. A few of these models have been studied in great detail in recent years~\cite{santos_rigol_10a, rigol_santos_10, santos_rigol_10b, kollath_roux_10, santos_borgonovi_12a, santos_borgonovi_12b, atas_bogomolny_13a, atas_bogomolny_13b}. Here, we show results for a prototypical lattice model of spinless (spin-polarized) fermions with nearest and next-nearest neighbor hoppings (with matrix elements $J$ and $J'$, respectively) and nearest and next-nearest neighbor interactions (with strengths $V$ and $V'$, respectively) \cite{santos_rigol_10a}. The Hamiltonian can be written as
\begin{eqnarray}
\hat{H} &=&\sum_{j=1}^{L}\left[-J\left(\hat{f}_j^{\dagger} \hat{f}^{}_{j+1} + \text{H.c.} \right)
+V \left(\hat{n}^{}_j -\frac{1}{2} \right) \left(\hat{n}^{}_{j+1} -\frac{1}{2}\right) \right.\nonumber \\
&&\left.\hspace{0.75cm} - J'\left( \hat{f}_j^{\dagger} \hat{f}^{}_{j+2} + \text{H.c.} \right)
+V' \left(\hat{n}^{}_{j} -\frac{1}{2}\right) \left(\hat{n}^{}_{j+2} -\frac{1}{2}\right) \right],
\label{eq:fermionHam}
\end{eqnarray}
where $\hat{f}^{}_j$ and $\hat{f}_j^{\dagger}$ are fermionic annihilation and creation operators at site $j$, $\hat{n}_j=\hat{f}_j^{\dagger} \hat{f}^{}_j$ is the occupation operator at site $j$, and $L$ is the number of lattice sites. Periodic boundary conditions are applied, which means that $\hat{f}^{}_{L+1}\equiv\hat{f}^{}_{1}$ and $\hat{f}^{}_{L+2}\equiv\hat{f}^{}_{2}$. A classical limit for this model can be obtained at very low fillings and sufficiently high energies. In the simulations presented below, the filling ($N/L$) has been fixed to $1/3$. Therefore, quantum effects are important at any value of the energy. In this example, we approach a dense energy spectrum, and quantum chaos, by increasing the system size $L$. The Hamiltonian~\eqref{eq:fermionHam} is integrable when $J'=V'=0$, and can be mapped (up to a possible boundary term) onto the well-known spin-1/2 $XXZ$ chain \cite{cazalilla_citro_11}. 

It is important to stress that Hamiltonian \eqref{eq:fermionHam} is translationally invariant. This means that when diagonalized in quasi-momentum space, different total quasi-momentum sectors (labeled by $k$ in what follows) are decoupled. In addition, some of those sectors can have extra space symmetries, for example, $k=0$ has reflection symmetry. Finally, if $J'=0$, this model exhibits particle-hole symmetry at half-filling. Whenever carrying out an analysis of the level spacing distribution, all those discrete symmetries need to be accounted for, that is, one needs to look at sectors of the Hamiltonian that are free of them. If one fails to do so, a quantum chaotic system may appear to be integrable as there is no level repulsion between levels in different symmetry sectors. All results reported in this review for models with discrete symmetries are obtained after properly taking them into account.

\begin{figure}[!t]
\includegraphics[width=0.99\textwidth]{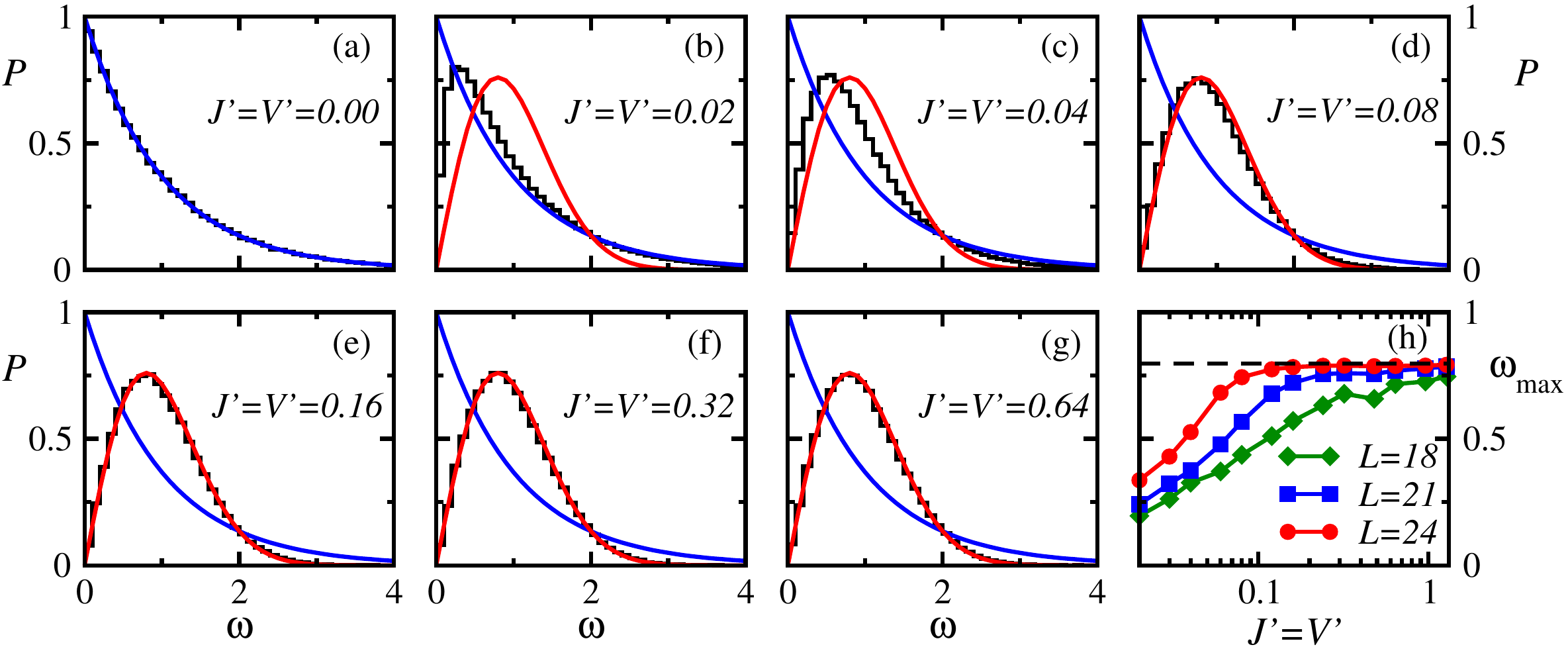}
\vspace{-0.15cm}
\caption{(a)--(g) Level spacing distribution of spinless fermions in a one-dimensional lattice with Hamiltonian \eqref{eq:fermionHam}. They are the average over the level spacing distributions of all $k$-sectors (see text) with no additional symmetries (see Ref.~\cite{santos_rigol_10a} for details). Results are reported for $L=24$, $N=L/3$, $J=V=1$ (unit of energy), and $J'=V'$ (shown in the panels) vs the normalized level spacing $\omega$. The smooth continuous lines are the Poisson and Wigner-Dyson (GOE) distributions. (h) Position of the maximum of $P(\omega)$, denoted as $\omega_\text{max}$, vs $J'=V'$, for three lattice sizes. The horizontal dashed line is the GOE prediction. Adapted from Ref.~\cite{santos_rigol_10a}.}
\label{fig:LDSF}
\end{figure}

In Fig.~\ref{fig:LDSF}(a)--\ref{fig:LDSF}(g), we show the level spacing distribution $P(\omega)$ of a system described by the Hamiltonian \eqref{eq:fermionHam}, with $L=24$ (see Ref.~\cite{santos_rigol_10a} for further details), as the strength of the integrability breaking terms is increased. Two features are immediately apparent in the plots: (i) for $J'=V'=0$, i.e., at the integrable point, $P(\omega)$ is almost indistinguishable from the Poisson distribution and (ii) for large values of the integrability breaking perturbation, $P(\omega)$ is almost indistinguishable from a Wigner-Dyson distribution [GOE in this case, as Eq.~\eqref{eq:fermionHam} is time-reversal invariant]. In between, as in Fig.~\ref{fig:level_hydrogen}, there is a crossover regime in which the distribution is neither Poisson nor Wigner-Dyson. However, as made apparent by the results in panel (h), as the system size increases the level spacing statistics becomes indistinguishable of the RMT prediction at smaller values of the integrability breaking parameters. This suggests that, at least for this class of models, an infinitesimal integrability breaking perturbation is sufficient to generate quantum chaos in the thermodynamic limit. Recent numerical studies have attempted to quantify how the strength of the integrability breaking terms should scale with the system size for the GOE predictions to hold in one dimension \cite{modak_mukerjee_14a,modak_mukerjee_14b}. These works suggest that the strength needs to be $\propto L^{-3}$ for this to happen, but the origin of such a scaling is not understood. Moreover, it is unclear how generic these results are. In particular, in disordered systems that exhibit many-body localization, it has been argued that the transition from the Poisson to the Wigner-Dyson statistics occurs at a finite value of the interaction strength. This corresponds to a finite threshold of the integrability breaking perturbation even in the thermodynamic limit (see Ref.~\cite{nandkishore_huse_14} and references therein).

\subsection{The Structure of Many-Body Eigenstates}\label{sec:manybodyeig}

\begin{figure}[!t]
\includegraphics[width=0.99\textwidth]{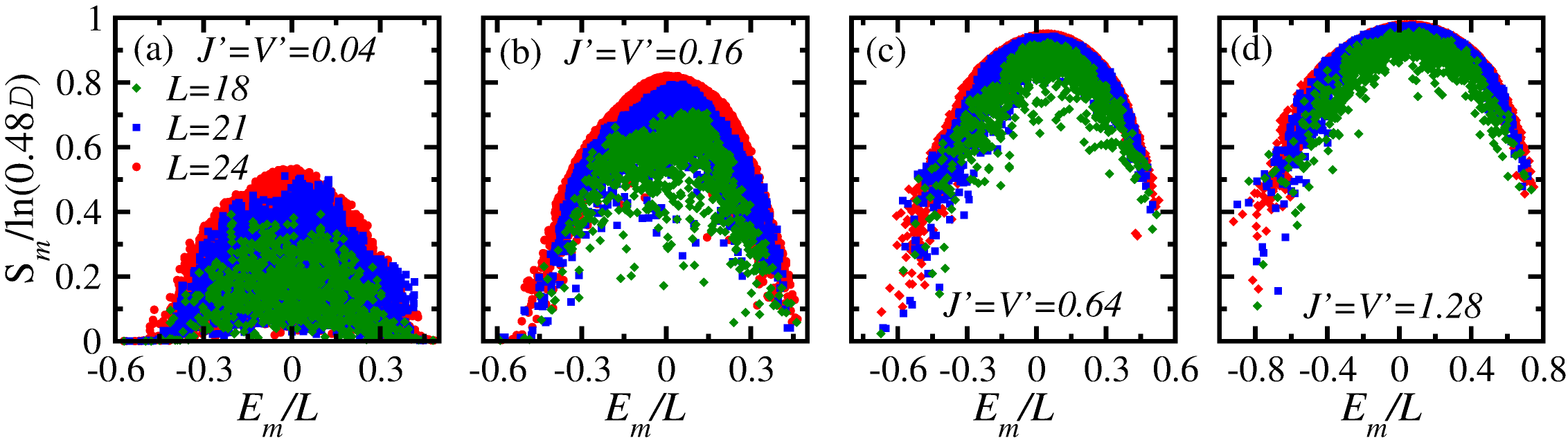}
\vspace{-0.15cm}
\caption{Information entropy (normalized using the GOE prediction) of the  eigenstates of spinless fermions in a one-dimensional lattice with Hamiltonian \eqref{eq:fermionHam}. Results are reported for $L=18$, 21, and 24, $N=L/3$,  $J=V=1$ (unit of energy), and $J'=V'$ (reported in the panels) vs the  energy of the eigenstates. The information entropy is calculated in the basis of the eigenstates of the integrable Hamiltonian  ($J=V=1$ and $J'=V'=0$), and in the $k=2$ quasi-momentum sector ($\mathcal D$ is the number of states in that sector). See also Ref.~\cite{santos_rigol_10a}.}
\label{fig:shannon}
\end{figure}

As we discussed in Sec.~\ref{sec:sec2}, RMT makes important predictions about the random nature of eigenstates in chaotic systems. According to Eq.~(\ref{eq:joprob}), any eigenvector of a matrix belonging to random matrix ensembles is a random unit vector, meaning that each eigenvectors is evenly distributed over all basis states. However, as we show here, in real systems the eigenstates have more structure. As a measure of delocalization of the eigenstates over a given fixed basis one can use the information entropy:
\begin{equation}
\mbox{S}_m \equiv -\sum_{i}  |c^i_m|^2 \ln |c^i_m|^2,
\label{entropy}
\end{equation}
where
\be
|m\rangle = \sum_i c^i_m |i\rangle
\ee
is the expansion of the eigenstate $|m\rangle$ over some fixed basis $|i\rangle$. For the GOE, this entropy, irrespective of the choice of basis, should be $\mbox{S}_{\text{GOE}} = \ln(0.48 \mathcal D) + {\cal O} (1/\mathcal D)$~\cite{kota_14},  where $\mathcal D$ is the dimensionality of the Hilbert space. However, numerical analyses of various physical systems indicate that $\mathrm S_m$ is only generically bounded from above by the RMT prediction~\cite{kota_sahu_98, santos_rigol_10a, santos_borgonovi_12b}. This situation is characteristic of both few-particle and many-particle systems. For concreteness, we will illustrate this using the eigenstates of the Hamiltonian~\eqref{eq:fermionHam} (see Ref.~\cite{santos_rigol_10a} for details). For the fixed basis $|i\rangle$, we use the eigenstates of the integrable limit of this Hamiltonian, corresponding to $J'=V'=0$. The results of the numerical simulations for the normalized Shannon entropy $S_m/S_{\rm GOE}$ are shown in Fig.~\ref{fig:shannon} [we note that $S_m$ and $S_{\rm GOE}$ were computed within a single quasi-momentum sector of the translationally invariant Hamiltonian \eqref{eq:fermionHam}]. It is clear from the figure that the entropy of the states in the middle of the spectrum approaches the RMT prediction as the strength of the integrability breaking perturbation and the system size increase, while the states near the edges of the spectrum remain ``localized''. The latter, namely, that the lowest and highest (if the spectrum is bounded) energy states are usually non-chaotic, is a generic feature of physical systems with few-body interactions and no randomness.

\begin{figure}[!t]
\begin{center}
 \includegraphics[width=0.82\textwidth]{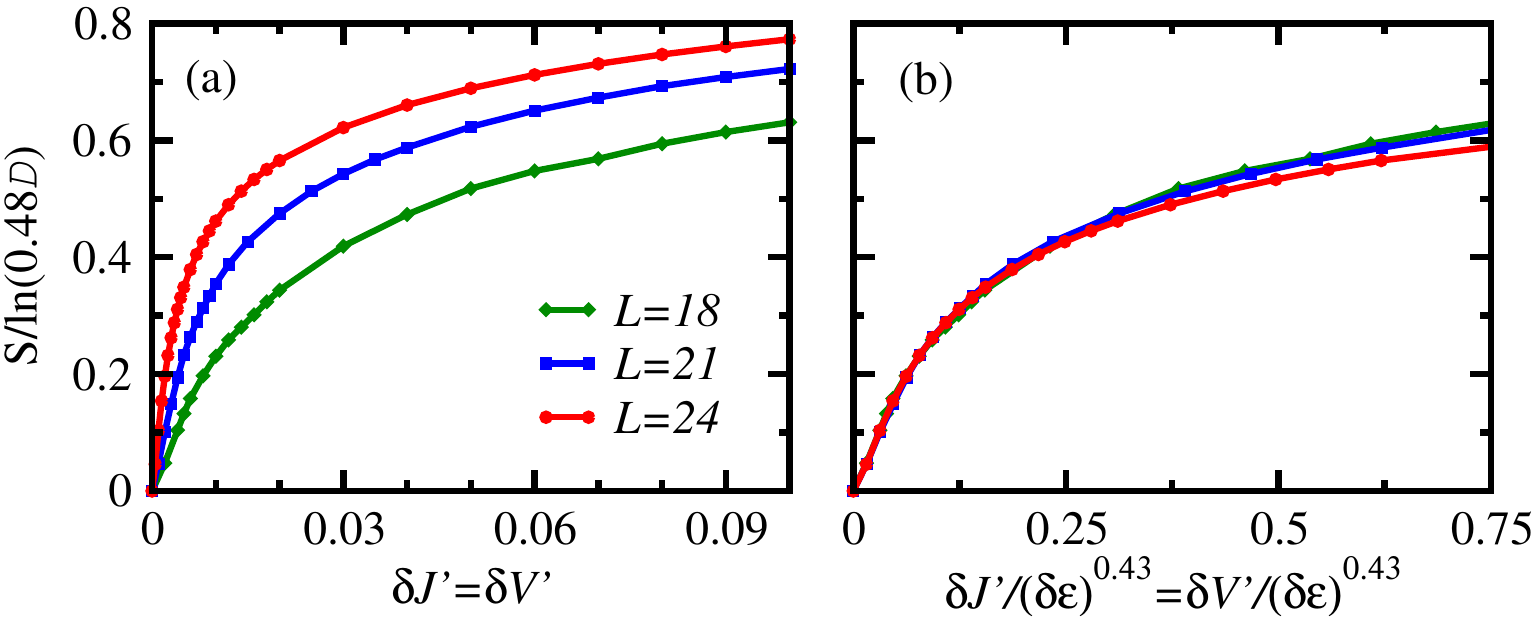}
\end{center}
\vspace{-0.35cm}
\caption{Average information entropy (normalized using the GOE prediction) of the eigenstates of spinless fermions in a one-dimensional lattice with Hamiltonian~\eqref{entropy}. The average is computed over the central  10\% of the energy spectrum. Results are reported for $L=18$, 21, and 24, $N=L/3$, $J=V=1$ (unit of energy), and $J'=V'$ as one departs from $J'=V'=0.5$. The information entropy is calculated in the basis of the  eigenstates of the nonintegrable Hamiltonian with $J=V=1$ and $J'=V'=0.5$, and in the $k=2$ quasi-momentum sector ($\mathcal D$ is the number of states in that sector). (a) The average information entropy is  reported as a function of $\delta J'\equiv J'-0.5=\delta V'\equiv V'-0.5$.  (b) The average information entropy is reported as a function of 
$\delta J'/(\delta \varepsilon)^{0.43}=\delta V'/(\delta\varepsilon)^{0.43}$, where $\delta\varepsilon$ is the average level spacing between the eigenstates used to compute the average entropy.}
\label{fig:sensitivity}
\end{figure}

Another implication of RMT is that the eigenstates of different Hamiltonians are essentially uncorrelated random vectors. This, of course, cannot be literary true in physical systems. Indeed, let us consider a family of Hamiltonians characterized by some continuous parameters like $\hat H(J',V')$ in Eq.~\eqref{eq:fermionHam}. If we change $J'\to J'+\delta J'$ and $V'\to V'+\delta V'$, then, obviously, for sufficiently small changes, $\delta J'$ and $\delta V'$, the eigenstates of the Hamiltonians $\hat H(J',V')$ and $\hat H(J'+\delta J', V'+\delta V')$ will be almost the same. However, one can anticipate that a very small parameter change, likely vanishing exponentially with the system size, is sufficient to mix different eigenstates of the original Hamiltonian with nearby energies such that new eigenstates look essentially random in the old basis. This is indeed what is seen in the numerical simulations. In Fig.~\ref{fig:sensitivity}(a), we show the scaled information entropy $S/S_{\rm GOE}$ of the eigenstates of $\hat H(J'+\delta J', V+\delta V')$ in the basis of $\hat H(J', V')$ as a function of $\delta J'=\delta V'$. These results show that at fixed values of $\delta J'$ and $\delta V'$, the information entropy rapidly increases with the system size. In Fig.~\ref{fig:sensitivity}(b), we show the same entropy plot vs the integrability breaking perturbation, but now scaled by a power of the mean level spacing $\delta J'/(\delta \varepsilon)^\alpha=\delta V'/(\delta\varepsilon)^\alpha$. We found numerically that there is good data collapse for $\alpha\approx 0.43$. While it is necessary to study much larger system sizes to determine the exponent $\alpha$ accurately, for the system sizes available to us it is already apparent that the relevant strength of the integrability breaking perturbation needed for a complete randomization of the energy levels is exponentially small in the system size. Indeed $\delta \varepsilon\propto \exp[-S(E)]$, where $S(E)$ is the thermodynamic entropy of the system, which scales linearly in the system size, and $E$ is the average energy of the eigenstates for which the information entropy is computed.

\subsection{Quantum Chaos and Entanglement}\label{sec:chaos_entanglement}

So far, we have discussed manifestations of quantum chaos in the statistics of level spacings and in the properties of many-body Hamiltonian eigenstates. At the same time, as we discussed in Sec.~\ref{sec:sec2}, classical chaotic systems do not have a well-defined analogue of stationary eigenstates because they do not have closed stationary orbits. Chaos in classical systems is usually defined as the exponential divergence in time of nearby trajectories. But this language does not apply to quantum systems, which do not have a well-defined notion of a trajectory. So, it seems that there is a fundamental discrepancy between the quantum and classical ways of defining chaos. Nevertheless, this discrepancy is somewhat superficial and one can treat quantum and classical chaos on the same footing by analyzing delocalization of the system either in phase space or in energy space, and using appropriate entropy measures to characterize this delocalization. Using such measures, it is possible to smoothly interpolate between quantum and classical regimes in chaotic systems and analyze various quantum to classical crossovers. However, some care is needed in defining such measures. To this end, here we first discuss the problem for classical systems and then extend the ideas to quantum systems.

Let us consider a setup in which the system is prepared in some initial state and is allowed to evolve according to some time-independent Hamiltonian $\hat H$. If the initial state is a stationary state (namely, a stationary probability distribution) of some initial Hamiltonian $\hat{H}_0\neq\hat H$ then this is what is usually called a quench. For example, one can consider a gas of particles in thermal equilibrium in a recipient with a given volume, and then one suddenly doubles the volume of the recipient, for example, by moving a piston. Alternatively, one can consider an equilibrium system of spins (classical or quantum) in which one suddenly changes a magnetic field or the coupling between the spins. A strong physical manifestation of chaos in classical systems is delocalization in the available phase space after the quench. A standard measure of this delocalization is the entropy, which is defined in phase space as
\be\label{eq:entclass}
S=-\int\int \frac{d{\bf x}d{\bf p}}{(2\pi\hbar)^D}\, \rho({\bf x},{\bf p})\ln[\rho({\bf x},{\bf p})],
\ee
where $\rho({\bf x},{\bf p})$ is the classical probability distribution, $D$ is the dimensionality of the phase space, and the usual factor of $(2\pi\hbar)^{D}$ is introduced to make the integration measure dimensionless. This entropy is maximized when $\rho({\bf x},{\bf p})$ is uniform in the available phase space. If the system is isolated, then according to Liouville's theorem the entropy \eqref{eq:entclass} is conserved in time~\cite{landau_lifshitz_5_80}. This is a consequence of the incompressibility of classical trajectories, which implies that the phase-space volume occupied by any closed system does not change in time. Liouville's theorem and the lack of the entropy increase was a topic of controversy for a long time, since Boltzmann introduced his H-theorem. 

To circumvent this problem and use entropy as a measure of delocalization in the available phase space, one can analyze the reduced probability distribution of $N_A$ particles obtained by averaging over the positions and momenta of the remaining $N-N_A$ particles,
\beq
&&\rho_A({\bf x}_1,\dots, {\bf x}_{N_A}, {\bf p}_1,\dots, {\bf p}_{N_A}, t)\nonumber\\&&\qquad=\int\int d{\bf x}_{N_A+1} d{\bf p}_{N_A+1}\dots d{\bf x}_N d{\bf p}_N\, \rho({\bf x}_1,\dots, {\bf x}_N,{\bf p}_1,\dots, {\bf p}_N,t) \;,
\eeq
and compute the entropy of this reduced probability distribution. This entropy is not restricted by Liouville's theorem and after a quench, for sufficiently large subsystems of an ergodic system (and for $N_A\ll N$), it is expected to increase in time to the maximum value given by the Gibbs distribution.  

In quantum systems, the situation is remarkably similar. Instead of a probability distribution one deals with a density matrix $\hat \rho$. A direct analogue of the classical (Liouville) entropy \eqref{eq:entclass} is the von Neumann entropy:
\be
S_{\rm vn}=-{\rm Tr}[\hat \rho\ln\hat\rho].
\ee
Similar to classical systems, the von Neumann entropy is conserved in time for isolated systems, which is a simple consequence of unitary evolution. Hence, extending the analogy to classical systems, we can define the reduced density matrix of a quantum system using a partial trace (typically, one traces over a region in real space):
\be
\hat \rho_A={\rm Tr}_B [\hat\rho]=\sum_{n_A, n_A'} |n_A\rangle \langle n_A'|\sum_{n_B} \langle n_A, n_B|\hat \rho| n_A', n_B\rangle,
\ee
where $|n_A\rangle$ and $|n_B\rangle$ are the complete basis states of the subsystems $A$ and $B$, respectively. One can then define the von Neumann entropy of the reduced density matrix
\be
S^{A}_{\rm vn}=-{\rm Tr}_A [\hat \rho_A\ln\hat \rho_A].
\ee
If the full density matrix is that of a pure state, that is, $\hat \rho=|\psi\rangle\langle \psi|$, then this entropy $S^A_{\rm vn}$ is also called the entanglement entropy. The entanglement entropy has been studied in the context of quenches and thermalization in clean interacting systems \cite{santos_polkovnikov_12, deutsch_li_13, mierzejewski_prosen_13, kim_huse_13, khlebnikov_kruczenski_14}, as well as in disordered systems in the context of many-body localization following quantum quenches and in the presence of a periodic drive~\cite{bardarson_pollman_12, vosk_altman_13, serbyn_papic_13, nanduri_kim_13, ponte_papic_14}. 

These ideas were recently tested in experiments with small systems involving superconducting qubits~\cite{neil_roushan_16} and ultracold atoms~\cite{kaufman_tai_16}. We will discuss the superconducting qubit experiment in the next section. Here, we review the results of the ultracold atom experiment. There, the authors prepared two identical chains each with six sites. The Hamiltonian describing the system is
\be
\hat H={U\over 2} \sum_{ij} \hat n_{i,j} (\hat n_{i,j}-1)-J_x \sum_{i,j} (\hat a_{i,j}^\dagger \hat a^{}_{i+1,j}+\text{H.c.})-J_y\sum_i (\hat a_{i,1}^\dagger \hat a^{}_{i,2}+\text{H.c.}),
\ee
where $i=1,\dots, 6$ is the site coordinate along the $x$-direction and $j=1,2$ is the site coordinate along the $y$-direction. As usual, $\hat a_{i,j}^\dagger$ and $\hat a^{}_{i,j}$ are the boson creation and annihilation operators, respectively, and $\hat n_{i,j}=\hat a_{i,j}^\dagger \hat a^{}_{i,j}$ is the site occupation number. 

The system was initialized in a Fock state with exactly one particle per site and both $J_x$ and $J_y$ being essentially equal to zero. At time $t=0$, the tunneling along the $x$-direction ($J_x$) was quenched to $J_x/U\approx 0.64$, with $J_y$ remaining negligible. The system was then allowed to evolve. This way, two identical copies of a many-body state were created. Implementing a swap operation~\cite{daley_pichler_12, islam_ma_15}, it was possible to measure the second Renyi entanglement entropy for each chain:
\be
S_2^A=-\ln\left[{\rm Tr}\hat \rho_A^2\right].
\ee 
The latter is very similar to the von Neumann entanglement entropy. If the system is quantum chaotic, $S_2^A$ is expected to coincide with the corresponding entropy in the thermal ensemble for $L_A/L<1/2$ (up to finite-size corrections). We note that, in general, $S^A_{\rm vn}$ bounds $S^A_2$ from above. The two entropies are equal for (maximally entangled) infinite temperature states and (non-entangled) product states.

\begin{figure}[!t]
\begin{center}
\includegraphics[width=0.47\textwidth,height=0.366\textwidth]{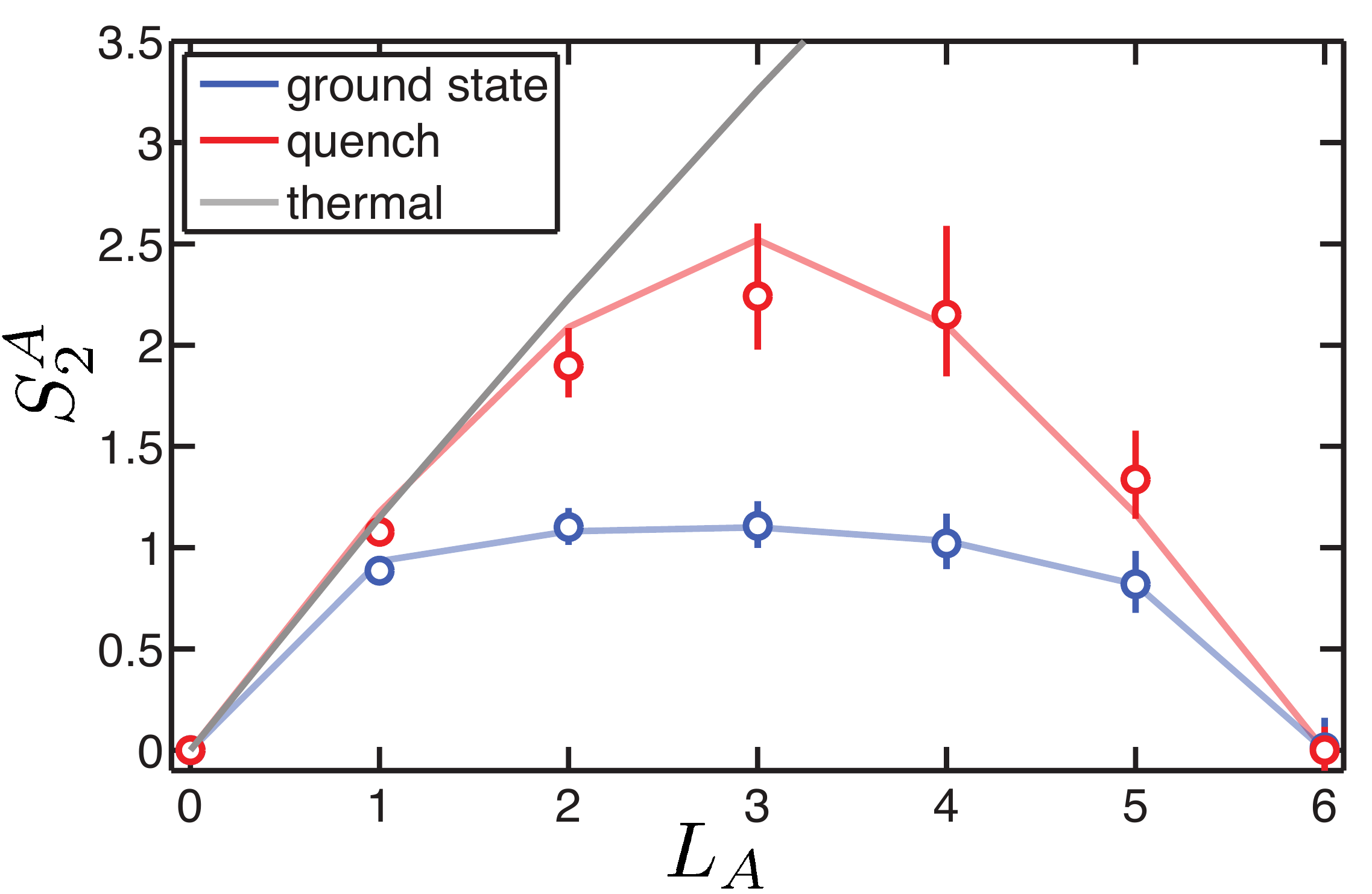}
\includegraphics[width=0.52\textwidth,height=0.37\textwidth]{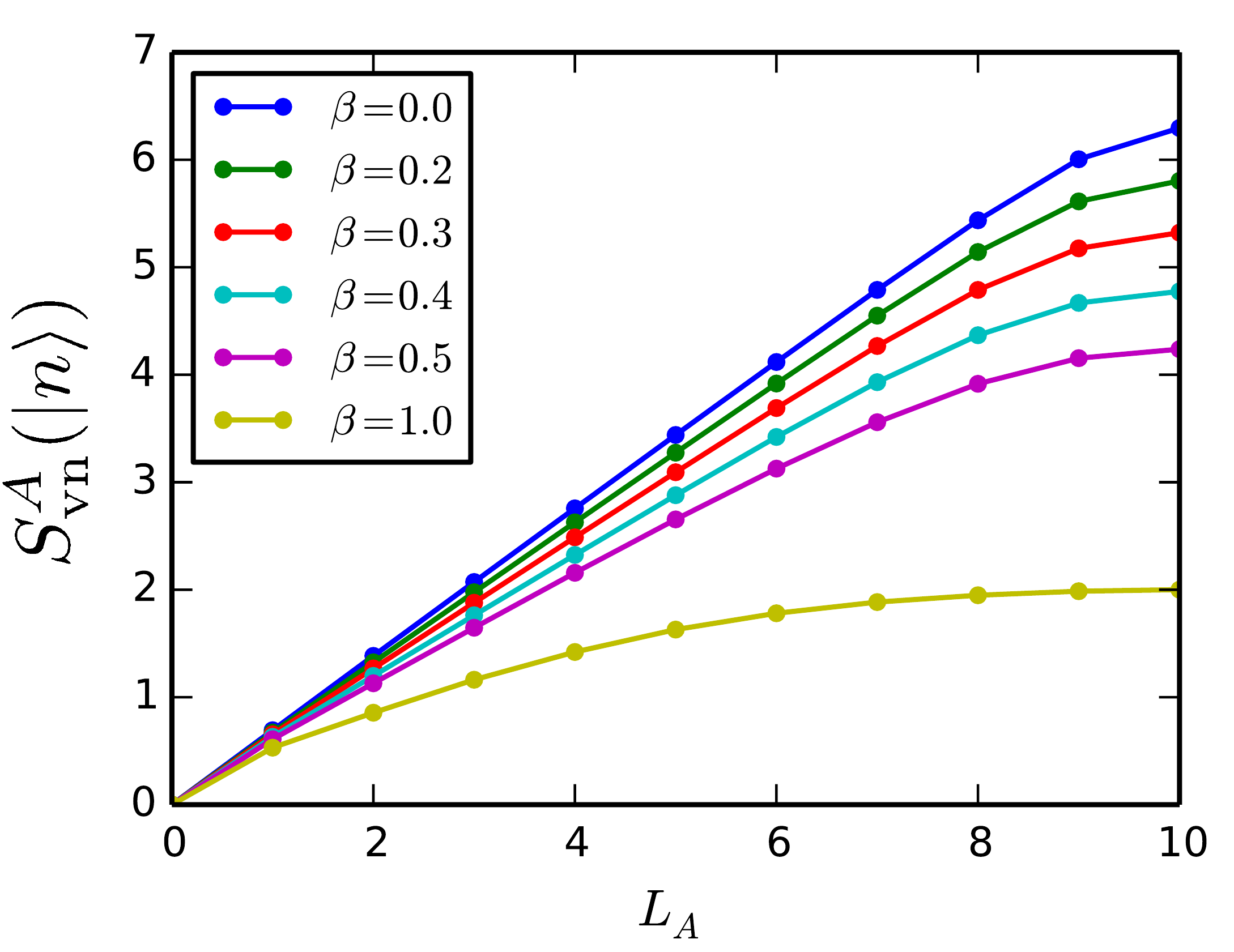}
\end{center}
\vspace{-0.15cm}
\caption{(Left panel) Second Renyi entanglement entropy vs subsystem size $L_A$ for a six-site Bose-Hubbard chain measured at long times after a quench (red symbols) and in the ground state (blue symbols). Red and blue lines following the data points depict the theoretical predictions, while the gray (straight) line depicts the theoretical prediction for the Renyi entropy of the system in thermal equilibrium. From Ref.~\cite{kaufman_tai_16}. (Right panel) Entanglement entropy as a function of a subsystem size for different representative eigenstates of the spin-1/2 Hamiltonian~\eqref{eq:hamiltonian_kim_huse} for $L=20$. The inverse temperature $\beta$ in both panels is obtained by matching the energies to those of systems in thermal equilibrium. The entanglement entropy grows linearly with $L_A$, when $L_A$ and $\beta$ are small, and coincides with the equilibrium entropy of the Gibbs ensemble. From Ref.~\cite{garrison_grover_15}.}
\label{fig:greiner1}
\end{figure}

In the left panel in Figure~\ref{fig:greiner1}, we show the measured long-time result of the second Renyi entanglement entropy after the quench as a function of the subsystem size. It is remarkable that, even for such a small system, $S_2^A$ is very close to the entropy of a Gibbs ensemble with the same mean energy for the smallest subsystem sizes. This experiment shows that, even in small quantum systems, one can see clear signatures of quantum chaotic behavior.

\begin{figure}[!t]
\begin{center}
 \includegraphics[width=0.6\textwidth]{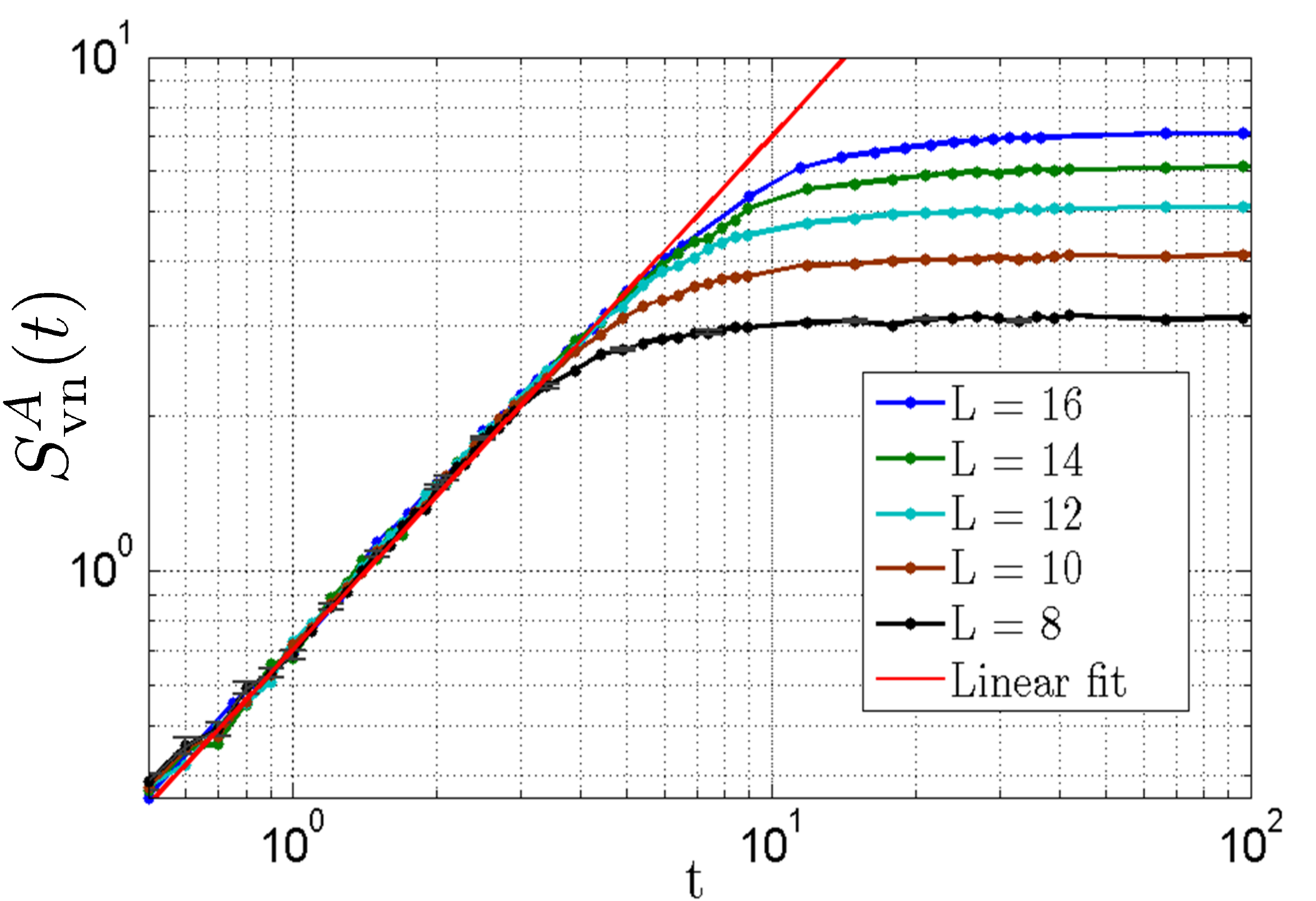}
\end{center}
\vspace{-0.3cm}
\caption{von Neumann's entropy of the reduced density matrix obtained after tracing one half of a spin-1/2 chain as a function of time (see text for details). The initial state corresponds to a product state of randomly polarized spins. The entropy grows linearly in time and saturates at a value which is very close to the maximum, corresponding to the infinite temperature state: $S_{\rm max}=L\ln{2}/2-1/2$. From Ref.~\cite{kim_huse_13}.}
\label{fig:kim_huse}
\end{figure}

Next, it is important to discuss theoretical predictions that closely follow the experimental findings. We focus on results for the spin-1/2 transverse field Ising chain, with Hamiltonian:
\be
\hat H=\sum_{j=1}^L g \hat\sigma^x_j+\sum_{j=2}^{L-1} h \hat \sigma^z_j+(h-J) (\hat \sigma^z_1+\hat \sigma^z_L)+J\sum_{j=1}^{L-1}\hat\sigma^z_j \hat\sigma^z_{j+1}.
\label{eq:hamiltonian_kim_huse}
\ee
This model exhibits quantum chaos in the parameter range studied in Figs.~\ref{fig:greiner1} (right panel) and \ref{fig:kim_huse}: $h=(\sqrt{5}+1)/4$, $g=(\sqrt{5}+5)/8$, and $J=1$.

In the right panel in Fig.~\ref{fig:greiner1}, we show the entanglement entropy for representative eigenstates of the Hamiltonian \eqref{eq:hamiltonian_kim_huse} as a function of the subsystem size \cite{garrison_grover_15}. Different curves are labeled according to the temperature of the Gibbs ensemble that has the same mean energy as the eigenstate. For small subsystem sizes, the entanglement entropy is clearly a linear function of the subsystem size (as in the experimental results for the second Renyi entanglement entropy shown in the left panel). Moreover, the slope is identical to the slope of the equilibrium entropy. As the subsystem size increases, the entanglement entropy deviates from the equilibrium result and the deviation increases as the effective temperature decreases. In Ref.~\cite{garrison_grover_15}, those deviations were argued to be subextensive in $L$ for any fixed ratio $L_A/L<1/2$. This means that, in the thermodynamic limit and for any nonvanishing effective temperature, the entanglement entropy of eigenstates in quantum chaotic systems is expected to have a triangular shape as a function of $L_A/L$, with a cusp at $L_A/L=1/2$. This is a result of the eigenstate thermalization phenomenon that we discuss in Sec.~\ref{sec:sec4}.

More directly related to the experiments in Ref.~\cite{kaufman_tai_16}, in Fig.~\ref{fig:kim_huse} we show the temporal evolution of the entanglement entropy obtained numerically after tracing out one half of the lattice in the transverse field Ising model \eqref{eq:hamiltonian_kim_huse}. The initial state corresponds to a product of spins with random orientations. Such a state has zero initial entropy.  As seen in Fig.~\ref{fig:kim_huse}, the entropy $S^A_{\rm vn}$ grows linearly in time and then saturates (as expected) close to that of a random pure state~\cite{page_93}: $S^\ast=L\ln 2/2-1/2$. This is exactly the result obtained in the right panel in Fig.~\ref{fig:greiner1} for $\beta=0$. Following the findings in Fig.~\ref{fig:kim_huse}, one can anticipate that if one studies a classical ergodic spin chain, instead of a quantum spin chain, and begins the dynamics from a factorized probability distribution one would get a similar increase of the Liouville entropy of one half of the system.

\subsection{Quantum Chaos and Delocalization in Energy Space}\label{sec:chaos_delocalization}

Another way to reveal delocalization of classical systems in available phase space is to study the time-averaged probability distribution over a time interval $t_0$
\be
 \rho_{t_0}({\bf x},{\bf p})={1\over t_0}\int_0^{t_0} \rho({\bf x},{\bf p},t)
\ee
and compute the entropy of this distribution. Because the negative logarithm is a convex function, using Jensen's inequality, it is straightforward to see that such an entropy can only increase as a function of $t_0$. For ergodic systems, it is expected that this entropy will increase to its maximally allowed value, that is, to the microcanonical entropy, because the system on average visits all points in phase space with equal probability. For non-ergodic systems, conversely, the system is expected to remain more localized in phase space even after time averaging, so that the entropy never reaches the microcanonical value. 

Continuing the analogy with classical systems, a second possibility to use entropy to quantify quantum delocalization is to study the entropy of the time-averaged density matrix. Assuming that there are no degeneracies, the off-diagonal matrix elements of the density matrix in the basis of the Hamiltonian oscillate in time according to~\cite{landau_lifshitz_3_81}: 
\be
\rho_{mn}(t)=\rho_{mn}(t_0)\exp[-i (E_m-E_n)(t-t_0)]
\ee
Therefore, in the quantum language, time averaging is equivalent to projecting the initial density matrix onto the diagonal subspace of the Hamiltonian, leading to what is known as the diagonal ensemble\footnote{The diagonal ensemble will play a crucial role throughout this review.} density matrix \cite{rigol_dunjko_08}:
\be\label{eq:denmatDE}
\hat{\rho}_\text{DE}\equiv\bar{\hat{\rho}}\equiv\lim_{t_0\to\infty} \frac{1}{t_0} \int_0^{t_0} \hat{\rho}(t) dt=\sum_m \rho_{mm} |m\rangle\langle m|.
\ee
Thus studying delocalization of the classical probability distribution in phase space at long times is equivalent, in the quantum language, to studying the spreading of the initial density matrix in the basis of the eigenstates of the Hamiltonian, or, simply, in energy space. From the discussion in Sec.~\ref{sec:manybodyeig}, one can expect that the diagonal density matrix will generically be delocalized for quantum chaotic systems. For integrable systems, on the other hand, the diagonal density matrix can be more (or less) localized depending on the initial state.

The analogy between delocalization in energy space and classical chaos was recently explored experimentally in a system of three coupled superconducting qubits~\cite{neil_roushan_16}, which effectively represent three 1/2 spins. The experiment was carried out in the sector where the effective total spin is $S=3/2$, and focused on periodic kicks with:
\be
\hat H(t)={\pi\over 2} \hat S_y+{\kappa\over 2 S }\hat S_z^2\sum_n \delta (t-n),
\ee
where $\hat S_y$ and $\hat S_z$ are spin operators. Like the kicked rotor model, this system in the $S\to\infty$ classical limit has a mixed phase space with both chaotic and regular trajectories. In the experiment, the quantum system was initialized in a coherent state centered around some point in the two-dimensional phase space. The system was then allowed to evolve under kicks and, after long times, both the entanglement entropy (of one qubit) and the diagonal entropy (the entropy of the time-averaged density matrix) were measured through quantum tomography. The results were contrasted with the phase-space dynamics of the $S\to\infty$ classical limit.

\begin{figure}[!t]
\begin{center}
 \includegraphics[width=15cm]{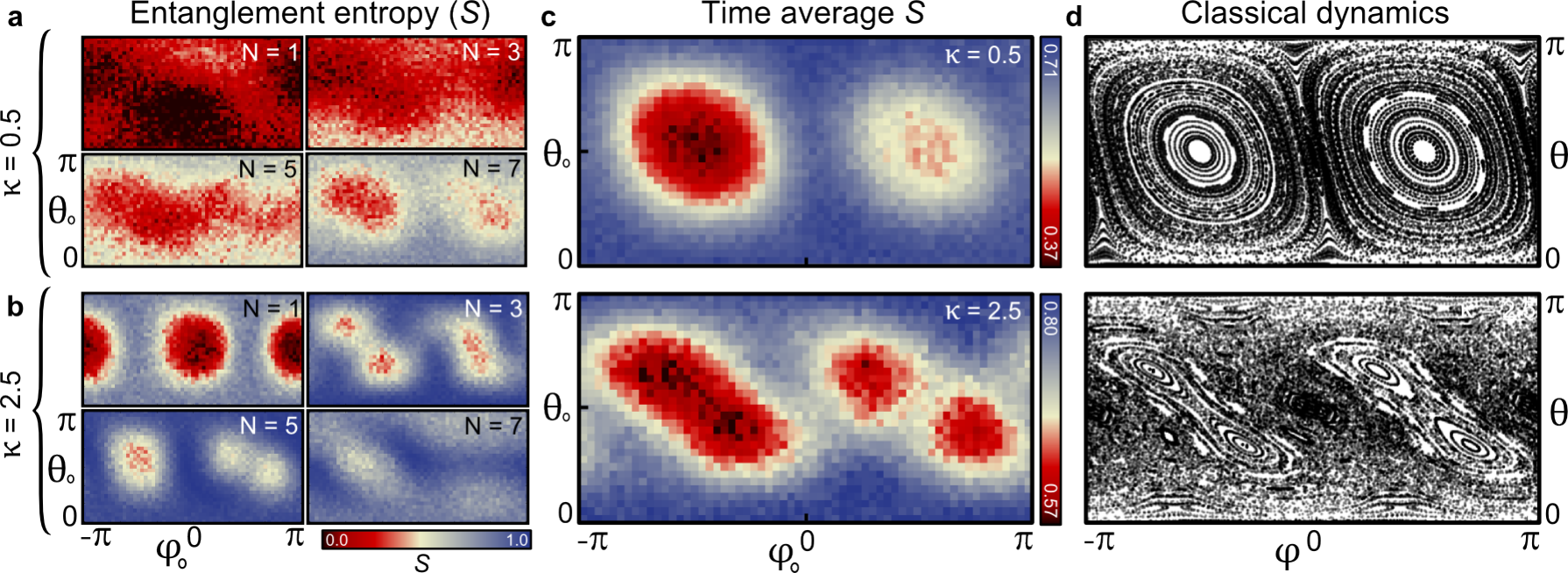}
\end{center}
\vspace{-0.3cm}
\caption{Left panels: entanglement entropy averaged over three different qubits after a different number of kicks $N$. In the color scale used, red signals a small entropy and blue signals the maximum entropy [ln(2)]. Different points on each panel correspond to different initial conditions, which are coherent states centered around the phase-space point $(\theta_0,\phi_0)$. Middle panels: Floquet diagonal entropy, which was computed as the entropy of the full density matrix averaged over 20 kicks (top) and 10 kicks (bottom).  Right panels: phase-space portrait of the classical system. The top and bottom panels depict results for different kick strengths: $\kappa=0.5$ and $\kappa=2.5$, respectively. From Ref.~\cite{neil_roushan_16}.}
\label{fig:roushan1}
\end{figure}

In Fig.~\ref{fig:roushan1}, we show the dynamics of the entanglement entropy (left panels) and the entropy of the density matrix averaged over several periods, which is equivalent to the Floquet diagonal entropy (middle panels). The entropies are reported (in a color scale) for different initial coherent states centered around spherical angles ($\theta_0,\phi_0$). The right panels show the phase space portraits of the corresponding classical systems. Both the entanglement entropy and the Floquet diagonal entropy show strong correlations with classical regions of chaotic and non-chaotic motion, with higher entropy corresponding to more chaotic behavior. Interestingly, these correlations persist deep in the quantum regime ($S=3/2$ is not particularly large). This experiment illustrates the ideas discussed in this and the previous section, namely, that quantum chaos results in delocalization of either the reduced density matrix of subsystems or the time-averaged density matrix of the full system.

To illustrate delocalization of an initial wave function among energy eigenstates in a larger quantum chaotic system, we follow Ref.~\cite{santos_polkovnikov_11}, which reported results for quantum quenches in one-dimensional periodic chains of interacting spinless fermions with Hamiltonian \eqref{eq:fermionHam}, and hard-core bosons with Hamiltonian
\begin{eqnarray}
\hat{H} &=&\sum_{j=1}^{L}\left[-J\left(\hat{b}_j^{\dagger} \hat{b}^{}_{j+1} + \text{H.c.} \right)
+V \left(\hat{n}^{}_j -\frac{1}{2} \right) \left(\hat{n}^{}_{j+1} -\frac{1}{2}\right) \right.\nonumber \\
&&\left.\hspace{0.75cm} - J'\left( \hat{b}_j^{\dagger} \hat{b}^{}_{j+2} + \text{H.c.} \right)
+V' \left(\hat{n}^{}_{j} -\frac{1}{2}\right) \left(\hat{n}^{}_{j+2} -\frac{1}{2}\right) \right],
\label{eq:HCBHam}
\end{eqnarray}
where $\hat{b}^{}_j$ and $\hat{b}_j^{\dagger}$ are hard-core bosons annihilation and creation operators at site $j$, $\hat{n}_j=\hat{b}_j^{\dagger} \hat{b}^{}_j$ is the occupation operator at site $j$, and $L$ is the number of lattice sites. Hard-core bosons satisfy the same commutation relations as bosons but have the constraints $\hat{b}^{2}_j=(\hat{b}_j^{\dagger})^2=0$, which preclude multiple occupancy of the lattice sites \cite{cazalilla_citro_11}. For $J'=0$, the hard-core boson Hamiltonian \eqref{eq:HCBHam} can be mapped onto the spinless fermion Hamiltonian \eqref{eq:fermionHam}, up to a possible boundary term \cite{cazalilla_citro_11}. Like the spinless fermion Hamiltonian \eqref{eq:fermionHam}, the hard-core boson one \eqref{eq:HCBHam} is integrable for $J'=V'=0$ and nonintegrable otherwise.

\begin{figure}[!t]
\begin{center}
 \includegraphics[width=0.95\textwidth]{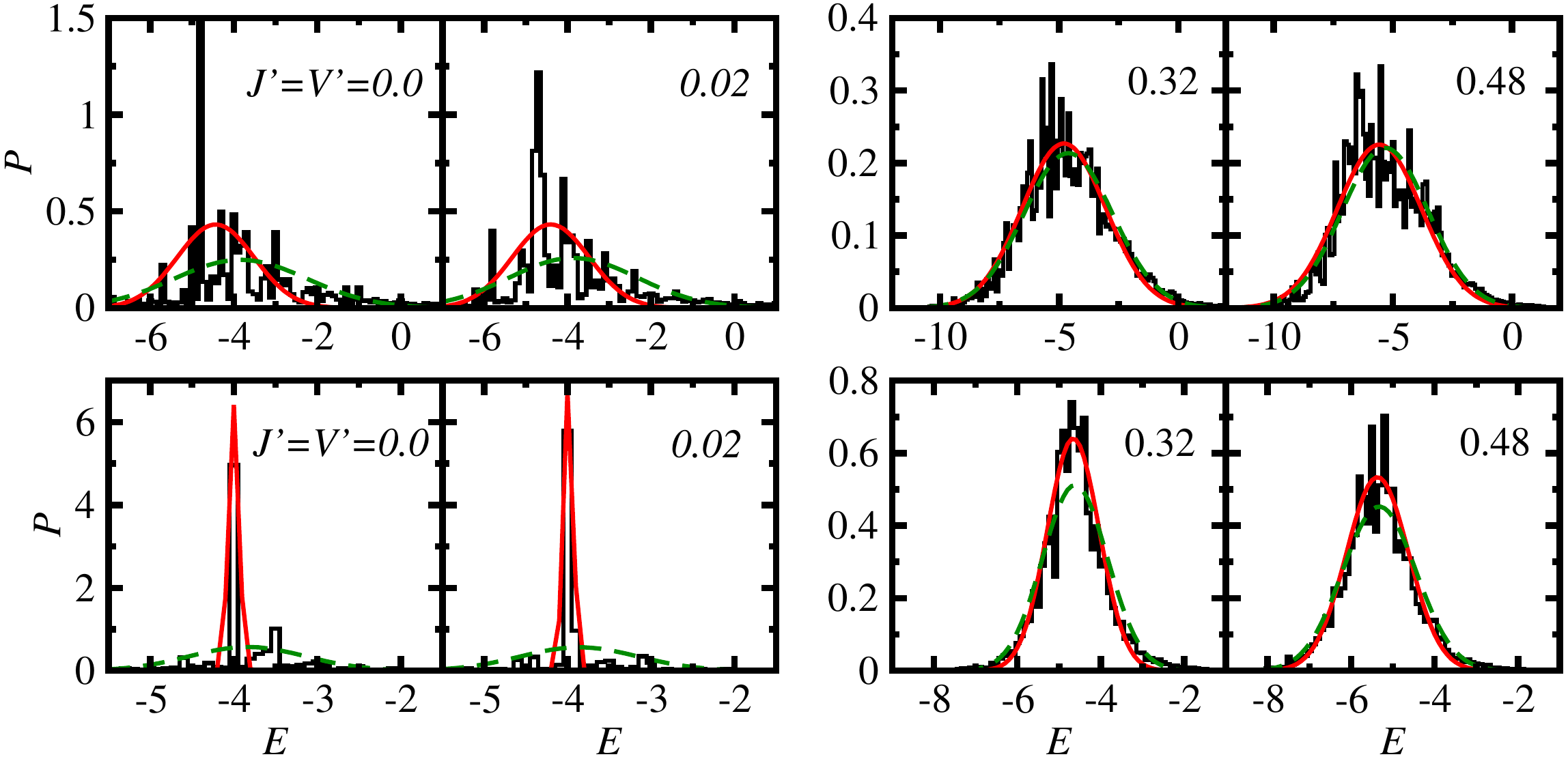}
\end{center}
\vspace{-0.3cm}
\caption{Normalized energy distribution function $P(E)$, see Eq.~\eqref{eq:normenegdist}, after quenches in hard-core boson chains. Results are presented for two different initial states $|\psi_I\rangle$ corresponding to eigenstates of Hamiltonian~\eqref{eq:HCBHam} with $J_I=0.5$, $V_I=2$ (top panels) and $J_I=2$, $V_I=0.5$ (bottom panels). The final parameters of the Hamiltonian are $J=V=1$. $J'=V'$ remain unchanged during the quench. Their values are indicated in the figure. Smooth solid lines: best Gaussian fit to $(\sqrt{2 \pi} a)^{-1} \exp[-(E-b)^2/(2 a^2)]$ for the parameters $a$ and $b$; dashed lines: $(\sqrt{2 \pi}\delta E)^{-1} \exp[-(E-\overline{E})^2/(2 \delta E^2)]$, where $\overline{E}=\langle\psi_I|\hat{H}|\psi_I\rangle$ and 
$\delta E^2=\langle\psi_I|\hat{H}^2|\psi_I\rangle-\langle\psi_I|\hat{H}|\psi_I\rangle^2$ is the energy variance after the quench. From Ref.~\cite{santos_polkovnikov_11}.}
\label{fig:santos}
\end{figure}

In Fig.~\ref{fig:santos}, we show the normalized energy distribution
\be\label{eq:normenegdist}
P(E)=\sum_m p_m \delta(E-E_m),
\ee
where
\be\label{eq:overlapsquared}
p_m=|\langle m|\psi_I\rangle|^2,
\ee
and $\langle m|\psi_I\rangle$ is the projection of the initial state $|\psi_I\rangle$ on eigenstate $|m\rangle$. The results presented are for quenches in the hard-core boson chain. The top and bottom panels correspond to different initial states. The parameters of the final Hamiltonian are $J=V=1$, and $J'=V'$ with the values indicated in the figure (increasing from left to right). Those plots make apparent that as the system becomes more quantum chaotic (larger $J'=V'$), the energy distribution becomes less sparse, which means that the initial state becomes more delocalized among the eigenstates of the final Hamiltonian. Another visible feature of the energy distribution is that in chaotic systems it rapidly approaches a Gaussian centered around the mean energy $\overline{E}=\langle\psi_I|\hat{H}|\psi_I\rangle$ and the width given by the variance of the energy in the initial state $\delta E^2=\langle\psi_I|\hat{H}^2|\psi_I\rangle-\langle\psi_I|\hat{H}|\psi_I\rangle^2$. Similar results were obtained for quenches in the spinless fermion chain \cite{santos_polkovnikov_11} and in other many-body Hamiltonians~\cite{neuenhahn_marquardt_12,santos_borgonovi_12a, santos_borgonovi_12b}.

For states that are eigenstates of some Hamiltonian $\hat{H}_I$ and are decomposed in the eigenstates of a new Hamiltonian $\hat{H}_F$ (as we did above), the normalized energy distribution is also known as the strength function \cite{casati_chirikov_93, casati_chirikov_96, flambaum_izrailev_00, santos_borgonovi_12a, santos_borgonovi_12b}. If $\hat{H}_F$ is taken to be $\hat{H}_I$ plus a perturbation, it has been shown that the normalized energy distribution (the strength function) evolves from a Breit-Wigner form to a Gaussian form as the strength of the perturbation is increased \cite{flambaum_izrailev_00, santos_borgonovi_12a, santos_borgonovi_12b}. The transition between those distributions, say, for a given state $|i\rangle$ (an eigenstate of $\hat{H}_I$), has been argued to occur as the average value of the nonzero off-diagonal matrix elements $|\langle i|\hat{H}_F|j\rangle|$ becomes of the same order of (or larger than) the average level spacing of the states $|j\rangle$ (also eigenstates of $\hat{H}_I$) for which $|\langle i|\hat{H}_F|j\rangle|\neq0$ \cite{santos_borgonovi_12a}. Hence, if $\hat{H}_F$ is quantum chaotic one expects that, provided that the system is large enough (the density of states increases exponentially fast with increasing system size), the normalized energy distribution will have a smooth Gaussian form after a quench independent of the nature of $\hat{H}_I$. It has been recently shown that this is not the case in generic (experimentally relevant) quenches to integrability. Namely, if $\hat{H}_F$ is integrable, the distribution of $p_m$ [see Eq.~\eqref{eq:overlapsquared}] ends up being sparse even in quenches whose initial states are thermal states of nonintegrable Hamiltonians \cite{rigol_16} (see Sec.~\ref{sec:XXZ}). The sparseness of $p_m$ at integrability is apparent in Fig.~\ref{fig:santos} and, for much larger system sizes, has also been explicitly shown in Refs.~\cite{cassidy_clark_11, he_santos_13, vidmar_rigol_16}.

To quantify the level of delocalization one can use the von Neumann entropy of the diagonal ensemble,\footnote{One could also use the inverse participation ratio: ${\rm IPR}=(\sum_m p_m^2)^{-1}$, with $p_m$ defined as in Eq.~\eqref{eq:overlapsquared}. ${\rm IPR}=1$ if only one state is occupied and it is maximized when all states are occupied with equal probability: ${\rm IPR}=\mathcal D$, where $\mathcal D$ is the dimension of the available Hilbert space.} which is known as the diagonal entropy~\cite{polkovnikov_11, santos_polkovnikov_11}:
\be
S_d=-\sum_m p_m \ln p_m.
\label{eq:Sd}
\ee
This entropy is the same as the information entropy of the initial state in the basis of the Hamiltonian governing the evolution, which we analyzed in Fig.~\ref{fig:sensitivity}. In particular, the entropy plotted in that figure is the diagonal entropy of the eigenstates of the Hamiltonian $H(J'+\delta J', V'+\delta V')$ in the basis of the Hamiltonian $H(J',V')$, averaged over eigenstates. 

\begin{figure}[!t]
\begin{center}
 \includegraphics[width=0.95\textwidth]{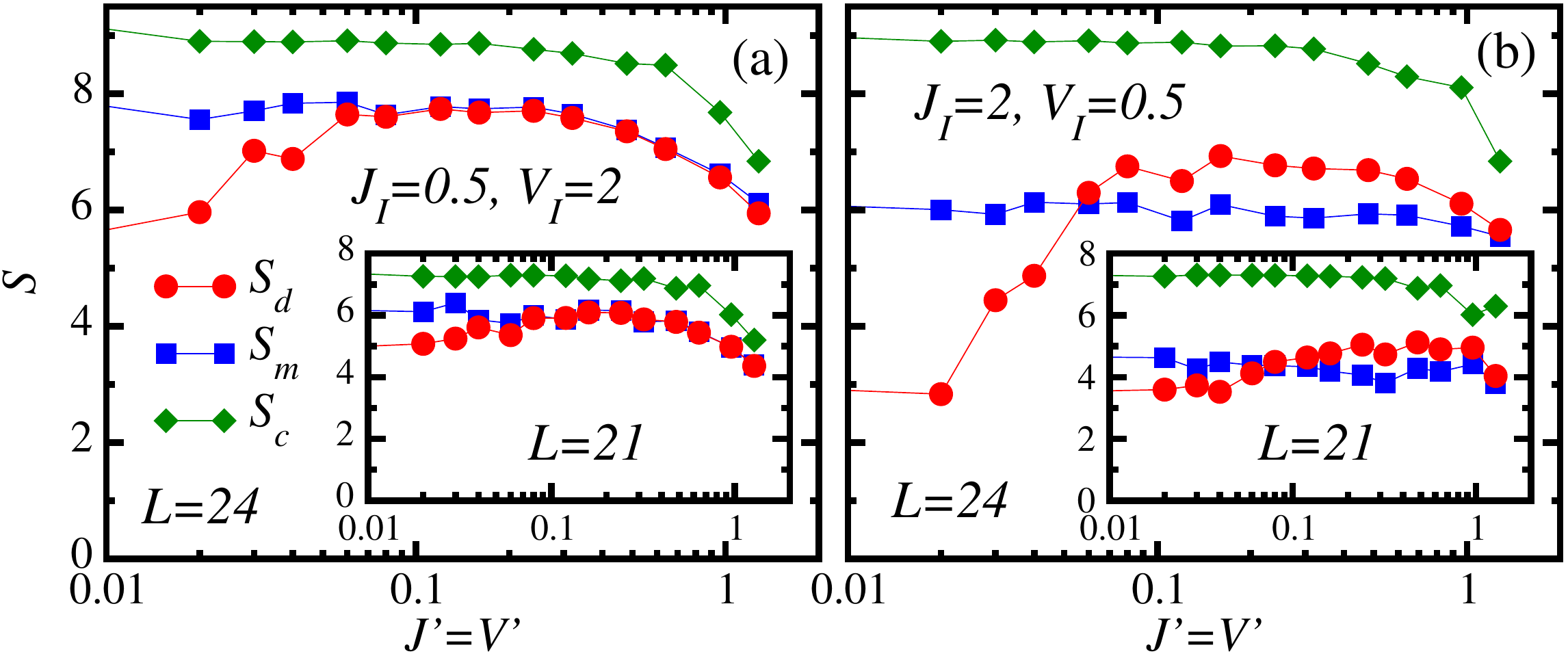}
\end{center}
\vspace{-0.3cm}
\caption{Entropies in hard-core boson chains with $L=24$ (main panels) and $L=21$ (insets). Results are reported for the diagonal entropy ($S_d$), the microcanonical entropy ($S_m$), and the canonical entropy ($S_c$), after quenches as those described in Fig.~\ref{fig:santos}, for initial states corresponding to eigenstates of Hamiltonian~\eqref{eq:HCBHam} with $J_I=0.5$, $V_I=2$ (a) and $J_I=2$, $V_I=0.5$ (b). The final parameters of the Hamiltonian are $J=V=1$. $J'=V'$ remain unchanged in the quench and their values are depicted in the $x$-axis. Thin lines joining the points are drawn to guide the eye. See also Ref.~\cite{santos_polkovnikov_11}.}
\label{fig:santos2}
\end{figure}

In Fig.~\ref{fig:santos2}, we plot the diagonal entropy vs $J'=V'$ for quenches as those in Fig.~\ref{fig:santos}, and systems with $L=21$ and $L=24$. We also plot the microcanonical entropy $S_m=\ln [\Omega(\bar E)\delta E]$, and the canonical entropy $S_c=-{\rm Tr}[\hat{\rho}_\text{CE}\ln\hat{\rho}_\text{CE}]$, where $\hat{\rho}_\text{CE}=\exp(-\beta\hat{H})/{\rm Tr}[\exp(-\beta\hat{H})]$ and the inverse temperature $\beta$ is fixed such that ${\rm Tr}(\hat{H}\hat{\rho}_\text{CE})=\overline{E}$. In all the quenches in Figs.~\ref{fig:santos} and~\ref{fig:santos2}, the initial state was selected such that $\beta^{-1}\approx3J$. Figure~\ref{fig:santos2} shows that, as one departs from the integrable point ($J'=V'=0$), the diagonal entropy becomes almost the same as the microcanonical entropy. This is up to finite-size effects, whose relevance to the results presented is made apparent by the differences between $S_m$ and $S_c$. As the system size increases, $S_m/L$ and $S_c/L$ must approach each other as they are equal up to subextensive corrections. Numerical evidence that the diagonal entropy after a quench and the thermal equilibrium entropy are identical in nonintegrable systems in the thermodynamic limit has been obtained in numerical linked cluster expansion studies~\cite{rigol_14a,rigol_16}. We will explain why $S_d/L$ agrees with the entropy in thermal equilibrium in Sec.~\ref{sec:fundamental_relation}, when we discuss the fundamental thermodynamic relation. On the other hand, in many numerical and analytical works it has been established that, in integrable systems, extensive differences generally occur between the diagonal entropy and the thermal entropy \cite{santos_polkovnikov_11, rigol_fitzpatrick_11, he_rigol_12, gurarie_13, kormos_bucciantini_14, collura_kormos_14, rigol_14a, rigol_16} (see Sec.~\ref{sec:XXZ}). In the spirit of our current discussion, this implies that integrable systems generally remain more localized in energy space.

%%%%%%%%%%%%%%%%%%%%%%%%%%%%%%%%%%%%%%%%%%%%%%%%%%%%%%%%%%%%%%%%%%%%%%%%%%%%%%%%%%%%%%%%%%
\section{Eigenstate Thermalization}\label{sec:sec4}
%%%%%%%%%%%%%%%%%%%%%%%%%%%%%%%%%%%%%%%%%%%%%%%%%%%%%%%%%%%%%%%%%%%%%%%%%%%%%%%%%%%%%%%%%%

\subsection{Thermalization in Quantum Systems}

In 1929, von Neumann wrote a remarkable paper in which he discussed how statistical mechanics behavior could emerge in quantum-mechanical systems evolving under unitary dynamics \cite{vonneumann_29}. As mentioned in the Introduction, one of von Neumann's crucial insights was to focus on macroscopic observables, as opposed to focusing on the wave function or the density matrix of the entire system. He proved what he named the quantum ergodic theorem, which has been recently discussed in detail by Goldstein et al.~in Ref.~\cite{goldstein_lebowitz_10}. In the words of the latter authors, the quantum ergodic theorem (or ``normal typicality'')\footnote{Not to be confused with canonical typicality \cite{tasaki_98,goldstein_lebowitz_06,popescu_06}, which makes statements about the reduced density matrix of typical states in the microcanonical energy shell. As discussed in the introduction, experimental out-of-equilibrium states are atypical as a result of the way they are created. Canonical typicality does not tell how typical states can be reached.} states that ``{\it for a typical finite family of commuting macroscopic observables, every initial wave function from a microcanonical energy shell evolves so that for most times in the long run, the joint probability distribution of these observables obtained from the unitarily time-evolved wave function is close to their microcanonical distribution}''. This theorem was a very important first step in the study of thermalization in quantum systems. However, some shortcomings are immediately apparent given our discussion so far. For example, the theorem makes no distinction between integrable and nonintegrable systems, as such, it leaves one wondering about the role of integrability. Also, typical observables in von Neumann's sense need not be relevant to experiments. As we discuss in Sec.~\ref{sec:normaltypical}, von Neumann's theorem is related to RMT. Hidden in it is the seed for eigenstate thermalization \cite{rigol_srednicki_12}, which is the topic of this section.

In the spirit of von Neumann's theorem, in this review thermalization refers to observables and is defined in a strong sense. Suppose that one prepares an isolated system in a nonstationary state with a well-defined mean energy, and subextensive energy fluctuations. An observable is said to thermalize if, during the time evolution of the system, it relaxes to the microcanonical prediction and remains close to it at most later times. Whether the isolated system is in a pure or mixed state is immaterial to the question of thermalization.

To understand the essential ingredients needed for thermalization to occur, let us consider a simple setup in which an isolated system is initially prepared in a pure state $|\psi_I\ra$\footnote{Everything we discuss in what follows can be straightforwardly generalized to mixed states.} and evolves under a time-independent Hamiltonian $\hat{H}$. We assume that the Hamiltonian has eigenvectors $|m\ra$ and eigenvalues $E_m$, that is, $\hat{ H}|m\ra=E_m|m\ra$. The time-evolving wave function can be written as
\be \label{eq:wftime}
|\psi(t)\ra=\sum_m C_m \mathrm e^{-i E_m t} |m\ra,
\ee
where $C_m=\la m | \psi_I \ra$ (notice that we set $\hbar\rightarrow 1$); and we are interested in $t\geq 0$. Obviously, the density matrix of the system will remain that of a pure state at all times [$\rho(t)^2=\rho(t)$], that is, it can never become a mixed (thermal) density matrix. Now, let us look at the time evolution of some observable $\hat{O}$, which in the basis of the eigenstates of the Hamiltonian can be written as
\beq
O(t)&\equiv&\la \psi(t)|\hat{O}|\psi(t) \ra
=\sum_{m,\,n} C_{m}^* C_{n}^{}
e^{i(E_{m}-E_{n})t} O_{mn}\nonumber\\
&=&\sum_{m} |C_{m}^{}|^2 O_{mm} 
+\sum_{m,\,n\neq m} C_{m}^* C_{n}^{}
e^{i(E_{m}-E_{n})t} O_{mn}\,
\label{eq:timeevolution}
\eeq
where $O_{mn}=\la m|\hat{O}|n \ra$. As stated before, we say that the observable $\hat O$ thermalizes if: (i) after some relaxation time, the average expectation value of this observable agrees with the microcanonical expectation value and (ii) temporal fluctuations of the expectation value about the microcanonical prediction are small at most later times. This implies that the long-time average accurately describes the expectation value of $\hat O$ at almost all times and agrees with the microcanonical prediction. 

The initial difficulties in reconciling these requirements with Eq.~(\ref{eq:timeevolution}) are obvious. In the long-time average, the second sum in Eq.~\eqref{eq:timeevolution} averages to zero (provided there are no degeneracies, or that there is a nonextensive number of them) and we are left with the sum of the diagonal elements of $\hat O$ weighted by $|C_m^{}|^2$. Some of the natural questions one can ask are: (i) Since the probabilities $|C_m|^2$ are conserved in time, how is it possible for $\sum_{m} |C_{m}^{}|^2 O_{mm}$ to agree with the microcanonical average? (ii) Moreover, in many-body systems, the eigenenergies are exponentially close to each other and therefore, to make sure that the second sum in Eq.~(\ref{eq:timeevolution}) averages to zero, one could potentially need to wait an exponentially (in system size) long time. Such a time, even for moderately small systems, could exceed the age of our universe, and therefore cannot be reconciled with the experimental observation that even large systems thermalize over much shorter time scales than the age of the universe (we observe them thermalize). 

Remarkably, if the Hamiltonian $\hat H$ was a true random matrix, then using the RMT prediction for observables [namely that $O_{mm}$ is independent of $m$ and that $O_{mn}$ for $m\neq n$ is exponentially small in the system size, see Eq.~\eqref{eth_rmt}] one finds that the observables thermalize in the sense specified above. This is because the first sum in Eq.~(\ref{eq:timeevolution}) becomes independent of the initial state
\be
\sum_{m} |C_{m}^{}|^2 O_{mm}\approx \bar O \sum_{m} |C_{m}^{}|^2=\bar O,
\ee
that is, it agrees with the microcanonical result. Note that within RMT, the microcanonical ensemble has no energy dependence and is thus formally equivalent to the infinite temperature ensemble.  It also becomes clear that exponentially long times may not be needed for relaxation. The off-diagonal matrix elements of $\hat O$ are exponentially small so, by destroying phase coherence between a finite fraction of the eigenstates with a significant contribution to the expectation value, it is possible to approach the infinite-time prediction with high accuracy in a time much shorter than the inverse (many-body) level spacing, which is required to destroy coherence between \textit{all} eigenstates. We will come back to this later. The relevance of RMT for understanding thermalization in many-body quantum systems was discussed by Deutsch in a seminal paper in the early 1990s \cite{deutsch_91}. There, he essentially extended Berry's conjecture to arbitrary quantum systems assuming that the eigenstates of ergodic Hamiltonians are essentially uncorrelated random vectors.

In order to describe observables in experiments, however, one needs to go beyond the RMT prediction. This because, in contrast to random matrices, in real systems: (i) thermal expectation values of observables depend on the energy density (temperature) of the system\footnote{For simplicity, we assume that the energy is the only conserved quantity in the system. If there are other conserved quantities, they have to be treated in a similar fashion.} and (ii) relaxation times are observable dependent. Hence, there is information in the diagonal and off-diagonal matrix elements of observables in real systems that cannot be found in RMT. In groundbreaking works throughout the 1990s, Srednicki provided the generalization of the RMT prediction that is needed to describe observables in physical systems \cite{srednicki_94,srednicki_96,srednicki_99}. Srednicki's ansatz is known as the ETH. It was first shown to apply to realistic quantum systems, where thermalization was observed for a strikingly small number of particles (5 bosons in 21 lattice sites), by Rigol et al.~\cite{rigol_dunjko_08}. We should mention that, in a remarkable discussion of numerical experiments with 7 spins, Jensen and Shankar \cite{jensen_shankar_85} advanced part of ETH [the first term on the RHS of Eq.~\eqref{eq:ETH}]. The smallness of the system they studied precluded them from observing a qualitatively different behavior between nonintegrable and integrable systems.

\subsection{The Eigenstate Thermalization Hypothesis (ETH)}
\label{sec:eth_def}

ETH can be formulated as an ansatz for the matrix elements of observables in the basis of the eigenstates of a Hamiltonian~\cite{srednicki_99}:
\be
O_{mn}=O\left(\bar E\right)\delta_{mn}+e^{-S\left(\bar E\right)/2}f_O\left(\bar E,\omega\right)R_{mn},
\label{eq:ETH}
\ee
where $\bar E\equiv(E_{m}+E_{n})/2$, $\omega\equiv E_{n}-E_{m}$, and $S(E)$ is the thermodynamic entropy at energy $E$. Crucially, $O\left(\bar E\right)$ and $f_O\left(\bar E,\omega\right)$ are smooth functions of their arguments, the value  $O\left(\bar E\right)$ is identical to the expectation value of the microcanonical ensemble at energy $\bar E$ and $R_{mn}$ is a random real or complex variable with zero mean and unit variance ($\overline{R_{mn}^2}=1$ or $\overline{|R_{mn}|^2}=1$, respectively). While there is no rigorous understanding of which observables satisfy ETH and which do not, it is generally expected that Eq.~\eqref{eq:ETH} holds for all physical observables, namely, observables for which statistical mechanics applies (see, e.g., discussion in Ref.~\cite{landau_lifshitz_5_80}). Specifically, ETH has been numerically verified for few-body observables in a variety of lattice models, no matter whether they are local or not (see Sec.~\ref{ss:ethlattice}). By few-body observables we mean $n$-body observables with $n\ll N$, where $N$ is the number of particles, spins, etc, in the system. This is the class of observables that can be experimentally studied in macroscopic systems. Projection operators to the eigenstates of the many-body Hamiltonian, $\hat P_m=|m\rangle\langle m|$, are operators for which Eq.~\eqref{eq:ETH}, as well as the predictions of statistical mechanics, do not hold. In a recent study of lattice systems, Garrison and Grover argued that ETH can hold for observables with support in up to 1/2 of the system size \cite{garrison_grover_15}.

The matrix elements of observables can be real or complex depending on the symmetries of the Hamiltonian and the basis used to diagonalize it. If the system obeys time-reversal symmetry, the eigenstates of the Hamiltonian can be chosen to be real and so will be the matrix elements of observables (Hermitian operators). This is not possible if the system does not obey time-reversal symmetry. By taking the Hermitian conjugate of Eq.~(\ref{eq:ETH}), we see that the function $f_O(\bar E,\omega)$ and the random numbers $R_{m n}$ must satisfy the following relations
\be
\begin{array}{ccc}
R_{nm}=R_{mn}, &\quad f_O(\bar E,-\omega)=f_O(\bar E,\omega) &\quad \text{(real matrix elements)}\\
R_{nm}^\ast=R_{mn}, &\quad f_O^\ast(\bar E,-\omega)=f_O(\bar E,\omega) &\quad \text{(complex matrix elements)}.
\end{array}
\label{eq:symmetry_f}
\ee

Srednicki's ansatz \eqref{eq:ETH} is similar to the RMT result in Eq.~(\ref{eth_rmt}). The differences are: (i) The diagonal matrix elements of observables $O(\bar E)$ are not the same in all eigenstates. Rather, they are smooth functions of the energy of the eigenstates. (ii) For the off-diagonal matrix elements, on top of the small Gaussian fluctuations, there is an envelope function $f_O(\bar E,\omega)$ that depends on the mean energy and the energy difference between the eigenstates involved. This ansatz is consistent with results obtained in the semi-classical limit of quantum systems whose classical counterpart is chaotic~\cite{feingold_peres_86, wilkinson_87, deutsch_91, srednicki_94, prosen_94, eckhardt_fishman_95, hortikar_srednicki_98}. 

The ETH ansatz reduces to the RMT prediction if one focuses on a very narrow energy window where the function $f_O(\bar E,\omega)$ is constant. In single-particle diffusive systems, this scale is given by the Thouless energy (see, e.g., Ref.~\cite{kota_14}), which is essentially equal to Planck's constant divided by the diffusion time~\cite{akkermans_montambaux_07}:
\be
E_T={\hbar {D}\over L^2},
\label{eq:Thouless}
\ee
where ${D}$ is the diffusion constant and $L$ is the linear size of the system. As we discuss in Sec.~\ref{ss:ethlattice}, the same appears to be true in generic diffusive many-body quantum systems. Namely, that if one focuses on an energy shell of width $\omega<E_T$ then $f_O(E,\omega)\approx {\rm const}$, so that the ETH ansatz is identical to RMT. In other words, there is no structure in the eigenstates of ergodic Hamiltonians in an energy window narrower than the Thouless energy. As this window vanishes in the thermodynamic limit, RMT has a very limited range of applicability. Note, however, that the level spacing vanishes much faster with the system size. Therefore, there is still an exponentially large number of energy levels in the region where RMT applies. The situation can be more subtle in systems with subdiffusive, for example, glassy dynamics. One can anticipate that $f(\omega)$ will saturate at $\omega<\hbar/\tau^\ast$, where $\tau^\ast$ is the slowest physical time scale in the system. As long as the corresponding energy window contains exponentially many energy levels, one expects that RMT will apply in this window.\footnote{In a very recent work by Luitz and Bar Lev [arXiv:1607.01012], the Gaussian ansatz for the off-diagonal matrix elements was found to hold for diffusive spin chains. At the same time it was found that, in sub-diffusive disordered chains, the distribution of off-diagonal matrix elements becomes non-Gaussian (it acquires long tails).} Conversely, the ETH ansatz does not have these RMT limitations and is believed to apply to arbitrary energies with the exception of the edges of the spectrum.  As we will see later, in Sec.~\ref{sec:fluctuation_dissipation}, the dependence of $f_O(E,\omega)$ on $\omega$  determines the decay of nonequal-time correlation functions. It also determines the relaxation time following a small perturbation about equilibrium (in the linear response regime) \cite{srednicki_99,khatami_pupillo_13}. In ergodic systems, it is expected that the diffusive time gives the slowest time scale in the system~\cite{lux_muller_14}. Within ETH, this follows from the fact that the function $f_O(E,\omega)$ becomes structureless (constant) for $\omega<E_T$.

\subsubsection{ETH and Thermalization}\label{sec:eth_thermalization}

The ETH ansatz (\ref{eq:ETH}) has immediate implications for understanding thermalization in many-body quantum systems. First, let us focus on the long-time average of observables. If there are no degeneracies in the energy spectrum, which is a reasonable assumption for generic quantum systems after removing all trivial symmetries, we obtain [using Eq.~\eqref{eq:timeevolution}]
\be\label{eq:mean&fluct}
\overline{O}\equiv \lim_{t_0\to\infty} {1\over t_0}\int_0^{t_0} dt\, O(t)
=\sum_{m} |C_{m}^{}|^2 O_{mm}={\rm Tr}\left[\hat{\rho}_{\text{DE}}\hat O\right], 
\ee
where $\rho_{\rm DE}$ is the density matrix of the diagonal ensemble, defined in Eq.~\eqref{eq:denmatDE}. On the other hand, statistical mechanics predicts
\be\label{eq:statmechlo}
O_\text{ME}={\rm Tr}\left[\hat{\rho}_{\text{ME}}\hat O\right],
\ee
where $\hat{\rho}_{\text{ME}}$ is the density matrix of the microcanonical ensemble (due to ensemble equivalence one can, of course, use a canonical, or any other equilibrium, density matrix instead). We then see that, independent of the actual values of $C_{m}^{}$, so long as energy fluctuations in the diagonal ensemble
\be
\delta E\equiv\sqrt{\langle\psi_I|\hat{H}^2|\psi_I\rangle-\langle\psi_I|\hat{ H}|\psi_I\rangle^2}
\ee
are sufficiently small (e.g., behaving like in traditional statistical mechanics ensembles), $\overline{O}$ will agree (to leading order) with the statistical mechanics prediction $O_\text{ME}$, provided that ${\rm Tr} \left[\hat{\rho}_{\text{ME}}\hat H\right]=\langle\psi_I|\hat{ H}|\psi_I\rangle\equiv\la E\ra$. This is because, using the ETH ansatz~\eqref{eq:ETH}, one can rewrite Eqs.~\eqref{eq:mean&fluct} and \eqref{eq:statmechlo} as
\be \label{eq:deeqsm}
\overline{O}\simeq O(\la E \ra)\simeq O_\text{ME}.
\ee
Furthermore, given Eq.~\eqref{eq:ETH}, one can quantify the difference between the two ensembles due to the fact that $\delta E$ is finite. Indeed, expanding the smooth function $O(E)$ into a Taylor series around the mean energy $\langle E\rangle$
\be
O_{mm}\approx O(\la E \ra) + (E_m-\la E \ra) \left.\frac{d O}{dE}\right|_{\la E \ra}
+{1\over 2}(E_m-\la E \ra)^2 \left.\frac{d^2 O}{dE^2}\right|_{\la E \ra},
\ee
and substituting this expansion into Eq.~\eqref{eq:mean&fluct}, we find
\be
\overline{O}\approx O(\la E \ra)+{1\over 2}(\delta E)^2 O''(\la E \ra)\approx O_{\rm ME}+
{1\over 2}\left[(\delta E)^2-(\delta E_\text{ME})^2\right] O''(\la E \ra),
\label{eq:barO_eth}
\ee
where $\delta E_\text{ME}$ are the energy fluctuations of the microcanonical ensemble, which are subextensive. If the energy fluctuations $\delta E$ in the time-evolving system are subextensive, which is generically the case in systems described by local Hamiltonians (see, e.g., the discussion in Sec.~\ref{ss:quencheth}), then the second term is a small subextensive correction to $O_{\rm ME}$, which is negligible for large system sizes. Moreover, the same Eq.~\eqref{eq:barO_eth} describes the difference between the equilibrium canonical and microcanonical expectation values of $\hat O$ if instead of $\delta E^2$ one uses energy fluctuations of the canonical ensemble.  It is remarkable that, using ETH, one can show that $\overline{O}\simeq O_\text{ME}$ without the need of making any assumption about the distribution of $C_{m}^{}$, beyond the fact that it is narrow. This is to be contrasted with the standard statistical mechanics statement about equivalence of ensembles, for which it is essential that the energy distributions are smooth functions of the energy.\footnote{The energy distributions after quenches to nonintegrable Hamiltonians are expected to be smooth, see Sec.~\ref{sec:chaos_delocalization}. However, because of ETH, thermalization in nonintegrable systems occurs independently of whether the energy distributions are smooth or not.}

Using the ETH ansatz, one can also calculate the long-time average of the temporal fluctuations of the expectation value of the observable $\hat O$
\beq\label{eq:fluct_expect_values}
\sigma_O^2&\equiv&\lim_{t_0\to \infty}{1\over t_0} \int_0^{t_0} dt \,[O(t)]^2\,-(\overline{O})^2\nonumber\\
&=& \lim_{t_0\to \infty} {1\over t_0}\int_0^{t_0} dt \sum_{m,n,p,q} O_{mn} O_{pq} 
C_m^\ast C_n C_p^\ast C_q \mathrm e^{i (E_m-E_n+E_p-E_q)t}
-(\overline{O})^2 \label{eq:sigma_O2} \label{eq:sigma_O} \\
&=& \sum_{m,n\neq m} |C_{m}^{}|^2 |C_{n}^{}|^2 |O_{mn}|^2
\leq \max|O_{mn}|^2\sum_{m,n} |C_{m}^{}|^2 |C_{n}^{}|^2= 
\max|O_{mn}|^2\propto \exp[-S(\bar E)].\nonumber
\eeq
Thus, the time fluctuations of the expectation value of the observable are exponentially small in the system size. These fluctuations should not be confused with the fluctuations of the observable that are actually measured in experiments, which are never exponentially small \cite{srednicki_99,rigol_dunjko_08}.  Instead, Eq.~\eqref{eq:sigma_O} tells us that at almost any point in time the expectation value of an observable $\hat O$ is the same as its diagonal ensemble expectation value. Thus the ETH ansatz implies ergodicity in the strong sense, that is, no time averaging is needed. In Sec.~\ref{sec:fluctuation_dissipation}, we show that ETH implies that temporal fluctuations of extensive observables satisfy standard fluctuation-dissipation relations. Let us point in passing that the results discussed so far are not restricted to pure states. They all straightforwardly generalize to mixed states by using the following substitutions in Eqs.~\eqref{eq:timeevolution}, \eqref{eq:mean&fluct}, and \eqref{eq:denmatDE}: $C_{m}^\ast C_{n}^{}\rightarrow \rho_{mn}$ and $|C_{m}^{}|^2\rightarrow \rho_{mm}$, where $\rho_{mn}$ are the matrix elements of the initial density matrix in the basis of the eigenstates of the Hamiltonian.

To contrast Eq.~\eqref{eq:fluct_expect_values} with the fluctuations of $\hat{O}$ seen in experiments, let us also show the expression for the latter:
\be
\overline{\delta O^2} = \lim_{t_0\to\infty} {1\over t_0}\int_0^{t_0} dt\, \la \psi(t) 
|(\hat{O}-\overline{O})^2|\psi(t)\ra=\sum_{m} |C_{m}^{}|^2 (O^2)_{mm}-\overline{O}^2.
\label{eq:fluct_O_DE}
\ee
This quantity is nonzero even if the initial state is an eigenstate of the Hamiltonian ($|\psi_I\ra=|m\ra$), while $\sigma_O$ is zero in that case. Assuming that $\delta E$ is sufficiently small, and using the ETH ansatz for $\hat{O}^2$, we find
\be\label{eq:flucethmic}
\overline{\delta O^2} \approx \delta O^2_{\rm ME}+{1\over 2}\left[{O^2}''(\la E \ra)-2O(\la E \ra)O''(\la E \ra)\right] \left[\delta E^2-(\delta E_\text{ME})^2\right].
\ee
And we see that the fluctuations of $\hat{O}$ scale as the equilibrium statistical fluctuations of $\hat O$. However, in this case, there is a second term which can be of the same order. We note that Eq.~\eqref{eq:flucethmic} describes the difference between the canonical and microcanonical fluctuations of $\hat O$ if instead of $\delta E^2$ one uses energy fluctuations of the canonical ensemble, that is, same order corrections to fluctuations also occur in equilibrium statistical mechanics. In generic cases, for example, for extensive observables in systems away from critical points, $\sqrt{\overline{\delta O^2}}/\overline{O}\simeq \sqrt{\delta O^2_{\rm ME}}/O_\text{ME}\simeq 1/\sqrt{V}$, where $V$ is the volume of the system.

An important question that we leave unaddressed here is that of relaxation times. Namely, how long it takes for an observable to reach the diagonal ensemble result. The answer to this question depends on the observable, the initial state selected, and the specifics of the Hamiltonian driving the dynamics. As we will show when discussing results from numerical experiments, the relaxation times of observables of interest in lattice systems are not exponentially large. They actually need not even increase with increasing system size.

Summarizing our discussion so far, we see that the language used to describe thermalization in isolated quantum systems is quite different from that in classical systems. Chaos, ergodicity, and thermalization are hidden in the nature of the Hamiltonian eigenstates. Relaxation of observables to their equilibrium values is nothing but the result of dephasing, as follows from the second term (in the last line) in Eq.~\eqref{eq:timeevolution}. Thus, the information about the eventual thermal state is encoded in the system from the very beginning, the time  evolution simply reveals it. In classical systems, one usually thinks of thermalization in very different terms using the language of particle collisions and energy redistribution between different degrees of freedom. It is important to realize that both approaches describe exactly the same processes. In Sec.~\ref{sec:kinetic_equations}, we will briefly discuss how one can understand relaxation in weakly nonintegrable quantum systems through the language of quantum kinetic equations. Kinetic equations, when justified, provide a unified framework to describe relaxation in both quantum and classical systems.

\subsubsection{ETH and the Quantum Ergodic Theorem}\label{sec:normaltypical}

Now that we have formulated ETH and seen its consequences for the dynamics of isolated quantum systems, let us come back to von Neumann's ergodic theorem and discuss how it relates to ETH (or, more precisely, to RMT)~\cite{rigol_srednicki_12}. As said before, von Neumann was interested in understanding what happens to observables during the unitary time evolution of {\it all} possible states drawn from the microcanonical shell. His theorem was then about the behavior of typical observables at most times. To state it, we follow the discussion by Goldstein {\it et al.} in Ref.~\cite{goldstein_lebowitz_10}. For a recent generalization of this theorem, see Ref.~\cite{reimann_15}. 

von Neumann considered a Hamiltonian $\hat{H}$ with eigenstates $|m\ra$ and eigenvalues $E_m$, that is, $\hat H |m\ra = E_m |m\ra$, and focused on a microcanonical energy window of width $\delta E$ around an energy $E$. This microcanonical energy window defines a Hilbert space $\mathscr{H}$ of dimension $\mathcal D$, which is spanned by $\mathcal D$ energy eigenstates $|m\ra$ with energies $E_m\in(E-\delta E/2,E+\delta E/2)$. For example, every state in the microcanonical energy window can be decomposed as $|\psi\ra=\sum_{m\in \mathscr{H}} C_m |m\ra$, where $C_m=\la m|\psi\ra$. The Hilbert space $\mathscr{H}$ is then decomposed into mutually orthogonal subspaces $\mathscr{H}_\nu$ of dimensions $d_\nu$, such that $\mathscr{H}=\bigoplus_\nu \mathscr{H}_\nu$ and $\mathcal D=\sum_\nu d_\nu$. Finally, the observables in $\mathscr{H}$ are written as $\hat{O}=\sum_\nu O_\nu \hat P_\nu$, where $\hat P_\nu$ is the projector onto $\mathscr{H}_\nu$. Here, both $\mathcal D$ and $d_\nu$ are assumed to be large. By definition, the expectation value of the observable at time $t$ is $O(t)=\la\psi|\exp[i\hat{H}t]\hat{O} \exp[-i\hat{H}t]|\psi\ra$ while its microcanonical average is $\la \hat{O}\ra_\text{ME}=\sum_{m \in \mathscr{H}}\la m|\hat{O}|m\ra/\mathcal D$. von Neumann's \textit{quantum ergodic theorem} states that: In the absence of resonances in $\hat{H}$, namely, if $E_m-E_n\neq E_m'-E_n'$ unless $m=m'$ and $n=n'$,  or $m=n$ and $m'=n'$, and provided that, for any $\nu$,  
\be\label{eq:GLTZ}
\max_m \left(\la m|\hat{P}_\nu|m\ra-\frac{d_\nu}{\mathcal D}\right)^2
+ \max_{m\neq n}|\la m |\hat{P}_\nu|n\ra|^2 
\text{ is exponentially small,} 
\ee
then
\be
|O(t)-\la \hat{O}\ra_\text{ME}|^2 < \epsilon\la\hat{O}^2\ra_\text{ME}
\ee
for all but a fraction $\delta$ of times $t$, where $\epsilon$ and $\delta$ are small numbers. It is easy to see that condition~\eqref{eq:GLTZ} guarantees that the eigenstate expectation value of $\hat{O}$ is identical to the microcanonical prediction~\cite{rigol_srednicki_12}. In fact:
\beq\label{eq:vN01}
\la m|\hat{O}|m\ra&=&\sum_{\nu} O_\nu \la m| \hat{P}_\nu |m\ra\approx \sum_\nu O_\nu \frac{d_\nu}{\mathcal D}\nonumber\\&=& \sum_{m\in \mathscr{H},\nu} O_\nu \frac{\la m|\hat P_\nu|m\ra}{\mathcal D} = \sum_{m\in \mathscr{H}} \frac{\la m|\hat O |m\ra}{\mathcal D} \equiv 
\la \hat{O}\ra_\text{ME}
\eeq
where the second equality holds up to exponentially small corrections, see Eq.~\eqref{eq:GLTZ}, and we have used that $\sum_{m\in \mathscr{H}} \la m|\hat P_\nu|m\ra = d_\nu$. Next, we have that
\be\label{eq:vN02}
\la m|\hat{O}|n\ra= \sum_\nu O_\nu \la m |\hat{P}_\nu|n\ra,
\ee
which is exponentially small if $\la m |\hat{P}_\nu|n\ra$ is exponentially small [as required in Eq.~\eqref{eq:GLTZ}] and if $O_\nu$ is not exponential in system size (as expected for physical observables).

We then see that Eqs.~\eqref{eq:vN01} and \eqref{eq:vN02} are nothing but the RMT predictions summarized in Eq.~\eqref{eth_rmt}, or, equivalently, the ETH ansatz restricted to the Thouless energy window $\delta E\sim E_T=\hbar D/L^2$, where the function $f(\bar E,\omega)$ is approximately constant. Without this condition, Eq.~\eqref{eq:GLTZ} cannot be satisfied. Ultimately, Eqs.~\eqref{eth_rmt}, \eqref{eq:ETH} and \eqref{eq:GLTZ} rely on the fact that the overlap between the energy eigenstates and eigenstates of the observables is exponentially small~\cite{rigol_srednicki_12}. It is important to note that RMT provides a wealth of information about the statistics of the level spacings and of the eigenstate components, which we have connected to quantum chaotic Hamiltonians, that was absent in von Neumann's (much earlier) theorem. ETH goes beyond RMT (and the quantum ergodic theorem), as we mentioned before, because it addresses what happens outside the featureless Thouless energy shell.

\subsection{Numerical Experiments in Lattice Systems}\label{ss:lattice}

\subsubsection{Eigenstate Thermalization}\label{ss:ethlattice}

Numerical evidence of the occurrence of eigenstate thermalization has been found in a number of strongly correlated nonintegrable lattice models in fields ranging from condensed matter to ultracold quantum gases. Such an evidence was first reported for a two-dimensional system of hard-core bosons \cite{rigol_dunjko_08}, and, since then, among others, it has been reported for a variety of models of hard-core bosons and interacting spin chains \cite{rigol_09a, rigol_09b, rigol_santos_10, santos_rigol_10b, khatami_pupillo_13, steinigeweg_herbrych_13, beugeling_moessner_14, kim_14, steinigeweg_khodja_14, khodja_steinigeweg_15, beugeling_moessner_15}, spinless and spinful fermions \cite{rigol_09b, neuenhahn_marquardt_12, khatami_rigol_12, genway_ho_12}, soft-core bosons \cite{biroli_kollath_10, roux_10, beugeling_moessner_14, sorg_vidmar_14, beugeling_moessner_15}, and the transverse field Ising model in two dimensions \cite{mondaini_fratus_16}. Below, we discuss the evidence for ETH separately for the diagonal and the off-diagonal matrix elements.

\paragraph{Diagonal matrix elements}
We begin by illustrating the behavior of the diagonal matrix elements of observables in the lattice hard-core boson model in Eq.~\eqref{eq:HCBHam}, which transitions between the integrable limit and the chaotic regime as $J'=V'$ departs from zero (see Sec.~\ref{sec:chaos_delocalization})~\cite{santos_rigol_10a}. 

\begin{figure}[!t]
\includegraphics[width=0.96\textwidth]{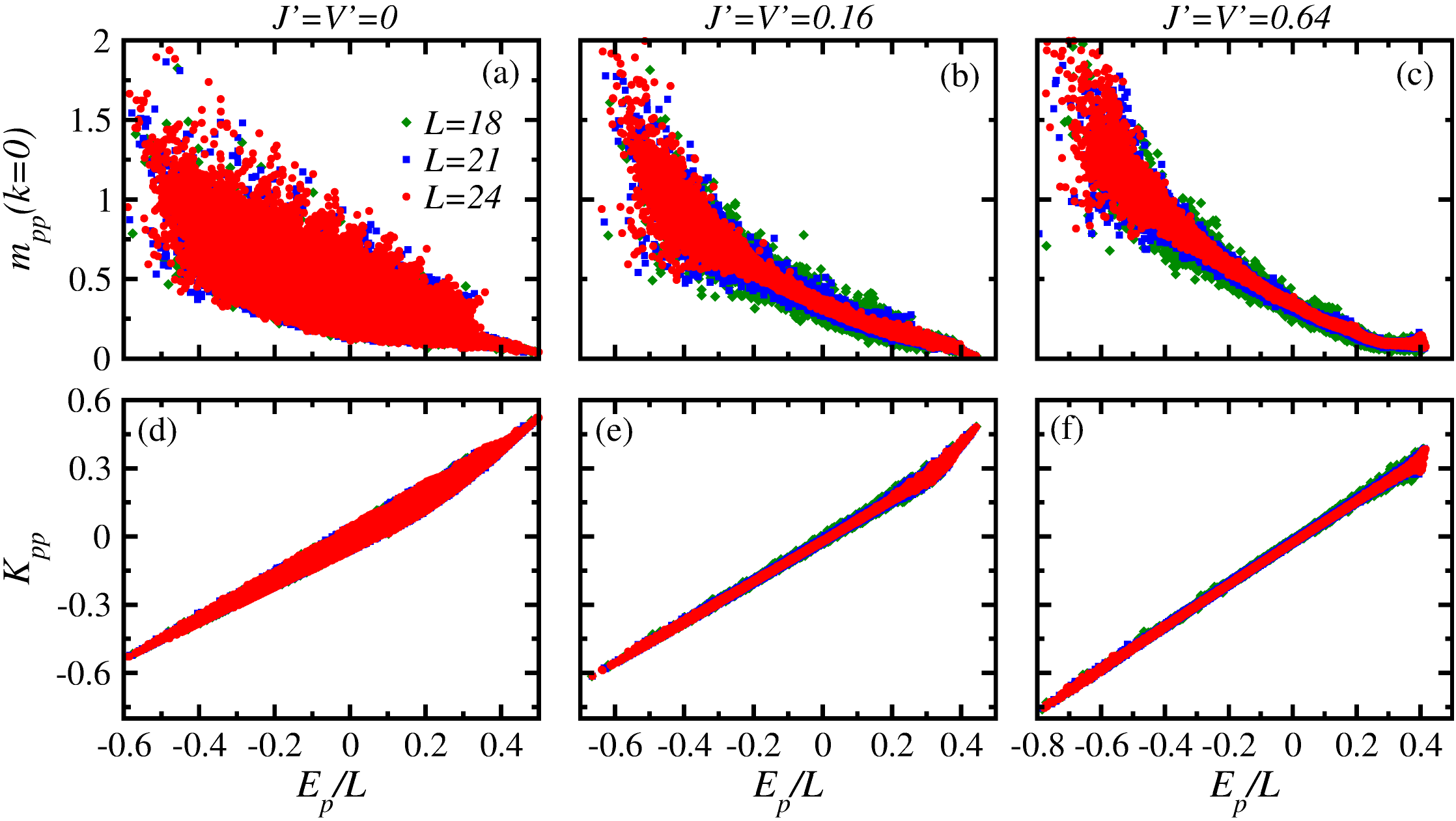}
\vspace{-0.15cm}
\caption{Eigenstate expectation values of the occupation of the zero momentum mode [(a)--(c)] and the kinetic energy per site [(d)--(f)] of hard-core bosons as a function of the energy per site of each eigenstate in the entire spectrum, that is, the results for all $k$-sectors are included. We report results for three system sizes ($L=18,$ 21, and 24), a total number of particles $N=L/3$, and for two values of $J'=V'$ [$J'=V'=0.16$ in panels (b) and (e) and $J'=V'=0.64$ in panels (c) and (f)] as one departs from the integrable point [$J'=V'=0$ in panels (a) and (d)]. In all cases $J=V=1$ (unit of energy). See also Ref.~\cite{rigol_09a}.}
\label{fig:nk_ETH}
\end{figure}

In Fig.~\eqref{fig:nk_ETH}, we show in panels (a)--(c) the energy eigenstate expectation values of the zero momentum mode occupation 
\be\label{eq:nkHCB}
\hat{m}(k)=\frac1L \sum_{i,j} e^{i k(i-j)} \hat{b}^{\dagger}_i\hat{b}^{}_j \;.
\ee
In panels (d)--(f), we show the kinetic energy per site
\be\label{eq:KHCB}
\hat{K}=\frac1L\sum_{j=1}^{L}\left[-J\left(\hat{b}_j^{\dagger} \hat{b}^{}_{j+1}+\text{H.c.}\right)
- J'\left( \hat{b}_j^{\dagger} \hat{b}^{}_{j+2} + \text{H.c.}\right)\right].
\ee
Eigenstate expectation values are plotted as a function of the eigenenergies per site ($E_p/L$), for three different system sizes as one increases $J'=V'$. The qualitative behavior of $m_{pp}(k=0)$ and $K_{pp}$ vs $E_p/L$, depicted in Fig.~\eqref{fig:nk_ETH}, has been observed in other few-body observables and models studied in the literature, and, as such, is expected to be generic. The main features to be highlighted are: (i) At integrability, $m_{pp}(k=0)$ and $K_{pp}$ can have quite different expectation values [see, particularly, $m_{pp}(k=0)$] in eigenstates of the Hamiltonian with very close energies. Moreover, the spread does not change with increasing system size and the variance (not shown) decreases as a {\it power law} of the system size. Similar results have been obtained in other integrable models for larger system sizes than those available from direct full exact diagonalization of the Hamiltonian \cite{cassidy_clark_11, he_santos_13, ikeda_watanabe_13, alba_15, vidmar_rigol_16}. (ii) As one departs from $J'=V'$, or as one increases the system size for any given value of $J'=V'\neq 0$, the spread (or maximal differences) between the eigenstate expectation values in eigenstates with very close energies decrease. This is true provided that the eigenstates are not too close to the edges of the spectrum.

\begin{figure}[!t]
\begin{center}
 \includegraphics[width=0.49\textwidth]{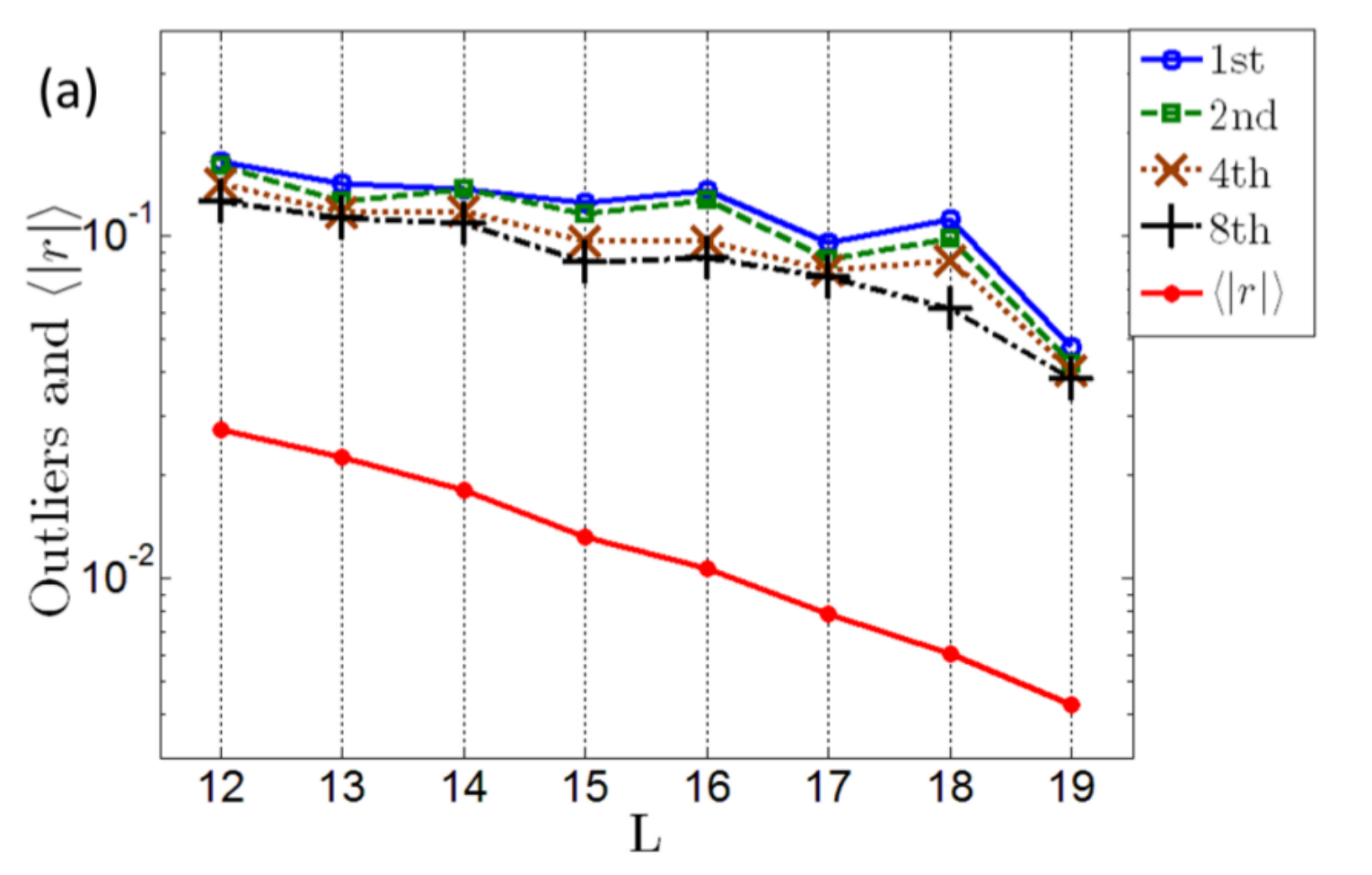}
 \includegraphics[width=0.49\textwidth]{Figures/Figure14a}
\end{center}
\vspace{-0.25cm}
\caption{The first, second, fourth, and eighth largest values of $r_p$ in the central half of the spectrum, as well as its mean value, are shown from top to bottom in both panels. (a) Results for the $x$-component of the magnetization in a nonintegrable transverse Ising Hamiltonian. (b) Results for the nearest neighbor density-density correlations in a nonintegrable hard-core boson Hamiltonian. From Ref.~\cite{kim_14}.}
\label{fig:kimhuseETH}
\end{figure}

Recently, Kim et al.~\cite{kim_14} studied the eigenstate-to-eigenstate fluctuations $r_p=O_{p+1p+1}-O_{pp}$ of both the $x$-component of the magnetization in a nonintegrable transverse Ising chain with a longitudinal field and of the nearest neighbor density-density correlations in the nonintegrable hard-core boson model \eqref{eq:HCBHam}. The results for the average value of $|r_p|$, and for some of the largest values of $r_p$ (in the central half of the spectrum), are shown in Fig.~\ref{fig:kimhuseETH} as a function of the system size. They support the ETH expectation that eigenstate-to-eigenstate fluctuations decrease exponentially fast with increasing system size (similar results were obtained in Ref.~\cite{mondaini_fratus_16} for the transverse field Ising model in two dimensions). Evidence that the variance of the eigenstate-to-eigenstate fluctuations of various observables decreases exponentially fast with increasing system size has also been presented in Refs.~\cite{steinigeweg_herbrych_13, beugeling_moessner_14, mondaini_fratus_16}.

The results discussed so far suggest that, away from the edges of the spectrum and for sufficiently large system sizes, any strength of an integrability breaking perturbation ensures that the first term in Eq.~\eqref{eq:ETH} describes the diagonal matrix elements of physical observables. By sufficiently large system sizes, we mean the same conditions that were discussed for the onset of quantum chaotic behavior in Sec.~\ref{sec:sec3}. Systems that exhibit a many-body localization transition do not conform with this expectation \cite{nandkishore_huse_14}.

\begin{figure}[!t]
\begin{center}
 \includegraphics[width=0.8\textwidth]{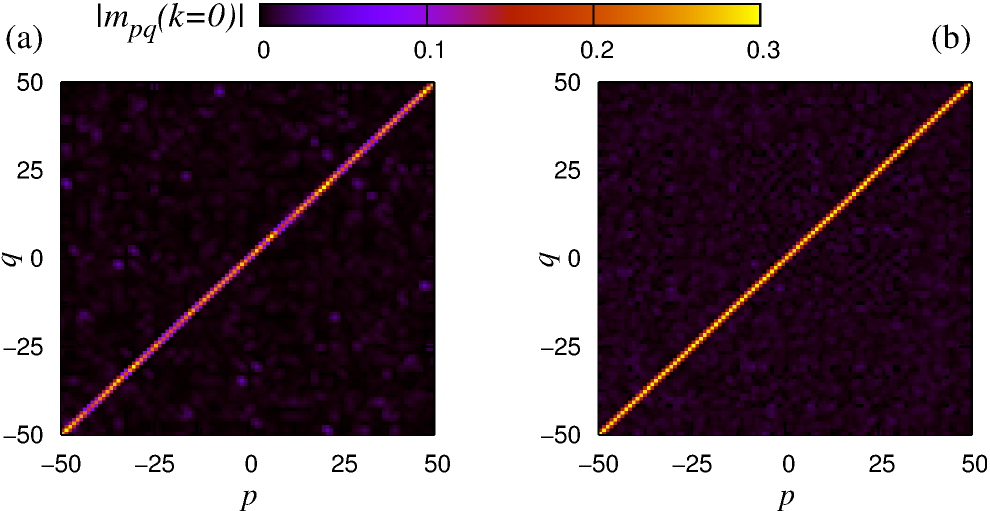}
\end{center}
\vspace{-0.15cm}
\caption{Off-diagonal matrix elements of $\hat{m}(k=0)$ in the eigenstates of the Hamiltonian for a system with $L=24$, $N=L/3$, and $J=V=1$ (unit of energy). (a) $J'=V'=0$ and (b) $J'=V'=0.32$. Results are shown for the matrix elements between the 100 eigenstates with energy closest to (a) $E/L=-0.16$ and (b) $E/L=-0.19$. Those energies were selected from canonical ensembles with $T=3$ in both systems. See also Ref.~\cite{rigol_09b}.}
\label{fig:nk_offdiag}
\end{figure}

\paragraph{Off-diagonal matrix elements}
In Fig.~\ref{fig:nk_offdiag}, we show the matrix elements of the zero momentum mode occupation $\hat{m}(k=0)$ between the 100 eigenstates whose energy is closest to the energy of the canonical ensemble with temperature\footnote{$T=3$ was selected so that the eigenstates considered are not too close to the ground state and do not correspond to infinite temperature either. These results are relevant to the quenches discussed in Sec.~\ref{ss:quencheth}.} $T=3$. Figure~\ref{fig:nk_offdiag}(a) and \ref{fig:nk_offdiag}(b) illustrates some of the most important properties of the off-diagonal matrix elements of few-body observables in integrable and nonintegrable systems. They have been discussed in Refs.~\cite{rigol_dunjko_08, rigol_09b, santos_rigol_10b} for various lattice models in one and two dimensions and, recently, systematically studied in Refs.~\cite{khatami_pupillo_13, steinigeweg_herbrych_13, beugeling_moessner_15}. The first obvious property, seen in Fig.~\ref{fig:nk_offdiag}(a) and \ref{fig:nk_offdiag}(b), is that no matter whether the system is integrable or not, the average value of the off-diagonal matrix elements is much smaller than the average value of the diagonal ones. In the integrable regime, Fig.~\ref{fig:nk_offdiag}(a), a few off-diagonal matrix elements can be seen to be relatively large, while many are seen to be zero \cite{khatami_pupillo_13, beugeling_moessner_15}. In the nonintegrable regime, Fig.~\ref{fig:nk_offdiag}(b), the (small) values of the off-diagonal matrix elements appear to have a more uniform distribution. Note that, in contrast to the integrable limit, no relatively large outliers can be identified among the off-diagonal matrix elements in the nonintegrable regime. In the latter regime, the values of the off-diagonal matrix elements have been shown to exhibit a nearly Gaussian distribution with zero mean \cite{steinigeweg_herbrych_13,beugeling_moessner_15}, and to be exponentially small in system size \cite{beugeling_moessner_15}. 

\begin{figure}[!t]
\begin{center}
 \includegraphics[width=0.94\textwidth]{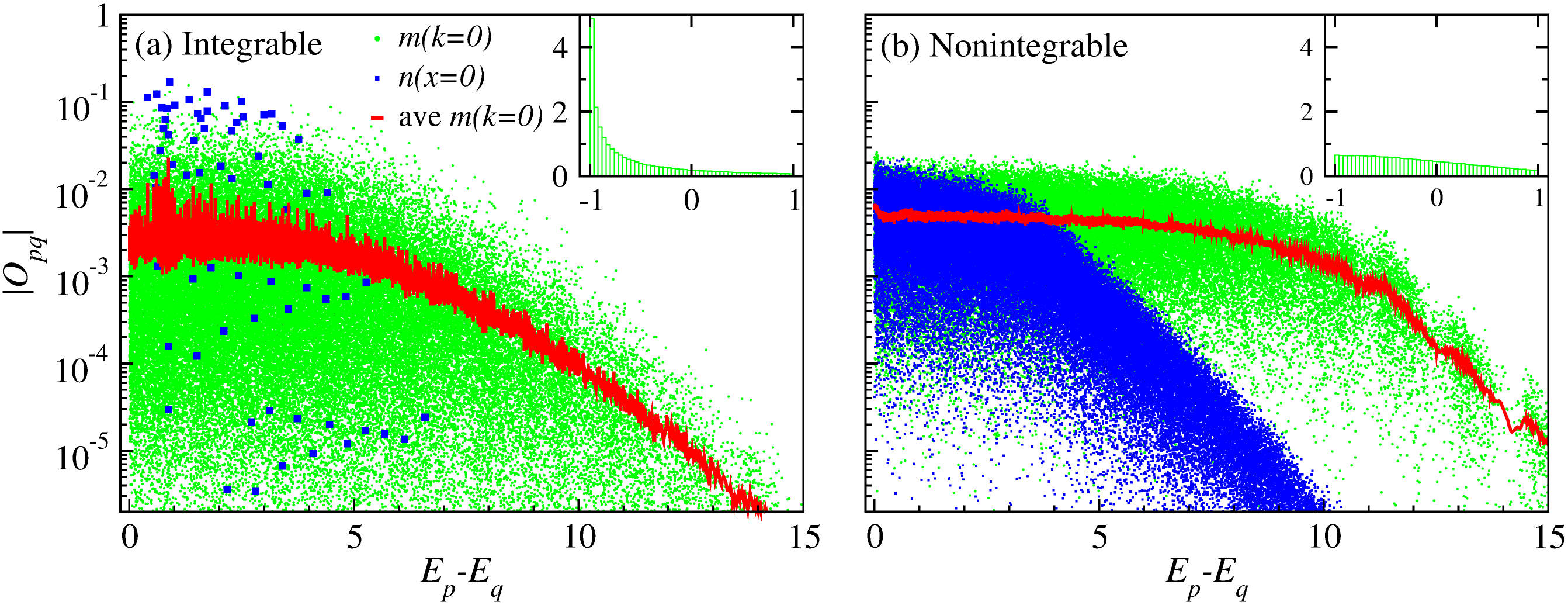}
\end{center}
\vspace{-0.15cm}
\caption{Absolute value of the off-diagonal matrix elements of the site occupation in the center of the system [$\hat{n}(x=0)$] and of the zero momentum mode occupation [$\hat{m}(k=0)$] in the eigenenergy basis vs the eigenenergy difference $E_p-E_q$, for a small window of energies $(E_p+E_q)/2$ (the center of the window was selected to be the energy of a canonical ensemble with $T=5$). Results are shown for an integrable (a) and a nonintegrable (b) system with $L=18$. Lines are running averages ($|O_{pq}|_\text{avg}$) for the matrix elements of $\hat{m}(k=0)$. The insets show histograms of the relative differences between the matrix elements of $\hat{m}(k=0)$ and the running averages. The relative difference is defined as ($|O_{pq}|-|O_{pq}|_\text{avg})/|O_{pq}|_\text{avg}$. The running averages were computed over 50 matrix elements for $L=15$ and over 200 matrix elements for $L=18$. Adapted from Ref.~\cite{khatami_pupillo_13}.}
\label{fig:khatami_offdiag}
\end{figure}

A better quantitative understanding of the behavior of the off-diagonal matrix elements of observables can be gained by plotting them as a function of $E_p-E_q$ for a small window of values $(E_p+E_q)/2$. This is done in Fig.~\ref{fig:khatami_offdiag} for a one-dimensional model of hard-core bosons with the Hamiltonian \cite{khatami_pupillo_13}
\beq
\hat{H}= - J \sum_{j=1}^{L-1} \left( \hat{b}^{\dagger}_j \hat{b}^{}_{j+1} + \text{H.c.} \right) +
V\sum_{j<l} \frac{\hat{n}^{}_j \hat{n}^{}_{l}}{|j-l|^3} +g\sum_{j} x_j^2\, \hat{n}^{}_j.
\label{eq:dhcbmodel}
\eeq
The number of bosons was set to be $L/3$. The three terms in this Hamiltonian describe, from left to right, hopping ($J=1$ sets the energy scale), dipolar interactions, and a harmonic potential ($x_j$ is the distance of site $j$ from the center of the trap). We note that $\hat{H}$ in Eq.~\eqref{eq:dhcbmodel} is not translationally invariant so that the thermodynamic limit needs to be taken with care~\cite{cazalilla_citro_11}. For $V=0$, this model is integrable (mappable to noninteracting spinless fermions) irrespective of the value of $J$ and $g$.

In Fig.~\ref{fig:khatami_offdiag}(a), we show results at integrability ($V=0$ and $g\neq0$), while, in Fig.~\ref{fig:khatami_offdiag}(b), we show results away from integrability ($V=2$ and $g\neq0$). For both cases, results are reported for two observables, the site occupation at the center of the trap and the zero momentum mode occupation. The off-diagonal matrix elements of both observables are qualitatively different in the integrable and nonintegrable regimes. In the integrable model, there is a small fraction of large outliers among the matrix elements (whose absolute value is orders of magnitude larger than that of the median of the nonvanishing absolute values). In addition, there is a large fraction of matrix elements that vanish. As a matter of fact, one can see in Fig.~\ref{fig:khatami_offdiag}(a) that only a few off-diagonal matrix elements of the site occupation are nonzero (this observable is the same for hard-core bosons and for the noninteracting fermions to which they can be mapped \cite{cazalilla_citro_11}). For the zero momentum mode occupation (which is not the same for hard-core bosons and noninteracting fermions \cite{cazalilla_citro_11}), the histogram of the differences between the absolute values and their running average [inset in Fig.~\ref{fig:khatami_offdiag}(a)] makes apparent that there is also a large fraction (increasing with system size \cite{khatami_pupillo_13}) of vanishing matrix elements. This demonstrates that, in the integrable model and for the observables shown, the off-diagonal matrix elements are not described by the ETH ansatz. In contrast, one does not find large outliers among the off-diagonal matrix elements in the nonintegrable model. In addition, the near flat histogram in the inset in Fig.~\ref{fig:khatami_offdiag}(b) shows that there is no large fraction of them that vanish as in the integrable case. One can then conclude that the running average of the off-diagonal matrix elements, that is, the absolute value of the function $f_O(E,\omega)$ in Eq.~\eqref{eq:ETH}, is a well-defined quantity in nonintegrable systems.

\begin{figure}
\includegraphics[width=1.0\textwidth]{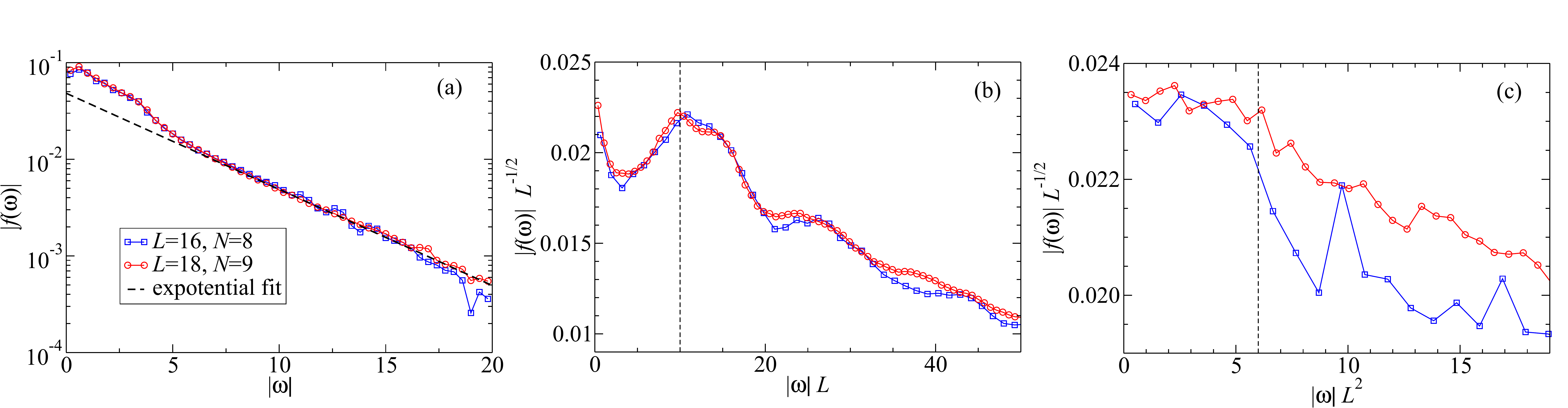}
\caption{Plot of the function $|f_O(\omega)|$ vs the eigenenergy difference $\omega=E_p-E_q$ for a system of $L/2$ bosons described by the Hamiltonian~\eqref{eq:dhcbmodel} with parameters $J=1$, $U=1$ and $g=16/(L-1)^2$. The observable $\hat O$ is the occupation at the  center of the trap (more precisely, the average of the occupation of the two central sites, as the calculations were done in the even sector when taking into account reflection symmetry and the center of the trap is in the middle of two sites) and the function $|f_O(\omega)|$ is obtained in a small energy window centered around the middle of the spectrum. (a) At large $\omega$, $|f_O(\omega)|$ decays exponentially. (b) At intermediate $\omega$, $|f_O(\omega)|$ is proportional to $L^{1/2}$ and has a broad peak whose position scales as $L^{-1}$. (c) At small $\omega$, $|f_O(\omega)|$ exhibits a plateau. $|f_O(\omega)|$ in the plateau is proportional to $L^{1/2}$, and the extension of the plateau is proportional to $L^{-2}$. }
\label{fig:FW}
\end{figure}

This function is studied in detail in Fig.~\ref{fig:FW} for the occupation at the center of the trap (see figure caption for a precise definition of the observable). Results are reported for the same system described above but with $L/2$ bosons. By selecting a narrow energy window in the center of the spectrum, and by comparing results for two different system sizes (including the largest we are able to solve numerically), it is possible to identify three qualitatively different regimes at large, intermediate, and small energy separation $\omega=E_p-E_q$ [in what follows, we drop ``$E$'' from $f_O(E,\omega)$, keeping in mind that $E$ is that in the center of the spectrum]. These three regimes are shown in panels (a), (b), and (c). (a) For $\omega\gg 1$, the function $|f_O(\omega)|$ decays exponentially and the curves corresponding to different system sizes show an excellent collapse supporting the ETH ansatz~\eqref{eq:ETH}. (b) At intermediate $\omega$, $|f_O(\omega)|$ is proportional to $L^{1/2}$ and, around the point marked with a vertical dashed line, one can see a broad peak whose position scales with $L^{-1}$. (c) For $\omega\ll 1$, $|f_O(\omega)|$ exhibits a plateau. Our results suggest that $|f_O(\omega)|$ in the plateau is proportional to $L^{1/2}$, and that its width is proportional to $L^{-2}$. The results in panel (c) are noisier than in panels (a) and (b) because of poor statistics, which is the result of having only few pairs of eigenstates in the center of the spectrum such that $\omega=E_p-E_q\ll1$.

The three regimes identified above, for large, intermediate, and small values of $\omega$, determine what happens to the observable at short, intermediate, and long times during the dynamics (c.f., Sec.~\ref{sec:fluctuation_dissipation}). In the fast, high-frequency, regime $|f_O(\omega)|$ is an exponentially decaying function independent of the system size. In Sec.~\ref{sec:fluctuation_dissipation}, we show that $|f_O(\omega)|^2$ is related to the spectral function of the observable $\hat O$ and to the dissipative part of the linear response susceptibility. Its exponential decay at high frequencies is expected on general grounds from perturbation theory, at least for systems with a bounded spectrum.\footnote{When the energy spectrum is bounded, in order to absorb energy $\omega\gg 1$, many-body processes are required. These processes appear only in high orders of perturbation theory, which leads to an exponential suppression of $|f_O(\omega)|$ for $\omega\gg 1$.} Such a high-frequency exponential tail was discussed, for example, in Ref.~\cite{mukerjee_oganesyan_06} for the conductivity, corresponding to the case where the observable $\hat O$ is the current operator. (ii) At intermediate times, the independence of $|f_O(\omega)|$ vs $\omega L$ on the system size indicates the existence of ballistic dynamics. (iii) At long times, the approximate collapse of $|f_O(\omega)|$ vs $\omega L^2$ for different system sizes indicates diffusive dynamics. Remarkably, at frequencies smaller than a characteristic frequency $\omega_c\sim 1/L^2$ (corresponding to times longer than the diffusive time $t_c\sim L^2$), the function $|f_O(\omega)|$ saturates at a constant value proportional to $L^{1/2}$. It is in this regime that the ETH ansatz~\eqref{eq:ETH} becomes equivalent to the RMT ansatz.\footnote{It is likely that $t_c\approx L^2/D$, where $D$ is the diffusion constant.} As the diffusive time is the longest relaxation time scale in the system, one expects that in this regime physical observables do not evolve. The fact that $|f_O(\omega)|$ at the plateau is proportional to $L^{1/2}$ can be understood as follow.  The function $|f_O(\omega)|^2$ is related to the nonequal-time correlation function of the observable $\hat{O}$, see Sec.~\ref{sec:fluctuation_dissipation}. In particular, when evaluated at $\omega=0$, we have 
\be
|f_O(\omega=0)|^2 \propto \int_0^{t_c}\, dt\,\la \hat{O}(t)\hat{O}(0) + \hat{O}(0)\hat{O}(t) \ra_c \propto \int_0^{t_c}\, \frac{dt}{\sqrt{t}} \propto \sqrt{t_c} \propto L,
\ee
where we have used that the diffusive time scale $t_c$ sets an upper bound for the time integral, and that, assuming diffusive behavior, the nonequal-time correlation function of $\hat{O}$ is expected to be $\propto t^{-1/2}$. It then follows that $f_O(\omega=0) \propto \sqrt{L}$. Remarkably, our results suggests that the scaling of $f_O(\omega)$ with $L^{1/2}$ is also valid at the intermediate frequencies that are relevant to ballistic transport [see Fig.~\ref{fig:FW}(b)].

\subsubsection{Quantum Quenches and Thermalization in Lattice Systems}\label{ss:quencheth}

Now, let us see what happens when systems such as those studied in Sec.~\ref{ss:ethlattice} are taken out of equilibrium. Among the most common protocols for taking systems out of equilibrium are the so-called sudden quenches or, simply, quenches. As explained in Sec.~\ref{sec:chaos_delocalization}, in a quench the system is assumed to be initially in equilibrium and then suddenly some parameter(s) is (are) changed. The dynamics proceeds without any further changes of parameters. For example, in ultracold gases experiments in optical lattices, one can suddenly change the depth of the optical lattice \cite{greiner_mandel_02, will_best_10, will_best_11, trotzky_chen_12, will_iyer_15}, displace the center of the  trapping potential \cite{ott_mirandes_04, fertig_ohara_05, strohmaier_takasu_07}, or turn off a trapping potential while keeping the optical lattice on \cite{schneider_hacke_12, ronzheimer_schreiber_13, xia_zundel_15}. Theoretically, one can think of a quench as a protocol in which one starts with a stationary state of a given Hamiltonian, often the ground state, and then suddenly changes some Hamiltonian parameter(s). The initial state is not stationary in the new (time-independent) Hamiltonian, as a result of which it has a nontrivial unitary dynamics. Quenches in which one changes parameters throughout the system are called global quenches, while quenches in which parameters are only changed in a finite region are called local quenches. In the former class of quenches, one generally adds an extensive amount of the energy to the system, while, in the latter class, the change in energy is subextensive.

A remarkable property of quantum quenches involving local Hamiltonians is that one can actually prove, under very general conditions, that the width $\delta E$ of the energy distribution after a quench scales with the square root of the volume (or of the number of particles) \cite{rigol_dunjko_08}. This behavior is expected from thermodynamics, and is essentially a consequence of the central limit theorem. This width sets the effective ``microcanonical window'' of the equivalent thermodynamic ensemble. It depends on the details of the initial state and the quench protocol.

To prove that after a global quench $\delta E\sim \sqrt{V}$, we consider, for concreteness, a lattice system prepared in an initial state $|\psi_I\rangle$ which is an eigenstate (not necessarily the ground state) of the initial Hamiltonian $\hat{H}_0$. After the quench, the Hamiltonian is $\hat{H}=\hat{H}_0+\hat{H}_1$, where $\hat{H}_1$ is a sum of local operators $\hat{H}_1=\sum_j \hat{h}_j$. One can then write \cite{rigol_dunjko_08}
\begin{eqnarray}
 \delta E \equiv \sqrt{\langle \psi_I | \hat{H}^2 | \psi_I \rangle
-\langle \psi_I | \hat{H} | \psi_I \rangle^2}=
 \sqrt{\langle \psi_I | \hat{H}_1^2 | \psi_I \rangle
-\langle \psi_I | \hat{H}_1 | \psi_I \rangle^2}\nonumber \\
=\sqrt{\sum_{j_1,j_2}\left[\langle \psi_I|\hat{h}_{j_1}\hat{h}_{j_2}|\psi_I\rangle  
-\langle\psi_I|\hat{h}_{j_1}|\psi_I\rangle \langle \psi_I|\hat{h}_{j_2}|\psi_I\rangle\right ]}.
\end{eqnarray}
From the expression above, one concludes that, in the absence of long-range {\it connected} correlations between $\hat h_j$ in the initial state, and if all matrix elements are finite,\footnote{This is guaranteed if the operator norm of each $\hat{h}_j$ is finite} the width $\delta E$ scales at most as the square root of the number of lattice sites in the system, that is, $\delta E\sim \sqrt{V}$. Because the energy itself is extensive in the volume of the system, we see that the relative energy fluctuations are inversely proportional to the square root of the volume $\delta E/E\sim 1/\sqrt{V}$ as expected from equilibrium thermodynamics. This result is a consequence of the locality of the Hamiltonian and, hence, is unrelated to whether the system is integrable or nonintegrable. This scaling of energy fluctuations, in combination with eigenstate thermalization, ensures that in generic systems with local interactions thermalization occurs after a quench. Generalizing this proof to continuous systems is straightforward.

Next, we address two important questions whose precise answer depends on the specifics of the system and of the observable of interest, but whose qualitative answer has been found to be quite similar for several strongly correlated lattice models and observables studied. The first question is how long it takes for experimentally relevant observables to relax to the diagonal ensemble predictions. The second one is how large the system sizes need to be for the relative difference between the diagonal ensemble and the statistical mechanics predictions to be small. These questions have been mainly addressed in numerical experiments. We reproduce some results of these numerical experiments below.

\paragraph{Dynamics}
We consider the dynamics of observables in the hard-core boson model \eqref{eq:HCBHam}. Some numerical results for this model were already discussed in Secs.~\ref{sec:chaos_delocalization} and \ref{ss:ethlattice}. For the quench dynamics discussed here, the initial states are taken to be eigenstates of the Hamiltonian with $J_I=0.5$, $V_I=2$, $J'=V'$, and the time evolutions are studied under final Hamiltonians with $J=V=1$ (unit of energy), and $J'=V'$ \cite{rigol_09a}. Hence, only the nearest neighbor parameters are changed during the quench. The strengths ($J'=V'$) of the integrability breaking terms remain unchanged. To characterize the dynamics of the entire momentum distribution function and of the kinetic energy (by comparing them to the diagonal ensemble results), the following relative differences are computed
\begin{equation}
  \delta m(t)=\dfrac{\sum_k|m(k,t)-m_\text{DE}(k)|}{\sum_k m_\text{DE}(k)},
  \quad \text{and}\quad 
  \delta K(t)=\dfrac{|K(t)-K_\text{DE}|}{|K_\text{DE}|},
\label{Eq:errortau}
\end{equation}
respectively. In these expressions, $t$ refers to time and the subscript ``DE'' refers to the diagonal ensemble prediction [recall Eq.~\eqref{eq:denmatDE}]. In order to be able to compare results for systems with different Hamiltonian parameters in a meaningful way, the initial state for each quench is selected to be the eigenstate of the initial Hamiltonian that, after the quench, has the closest energy to that of a system with temperature $T$, namely, $\langle\psi_I|\hat{H}|\psi_I\rangle=\text{Tr}[\hat{H}\exp(-\hat{H}/T)]/\text{Tr}[\exp(-\hat{H}/T)]$, where the Boltzmann constant is set to unity. For the quenches discussed in what follows, $T=3$ as in Fig.~\ref{fig:nk_offdiag}. This temperature is such that eigenstate thermalization can be seen in these small systems and $O(\bar E)$ is not featureless as expected in the center of the spectrum (i.e., at ``infinite temperature").

\begin{figure}[!t]
\includegraphics[width=0.96\textwidth]{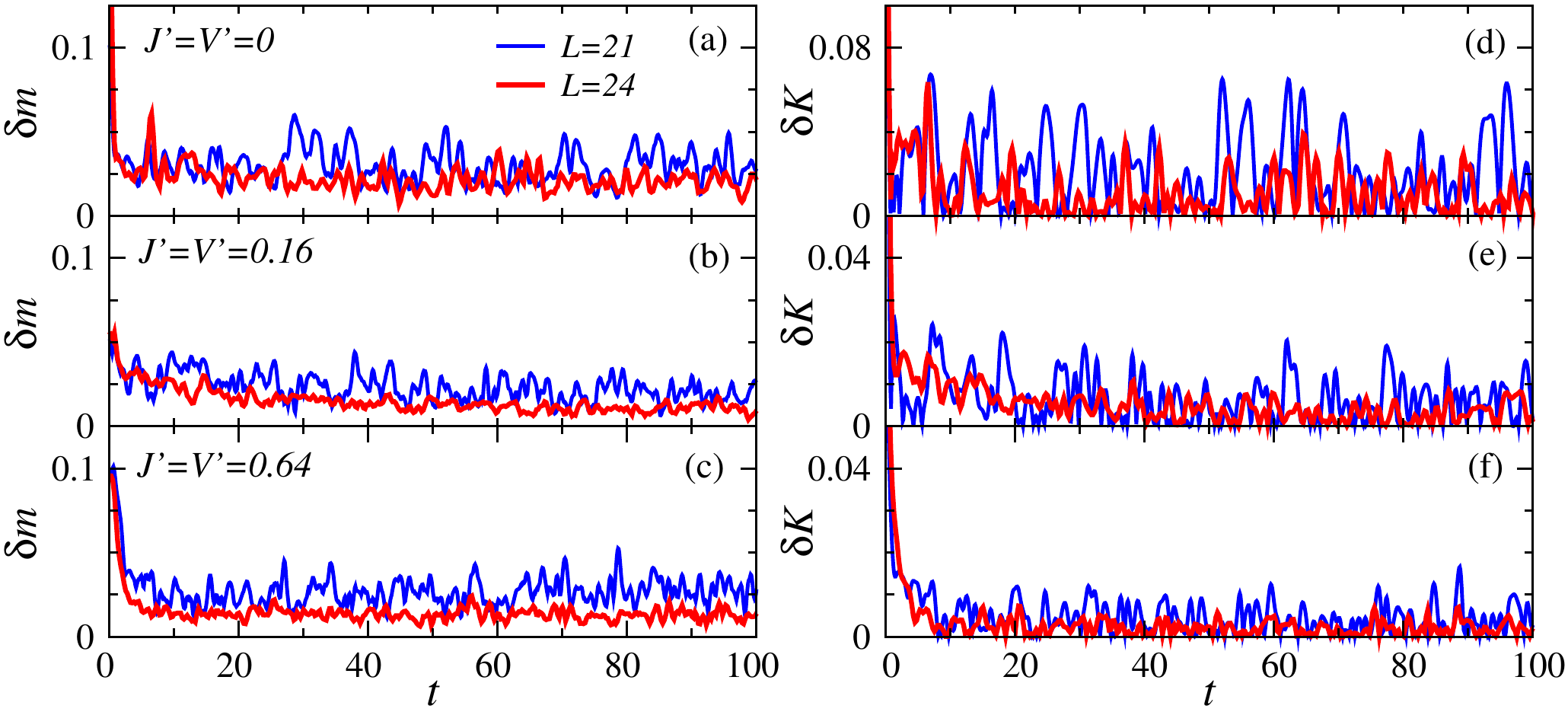}
\vspace{-0.12cm}
\caption{(a)--(c) Relative difference between the instantaneous momentum distribution function and the diagonal ensemble prediction (see text) as a function of time. (d)--(f) Relative difference between the instantaneous kinetic energy and the diagonal ensemble prediction (see text) as a function of time. The strength of the integrability breaking terms ($J'=V'$) increases from top to bottom: (a),(d) $J'=V'=0$ (integrable point); (b),(e) $J'=V'=0.16$; and (c),(f) $J'=V'=0.64$. Results are reported for two system sizes ($L=21$ and 24) and $N=L/3$. In all cases, $J=V=1$ (unit of energy). Time is given in units of $\hbar/J$. See also Ref.~\cite{rigol_09a}.}
\label{fig:TimeEvol}
\end{figure}

In Fig.~\ref{fig:TimeEvol}(a)--\ref{fig:TimeEvol}(c), we show results for $\delta m(t)$ vs $t$ for systems with $L=21$ (blue lines) and $L=24$ (red lines), and for three values of $J'=V'$. The behavior of $\delta m(t)$ vs $t$ is qualitatively similar for all values of $J'=V'$. Namely, at $t=0$, one can see that $\delta m$ is large ($\gtrsim10\%$, except for $J'=V'=0.16$) and then it quickly decreases (in a time scale of the order of $\hbar/J$) and starts oscillating about a small nonzero value ($\sim2\%$ for $L=24$). With increasing system size, the value about which $\delta m(t)$ oscillates, as well as the amplitude of the oscillations, decrease. A qualitatively similar behavior, though with a significantly smaller mean and amplitude of the oscillations about the mean, can be seen during the time evolution of $\delta K(t)$ [Fig.~\ref{fig:TimeEvol}(d)--\ref{fig:TimeEvol}(f)]. Comparable results have been obtained for other nonintegrable models and observables \cite{kollath_lauchli_07, manmana_wessel_07, rigol_dunjko_08, cramer_flesch_08a, cramer_flesch_08b, rigol_09a, rigol_09b, eckstein_kollar_09, rigol_santos_10, banuls_cirac_11, khatami_pupillo_13, zangara_dente_13, sorg_vidmar_14}. All that numerical evidence makes clear that, despite the exponentially small (in system size) level spacing in many-body quantum systems, the relaxation of physically relevant observables to the diagonal ensemble results does not take exponentially long time. Furthermore, in accordance with our expectations based on the ETH ansatz, numerical experiments have also shown that the scaling of the variance of the time fluctuations of expectation values of observables is consistent with an exponential decrease with increasing system size \cite{zangara_dente_13}.

Note that the results reported in Fig.~\ref{fig:TimeEvol} were obtained in systems in which there are only seven and eight hard-core bosons, for $L=21$ and $L=24$, respectively. Namely, the time fluctuations of the expectation values of observables can be very small even for systems with a very small number of particles (see Ref.~\cite{rigol_dunjko_08} for an analysis of a two-dimensional system with only five hard-core bosons that exhibits a qualitatively similar behavior).

\paragraph{Post relaxation} 
After showing that even small finite systems relax to the predictions of the diagonal ensemble and remain close to them, we need to check how close the diagonal ensemble predictions are to those made by standard statistical mechanics. This is the final step needed to know whether thermalization takes place. Since we are dealing with small systems, which ensemble is taken among the microcanonical, canonical, and grand canonical ensembles makes a difference. Considering that the systems of interest here are isolated, the most appropriate statistical ensemble is the microcanonical ensemble \cite{rigol_dunjko_08,rigol_09a}. Therefore, we compute the following relative differences to characterize whether the system thermalizes or not
\begin{equation}
  \Delta m =\dfrac{\sum_k|m_\text{DE}(k)-m_\text{ME}(k)|}{\sum_k m_\text{DE}(k)},
  \quad \text{and}\quad 
  \Delta K =\dfrac{|K_\text{DE}-K_\text{ME}|}{|K_\text{DE}|}.
\label{Eq:errorn}
\end{equation}
In these expressions, the subscripts ``DE'' and ``ME'' refer to the diagonal and microcanonical ensemble predictions, respectively. 

\begin{figure}[!t]
\includegraphics[width=1\textwidth]{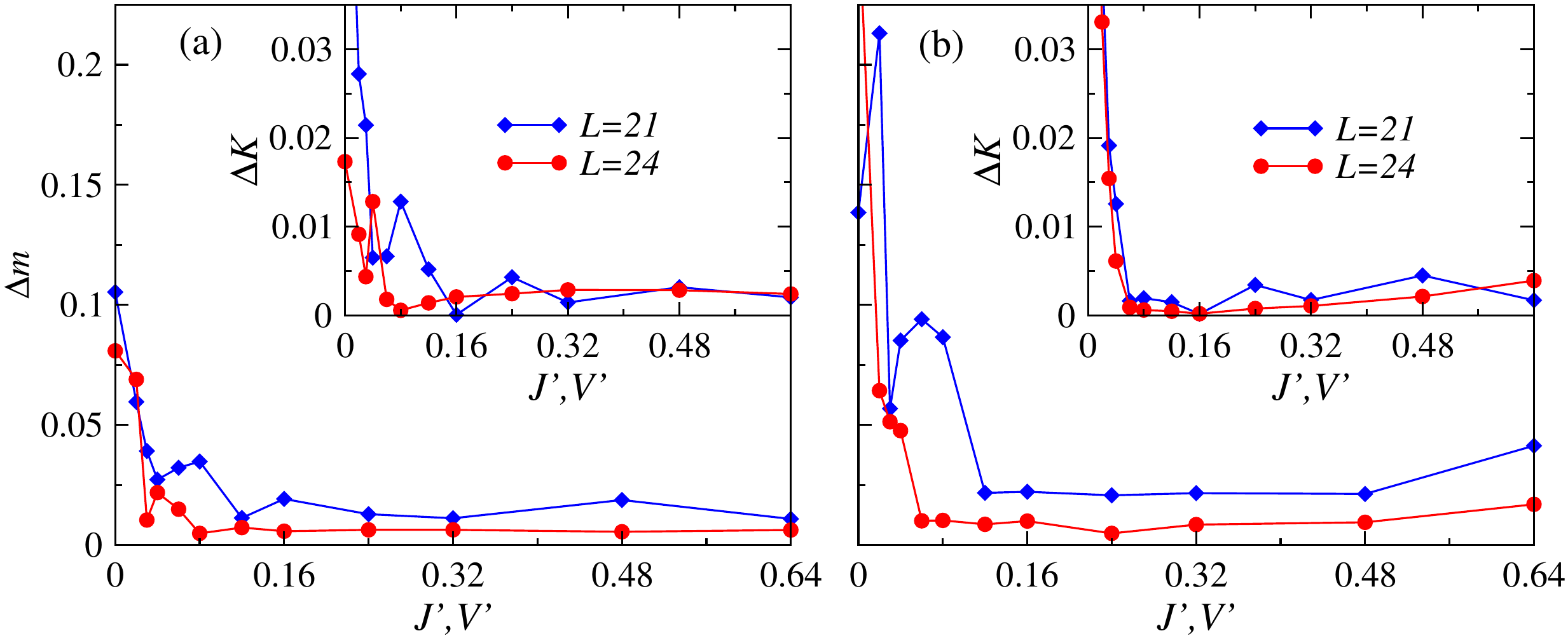}
\vspace{-0.15cm}
\caption{(Main panels) Relative difference between the predictions of the diagonal and microcanonical ensembles for the momentum distribution function (see text) as a function of the strength of the integrability breaking terms. (Insets) The same as the main panels but for the kinetic energy. Results are reported for hard-core bosons (a) and for spinless fermions (b), and for two system sizes $L=21$ (blue lines) and $L=24$ (red lines) with $N=L/3$. In all cases, $J=V=1$ (unit of energy). See also Refs.~\cite{rigol_09a,rigol_09b}.}
\label{fig:DEvsME}
\end{figure}

The main panel (inset) in Fig.~\ref{fig:DEvsME}(a) depicts results for $\Delta m$ ($\Delta K$) in the hard-core boson systems whose dynamics was reported in Fig.~\ref{fig:TimeEvol}. The corresponding results when the hard-core bosons are replaced by spinless fermions are shown in Fig.~\ref{fig:DEvsME}(b). One can see that the behavior of $\Delta m$ ($\Delta K$) is qualitatively similar for hard-core bosons and spinless fermions. The largest differences between the predictions of the diagonal and microcanonical ensembles are seen at (and close to) the integrable point. As one departs from the integrable point, the differences decrease. After a fast decrease, there is an interval of values of $J'=V'$ at which $\Delta m$ ($\Delta K$) becomes almost independent of the exact value of $J'=V'$ (up to finite-size fluctuations). In that interval, $\Delta m$ ($\Delta K$) can be seen to decrease as one increases system size (up to finite-size fluctuations). This is consistent with the expectation that those differences vanish in the thermodynamic limit. Numerical evidence that, in the thermodynamic limit, the predictions of the diagonal ensemble for observables after a quench to a nonintegrable model are identical to those from traditional statistical mechanics ensembles has been obtained in numerical linked cluster expansion studies~\cite{rigol_14a, rigol_16}. On the other hand, in quenches to integrable points in the thermodynamic limit, it was found that lack of thermalization is ubiquitous \cite{rigol_14a, rigol_16} (see Sec.~\ref{sec:XXZ}). For finite systems, it is striking that the differences between the diagonal and the microcanonical ensembles can be a fraction of a percent even for systems with less than 10 particles (which makes their experimental detection unlikely). Similar results have been obtained for other observables in Refs.~\cite{rigol_dunjko_08, rigol_09a, rigol_09b, rigol_santos_10, khatami_pupillo_13, sorg_vidmar_14}.

The results presented in this section support the expectation that nonintegrable quantum systems exhibit eigenstate thermalization and therefore thermalize in the strong sense as defined in this review.

%%%%%%%%%%%%%%%%%%%%%%%%%%%%%%%%%%%%%%%%%%%%%%%%%%%%%%%%%%%%%%%%%%%%%%%%%%%%%%%%%%%%%%%%%%
%%%%%%%%%%%%%%%%%%%%%%%%%%%%%%%%%%%%%%%%%%%%%%%%%%%%%%%%%%%%%%%%%%%%%%%%%%%%%%%%%%%%%%%%%%
%%\chapter{PART 2: from Quantum Chaos to Thermodynamics}
%%%%%%%%%%%%%%%%%%%%%%%%%%%%%%%%%%%%%%%%%%%%%%%%%%%%%%%%%%%%%%%%%%%%%%%%%%%%%%%%%%%%%%%%%%
%%%%%%%%%%%%%%%%%%%%%%%%%%%%%%%%%%%%%%%%%%%%%%%%%%%%%%%%%%%%%%%%%%%%%%%%%%%%%%%%%%%%%%%%%%

%%%%%%%%%%%%%%%%%%%%%%%%%%%%%%%%%%%%%%%%%%%%%%%%%%%%%%%%%%%%%%%%%%%%%%%%%%%%%%%%%%%%%%%%%%
\section{Quantum Chaos and the Laws of Thermodynamics\label{sec:sec5}}
%%%%%%%%%%%%%%%%%%%%%%%%%%%%%%%%%%%%%%%%%%%%%%%%%%%%%%%%%%%%%%%%%%%%%%%%%%%%%%%%%%%%%%%%%%

If one assumes that a system is prepared in a Gibbs, or other equivalent ensemble, then one does not need assumptions about chaos and ergodicity to prove various statements of statistical mechanics and thermodynamics. For example, the fluctuation-dissipation relation can be straightforwardly proved using standard perturbation theory. Quantum chaos and ETH allow one to prove all the statements for individual eigenstates of chaotic Hamiltonians, and therefore for arbitrary stationary ensembles (with subextensive energy fluctuations). This distinction is at the heart of the importance of quantum chaos for the proper understanding of thermodynamics in isolated systems. In the earlier sections, we argued that eigenstate thermalization is generally needed for isolated quantum systems taken far from equilibrium to thermalize. Likewise, in the following sections, we will show that the same assumptions of quantum chaos together with ETH are sufficient for establishing thermodynamic relations in such isolated systems.

\subsection{General Setup and Doubly Stochastic Evolution\label{sec:DS}}

Equilibrium thermodynamics studies transformations between equilibrium states of macroscopic systems. During such a transformation, thermodynamic quantities (such as the free energy, magnetization, and pressure) evolve in time. These changes are usually induced by either heat exchange with another macroscopic system or the work done on the system via changing some macroscopic parameters in time (like its volume or the applied magnetic field), or both. For example, consider a phase transformation from a solid to a liquid as temperature is changed, or the exchange of energy, in the form of heat and work, in heat engines. The laws of thermodynamics dictate which process are possible and which are not. They give bounds for engine efficiencies and provide relations between superficially different quantities (e.g., the Onsager relations, which will be discussed in the next section).

Since equilibrium thermodynamics provides relations between different equilibrium states, the concept of a quasi-static process is central to the development of the theory. A quasi-static process is one in which the state of the system is changed very slowly through a sequence of equilibrium states. However, it is important to stress that thermodynamics is not limited to quasi-static processes. For example, in one formulation of the second law of thermodynamics one considers an equilibrated isolated system that undergoes a dynamical process (which need not be quasi-static). As a result, the entropy difference between the final equilibrium state of the system and the initial equilibrium state is positive or zero independent of how rapidly the process is carried out. Moreover, this entropy difference is uniquely determined by the total energy change in the system, no matter how fast or slow the process of energy exchange is. Another remarkable example of thermodynamic relations are the recently discovered fluctuation theorems~\cite{jarzynski_97, crooks_98,  campisi_hanggi_11,seifert_12}, which make exact statements about work, heat, and free energy changes in arbitrary nonequilibrium processes.

\begin{figure}[!t]
\begin{center}
 \includegraphics[width=0.5\textwidth]{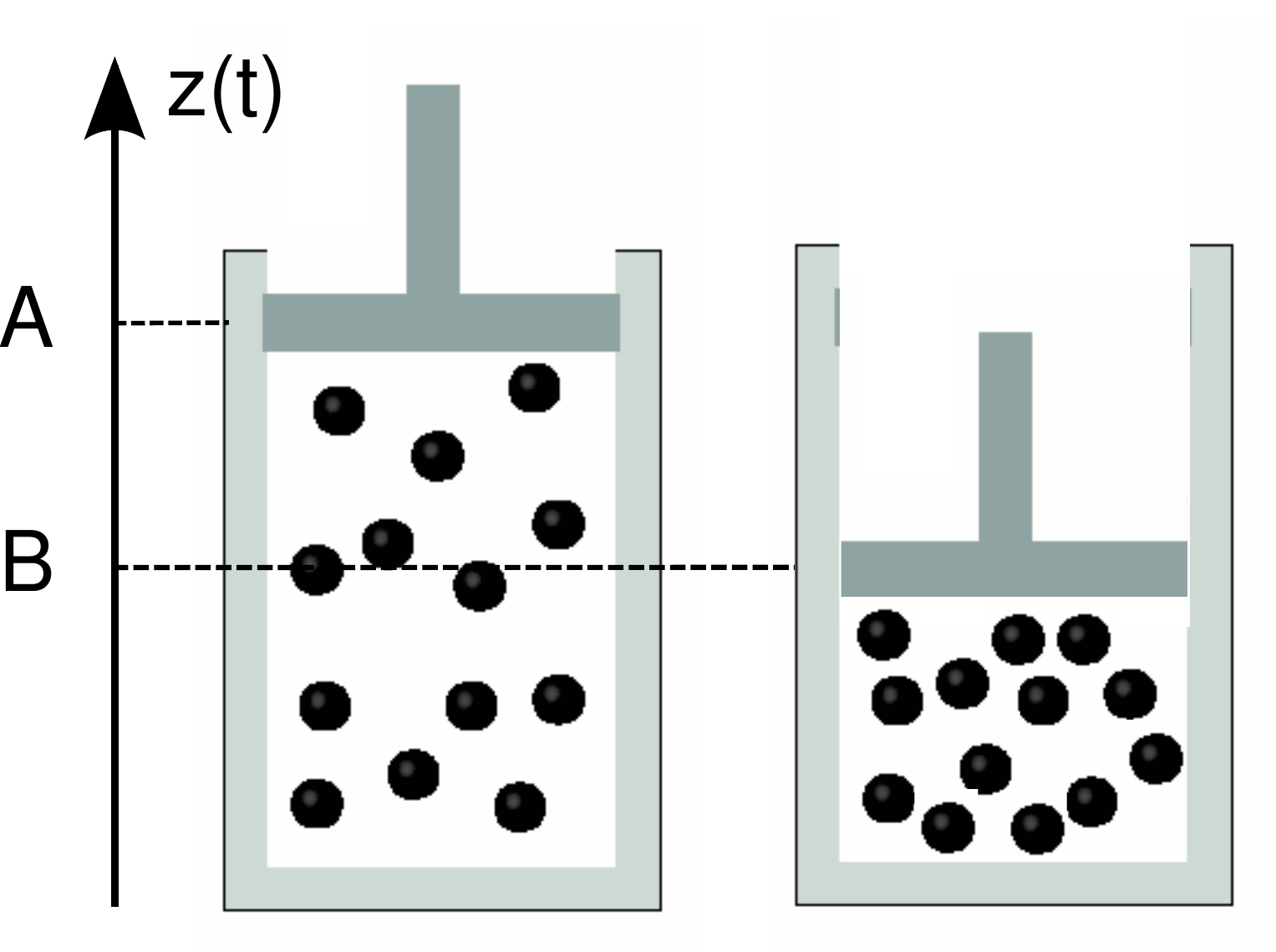}
\end{center}
\caption{Schematic representation of a gas undergoing a compression}
\label{fig:piston}
\end{figure}

To derive thermodynamic relations one needs to consider dynamical processes that start from a stationary state. To this end, we focus on an isolated system initially prepared in a stationary state, which undergoes a unitary evolution in response to an external change. The latter is modeled by a change in time of macroscopic parameters in the Hamiltonian according to a prescribed protocol. The protocol considered is such that the parameters are changed during a finite time, after which the Hamiltonian is time independent and the system is allowed to relax to equilibrium. As an example, one can think of a gas confined in a container by a piston (see Fig.~\ref{fig:piston}). In this case, the macroscopic parameter is the position of the piston, which is changed in time from position $z=A$ to $z=B$ according to a protocol $z(t)$. At the end of the process, the piston is kept fixed at position $z=B$ and the gas is allowed to equilibrate. While we focus on isolated systems, our setup can also describe open systems. This is because, if the dynamical process involves changing parameters only in a part of the system (a local operation), the rest of the system can act as a thermal bath. If the bath is much larger than the part of the system in which the dynamical process is implemented, the intensive quantities characterizing the latter (such as the energy density defining the temperature or the pressure) after it relaxes to equilibrium will be identical to their initial values.

Mathematically, the assumption that the systems start in a stationary (i.e., a state that is translationally invariant in time) amounts to taking the initial density matrix to be diagonal in the basis of the initial Hamiltonian (if there are degeneracies in the spectrum, one just needs to find a basis where the density matrix is diagonal)
\be
\rho_{nm}^{(0)}=\rho_{nn}^{(0)}\,\delta_{nm}.
\ee
It is important to note that, at the moment, we make no assumptions on the structure of the density matrix beyond stationarity. After the system undergoes a dynamical process, the density matrix in the basis of the final Hamiltonian, that is, the Hamiltonian at the end of the dynamical process, is not diagonal anymore. As discussed previously, the off-diagonal matrix elements dephase and, at long times, the expectation values of observables are solely determined by the diagonal matrix elements. The latter are given by
\be
\rho_{\tilde m \tilde m}=\sum_n U_{\tilde mn} \rho_{nn}^{(0)} U^\dagger_{n \tilde m},
\label{eq:evolution_rho}
\ee
where $U_{nm}$ are the matrix elements of the evolution operator
\be
\hat U=T_t\exp\left[-i\int_0^t dt' \hat H(t')\right],\; U_{\tilde m n}=\langle \tilde m|\hat U|n\rangle.
\ee
Here, $T_t$ denotes time ordering, and the ``tilde'' indicates that the states $|\tilde m\rangle$ are eigenstates of the Hamiltonian after the evolution, while $|n\rangle$ are the eigenstates of the initial Hamiltonian. The two basis sets coincide only in the special case of cyclic processes. 

Equation~(\ref{eq:evolution_rho}) can be rewritten as a master equation for the occupation probabilities of the microstates
\be
\rho_{\tilde m \tilde m}^{(1)}=\sum_n \rho_{nn}^{(0)}\,\left( U_{\tilde{m} n}U^\dagger_{n\tilde{m}}\right) \equiv \sum_n \rho_{nn}^{(0)}\, p_{n\to\tilde m},
\label{eq:master}
\ee
where we have defined the transition probabilities between states $|n\rangle$ and $|\tilde m\rangle$ associated with the dynamical process to be\footnote{Note that the transition probabilities are conditional, i.e., they define the probability of a transition from state $|n\rangle$ to state $|\tilde m\rangle$ if the system is initially prepared in state $|n\rangle$.} 
\be
p_{n\to\tilde m}=U_{\tilde{m} n}U^\dagger_{n\tilde{m}} =|U_{\tilde m n}|^2 \;.\label{eq:def_pnm}
\ee
The last equality trivially follows from the identity $U^\dagger_{n\tilde m}=(U_{\tilde m n})^\ast$, where the star indicates complex conjugation.

From the unitarity of the evolution operator it follows that
\be
\sum_{\tilde m} U_{\tilde m n} U^\dagger_{k \tilde m} = \delta_{nk},
\quad \text{and} \quad
\sum_{n} U^\dagger_{n \tilde k} U_{\tilde m n}=\delta_{\tilde{k}\tilde{m}}\;.
\ee
Setting $k=n$ and $\tilde k=\tilde m$, we immediately see that
\be
\sum_{\tilde m} p_{n\to \tilde m}=1, \quad \sum_{n} p_{n\to \tilde m}=1.
\label{eq:double_stochastic}
\ee
These conditions allow for a simple physical interpretation if we rewrite it changing the dummy indices $n$ and $m$ in the second sum, namely
\be
\sum_{\tilde m \neq  \tilde n} p_{n\to \tilde m}=\sum_{m \neq n} p_{m\to \tilde n}.
\ee
In other words, the sum of incoming probabilities to any given state $|\tilde n \rangle$ of the final Hamiltonian is equal to the sum of the outgoing probabilities from an equivalent, for example, adiabatically connected state $|n\rangle$ of the initial Hamiltonian. For a cyclic process, one can remove the tildes and simply say that the sum of incoming probabilities to any eigenstate is equal to the sum of outgoing probabilities from the same state.

The transition probabilities $p_{n\to\tilde m}=|U_{n \tilde m}|^2$ are positive semi-definite for any pair of states. This, combined with the constraints above, leads to
\be
0\le p_{n\to\tilde m} \le 1 \;, \label{eq:bound}
\ee 
as expected. Any semi-positive matrix $p$ satisfying the constraints~\eqref{eq:double_stochastic} and \eqref{eq:bound} is called doubly stochastic~\cite{thirring_book}. The corresponding evolution described by the master equation~\eqref{eq:master} is called a doubly stochastic evolution. 

As it has been known for a long time, the constraints~\eqref{eq:double_stochastic} have far-reaching consequences and, for example, play a prominent role in the formulation of the kinetic theory of gases~\cite{lifshitz_pitaevskii_06}. Moreover, doubly stochastic evolution is the proper framework to discuss thermodynamic processes in isolated quantum systems since it emerges naturally based on the assumptions that: (i) the system starts from a stationary state, (ii) the system evolves unitarily, and (iii) the long-time behavior of the observables is determined only by the diagonal elements of the density matrix in the basis of the final Hamiltonian. With this in mind, we review some general properties of the master equation, and those associated with doubly stochastic matrices in particular.

\subsubsection{Properties of Master Equations and Doubly Stochastic Evolution\label{sec:prop}}

Equation~\eqref{eq:master} is a discrete-time master equation. The matrix ${\bf p}$, with elements $p_{n,\tilde m}=p_{n \to \tilde m}$ satisfying the first of the two conditions in Eq.~\eqref{eq:double_stochastic}, is known as a Markov matrix or, equivalently, as a stochastic matrix. The action of ${\bf p}$ on a probability vector, in our case $\rho^{(0)}$ (with elements $\rho_{nn}^{(0)}$), gives a new probability vector $\rho^{(1)}$ (with elements $\rho_{\tilde m \tilde m}^{(1)}$), which is the result of stochastic transitions between the different states of the system. For completeness, we briefly review some of the properties of a Markov matrix that will be used in our discussion. More details and complete proofs can be found, for example, in Ref.~\cite{VanKampen}. 

We first note that the conservation of probability implies that the \textit{outgoing} transition probabilities from any state must sum to one [first condition in Eq.~\eqref{eq:double_stochastic}], so that the sum over each column of the Markov matrix is $1$. This holds for any master equation. Indeed, from Eq.~\eqref{eq:master},
\be
\sum_{\tilde m} \rho_{\tilde m \tilde m}^{(1)}=\sum_{\tilde m} \sum_n \rho_{nn}^{(0)}\, p_{n\to\tilde m}=\sum_n \rho_{nn}^{(0)}=1.
\ee
In general, the matrix ${\bf p}$ is not symmetric and therefore admits separate left and right eigenvectors (the spectrum associated with left and right eigenvectors is the same). If one applies the matrix ${\bf p}$ many times on a probability vector one expects that the probability distribution relaxes to a steady state. This implies that the matrix ${\bf p}$ has one eigenvalue $\lambda_0=1$ whose corresponding right eigenvector is the steady-state probability distribution~\cite{VanKampen}. The existence of the eigenvalue $\lambda_0=1$ is straightforward to prove as the left vector $(1,1,1 \ldots)$ is always, by conservation of probability, the corresponding left eigenvector. One can show that if the Markov matrix does not have a block diagonal form, which implies that some states cannot be reached from others, then the right eigenvector corresponding to the eigenvalue $1$ is unique \cite{VanKampen}. Finally, we note that the relaxation to the steady-state following many applications of ${\bf p}$ is dictated by the other eigenvectors and their corresponding eigenvalues. One then expects that the eigenvalues $\lambda_i$ satisfy $|\lambda_i|\leq 1$ (this can be proved rigorously \cite{VanKampen}).   

\begin{figure}[!t]
\begin{center}
 \includegraphics[width=0.55\textwidth]{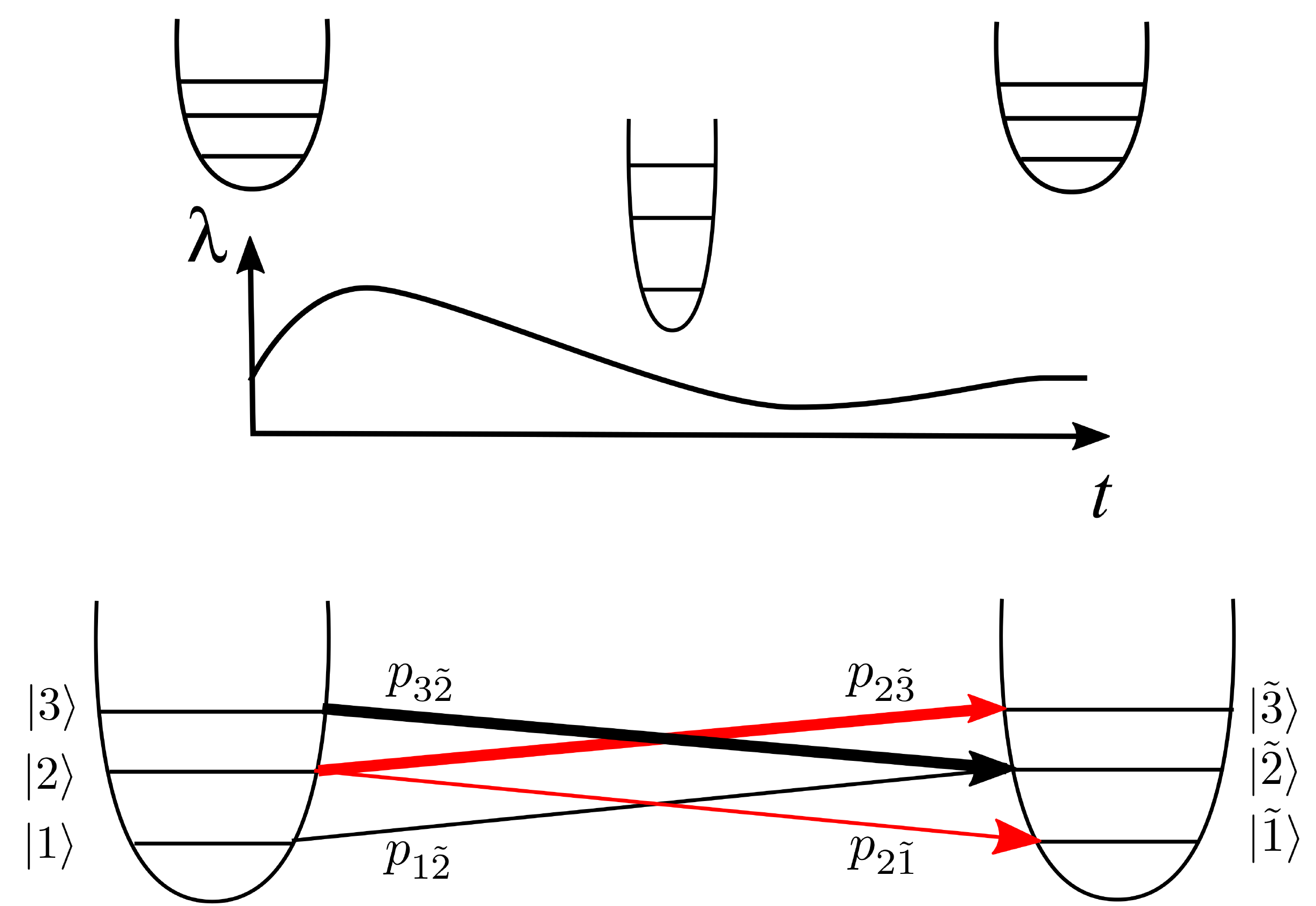}
\end{center}
\caption{Schematic representation of a doubly stochastic evolution for a cyclic process. An isolated system undergoes some dynamical process where the control parameter $\lambda$ changes in time in some arbitrary way (top). The cumulative effect of the evolution is given by the transition probabilities (bottom). Red arrows indicate the outgoing transition probabilities from the level $|2\rangle$ of the initial Hamiltonian to levels $|\tilde 1\rangle$ and $|\tilde 3\rangle$ of the final Hamiltonian (which in this case are the same as $|1\rangle$ and $|3\rangle$ because we consider a cyclic process). Black arrows describe the incoming transition probabilities to the level $|\tilde 2\rangle$ of the final Hamiltonian: $|1\rangle\to |\tilde 2\rangle$ and $|3\rangle\to|\tilde 2\rangle$.  Doubly stochastic evolution implies that $p_{2\tilde 1}+p_{2\tilde 3}=p_{1\tilde 2}+p_{3\tilde 2}$.}
\label{fig:double_stochastic}
\end{figure}

Next, we turn to the second condition in Eq.~\eqref{eq:double_stochastic}, which is associated with the doubly stochastic nature of ${\bf p}$. It states that the \textit{incoming} transition probabilities to any state sum to one. This constraint is less trivial and does not generally hold for non-unitary evolution.\footnote{For example, if the Markov process admits an absorbing state, this condition is violated since the sum of the incoming transition probabilities into the absorbing state is larger than one.} At the same time doubly stochastic evolution is more general than unitary evolution. In particular, the product of two doubly stochastic matrices is a doubly stochastic matrix [see Eq.~\eqref{eq:doubly_stochastic_prod}]. This implies that any projective measurement performed during the evolution, which breaks unitarity, keeps the transition matrix doubly stochastic. Moreover, any statistical mixture of doubly stochastic matrices is doubly stochastic. This implies that if, for example, one repeats slightly different dynamical protocols starting with the same density matrix and ending with the same final Hamiltonian, then the transition matrix describing the average effect of these dynamical protocols is still doubly stochastic. Because of this, various dephasing mechanisms (e.g., the presence of external noise or fluctuating waiting times between different pulses) keep the evolution doubly stochastic, even if they generally break its unitarity. The second condition in Eq.~\eqref{eq:double_stochastic} is a direct consequence of the fact that any doubly stochastic matrix can be represented as $p_{n\to \tilde m}= |\langle \tilde m | \hat{U} |n\rangle|^2$ for \textit{some} (maybe fictitious) unitary operator $\hat{U}$, see Ref.~\cite{marshall_olkin_79}. For a unitary process, one can always define its inverse. Therefore, the role of the initial and final states is interchangeable, and the same sum rule applies to both summations over $n$ and $\tilde m$. The doubly stochastic condition is schematically illustrated in Fig.~\ref{fig:double_stochastic}, where it is shown that the sum of the outgoing rates \textit{from} a state $|2\rangle$ (red lines) is equal to the sum of the incoming probabilities \textit{into} the state $|\tilde 2\rangle$ (black lines).

The easiest way to satisfy the doubly stochastic constraint is to have identical transition probabilities between any two pair of energy levels, $p_{n\to \tilde m}=p_{\tilde m\to n}$. This condition is known as detailed balance for an isolated system.\footnote{For systems in contact with a thermal bath at temperature $T$, the detailed balance condition is $p_{n\to \tilde m} =p_{\tilde m\to n} \exp\left[\beta (E_n-E_{\tilde m})\right]$, where $\beta=(k_B T)^{-1}$ is the inverse temperature. In Sec.\ref{detailed_balance_open_systems}, we show how this condition follows from the doubly stochastic transition rates.} Detailed balance is satisfied in: (i) two level systems, (ii) systems with more than two energy levels within first order of perturbation theory (e.g., a Fermi golden rule)~\cite{polkovnikov_08}, and (iii) systems with real Hamiltonians, which satisfy instantaneous time-reversal symmetry, subjected to symmetric cyclic protocols such that $\hat H(t)=\hat H(T-t)$, where $T$ indicates the total duration of the dynamical process~\cite{bunin_dalessio_11}. In general, however, pairwise transition probabilities are not the same, that is, the detailed balance condition $p_{n\to \tilde m}= p_{\tilde m\to n}$ is violated and only the sum rules~(\ref{eq:double_stochastic}) are satisfied. 

In passing, we note that doubly stochastic matrices are intimately related to permutation matrices, as stated by Birkhoff's theorem, which is presented in Appendix~\ref{appendix:birkhoff}. This theorem allows one to make physical predictions for arbitrary doubly stochastic evolution of systems with many degrees of freedom. For example, it allows one to rigorously bound the maximum amount of work that can be extracted from a microcanonical bath~\cite{allahverdyan_hovhannisyan_11}.
  
\textit{Repeated processes.} Next, we show that doubly stochastic matrices form a group under multiplication, that is, the product of two doubly stochastic matrices $p$ and $q$ is a doubly stochastic matrix $s$ (the remaining group properties follow trivially):
\be
s_{n\to k}\equiv \sum_{m} p_{n\to m}q_{m\to k} \Rightarrow\left\{ \begin{array}{c}
\sum_n s_{n\to k} = \sum_m q_{m\to k}=1\\
\sum_k s_{n\to k} = \sum_m p_{n\to m}=1
\end{array}\right.,
\label{eq:doubly_stochastic_prod}
\ee
where, to simplicity the notation, we dropped the tilde over the state labels. Physically, Eq.~\eqref{eq:doubly_stochastic_prod} tells us that performing a sequence of two (or more) doubly stochastic processes on a system is again a doubly stochastic process. This property allows one to split any doubly stochastic process in a sequence of arbitrary many doubly stochastic processes. We now apply this result to a concrete setup in which an initially stationary density matrix $\rho^{(0)}_{nn}$ undergoes an arbitrary dynamical process that is interrupted by a sequence of ideal projective measurements, i.e., we consider the sequence:
\be
\rho^{(0)}_{nn} \underbrace{\longrightarrow}_{U_1} M_1 \underbrace{\longrightarrow}_{U_2} M_2 \dots \underbrace{\longrightarrow}_{U_N} M_N 
\ee
where $U_j$ represents an arbitrary dynamical process and $M_j$ an arbitrary projective measurement. Immediately after each projective measurement, the density matrix is diagonal in the basis of the projection operator $M_j$. Hence the previous sequence is equivalent to:
\be
\rho^{(0)}_{nn} \underbrace{\longrightarrow}_{U_1} \rho^{(1)}_{ll} \underbrace{\longrightarrow}_{U_2} \rho^{(2)}_{kk} \dots \underbrace{\longrightarrow}_{U_N} \rho^{(N)}_{mm}  
\ee
where $\rho^{(j)}$ is a diagonal matrix in the basis of the projection operator $M_j$.\footnote{To simplify the presentation, here we denote all basis states using the same notation. One needs to keep in mind that the initial state is always diagonal in the basis of the initial (measurement) Hamiltonian and the final state is always diagonal in the basis of the final (measurement) Hamiltonian.} Each fundamental block is a doubly stochastic process and is represented by a proper doubly stochastic matrix. For example:
\be
 \rho^{(j-1)}_{ll} \underbrace{\longrightarrow}_{U_j} \rho^{(j)}_{kk} \quad \Leftrightarrow \quad \rho^{(j)}_{kk} = \sum_l \rho^{(j-1)}_{ll}\,p^{(j)}_{l\to k}.
\ee
Using the group property discussed above, the sequence is described by a single doubly stochastic matrix $s$ obtained as the (matrix) product of the doubly stochastic matrices of the fundamental processes:
\be
\rho^{(N)}_{mm} = \sum_n \rho^{(0)}_{nn}\,s_{n\to m},\quad  {\bf s} =  {\bf p}^{(N)}\dots {\bf p}^{(2)} {\bf p}^{(1)} \;,
\label{combined}
\ee 
where the bold symbols indicate matrices.

One can generally think of projective measurements as quenches to a measurement Hamiltonian and dephasing. For example, let us imagine we have a system of (possibly interacting) spins and we want to do a projective measurement of the $z$-magnetization of a given spin. Formally, we can simply say that we project this spin to the $z$-axis and read probabilities. Alternatively, one can think about the same process as a quench to a very strong local magnetic field along the $z$-axis, such that the remaining part of the Hamiltonian does not matter, and then dephasing (or time averaging) of the density matrix. Thus, this combination of quench and dephasing projects the local spin density matrix into a statistical mixture of ``up'' and ``down'' states. For this reason, the factorization property of the transition matrix~\eqref{combined} holds if we have a series of quenches with long random waiting times in between. These random waiting times are equivalent to the projection of the density matrix to the basis of the Hamiltonian after each quench or, formally equivalent, to the projective measurement of the energy of this intermediate Hamiltonian. 

In ergodic systems, random waiting times are not needed, it is sufficient to wait times that are longer than the relevant relaxation time. As we discussed in the previous section, apart from small fluctuations, from the point of view of observables the density matrix is effectively dephased (see Ref.~\cite{ikeda_sukumichi_15} for a more formal discussion of this point, and Ref.~\cite{ji_fine_2011} for caveats). If the waiting time between quenches (random or not) is shorter than the relaxation time, then the transition matrix describing the whole dynamical process is doubly stochastic but it is not the product of the transition matrices corresponding to the individual quenches. For example, if one considers a large periodically driven ergodic system, one can anticipate that, if the driving period is longer than the relaxation time, the exact periodicity of the driving protocol is not important and the transition probability factorizes (in small ergodic systems the factorization can be violated even for long driving periods~\cite{ji_fine_2011}). If the period is short compared to the relaxation time, one has to use the Floquet formalism (see, e.g., Ref.~\cite{bukov_dalessio_14} for review) to accurately describe the time evolution after many periods. In this case, the dynamics between periods is coherent and the factorization property of the transition probability~\eqref{combined} does not apply. Still, the overall evolution remains doubly stochastic.

Let us now discuss the implications of doubly stochastic evolution for time-reversed processes. The two conditions~\eqref{eq:double_stochastic} imply that one can define the transpose transition rate matrix $p^T_{\tilde m\to n}=p_{n\to \tilde m}$, which is also a doubly stochastic matrix that corresponds to a reversed process in which the role of initial and final states is swapped. For a unitary process (i.e., a process without projective measurements or dephasing), the time-reversal process corresponds to the transition matrix ${\bf p}^T$, that is, ${\bf p}^{\rm t.r.}={\bf p}^T$. Indeed, for the time-reversal process, the evolution operator is given by $\hat U^{-1}=\hat U^\dagger$. Therefore 
\be
p^\text{t.r.}_{\tilde m\to n}\equiv |U^\dagger_{n \tilde m}|^2=|U_{\tilde m n}|^2=p_{n\to \tilde m}.
\label{eq:time_reversal}
\ee
In practice, time-reversal processes are very difficult to realize. They require either an overall change of sign of the Hamiltonian or taking the complex conjugate of the wave function, which in the classical language is equivalent to changing the sign of the velocities of all particles. As noted in Ref.~\cite{campisi_hanggi_11}, the dynamics leading to the transition probabilities $p^\text{t.r.}_{\tilde m\to n}$ can be achieved much more easily by using the so-called reversed protocol. To see the difference between the time-reversal and reversed processes, consider again a unitary evolution. The evolution operator and its time inverse are given by the time-ordered exponentials (such that later times $t$ appear on the left):
\be
\hat U=T_t \exp\left[-i \int_{0}^{T} \hat H(t) dt\right],\quad \hat U^\dagger=T_t \exp\left[i \int_{0}^{T} \hat H(T-t) dt\right]
\ee
Let us now define the reverse protocol as the forward time evolution with the Hamiltonian $H(T-t)$, that is, the Hamiltonian for which we simply reverse the dynamical protocol. The corresponding evolution operator is given by
\be
\hat U^\text{r}=T_t \exp\left[-i \int_{0}^{T} \hat H(T-t) dt\right].
\ee
Note that $\hat U^{\rm r}$ and $\hat U^\dagger$ are very different. For example, $\hat U \hat U^\dagger$ is the identity while $\hat U \hat U^r$ is not. Nevertheless, if the Hamiltonian $\hat H(t)$ is real at each moment of time, that is, satisfies instantaneous time-reversal symmetry, then $\hat U^{\rm r}$ and $\hat U^\dagger$ lead to the same transition probabilities. Using this fact, the eigenstate $|n\rangle$ and $|\tilde m\rangle$ can be chosen to be real in that case, we find
\be
\langle n | \hat U^r|\tilde m\rangle^\ast=\langle n^\ast| \left(\hat U^r\right)^\ast |\tilde m^\ast\rangle=\langle n| \hat U^\dagger |\tilde m\rangle .
\ee 
Therefore
\be
p^{\rm r}_{\tilde m\to n}=|U^{\rm r}_{n\tilde m}|^2=|\langle n| U^\dagger |\tilde m\rangle|^2=p^{\rm t.r.}_{\tilde m\to n}.
\label{detailed_bal}
\ee
Unlike the time-reversal process, which generally exists only for unitary evolution, the reverse process is defined even if the forward protocol is not unitary. For example, if it involves projection measurements along the way. As we discussed, in this case the transition probability matrices factorize into products of transition probability matrices corresponding to processes between measurements. It is then straightforward to see that for the reversed process, which involves exactly the same sequence of measurement performed in the opposite order, one still has $p^r_{\tilde m\to n}=p_{n\to \tilde m}$. If the protocol is time symmetric, that is, $\hat H(T-t)=\hat H(t)$, then $\hat U=\hat U^{\rm r}$ and hence $p^{\rm r}_{m\to n}=p_{m\to n}$. Combining this condition with $p^r_{m\to n}=p_{n\to m}$, we see that for such symmetric protocols, detailed balance is automatically satisfied, i.e., $p_{n\to m}=p_{m\to n}$.

\subsection{General Implications of Doubly Stochastic Evolution\label{sec:DS_only}}

We now derive the physical implications of doubly stochastic evolution. The results in this subsection rely {\it only} on doubly stochasticity and are therefore valid for both nonintegrable and integrable systems.

\subsubsection{The Infinite Temperature State as an Attractor of Doubly-Stochastic Evolution\label{sec:inf_T}}

First, let us consider a cyclic process. In this case, the basis $\tilde m$ and $m$ are identical and the master equation~\eqref{eq:master} becomes:
\be
\rho^{(1)}=\boldsymbol{p} \rho^{(0)},
\ee
where on the RHS we have a matrix vector multiplication. If we repeat the process $N$ times, we obtain
\be
\rho^{(N)}=\boldsymbol{p}^N \rho^{(0)}. \label{master2}
\ee
On physical grounds one expects that, after many applications of a dynamical process, all eigenstates of the Hamiltonian should have an equal occupation:
\be
\lim_{N\rightarrow\infty} \rho^{(N)}_{m m} =\text{const.}=\frac{1}{\mathcal D},
\ee
where $\mathcal D$ is the dimensionality of the Hilbert space. The state characterized by $\rho_{m m} =\text{const.}$ is often called an ``infinite temperature state" since it is formally identical to a Gibbs distribution, $\rho_{nn} = e^{-\beta E_n}/Z$, in the limit $\beta\rightarrow 0$. The invariance of the infinite temperature state under doubly stochastic evolution trivially follows from the master equation. By substituting the infinite temperature state (which is the right eigenvector of ${\bf p}$ corresponding to $\lambda_0=1$) in the master equation, we obtain:
\be
\rho^{(N+1)}_{mm}=\sum_n p_{n\to m} \rho^{(N)}_{nn}=\frac{1}{\mathcal D}\sum_n p_{n\to m}=\frac{1}{\mathcal D}=\rho^{(N)}_{nn}.
\ee
In Appendix~\ref{app:inftematt}, we prove that the infinite temperature state is an attractor of the doubly stochastic evolution. The approach to the steady state is controlled, as discussed in Sec. \ref{sec:prop}, by the eigenvalues of ${\bf p}$ whose absolute value is smaller than one. 

Let us discuss in detail the three-level system depicted in Fig.~\ref{fig:double_stochastic}. Besides providing a concrete example of the approach to infinite temperature, this example clarifies under which conditions the system always relaxes to the infinite temperature state. Instead of considering the most general doubly stochastic evolution (which is discussed in appendix~\ref{appendix:birkhoff} in connection with Birkhoff's theorem), we assume that: (i) the process is cyclic (therefore we can drop tilde signs over eigenstate labels of the final Hamiltonian), (ii) the transition probabilities satisfy the detailed balance condition $p_{n\to m}=p_{m\to n}$, and (iii) the only nonzero transition probabilities are between states $1$ and $2$ ($p_{12}=\gamma_{12}$) and states $2$ and $3$ ($p_{23}=\gamma_{23}$). From probability conservation, we must have $p_{11}=1-\gamma_{12}$, $p_{22}=1-\gamma_{12}-\gamma_{23}$ and $p_{33}=1-\gamma_{23}$
\be
{\bf p} = \left(\begin{array}{ccc}
1-\gamma_{12}  & \gamma_{12} & 0 \\
\gamma_{12} & 1-\gamma_{12}-\gamma_{23} & \gamma_{23}  \\
0 & \gamma_{23}  & 1-\gamma_{23}.
\end{array}\right)
\ee
Note that ${\bf p}$ is symmetric because of the detailed balance condition. Its eigenvalues are:
\be
\lambda_0=1,\; \lambda_1= 1-\gamma_{12}-\gamma_{23}+\sqrt{\gamma_{12}^2+\gamma_{23}^2-\gamma_{12}\gamma_{23}},\; \lambda_2= 1-\gamma_{12}-\gamma_{23}-\sqrt{\gamma_{12}^2+\gamma_{23}^2-\gamma_{12}\gamma_{23}}
\ee
One can then see that unless either $\gamma_{12}=0$ or $\gamma_{23}=0$, that is, unless the transition matrix is block diagonal, $|\lambda_1|,|\lambda_2|<1$. As a result, for a repeated process, any probability distribution will relax to the eigenstate corresponding to the eigenvalue $\lambda_0=1$, which is nothing but the uniform probability distribution $(1/3, 1/3, 1/3)$.

\subsubsection{Increase of the Diagonal Entropy Under Doubly Stochastic Evolution\label{sec:diag_S}}

As shown above, any initial state evolving under a repeated doubly stochastic process approaches the ``infinite temperature state". This state is the one with the maximal spread in the eigenstates of any final Hamiltonian. As we discussed in Sec.~\ref{sec:chaos_delocalization}, a natural measure of the spreading of states, in the basis of a given Hamiltonian, is the diagonal entropy:
\be
S_d=-\sum_n \rho_{nn}\ln \rho_{nn}.
\ee
This entropy is maximized for the uniform occupation probability (which, as shown in Appendix~\ref{app:inftematt}, is an attractor) so one can anticipate that $S_d$ can only increase under doubly stochastic evolution.

The diagonal entropy has many interesting properties. For example, it coincides with the usual von Neumann entropy for stationary density matrices. In addition, the diagonal entropy can be viewed as the entropy of the time averaged density matrix. The diagonal entropy also sets a natural ``distance" between the density matrix $\rho$ and the infinite temperature density matrix. Indeed given two discrete distributions $P$ and $Q$ a natural distance between them, also known as the Kullback-Leibler (KL) divergence~\cite{kullback_leibler_51}, is\footnote{Some caution is needed here as the Kullback-Leibler divergence is not symmetric and does not satisfy the triangular inequalities. Therefore, it is not a distance in the metric sense.}
\be
D_{KL}(P||Q)=\sum_n P_n \ln(P_n/Q_n).
\ee
It is straightforward to see that this distance is non-negative and that it is zero only when the two distributions coincide, that is, only when $P_n=Q_n$ for all values of $n$. If we substitute $P_n\rightarrow\rho_{nn}$ and $Q_n\rightarrow{1/\mathcal D}$ then 
\be
D_{KL}(\rho_{nn}||\rho^{\infty})=S_\infty-S_d \ge 0,
\ee
where $S_\infty=\ln(\mathcal D)$ is the entropy of the infinite temperature state (the highest possible entropy). Therefore, an increase of the diagonal entropy is equivalent to decreasing the distance between the actual and the infinite temperature energy distributions. 

We prove next that doubly stochastic evolution leads to an increase of the diagonal entropy. First, recall that if a function is convex in a given interval then 
\be
f(x)\ge f(y)+(x-y)f'(y)
\ee for any $x,y$ in that interval. In particular, if we chose the function $f(x)=x\ln(x)$, which is convex for any $x\ge0$, we obtain 
\be
x\ln(x)- y\ln(y) \ge  (x-y)[\ln(y)+1].
\ee 
By replacing $x\rightarrow \rho_{nn}^{(0)}$ and $y\rightarrow \rho_{\tilde m \tilde m}^{(1)}$, we obtain
\be
\rho_{nn}^{(0)}\ln\rho_{nn}^{(0)}- \rho_{\tilde m \tilde m}^{(1)}\ln\rho_{\tilde m \tilde m}^{(1)} \ge \left(\rho_{nn}^{(0)}-\rho_{\tilde m\tilde m}^{(1)}\right) \left(\ln\rho_{\tilde m\tilde m}^{(1)}+1\right).
\ee
Multiplying both sides of the equation above by $p_{n\rightarrow \tilde m}$, and summing over $n$, leads to
\be
\sum_n \left(p_{n\rightarrow \tilde m}\,\rho_{nn}^{(0)}\ln\rho_{nn}^{(0)}\right) - \rho_{\tilde m \tilde m}^{(1)}\ln\rho_{\tilde m \tilde m}^{(1)} \ge 0,
\ee
where we have used that $\sum_n p_{n\rightarrow \tilde m}=1$ and $\sum_n \rho_{nn}^{(0)}\,p_{n\rightarrow \tilde m} = \rho_{\tilde m\tilde m}^{(1)}$. Finally, summing this inequality over $\tilde m$ and using that $\sum_{\tilde m} p_{n\rightarrow \tilde m} =1$, one obtains
\be
\Delta S_d \equiv S_d^{(1)}-S_d^{(0)} = \sum_n \rho_{nn}^{(0)}\ln\rho_{nn}^{(0)}- \sum_{\tilde m} \rho_{\tilde m \tilde m}^{(1)}\ln\rho_{\tilde m \tilde m}^{(1)} \ge 0.
\label{eq:DSge0}
\ee
This implies that, under \textit{any} doubly stochastic evolution, the diagonal entropy can only increase or stay constant. Hence, the distance from the uniform or infinite temperature distribution monotonically decreases or stays constant. It is interesting that this statement is not tied in any way to quantum chaos. For example, if we take an arbitrary two level system in a stationary state and apply any sequence of pulses then the diagonal entropy cannot decrease. This statement, however, does not hold if the initial state is nonstationary. In that case, the evolution is not doubly stochastic and the diagonal entropy can decrease.

The increase of the diagonal entropy under doubly stochastic evolution should be contrasted with the exact conservation of von Neumann's entropy $S_{\rm vn}=-{\rm Tr}\left[\hat \rho\ln\hat \rho\right]$ under \textit{any} unitary evolution. These two results do not contradict each other. In fact, the relation between these two results can be understood by considering a \textit{unitary} evolution of an initially \textit{stationary} density matrix. Then the following chain of relations hold:
\be
S_d(0)= S_\text{vn}(0)=S_\text{vn}(t)\le S_d(t) \;.
\ee
The first equality follows from the fact that, for a stationary density matrix, the von Neumann and the diagonal entropy are identical. The second equality reflects the obvious fact that under unitary evolution the von Neumann entropy is conserved. Finally, the last inequality follows from Eq.~\eqref{eq:DSge0}. This has a direct analogy in classical systems where Liouville's theorem conserves the volume in phase space while the total entropy of an isolated system increases or stays constant.

The fact that
\be
S_d(t)\ge S_d(0),
\label{eq:sd_unitary}
\ee
means that, under unitary evolution starting from a stationary state, the diagonal entropy at any time $t>0$ is larger than (or equal to) the initial diagonal entropy. This \textit{does not} mean that the diagonal entropy increases continuously in time, that is, in general it is not true that $S_d(t_2)\ge S_d(t_1)$ for $t_2>t_1>0$ because at intermediate times the system might retain coherence. A monotonic increase occurs if we consider repeated doubly stochastic processes, as discussed in Sec.~\ref{sec:prop}. One can also prove a more general statement without assuming any dephasing, namely, that if one waits for a fixed long time between two pulses the probability that the diagonal entropy increases in time is higher (exponentially higher for many particles) than the probability that it decreases. The proof of this statement is beyond the scope of this review and can be found in Ref.~\cite{ikeda_sukumichi_15}.

\subsubsection{The Second Law in the Kelvin Formulation for Passive Density Matrices\label{sec:thompson}}

As shown above, under repeated doubly stochastic evolution that starts from a stationary density matrix, the diagonal entropy increases until it reaches its maximum value, corresponding to an ``infinite temperature state".

Now, we take a step further and assume that the initial probabilities decrease monotonically in energy, that is, for any $n$ and $m$
\be
\left(\rho_{nn}^{(0)}-\rho_{mm}^{(0)}\right)(E_n-E_m)\leq 0,
\label{eq:passive}
\ee
where $E_n$ and $E_m$ are eigenenergies of the system. Relying on this assumption, one can prove that, for any doubly stochastic \textit{cyclic} evolution (in particular, for any cyclic unitary process), the energy of the system can only increase or stay constant~\cite{thirring_book, lenard_78}:
\be
\sum_n E_n \rho^{(1)}_{nn} \geq \sum_n E_n\rho_{nn}^{(0)}.
\ee
By energy conservation, this difference must be equal to the average work done on the system during the cyclic process
\be
\langle W \rangle=\sum_n  E_n \rho_{nn}^{(1)}-\sum_n E_n\rho_{nn}^{(0)}\geq 0 \;.
\label{eq:energy_change_ds}
\ee
Diagonal density matrices satisfying the inequality~\eqref{eq:passive} are termed passive~\cite{thirring_book} and are common. The Gibbs distribution for systems in thermal equilibrium, $\rho_{nn}=e^{-\beta E_n}/Z$, is a passive density matrix. Therefore, condition \eqref{eq:energy_change_ds} is quite general and can be interpreted as a manifestation of the second law of thermodynamics in Kelvin's formulation -- one cannot extract work from a closed equilibrium system by carrying out a cyclic process. As all the results in Sec.~\ref{sec:DS_only}, this statement is solely based on  doubly stochastic evolution and on the passivity of the initial density matrix (and therefore, applies to both integrable and nonintegrable systems). In Sec.~\ref{sec:sec7}, we show explicitly how it works for a single particle driven in a chaotic cavity.

The proof of Eq.~\eqref{eq:energy_change_ds} relies on the fact that any doubly stochastic evolution tends to make the occupation probabilities uniform. In the case of an initial passive density matrix, this process requires a transfer of probability from low- to high-energy states causing the energy of the system to increase. If a stronger detailed balance condition is satisfied, that is, $p_{n\to m}=p_{m\to n}$ for any $m,n$, then the proof becomes particularly simple~\cite{polkovnikov_08}:
\be
\langle W\rangle =\sum_{n,m} E_n p_{n\to m} [\rho_{mm}^{(0)}-\rho_{nn}^{(0)}]={1\over 2}
\sum_{n,m}  p_{n\to m} (E_n-E_m)[\rho_{mm}^{(0)}-\rho_{nn}^{(0)}]\geq 0.
\ee
The second equality follows from symmetrizing with respect to $n$ and $m$ and using the detailed balance condition. However, in general, pairwise transition probabilities are not the same and only the sum rule~(\ref{eq:double_stochastic}) is satisfied. In this case, the proof is more complicated but still straightforward~\cite{thirring_book}. For completeness, it is presented in Appendix~\ref{appendix:kelvin}.

\subsection{Implications of Doubly-Stochastic Evolution for Chaotic Systems\label{sec:ETH+DS}}

In the previous three subsections we discussed three important results that are all manifestations of the second law of thermodynamics. In Sec.~\ref{sec:inf_T} (and Appendix~\ref{app:inftematt}), we showed that the ``infinite temperature state" is the only generic attractor of a doubly stochastic evolution. In Sec.~\ref{sec:diag_S}, we proved that under a repeated doubly stochastic evolution the diagonal entropy increases until it reaches its maximum value, which corresponds to that of the ``infinite temperature state". Finally, in Sec.~\ref{sec:thompson}, we proved that any cyclic doubly stochastic evolution leads to an increase of the (average) energy of the system (provided the initial density matrix is passive). This statement is equivalent to the second law of thermodynamics in the Kelvin form.

All these statements rely only on doubly stochastic evolution and therefore apply to both integrable (and in particular noninteracting) and chaotic systems. In this section, we take a step further and assume that the system undergoing the dynamical process is chaotic. 

\subsubsection{The Diagonal Entropy and the Fundamental Thermodynamic Relation}\label{sec:fundamental_relation}

The entropy of a system in thermal equilibrium is a unique function of its energy and other relevant extensive variables (denoted by $\lambda$). This fact implies the fundamental thermodynamic relation
\begin{equation} \label{eq:fundamental_relation}
	dS=\frac{1}{T}\left( dE + F_\lambda d \lambda \right).
\end{equation}
Here, $F_\lambda$ is the generalized force conjugate to $\lambda$. Since this expression is directly derived from $S(E,\lambda)$, it applies to both reversible and irreversible processes. In the case of a reversible transformation, we can identify $TdS$ as the heat transferred and $F_\lambda d \lambda$ as the work done by the system. In contrast, for an irreversible transformation, one cannot make these identifications. Equation~\eqref{eq:fundamental_relation} is then taken as a mathematical relation between thermodynamic functions. For example, if an isolated gas in a container is expanded by a volume $d \lambda$ by moving a partition very quickly (a Joule experiment), there is no work done and the energy change in the system is zero. The fundamental relation then implies that the change in entropy is given by $dS=F_\lambda d \lambda /T$ with $F_\lambda$ being the pressure before the expansion. If one insists on giving an interpretation to the equation as describing a dynamical process, it can be thought of as a fictitious reversible process (expansion with work and heat exchange) that is not related to the actual (irreversible) process that has taken place. As we show in this and the following subsection, thinking about the fundamental relation from a microscopic point of view is illuminating. The changes in the entropy can actually be assigned to underlying, in general irreversible, physical processes (which may result from work and/or heat exchange). The entropy change is then simply related to transitions between the energy levels of the system during the dynamical process.  

From the microscopic point of view, the fundamental relation is not at all trivial (see, e.g., Ref.~\cite{balian_06}). The energy and its change are uniquely defined by the density matrix and the energy eigenstates. The generalized force is also expressed through the density matrix and the Hamiltonian. Thus, for the fundamental relation to apply microscopically, we need to define an object, the entropy, which can also be expressed through the density matrix and possibly the Hamiltonian and ensure that Eq.~(\ref{eq:fundamental_relation}) holds for any dynamical process both for open and isolated systems. Let us show that the diagonal entropy, which we defined earlier as the measure of delocalization in the energy space, satisfies the fundamental relation in chaotic systems~\cite{polkovnikov_11,santos_polkovnikov_11}. As we will see, $S_d$ satisfies
\be
dE=TdS_d-F_\lambda d\lambda,
\label{eq:fund_rel_diag_ent}
\ee
for both reversible and irreversible processes. Once we identify $S_d$ with the entropy, this constitutes the fundamental relation. 

To derive the fundamental relation, let us first use standard statistical mechanics and come back to the role of quantum chaos later. We assume that the initial density matrix is described by a Gibbs distribution (the extension to other ensembles is straightforward)
\be
\rho_{nm}(\lambda)={1\over Z(\lambda)} e^{-\beta E_n(\lambda)}\delta_{nm}.
\label{rho_gibbs}
\ee
Using that the energy of the system is given by $E(\lambda)=\sum_n \rho_{nn} E_n(\lambda)$, and calculating its change for an arbitrary `small' dynamical process (we are not assuming here that the system is isolated or that the process is unitary), we find
\be
dE(\lambda) =d\left(\sum_n \rho_{nn} E_n(\lambda)\right)=\sum_n \left[E_n(\lambda)  d\rho_{nn}+\rho_{nn} {dE_n\over d\lambda} d\lambda\right]\\
=\sum_n E_n (\lambda)d\rho_{nn}-F_\lambda d\lambda,
\label{eq:energy_change}
\ee
where $F_\lambda=-\sum_n\rho_{nn}dE_n(\lambda)/d\lambda$. Next, we compute the change in the diagonal entropy for the same process. This gives
\be\label{eq:entropy_change}
dS_d=-d\left[\sum_n \rho_{nn}\ln(\rho_{nn})\right]=-\sum_n d\rho_{nn} \ln(\rho_{nn})-\sum_n d\rho_{nn}=\beta\sum_n E_n(\lambda) d\rho_{nn},
\ee
where we used that, by conservation of probability, $\sum_n d\rho_{nn}=0$. Comparing Eqs.~\eqref{eq:energy_change} and \eqref{eq:entropy_change}, and noting that the generalized force can be also written as
\be\label{eq:genforce}
F_\lambda=-{\partial E(\lambda)\over \partial \lambda}\biggr|_{S_d},
\ee
we recover that the diagonal entropy indeed satisfies the fundamental thermodynamic relation, Eq.~\eqref{eq:fund_rel_diag_ent}, for any dynamical process. Remarkably, under the assumption that the system is initially described by the Gibbs distribution, the fundamental relation applies exactly for both large and small systems whether they are open or closed during the dynamical process. Moreover, it applies to integrable and nonintegrable systems alike.

To see where quantum chaos enters, assume that an isolated system undergoes a quench (or any other dynamic process) protocol. Then, according to ETH, physical observables after relaxation are described by an equilibrium thermal ensemble. This is true despite the fact that the density matrix of the entire system is not that of the Gibbs ensemble. With this in mind, we want to prove that if the system is chaotic then the fundamental relation holds up to possible subextensive corrections. The easiest way to prove it without assuming a standard equilibrium density matrix is to show that the diagonal entropy, which is a function of the density matrix, coincides with the thermodynamic entropy up to subextensive corrections. Then, the fundamental relation and its generalization immediately follows. Recall that we already presented numerical evidence that the diagonal entropy coincides with the thermodynamic entropy in Sec.~\ref{sec:chaos_delocalization}, when discussing implications of quantum chaos and RMT to delocalization in energy space (see Fig.~\ref{fig:santos2}).

We start our discussion by noticing that, for large system sizes (no matter whether they are in a pure or in a mixed state), the diagonal entropy can written as an integral over energies
\be
S_d\simeq-\int dE\, \Omega(E)\, \rho(E)\,\ln[\rho(E)],
\ee
where $\rho(E)$ is an interpolating function satisfying $\rho(E_n)=\rho_{nn}$ and $\Omega(E)$ is the smoothed many-body density of states. We note that $\rho(E)\Omega(E)=P(E)$ is the energy distribution function in the system [see Eq.~\eqref{eq:normenegdist}], from which all moments of the energy can be computed. For example, the average energy is given by $\la E\ra=\int dE\, E\, P(E)$. One can rewrite the diagonal entropy as
\be
S_d=-\int dE P(E) \ln\left[\frac{P(E)\delta E}{\Omega(E)\delta E}\right]=\int dE P(E) S_m(E) -\int dE P(E)\ln [P(E)\delta E],
\label{eq:Sd_int}
\ee
where $S_m(E)=\ln[ \Omega(E)\delta E]$ is the microcanonical entropy at energy $E$ ($\delta E$ is the width of the microcanonical energy window).

The last term in Eq.~\eqref{eq:Sd_int} is the one that exhibits a qualitatively different behavior in integrable and chaotic systems \cite{santos_polkovnikov_11}. In nonintegrable systems, one expects $P(E)$ to be a smooth function of the energy (see, Sec.~\ref{sec:chaos_delocalization}). As a result, $\int dE P(E)\ln P(E)$ is not extensive because $\int dE P(E)$ is normalized to one and the width of the energy distribution is not exponentially large in the system size (see Sec.~\ref{ss:quencheth}). In integrable systems, on the other hand, $P(E)$ after a dynamical process (such as a quench) generally exhibits large fluctuations (see, e.g., left panels in Fig.~\ref{fig:santos}). As a result, the last term in Eq.~\eqref{eq:Sd_int} can be extensive and, therefore, comparable to the contribution of the first term, that is, $S_d$ can differ from the thermodynamic entropy. Numerical studies have indeed found that $S_d$ agrees (disagrees) with the thermodynamic entropy in quenches in nonintegrable (integrable) systems \cite{santos_polkovnikov_11, rigol_14a, rigol_16} (for results at integrability, see Sec.~\ref{sec:XXZ}).

Actually, if $P(E)$ after a dynamical process is well approximated by a smooth Gaussian (expected for sufficiently large nonintegrable systems, see Sec.~\ref{sec:chaos_delocalization})
\be
P(E)\approx {1\over \sqrt{2\pi\sigma^2}} \exp\left[-\frac{(E-\la E\ra)^2}{2\sigma^2}\right],
\label{eq:PE_gaussian}
\ee
with $\sigma^2$ being, at most, extensive (for a discussion of $\sigma^2$ after a quench, see Sec.~\ref{ss:quencheth}), then the fact that $S_d$ agrees with the thermodynamic entropy follows straightforwardly. To show that, let us expand $S_m(E)$ around the mean energy $\la E\ra$
\be
S_m(E) \approx S_m(\la E\ra) + \left.\frac{\partial S_m(E)}{\partial E}\right|_{\la E\ra} (E-\la E\ra) +\frac{1}{2} \left.\frac{\partial^2 S_m(E)}{\partial E^2}\right|_{\la E\ra} (E-\la E\ra)^2 +\dots \ .
\label{eq:Sm_expansion}
\ee
By substituting Eqs.~\eqref{eq:PE_gaussian} and \eqref{eq:Sm_expansion} into Eq.~\eqref{eq:Sd_int}, and computing the Gaussian integrals, we obtain
\be
S_{d}\approx S_m(\la E\ra)-{1\over 2}\left({\sigma^2\over \sigma_c^2}-1\right),
\label{eq:result}
\ee
where $S_m(\la E\ra)=\ln [\Omega(\la E\ra)\sqrt{2\pi}\sigma]$ is the von Neumann entropy of a microcanonical distribution with mean energy $\la E\ra$ and energy width $\delta E=\sqrt{2\pi}\sigma$. In the expression above
\be
\sigma_c^{-2}=-\left.\frac{\partial\beta(E)}{\partial E}\right|_{\la E\ra},\quad \text{with} \quad \beta(E)= \frac{\partial S(E)}{\partial E}.
\ee
We note that here the inverse temperature $\beta(E)$ is defined solely by the density of states at energy $E$, and that $\sigma_c^2$ is the variance of the energy in a canonical ensemble with inverse temperature $\beta(\la E\ra)$.\footnote{We assume that we are not at a phase transition, at which $\sigma_c^{2}$ might diverge.} Since $\sigma_c^{2}$ is extensive, the last term in Eq.~\eqref{eq:result} is clearly non-extensive and can be ignored in large systems. We then see that, in chaotic systems, one can define a functional of the density matrix (the diagonal entropy) that coincides with the thermodynamic entropy both in open and closed systems after they are driven from equilibrium and allowed to relax. For closed systems, this is a nontrivial statement that relies on the assumption that the final Hamiltonian (the one after the dynamical process) is quantum chaotic (nonintegrable). From this result, the fundamental thermodynamic relation for chaotic systems follows without the assumption that the system is in thermal equilibrium.

To conclude this section, let us mention an apparent paradox that is frequently raised to argue that there is a deficiency in the diagonal entropy (or, for that matter, von Neumann's entropy) for quantum systems. A similar ``paradox'' can be argued to occur for Liouville's entropy for classical systems. If one starts a cyclic process from an eigenstate of an ergodic Hamiltonian, where von Neumann's entropy is zero by definition, after reaching the new equilibrium the energy change can be made arbitrarily small while the entropy change cannot. The latter will be the thermodynamic entropy. Hence, the equality $dE=TdS$ seems to be violated. There is, in fact, no paradox. The entropy of a single eigenstate is a singular quantity. Any arbitrarily small perturbation will immediately lead to mixing exponentially many eigenstates of the Hamiltonian and lead to the thermodynamic entropy (see, e.g., the discussion in Ref.~\cite{santos_polkovnikov_12}). In particular, any attempt to measure the temperature of the eigenstate, which is necessary to test the fundamental relation,  will introduce an extensive thermodynamic entropy and thus the paradox is immediately removed.

\subsubsection{The Fundamental Relation vs the First Law of Thermodynamics}

It is interesting to put the results presented in the previous subsection in the context of the first law of thermodynamics:
\be
dE=dQ+dW.
\ee
This law is a statement about energy conservation and implies that the energy of the system can change only due to heat (defined as the energy flow from one system to another at fixed macroscopic couplings) and work (defined as the energy change in the system due to a dynamical change of these couplings). Note that, as previously stressed, the fundamental relation~\eqref{eq:fundamental_relation} is a mathematical expression relating equilibrium quantities, while the first law only deals with the conservation of energy.

From the microscopic stand point it is convenient to split an infinitesimal energy change into two contributions [see Eq.~\eqref{eq:energy_change}]:
\be
dE=d\tilde Q+ dW_{\rm ad}.
\label{eq:tildeQ_def}
\ee
The first one, $d\tilde Q=\sum_n E_n d\rho_{nn}$, results from changes in occupation numbers of the microscopic energy levels (and is not to be confused with the common definition of heat in the first law) and the second one, $d W_{\rm ad}=\sum_n dE_n \rho_{nn}$, results from the changes of the energy spectrum at fixed occupation numbers. This last term, as we discussed, can be written as the full derivative of the energy (assuming that the energy spectrum is differentiable) and thus represents the adiabatic work done on the system, $dW_{\rm ad}= -F_\lambda d\lambda$ [see Eq.~\eqref{eq:genforce}]. If the dynamical process is infinitesimally slow, changing the macroscopic parameter of the system does not change occupation probabilities. Formally, one can prove this statement using adiabatic perturbation theory, similarly to what was done for energy in Ref.~\cite{dalessio_polkovnikov_14}. We thus see that for infinitesimally slow processes $d\tilde Q=T dS=d Q$ and $dW=dW_{\rm ad}$, which is well known from thermodynamics. We note that in large systems the strict quantum-mechanical adiabatic limit requires exponentially slow processes in order to suppress transitions between many-body eigenstates. This is another way of saying that isolated eigenstates are very fragile. Thermodynamic adiabaticity on the other hand requires that the dynamical process is slow with respect to physical time scales, which are much shorter than the inverse level spacing. This is of course consistent with Eq.~\eqref{eq:tildeQ_def} as transitions between nearest eigenstates lead to exponentially small heating and essentially do not contribute to $d\tilde Q$ and hence to $dE$. So the only way to have significant heating is to introduce transitions across exponentially many energy levels, which requires much faster dynamics.

The situation becomes somewhat different in a setup where the process is not infinitesimally slow, even if it is still effectively quasi-static. For example, one can imagine a compression and expansion of a piston containing a gas (see Fig.~\ref{fig:piston}) at some finite rate. At the end of the process, when the piston is back at its original position, the energy of the gas is higher than its initial energy. This is essentially the way microwave ovens function. There, the food heats up because of the non-adiabatic work performed by the time-dependent electromagnetic field. This process can still be quasi-static because if in each cycle the energy of the system increases by a small amount then it can be approximately described by local equilibrium. This ``microwave heating'' in ergodic systems is indistinguishable, at the end of the process, from conventional heating resulting from connecting the system to a thermal reservoir.  

Therefore, from a microscopic standpoint, it is more natural to define $d\tilde Q$ (and not $d Q$) as heat. In the literature, $d\tilde Q$ has been called heat \cite{polkovnikov_08}, excess heat \cite{kolodrubetz_clark_12}, excess energy \cite{bonnes_essler_14}, non-adiabatic work \cite{bunin_dalessio_11}, and others. We will not argue one way or another in terming $d\tilde Q$. We only note that, physically, it represents the heating of the system, that is, the energy change in the system caused by transitions between levels, independent of whether those transitions are induced by contact with another system or by changing non-adiabatically some coupling $\lambda$, or both. As we proved earlier [see Eq.~\eqref{eq:fund_rel_diag_ent}], for small changes in ergodic systems, one always has $d\tilde Q=TdS_d$, so this energy change is uniquely associated with the entropy change. If an isolated system starts in the stationary state then, as we proved, the entropy change is always non-negative and thus one always has $d\tilde Q\geq 0$. This, in turn, implies that $dW\geq dW_{\rm ad}$ (in the latter $dQ=0$), in agreement with the results in Sec.~\ref{sec:thompson}.\footnote{For processes that are not infinitesimally slow, in the presence of unavoidable level crossings, there can be exceptions where $dW<dW_{\rm ad}$ \cite{allahverdyan_nieuwenhuizen_05}} If the system is not closed, then, by definition,
\be
d\tilde Q=d Q+(dW-dW_{\rm ad}).
\ee
As a result, $d\tilde Q\geq dQ$ so that $TdS_d\geq dQ$, as expected from thermodynamics.

%%%%%%%%%%%%%%%%%%%%%%%%%%%%%%%%%%%%%%%%%%%%%%%%%%%%%%%%%%%%%%%%%%%%%%%%%%%%%%%%%%%%%%%%%%
\section{Quantum Chaos, Fluctuation Theorems, and Linear Response Relations\label{sec:sec6}}
%%%%%%%%%%%%%%%%%%%%%%%%%%%%%%%%%%%%%%%%%%%%%%%%%%%%%%%%%%%%%%%%%%%%%%%%%%%%%%%%%%%%%%%%%%

The recently discovered fluctuation theorems are remarkable \textit{equalities} involving thermodynamic variables. They are valid for systems initially prepared in equilibrium and then driven far from equilibrium in an arbitrary way (see, e.g., Refs.~\cite{campisi_hanggi_11,seifert_12} for reviews). These theorems effectively replace many thermodynamic inequalities by equalities (e.g., the second law of thermodynamics in the Kelvin form discussed previously). In many cases, the proof of the fluctuation theorems, as previously done for the fundamental relations, assumes that the initial state is described by a Gibbs distribution. When this is the case, one does not need any additional assumptions, such as quantum chaos. However, if the system is not weakly coupled to an equilibrium bath, then the assumption of a Gibbs distribution is often not justified and one has to rely on quantum chaos and ETH to prove these relations.

In this section, we derive the fluctuation theorems for individual eigenstates and hence extend them to arbitrary stationary distributions that are narrow in energy. Based on these fluctuation theorems, we derive energy drift-diffusion relations for both isolated and open systems, and discuss how they lead to nontrivial asymptotic energy distributions for driven isolated systems. For clarity, we derive these fluctuation relations in two ways. First we show a standard derivation for an initial Gibbs ensemble, and then, for quantum chaotic systems, we generalize this derivation to systems prepared in individual eigenstates. The latter approach clarifies in which situations fluctuation theorems apply to isolated systems. It also allows us to derive finite-size corrections and to extend them to open systems that are strongly coupled to a bath.

\subsection{Fluctuation Theorems\label{sec:JE_CT}}

Particularly simple proofs of fluctuation theorems are obtained by considering isolated quantum systems initially prepared in contact with a thermal bath at temperature $T$. The bath is then disconnected from the system which undergoes a unitary (or, more generally, doubly stochastic) evolution in response to an external protocol that changes some macroscopic parameter in time. The protocol has a specified duration after which the parameter is kept constant and the system is allowed to relax back to equilibrium.

Thermodynamics tell us that the average external work, $W$, done on the system during a thermodynamic protocol is bounded from below by the difference in the equilibrium free-energies (at the same temperature $T$) evaluated at the initial (A) and final (B) value of the control parameters. Specifically,
\be
\langle W\rangle \ge\Delta F\equiv F_{B,T}-F_{A,T}.\label{eq:clausius}
\ee
Because the system is isolated, there is no heat flowing to the system, and according to the first law of thermodynamics $\langle W\rangle=W_{\rm ad}+\tilde Q$, where $\tilde Q$  [introduced in Eq.~\eqref{eq:tildeQ_def} in the previous section] is the irreversible work or, microscopically, the energy change associated with the transitions between different energy levels. Then Eq.~\eqref{eq:clausius} becomes
\be
\tilde Q\geq 0.
\label{eq:meaning}
\ee

For a cyclic process, $W_{\rm ad}=0$. Therefore, $\langle W\rangle=\tilde Q$, and this inequality reduces to Kelvin's formulation of the second law. For an adiabatic process, $\tilde Q=0$, and the inequality~\eqref{eq:meaning} becomes an equality.

By properly taking into account the fluctuations, the Jarzynski equality turns the inequality \eqref{eq:clausius} into an equality even if the protocol drives the system far from equilibrium. This equality reads:
\be
\langle e^{-W/k_{B}T}\rangle=e^{-\Delta F/k_{B}T},\label{eq:JE}
\ee
where the angular brackets denote the average over many experimental realizations of the same protocol. In particular, for cyclic processes (for which $A=B$), the Jarzynski equality reduces to 
\be\label{eq:BK}
\langle e^{-W/k_{B}T}\rangle=1,
\ee
which was first discovered by Bochkov and Kuzovlev~\cite{bochkov_kuzovlev_77, bochkov_kuzovlev_81} as a nonlinear generalization of the fluctuation-dissipation theorem. Equation~\eqref{eq:BK} is frequently referred to as the Bochkov-Kuzovlev work-fluctuation theorem~\cite{campisi_hanggi_11}.

To clarify the meaning of Eq.~\eqref{eq:JE}, it is better to refer to a concrete example, say, the compression of a gas by a moving piston (see Fig.~\ref{fig:piston}). Initially, the gas is assumed to be in thermal equilibrium connected to a bath, with the piston at position $z(0)=A$. We assume that, during the protocol $z(t)$, the system is not connected the bath anymore, i.e., it can be regarded as isolated. At the end of the protocol, the control parameter reaches the value $z(t)=B$. We record the external work $W$, which is formally defined as the energy change of the piston. Note that the free energy $F_{B,T}$ is not the free energy of the system after the protocol. It is rather the equilibrium free energy evaluated at the initial temperature and the final value of the control parameter.\footnote{It would be the free energy of the system if it is reconnected to the same bath and allowed to re-equilibrate. Since no work is done on the system during its re-equilibration, such a process does not affect the Jarzynski equality.} Upon repeating the protocol many times, the work will fluctuate around some average value. The Jarzynski equality \eqref{eq:JE} states that the exponential of the work done, averaged over many realizations of the same experiment, is equal to the exponential of the equilibrium free energy difference. Hence, the Jarzynski equality connects a dynamical quantity, work, which depends on the details of the protocol, and an equilibrium quantity, the free energy difference, which only depends on the initial and final values of the control parameter. In particular, this relation can be used to measure free energy differences in small systems by measuring the work. In large systems, the Jarzynski relation is generally not very useful unless $W$ is small. This because the average of the function $\exp[-\beta W]$ will be dominated by rare events in which the work is negative. The assumption that the system is not connected to the bath during the protocol, which was present in the original work of Jarzynski \cite{jarzynski_97} and which caused some confusion, is not necessary (see Ref.~\cite{jarzynski_11} and the proof below). 

Equation~\eqref{eq:JE} can be understood as a constraint on the work distribution $P\left(W\right)$. This constraint is independent of the details of the protocol $z(t)$, and depends only on the initial and final values of the macroscopic parameter, $A$ and $B$, respectively, through the free energy difference $\Delta F\equiv F_{B,T}-F_{A,T}$. Note that the full distribution $P(W)$ depends on the details of $z(t)$.  If we take the logarithm of Eq.~\eqref{eq:JE}, and perform the cumulant expansion, we obtain the expression:
\be
\sum_{n\geq 1} {(-1)^n\over n! }{ \langle W^n\rangle_c\over (k_B T)^n}=-{\Delta F\over k_B T}.
\label{eq:jarzynski_cumulant}
\ee
We therefore see that the Jarzynski equality constraints different cumulants of the work. If the work is small, or if the temperature is high, then only a few cumulants effectively enter the sum. When this happens, the constraint has an important consequence and leads to standard  thermodynamic relations. However, if many cumulants contribute to the expansion (as expected when $W$ is large), then the constraint does not place strong restrictions to the moments of the work. 

Note that, by combining the Jarzynski equality \eqref{eq:JE} and Jensen's inequality $\langle\exp[ x]\rangle\ge\exp[\langle x\rangle]$, we recover the Clausius inequality~\eqref{eq:clausius}. Let us emphasize that this inequality only applies to the average work. The work carried out in a single realization of the experiment can be smaller than $\Delta F$. This can be the case especially in small systems, where fluctuations are large. The Jarzynski equality allows one to estimate the likelihood of such rare events~\cite{jarzynski_11}.

Closely related to the Jarzynski equality is the Crooks theorem, which was originally formulated for classical systems~\cite{crooks_98} and then extended to quantum systems~\cite{kurchan_00, tasaki_00}. The Crooks theorem relates the probability of performing a work $W$ during the forward process, $P_{F}(W)$ ($A\rightarrow B$), with the probability of performing a work $-W$ during the reverse process, $P_{R}(-W)$ ($B\rightarrow A$, in a time-reversed manner):
\be
\frac{P_{F}\left( W\right)}{P_{R}\left(- W\right)}=e^{\left(W-\Delta F\right)/k_B T}.\label{eq:Crooks}
\ee
The expression above can be interpreted as a symmetry of the distribution function between the forward and the reverse process. Note that this symmetry is with respect to $W=0$ and not with respect to the average work. Rewriting the Crooks relation as
\be
P_{F}\left(W\right)e^{-W/k_B T}=P_{R}\left(-W\right)e^{-\Delta F/k_B T},
\ee
and integrating over $W$, one recovers the Jarzynski equality.

\subsubsection{Fluctuation Theorems for Systems Starting from a Gibbs State\label{sec:fluct_gibbs}}

Here we derive the Jarzynski equality \eqref{eq:JE} and the Crooks theorem~\eqref{eq:Crooks} for a system that is initially in a Gibbs state and is not coupled to a bath during its evolution. In this case, neither quantum chaos nor the limit of a large system size need to be invoked. The derivation relies exclusively on the symmetry of the doubly stochastic evolution between the forward and the reverse process [see Eq.~\eqref{detailed_bal}]:
\be
p^r_{\tilde m\to n}=p_{n\to \tilde m}.
\ee
We recall, that a doubly stochastic evolution describes unitary dynamics of systems starting from stationary states and extends to some non-unitary processes involving projective measurements or dephasing.

Let us consider a system prepared in a state characterized by an initial energy $E_{A}$, corresponding to an initial value of the control parameter $\lambda(0)\equiv \lambda_A$, drawn from the Gibbs ensemble of the initial Hamiltonian. Then the system undergoes an arbitrary dynamical process described by a doubly stochastic evolution. At the end of the process, the system has an energy $E_B$, which is a random variable. The fluctuating work\footnote{There is an active discussion on how to define work in quantum systems, with many conflicting definitions. We use the definition due to  Kurchan~\cite{kurchan_00} and Tasaki \cite{tasaki_00}, which has a transparent physical meaning, namely, the energy change in the system. By energy conservation, $W$ is also the energy change of the macroscopic degree of freedom associated with the control parameter.} is formally defined as $W=E_B-E_A$~\cite{kurchan_00, tasaki_00}. This work is characterized by the probability distribution:
\be
P_F(W)=\sum_{n,\tilde m}\,\rho_{nn}^{(0)}\,p_{n\to \tilde m}\delta(\tilde{E}_{\tilde{m}}-E_n-W) = \sum_{n,\tilde m}\,\frac{{\rm e}^{-\beta E_n}}{Z_A}\,p_{n\to \tilde m}\,\delta(\tilde E_{\tilde{m}}-E_n-W),
\label{eq:forward1}
\ee
where $n$ and $E_n$ ($\tilde m$ and $\tilde E_{\tilde m}$) refer to states and the spectrum of the initial (final) Hamiltonian, and $Z_A$ is the partition function associated with the initial value of the control parameter $\lambda_A$. The probability of performing work $-W$ during the reverse process, starting from a Gibbs distribution, is
\beq
P_R(-W)&=&\sum_{n,\tilde m}\,{{\rm e}^{-\beta \tilde E_{\tilde m}}\over Z_B} p^r_{\tilde m\to n}\,\delta(E_n-\tilde E_{\tilde m}+W)\nonumber\\ &=& \sum_{n,\tilde m}\,\frac{{\rm e}^{-\beta (E_n+ W)}}{Z_B}\,p_{n \to \tilde m}\,\delta(\tilde E_{\tilde m}-E_n-W)
= P_F(W)\,e^{-\beta W}\,\frac{Z_A}{Z_B},\label{eq:dct}
\eeq
where $Z_B$ is the partition function associated with the thermal equilibrium distribution at the final value of the control parameter, that is, $\lambda_B$. Writing the free energy difference as $\Delta F=F_B(T)-F_A(T)$, and using the relation between free energy and partition function, that is, $Z_{A,B}=e^{-\beta F_{A,B}}$, one sees that Eq.~\eqref{eq:dct} is nothing but the Crooks theorem~\eqref{eq:Crooks}. 

For cyclic symmetric protocols, for which the reverse process is identical to the forward process, the Crooks theorem simplifies to
\be\label{eq:symmprocro}
P(W) e^{-\beta W}= P(-W),
\ee
where we suppressed the indexes $F$ and $R$ since, in this case, they are redundant. In turn, this relation can be recast in the form of a symmetry relation for the cumulant generating function $G(\zeta)$:
\be
G(\zeta)=\ln\left[\int dW P(W) e^{-\zeta W}\right]=\sum_{n=1}^\infty \langle W^n\rangle_c {(-\zeta)^n\over n!}.
\ee
To see this we multiply both sides of Eq.~\eqref{eq:symmprocro} by $e^{\zeta W}$:
\be
P(W) e^{-\beta W} e^{\zeta W}= P(-W) e^{\zeta W},
\ee
and integrate over $W$ to obtain 
\be
G(\beta-\zeta)=G(\zeta).
\label{eq:symmetry_G}
\ee
The generating function formalism is a convenient tool for deriving various linear response relations, for example, Onsager relations and their nonlinear generalizations (see Ref.~\cite{andrieux_gaspard_07} and the discussion below).

\subsubsection{Fluctuation Theorems for Quantum Chaotic Systems\label{sec:Jarzynski_equality_chaos}}

Now, let us focus on eigenstates of many-body chaotic Hamiltonians and derive the corresponding fluctuation relations for isolated systems. The applicability of fluctuation relations to individual eigenstates allows one to extend them to arbitrary initial stationary distributions so long as they are sufficiently narrow. The approach based on individual eigenstates also allows us to derive the leading finite-size corrections to the cumulant expansion of these relations and prove these relations for open systems, even if they are strongly coupled to the bath throughout the dynamical process.

Let us analyze the probability of doing work $W$ during the forward process starting from a given many-body energy eigenstate $|n\rangle$. By definition, this is given by 
\beq
P_F(E_n\to E_n+W)&\equiv& P_F(E_n, W)=\sum_{\tilde m}\,p_{n\to \tilde m}\,\delta(\tilde E_{\tilde m}-E_n-W) \nonumber\\&=&\int\,d \tilde E\,\Omega_{B}(\tilde E)\,p (E_n\rightarrow \tilde E)\,\delta(\tilde E-E_n-W),
\eeq
where $\Omega_B$ is the density of states at the final value of the control parameter and we used the fact that, for chaotic systems, the probability $p_{n\to \tilde m}\approx p(E_n\rightarrow \tilde E)$ is a smooth function of the energy $\tilde E_{\tilde m}$, up to a small Gaussian noise (c.f., Sec.~\ref{sec:chaos_delocalization}). In non-chaotic systems, the transition probability $p_{n\to \tilde m}$ can fluctuate strongly between states that are close in energy, that is, changing the summation over $\tilde m$ by an integration over $\tilde E$ is not justified and, in general, is not valid. Integrating the expression above over the energy, we find
\be
P_F(E_n\to E_n+W) =  p (E_n\rightarrow E_n+W)\, \Omega_B(E_n+W).
\label{eq:forward}
\ee

\begin{figure}[!t]
\begin{center}
 \includegraphics[width=0.5\textwidth]{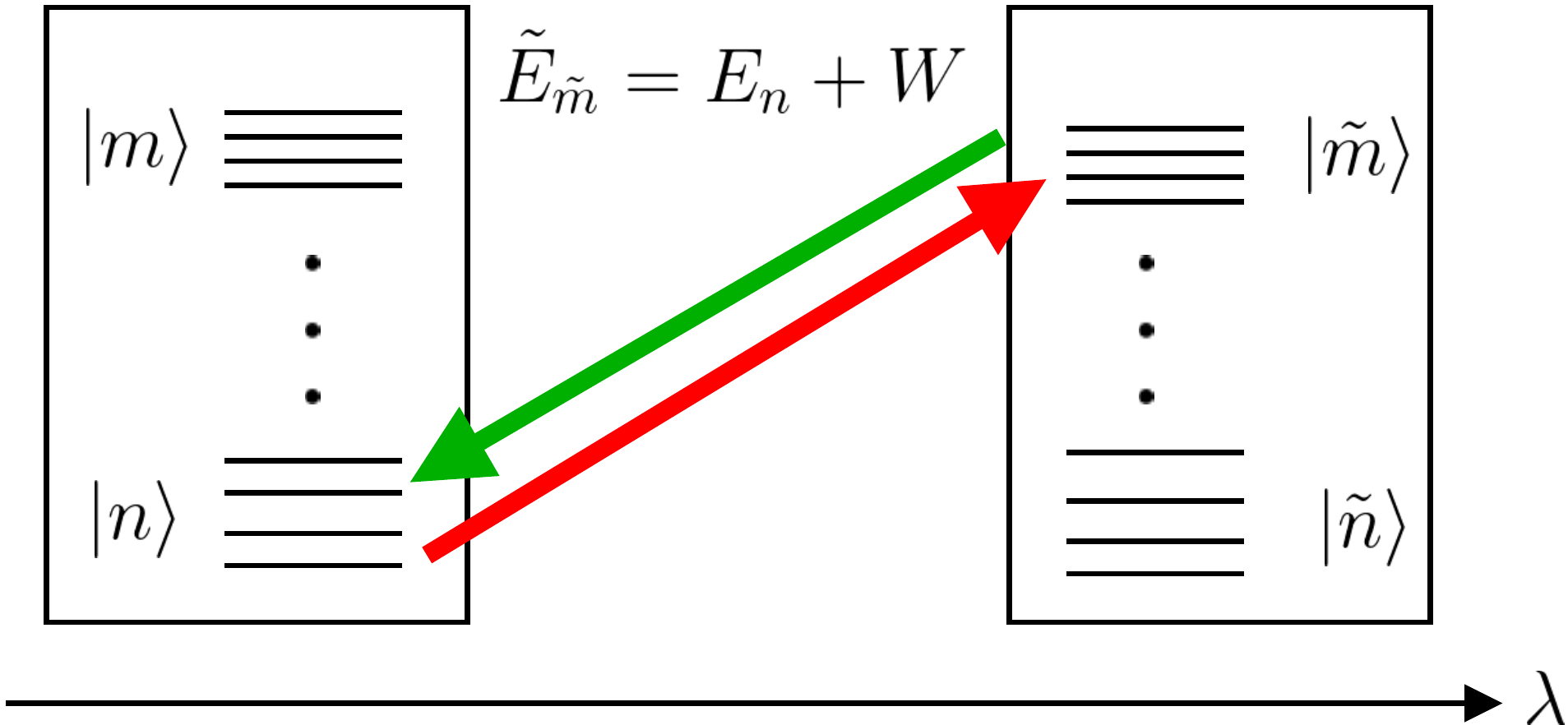}
\end{center}
\vspace{-0.15cm}
\caption{Illustration of relation~\eqref{eq:raw_result}. A system is composed of many closely spaced energy levels. In response to some dynamic process, during the forward process (red arrow) the system undergoes a transition from an initial state $|n\rangle$ to a final state $|\tilde m\rangle$ with the energy $\tilde E_{\tilde m}=E_n+W$. During the reverse process (green arrow), the opposite transition happens. The ratio of probabilities of doing work $W$ for the red process and $-W$ for the green process is given by the ratio of the density of states of the corresponding final states [see Eq.~\eqref{eq:raw_result}].}
\label{fig:Crooks}
\end{figure}

Using similar considerations, we find that the transition probability for doing work $-W$ during the reverse process starting from state $|\tilde m\rangle$ is:
\beq
&&P_R(\tilde E_{\tilde m }\rightarrow \tilde E_{\tilde m }-W)\equiv P_R(\tilde E_{\tilde m},-W)=\sum_n\,p^r_{\tilde m\to n}\,\delta(E_n-\tilde E_{\tilde m}+W)\nonumber\\ &&\qquad\qquad= \sum_n\,p_{n\to \tilde m}\,\delta(E_n-\tilde E_{\tilde m}+W) 
  =\int\,d E\,\Omega_A(E)\,p (E\rightarrow \tilde E_m)\,\delta(E-\tilde E_{\tilde m}+W)\nonumber\\ &&\qquad\qquad= p(\tilde E_{\tilde m}-W\to \tilde E_{\tilde m}) \Omega_A(\tilde E_{\tilde m}-W).
\label{eq:all_steps}
\eeq
Comparing the expressions for the forward and backward processes, and substituting $E_n\rightarrow E$ and $\tilde E_{\tilde m}\rightarrow E+W$, we obtain
\be
{P_F(E,W)\over P_R(E+W,-W)}={\Omega_B(E+W)\over \Omega_A(E)}\equiv \mathrm e^{S_B(E+W)-S_A(E)},
\label{eq:raw_result}
\ee
which is known as the Evans-Searles fluctuation relation \cite{evans_searles_94} (for classical derivations of a similar nature, see Ref.~\cite{pradhan_kafri_levine_08}). This result tells us that the ratio of probabilities of doing work $W$ in the forward process and $-W$ in the reverse process is simply equal to the ratio of the final and initial densities of states, that is, the ratio of the number of available microstates. Typically, as schematically illustrated in Fig.~\ref{fig:Crooks}, the density of states is an exponentially increasing function of the energy (corresponding to positive temperature states). Hence, the number of available microstates is larger for processes corresponding to an energy increase. This asymmetry is the microscopic origin of the higher probability of doing positive work, despite the equivalence of the forward and backward microscopic rates. Note that these considerations are only valid for chaotic systems. For integrable systems, not all microstates might be accessible and one needs to refine the argument.

Relation~\eqref{eq:raw_result} is very general. In particular, it contains fluctuation theorems and, in a sense, generalizes them. To see this, consider a total system with $N$ particles and let us assume that a dynamical process is applied only to a small subsystem with $N_1$ particles, such that $N_1$ is kept fixed as $N\to\infty$ ($N_1$ can be arbitrarily small, e.g., just one particle, or can be macroscopic). If the subsystem is weakly coupled to the rest of the system, it is expected that the rest of the system will act as a thermal bath and, as discussed previously, from ETH we expect that the small subsystem is described by the Gibbs ensemble. In this case, we are back to the standard setup considered in Sec.~\ref{sec:fluct_gibbs}. On the contrary, when the subsystem is strongly coupled to the rest of the system the assumption about the Gibbs distribution is not justified. However, even in this case, the Crooks theorem~\eqref{eq:Crooks} (and, hence, the Jarzynski equality) still applies. The proof is straightforward. Since $W$ is at most proportional to $N_1$ and it is therefore non-extensive in $N$, we can expand the entropy in Eq.~\eqref{eq:raw_result} to the leading order in $W$:
\begin{multline}
S_B(E+W)-S_A(E)=\frac{\partial S_B}{\partial E}W+S_B(E)-S_A(E)+\mathcal O(N_1/N) =\beta W-\beta \Delta F+\mathcal O(N_1/N),
\end{multline}
where we used the standard thermodynamic result that at constant temperature
\be
-\beta \Delta F =S(E,\lambda_B)-S(E,\lambda_A)\equiv S_B(E)- S_A(E).
\ee
Therefore, up to $N_1/N$ corrections, Eq.~\eqref{eq:raw_result} implies that
\be
{P_F(E,W)\over P_R(E+W,-W)}= \mathrm e^{\beta (W-\Delta F)},
\ee
irrespective of whether the subsystem is described by the Gibbs ensemble or not. Note that the free energy difference here is that of the whole system. It becomes the free energy difference in the subsystem only for a weak coupling between the subsystem and the rest of the system.

If the work $W$ is small but extensive, for example, coming from a global protocol, or if one does not take the thermodynamic limit, one can still perform a Taylor expansion of the entropy in Eq.~\eqref{eq:raw_result} in powers of $W$. Then, corrections to the Crooks relation will be generally finite. In particular, in Sec.~\ref{linear_response}, we will discuss how such corrections affect the Einstein drift-diffusion relation for the energy current (as occurs, for example, in microwave heating) of a driven isolated system, and to the Onsager relations. In what follows, we discuss various implications of Eq.~\eqref{eq:raw_result} to setups involving both isolated and open systems.

\subsection{Detailed Balance for Open Systems\label{detailed_balance_open_systems}}

\begin{figure}[!t]
\begin{center}
 \includegraphics[width=0.6\textwidth]{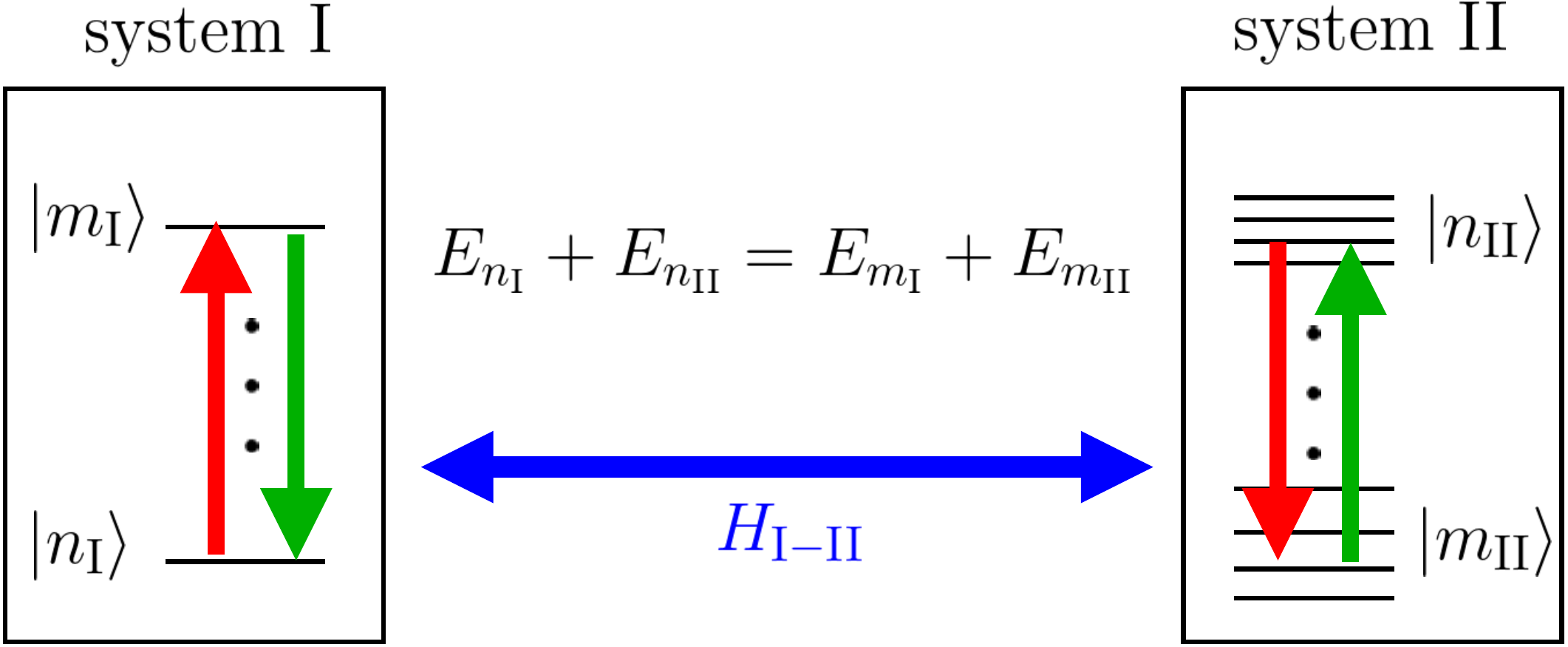}
\end{center}
\vspace{-0.15cm}
\caption{Schematic representation of the setup considered to prove the detailed balance condition, Eq.~\eqref{eq:detailed_balance}. A system $\rm I$, which can be either quantum chaotic or not is coupled to another system $\rm II$ (bath), which is quantum chaotic. The two systems are weakly coupled and are allowed to exchange energy. The microscopic transition probabilities from a pair of states $|n_{\rm I}, n_{\rm II}\rangle$ to another pair of states $|m_{\rm I},m_{\rm II}\rangle$ (red arrows) is the same as the probability of the reverse process (green arrows). However, if one is only interested in transition probabilities in system $\rm I$, irrespective of the outcome in the bath, this symmetry is broken and one obtains the standard equilibrium detailed balance.}
\label{fig:detailed_balance}
\end{figure}

Let us use the general relation~\eqref{eq:raw_result} to derive the familiar detailed balance condition for open systems. Imagine we have two weakly coupled systems $\rm I$ and $\rm II$. We do not need to make any assumption about system $\rm I$. It can be arbitrarily small, for example, consisting of one spin, and can be either integrable or ergodic. We assume that system $\rm II$ is quantum chaotic. For convenience, we call system $\rm II$ a bath, but we emphasize that we do not assume that system $\rm II$ is much bigger than system $\rm I$. The weak coupling assumption is made explicit by writing the Hamiltonian of the entire system as
\be
H_\text{tot}=H_{\rm I}+H_{\rm II}+\gamma H_{I-II},
\ee
with $\gamma$ small. Initially, the two systems are prepared in a stationary state with respect to the uncoupled Hamiltonian $H_{\rm I}+H_{\rm II}$ and then they are allowed to interact. The initial state can be the product of two Gibbs states with different temperatures or can be the product of two eigenstates. Because of the coupling, energy is allowed to flow between the two systems. If the coupling is weak, then the sum of the energies of systems I and II is (approximately) conserved. This implies that only microscopic transitions between states $|n_{\rm I}, n_{\rm II}\rangle$ and $|m_{\rm I}, m_{\rm II}\rangle$ satisfying the energy conservation $E^{\rm I}_{n_{\rm I}}+E^{\rm II}_{n_{\rm II}}=E^{\rm I}_{m_{\rm I}}+E^{\rm II}_{m_{\rm II}}$ are allowed (see the red arrows in Fig.~\ref{fig:detailed_balance}). This setup is formally equivalent to a quench in the coupling $\gamma$. Therefore, the transition probabilities are doubly stochastic. Moreover, as we discussed in the previous section, within the Fermi golden rule (or because this quench can be viewed as a time-symmetric protocol) these probabilities satisfy the stronger global detailed balance condition
\be
p_{n_{\rm I},n_{\rm II}\to m_{\rm I}, m_{\rm II}}=p_{m_{\rm I},m_{\rm II}\to n_{\rm I}, n_{\rm II}}.
\label{eq:global_detailed_balance}
\ee
Suppose that we are interested in the transition between microstates only in system $\rm I$, irrespective of the outcomes in the bath $\rm II$. Hence, we have to sum over all final states of the bath. Using that in quantum chaotic systems the transition probabilities are the same for nearby eigenstates
\be
p_{\rm I}(n_{\rm I}\to m_{\rm I})=\sum_{m_{\rm II}} p_{n_{\rm I},n_{\rm II}\to m_{\rm I}, m_{\rm II}}= p_{n_{\rm I},n_{\rm II}\to m_{\rm I}, m_{\rm II}} \Omega_{\rm II} (E^{\rm II}_{m_{\rm II}}).
\ee
This equation is a direct analogue of Eq.~\eqref{eq:forward}, with the solely difference being that only the density of states of the bath enters the RHS, as the sum is carried out over the final states of the bath. For the reverse process (green arrows in Fig.~\ref{fig:detailed_balance}), using the same arguments, we find
\be
p_{\rm I}(m_{\rm I}\to n_{\rm I})= p_{m_{\rm I},m_{\rm II}\to n_{\rm I}, n_{\rm II}} \Omega_{\rm II} (E^{\rm II}_{n_{\rm II}})
\ee
Comparing these two results, using the conservation of energy together with the global detailed balance condition, and simplifying notations $E_{n_{\rm II}}^{\rm II}\to E^{\rm II}$~\eqref{eq:global_detailed_balance}, we find
\be
{p_{\rm I}(n_{\rm I}\to m_{\rm I})\over p_{\rm I}(m_{\rm I}\to n_{\rm I})}={\Omega_{\rm II} (E^{\rm II}+E^{\rm I}_{n_{\rm I}}-E^{\rm I}_{m_{\rm I}})\over \Omega_{\rm II}(E^{\rm II})}=\mathrm e^{S_{\rm II} (E^{\rm II}+E^{\rm I}_{n_{\rm I}}-E^{\rm I}_{m_{\rm I}})-S_{\rm II} (E^{\rm II})}.
\ee
If the bath is large, or if the energy change $\delta E^{\rm I}_{m n}\equiv E^{\rm I}_{m_{\rm I}}-E^{\rm I}_{n_{\rm I}}$ is small compared to $E^{\rm II}$, as before, one can expand the entropy difference in the energy change. By doing that, one recovers the detailed balance condition in its most familiar form \cite{Krapivsky_2010}:
\be
{p_{\rm I}(n_{\rm I}\to m_{\rm I})\over p_{\rm I}(m_{\rm I}\to n_{\rm I})}\approx \mathrm e^{-\beta_{\rm II}(E^{\rm II}) \, \delta E^{\rm I}_{m n}},
\label{eq:detailed_balance}
\ee
where $\beta_{\rm II}(E^{\rm II})$ is the temperature of the bath corresponding to the energy $E^{\rm II}$.

\subsection{Einstein's Energy Drift-Diffusion Relations for Isolated and Open Systems\label{linear_response}}

In Sec.~\ref{sec:JE_CT}, we showed that the Jarzynski equality can be viewed as a constraint on the cumulant expansion of the work generating function [see Eq.~\eqref{eq:jarzynski_cumulant}]. Consider, for example, the Jarzynski equality for a cyclic process applied to a small subsystem of a large system or to a system initialized in a Gibbs state. In these cases, the Jarzynski equality is exact and, given the fact that the process is cyclic, $\Delta F=0$, so that 
\be
\langle e^{-\beta W} \rangle = 1.
\label{eq:JE2}
\ee
Next, let us perform a cumulant expansion of this equality:
\begin{equation}
0=\ln \langle e^{-\beta W} \rangle \; \; = \; \; -\beta \langle W \rangle +\frac{1}{2}\beta^2 \langle W^2 \rangle_c+\frac{1}{6}\beta^3 \langle W^3 \rangle_c + \ldots\;\;.
\label{eq:JE2a}
\end{equation}
where $\langle W^2 \rangle_c=\langle W^2 \rangle-\langle W \rangle^2$ and  $\langle W^3\rangle_c=\langle W^3\rangle - 3 \langle W^2\rangle \langle W\rangle+2\langle W\rangle^3$. If the average work performed is small and its distribution is close to Gaussian or if temperature is high, then cumulants of order three and higher in the expansion above can be neglected leading to:
\be
0 \approx -\beta \langle W \rangle +\frac{1}{2}\beta^2 \langle W^2 \rangle_c \Rightarrow \langle W \rangle = {\beta\over 2} \langle W^2 \rangle_c \,.
\label{eq:fdis}
\ee
The relation above can be interpreted as a fluctuation-dissipation relation for a system coupled to an external noise. $\langle W\rangle$ is the average work characterizing energy dissipation in the system and $\langle W^2\rangle_c$ characterizes the uncertainty in the work. Equation~\eqref{eq:fdis} can also be interpreted as an analogue of Einstein's drift-diffusion relation for the energy. Indeed, $\langle W\rangle$ is the average work done on the system during the cyclic process, which is the energy drift, and $\langle W^2\rangle_c$ represents the work fluctuations, which is the energy diffusion.

Next we consider a system prepared in a single eigenstate (hence, the results equally apply to a setup where we start from an arbitrary narrow stationary distribution). We focus on the general expression for the work probability distribution, Eq.~\eqref{eq:raw_result}. For simplicity, we focus once again on cyclic\footnote{If the protocol is not cyclic the total work $W$ is the sum of the adiabatic work $W_{ad}$, which is not fluctuating, and the non-adiabatic work $\tilde Q$ which is fluctuating. Then, the following derivation remains valid if one identifies $W$ with $\tilde Q$.} and symmetric protocols.\footnote{If the protocol is not symmetric one has to distinguish between forward and reverse processes.} Then 
\be
P(E, W)\, \mathrm e^{-S(E+W)+S(E)}  = P(E+W,-W). 
\label{eq:raw_cyclic}
\ee
In order to proceed further, let us assume that the work $W$ is small (though it can be extensive).  Then, the probability $P(E,W)$ is a slow function of the first argument and a fast function of the second argument $W$. Similarly, the entropy is a slow function of $W$. Formally, the RHS of the equation above can be rewritten as 
\be
P(E+W,-W)=\mathrm e^{W\partial_E} P(E,-W),
\ee
where we used the notation $\partial_E=\partial/\partial E$. Finally, expanding the entropy $S(E+W)$ in Eq.~\eqref{eq:raw_cyclic} to second order in $W$, and integrating over $W$, we find
\be
\langle \mathrm e^{-\beta W-\frac{W^2}{2} \frac{\partial \beta}{\partial E}} \rangle_{E}\approx \langle \mathrm e^{-W \partial_E} \rangle_E,
\ee
where $\langle\dots\rangle_E$ means that an average is taken with respect to the probability distribution $P(E,W)$ at fixed $E$. Taking the logarithm of both sides of this relation, and performing the cumulant expansion to the second order, one finds
\be\label{eq:cumfirsteq}
-\beta \langle W \rangle_E + \frac{1}{2}{\beta^2}\langle W^2\rangle_{E,c}-\frac{1}{2}{\partial_E\beta}\,\langle W^2\rangle_E\approx -\partial_E \langle W\rangle_E+{1\over 2} \left[\partial^2_{EE} \langle W^2\rangle_E-(\partial_E \langle W\rangle_E)^2\right].
\ee
Note that $\langle W^2 \rangle_E=\langle W^2\rangle_{E,c}+\langle W\rangle_E^2$, where the first term on the RHS is linear and the second is quadratic in cumulants. By equating all linear terms in the cumulants\footnote{\label{quadratic}Formally, this can be justified by introducing some parameter $\varepsilon$ such that all cumulants are linear in $\varepsilon$. This parameter can be, for example, duration of the pulse $\varepsilon=dt$ or the size of the subsystem that is coupled to the driving term. Since the relation~\eqref{eq:raw_cyclic} is valid in all orders in $\varepsilon$, it is sufficient to verify it only to linear order.} Eq.~\eqref{eq:cumfirsteq} reduces to:
\be
-\beta \langle W \rangle_E + \frac{1}{2} \beta^2\langle W^2 \rangle_{E,c} -\frac{1}{2} \partial_E \beta\, \langle W^2 \rangle_{E,c}  \approx -\partial_E \langle W \rangle_E + \frac{1}{2} \partial^2_{EE} \langle W^2 \rangle_{E,c}\,.
\label{eq:correction}
\ee
Using that 
\be
{\partial_E\beta}\, \langle W^2\rangle_{E,c}=\partial_E (\beta \langle W^2\rangle_{E,c})-\beta \partial_E \langle W^2\rangle_{E,c}\,,
\ee
and regrouping all the terms, one finds that the equation above simplifies to
\be
-\beta \left(\langle W\rangle_E-{\beta \over 2}\langle W^2\rangle_{E,c}-{1\over 2} \partial_E \langle W^2\rangle_{E,c}\right)+\partial_E \left(\langle W\rangle_E-{\beta\over 2} \langle W^2\rangle_{E,c}-{1\over 2}\partial_E \langle W^2\rangle_{E,c}\right)=0.
\label{eq:intermediate}
\ee
Therefore, the first and the second cumulant have to satisfy the relation:
\be
\langle W\rangle_E={\beta \over 2}\langle W^2\rangle_{E,c}+{1\over 2}\partial_E \langle W^2\rangle_{E,c}\,.
\label{eq:fdis1}
\ee

Equation~\eqref{eq:fdis1} extends Einstein's drift-diffusion relation~\eqref{eq:fdis} connecting work and work fluctuations for a system prepared in a single eigenstate (and hence to a microcanonical shell). In large systems, the last term in~\eqref{eq:fdis1} is a subleading correction since the energy is extensive. So, in the thermodynamic limit, this term can be dropped and one is back to Eq.~\eqref{eq:fdis}. However, if one is dealing with a small chaotic system, then the last term cannot be neglected. As we will show in the next section, it has important implications for determining the correct asymptotic distribution of driven chaotic systems. Relation~\eqref{eq:fdis1} was first obtained by C. Jarzynski for a single classical particle moving in a shaken chaotic cavity~\cite{jarzynski_92,jarzynski_93}, and then extended to arbitrary quantum or classical systems along the lines of our derivation in Ref.~\cite{bunin_dalessio_11}.

The exact same analysis can be carried out if, instead of a single driven system, one considers two weakly coupled systems $\rm I$ and $\rm II$ as illustrated in Fig.~\ref{fig:detailed_balance}. For the purposes of this discussion, we will assume that both systems are ergodic. Let us assume that the two systems are initialized in eigenstates with energies $E_{\rm I}$ and $E_{\rm II}$ and then  coupled weakly. This coupling leads to an energy exchange between the two systems. Since the dynamics is unitary, and after the assumption that each system is quantum chaotic, we can apply Eq.~\eqref{eq:raw_cyclic} to this setup. The only new ingredient is that the entropy $S$ is replaced by the sum of entropies of the systems $\rm I$ and $\rm II$ (corresponding to the factorization of the densities of states of the uncoupled systems). Thus
\begin{multline}
P(E_{\rm I}\to E_{\rm I}+W, E_{\rm II}\to E_{\rm II}-W)\equiv P(E_{\rm I}, E_{\rm II},W)\\
=\mathrm e^{S_{\rm I}(E_{\rm I}+W)+S_{\rm II}(E_{\rm II}-W)-S_{\rm I}(E_{\rm I})-S_{\rm II}(E_{\rm II})} P(E_{\rm I}+W, E_{\rm II}-W,-W).
\end{multline}
Note that, here, we use $W$ to denote the energy exchange between systems $\rm I$ and $\rm II$ (even though now it means heat). As we will see shortly, for isolated chaotic systems an external drive is equivalent to a coupling to an infinite temperature bath $\beta_{\rm II}$. Therefore, we prefer to keep the same notation for both coupled and isolated driven systems. Repeating exactly the same steps as before, that is, expanding the two entropies and $P$ with respect to $W$ up to the second order and integrating over the distribution function $P(E_{\rm I}, E_{\rm II},W)$, one obtains (see Appendix~\ref{Appendix_FLUCTUATIONS}):
\be
\langle W\rangle_{E_{\rm I},E_{\rm II}}=\frac{\beta_{\rm I}-\beta_{\rm II}}{2}\langle W^2\rangle_{E_{\rm I},E_{\rm II},c}+{1\over 2}\left(\partial_{E_{\rm I}}-\partial_{E_{\rm II}}\right) \langle W^2\rangle_{E_{\rm I},E_{\rm II},c},
\label{eq:fdis2}
\ee
where the symbol $\langle\dots\rangle_{E_{\rm I},E_{\rm II}}$ means that an average is taken with respect to the probability distribution $P(E_{\rm I},E_{\rm II},W)$ at fixed $E_{\rm I}$ and $E_{\rm II}$, the suffix ``c" means connected, and $\beta_{\rm I}-\beta_{\rm II}$ is the difference in temperature between the two systems. If both systems are large, then the last term is again subextensive and can be dropped. It is interesting to note that Einstein's relation for coupled systems~\eqref{eq:fdis2} reduces to the one for isolated systems~\eqref{eq:fdis1} if the temperature of system $\rm II$ is infinite $\beta_{\rm II}=0$ and $\la W^2\ra_{E_{\rm I},E_{\rm II},c}$ is either negligible or independent of $E_{\rm II}$. Hence, from the point of view of energy flow, driving an isolated system by means of an external cyclic perturbation (like it happens, e.g., in microwave ovens) is equivalent to coupling it to an infinite temperature reservoir. 

\subsection{Fokker-Planck Equation for the Heating of a Driven Isolated System}\label{sec:FP}

The drift-diffusion relation~\eqref{eq:fdis1} can be used to understand energy flow (or heating) in a driven ergodic system not coupled to a bath. Let us imagine that the process consists of many pulses, well separated in time, such that the system can relax between pulses to the diagonal ensemble. Then, after coarse-graining, one can view this as a continuous (in number of pulses) process. If the mean energy deposited in each pulse is small then both $\langle W\rangle$ and $\langle W^2 \rangle_c$ (as well as other cumulants of $W$) scale linearly with the number of pulses and, hence, with the coarse-grained time. Next, instead of a series of discrete processes, one can consider a single continuous process. Then, as long as the relaxation time is faster than the characteristic time for the energy change in the system, $\langle W\rangle$ and $\langle W^2\rangle_c$ are approximately linear in time. It is well known that under such assumptions transport (energy transport in our case) can be described by the Fokker-Planck equation. The derivation of the Fokker-Planck equation is fairly standard (see, e.g., Ref.~\cite{VanKampen}) but we repeat it here for completeness. We start from the microscopic master equation~\eqref{eq:master}. For simplicity, assuming a cyclic process and dropping the tildes:
\be
\rho_n(t+\delta t)=\sum_m p_{m\to n}(\delta t) \rho_m(t),
\label{eq:master_FP}
\ee
where $\delta t$ is a small interval of time. By assumption, $p_{m\to n}(\delta t)\propto \delta t$. Let us define the energy distribution function 
\be
\Pi(E_n,t)=\rho_n(t) \Omega(E_n).
\ee
Now, replacing the summation over $m$ in the master equation by an integration over $W$ (and multiplying by the density of states), recalling that the probability of performing work $W$ is $P(E_n,W)=p_{n\to m} \Omega(E_m)$, where $E_m=E_n+W$, and dropping the index $n$ in $E_n$, we can rewrite Eq.~\eqref{eq:master_FP} as
\be
\Pi(E,t+\delta t)=\int dW P(E-W,W)  \Pi(E-W, t).
\label{eq:master_FP1}
\ee
Note that $P(E-W,W)$ is a fast function of the second argument with the width $\delta W^2\equiv \langle W^2\rangle_c\propto \delta t$, but a slow function of the total energy of the system, that is, of the first argument. Similarly, the probability distribution $\Pi(E,t)$ is expected to be a slow function of $E$ on the scale of $\delta W$. Hence, we can use the Taylor expansions:
\beq
\Pi(E-W,t)&=& \Pi(E,W)-W\partial_E \Pi(E,W)+\dots,\nonumber\\ 
P(E-W,W)&=& P(E,W)-W \partial_E P(E,W)+\dots\nonumber
\eeq
By substituting these expansions in Eq.~\eqref{eq:master_FP1}, and expanding to second order in the work $W$, we find:
\beq
\Pi(E,t+\delta t)-\Pi(E,t)&=&-\langle W\rangle_E \partial_E \Pi(E,t)-\partial_E \langle W\rangle_E \Pi(E,t)
+\frac{1}{2}{\langle W^2\rangle_E} \partial^2_{EE} \Pi(E,t)\nonumber\\ &&+\partial_E \langle W^2\rangle_E \partial_E \Pi(E,t)+{1\over 2}\partial^2_{EE}\langle W^2\rangle_E \Pi(E,t).
\eeq
Dividing the equation above by $\delta t$, and using the notation 
\be
J_E={\langle W\rangle_E\over \delta t},\; D_E={\langle W^2\rangle_{E,c}\over \delta t},
\ee
we find the Fokker-Planck equation
\be
{\partial \Pi(E,t)\over \partial t}=-\partial_E [J_E \Pi(E,t)]+{1\over 2}\partial^2_{EE} [D_E \Pi(E,t)].
\label{eq:fokker-planck}
\ee
To obtain this result, we notice that the cumulants of the work are linear in transition probabilities and, hence, they are linear in $\delta t$. So, in the limit $\delta t\to 0$, one can substitute $\langle W^2\rangle_E\to \langle W^2\rangle_{E,c}$. 

This Fokker-Planck equation for the energy drift and diffusion is completely general. Effectively, it describes the evolution of the energy distribution function under many small uncorrelated pulses or under a continuous slow driving such that, at each moment of time, the system is, approximately, in a stationary state. In general, the drift $J_E$ and diffusion $D_E$ coefficients are independent. However, in ergodic systems, they are related to each other by Eq.~\eqref{eq:fdis1}, which, after dividing by $\delta t$, reads~\cite{bunin_dalessio_11}
\be
J_E={\beta \over 2}D_E+{1\over 2}\partial_E D_E.
\label{eq:AB1}
\ee
Likewise, for open systems, Eq.~\eqref{eq:fdis2} implies that~\cite{jarzynski_92, jarzynski_93, bunin_kafri_13}
\be
J_E={\Delta \beta\over 2} D_E+{1\over 2} \partial_E D_E
\label{eq:AB2}
\ee
In the next section, we will see how powerful this result is for finding universal energy distributions in driven isolated and open systems. We note that the Fokker-Planck equation can be derived using a different approach, as done in Refs.~\cite{ott_79,cohen_00}.

Interestingly, one can derive Eqs.~\eqref{eq:AB1} and \eqref{eq:AB2} from a very simple argument~\cite{bunin_dalessio_11}. Let us discuss it here only for an isolated setup, which is relevant to Eq.~\eqref{eq:AB1}. It is straightforward to extend the discussion to open systems, relevant to Eq.~\eqref{eq:AB2}. According to the discussion in Sec.~\ref{sec:inf_T}, the only attractor for the probability distribution of a driven system is the infinite temperature distribution: $\rho_n^\ast=C_1$, where $C_1$ is a constant. This implies that $\Pi^\ast(E)=C_2\exp[S(E)]$, with $C_2$ another constant, should be stationary under the Fokker-Planck equation or, in other words, that
\be
-\partial_E [J_E \Pi^\ast (E)]+{1\over 2}\partial^2_{EE} [D_E \Pi^\ast(E)]=0,
\ee
which, in turn, implies that
\be\label{eq:simpargdiff}
-J_E \mathrm e^{S(E)}+{1\over 2} \partial_E [D_E \mathrm e^{S(E)}]=C_3=0.
\ee
The integration constant $C_3$ has to be equal to zero because one can go to (typically, low) energies where there are no states and, hence, $J_E=D_E=0$. From Eq.~\eqref{eq:simpargdiff}, one immediately recovers the drift-diffusion relation~\eqref{eq:AB1}.

\subsection{Fluctuation Theorems for Two (or More) Conserved Quantities\label{Crooks_2}}

\begin{figure}[!t]
\includegraphics[width=0.33\textwidth]{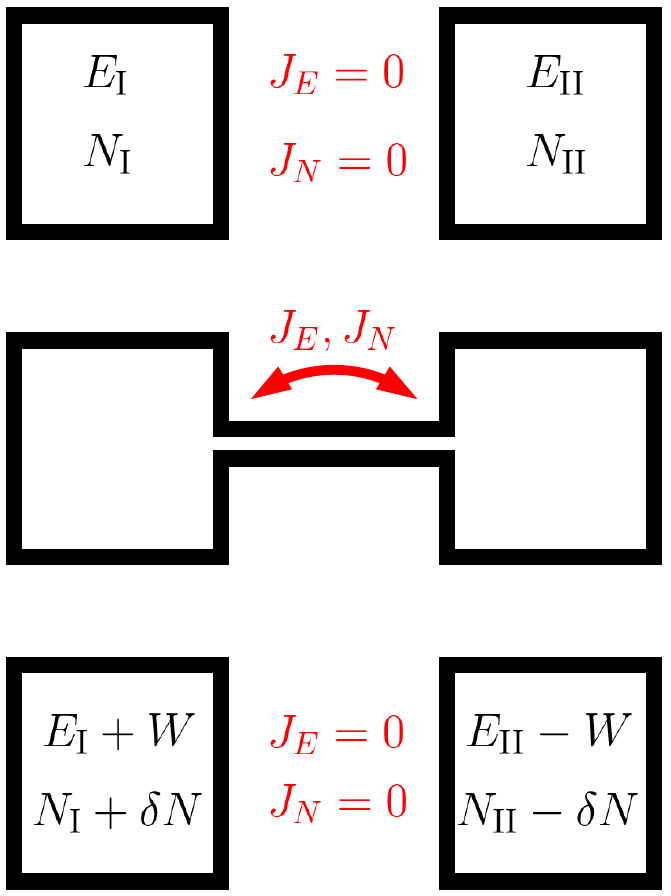}
\centering
\caption{Two systems are connected and exchange energy and particles. (Top) Initially, the two systems have a well-defined energy and number of particles. (Middle) Then they are connected and allowed to exchange energy and particles. (Bottom) Finally they are disconnected and each system is allowed to equilibrate at the new fixed values of energy and particle numbers.}
\label{fig:two-systems} 
\end{figure}

Until now we focused on systems with only one conserved quantity, the energy. Quite often one deals with situations where, in addition to the energy, there are other conserved quantities, such as the number of particles, magnetization, momentum, charge, and volume. In this section, we extend the fluctuation relations and their implications to such setups. To simplify our derivations, we will focus on systems with two conserved quantities: energy $E$ and the number of particles $N$. At the end of Sec.~\ref{onsager}, we comment on how to extend our results to an arbitrary (non-extensive) number of conserved quantities. We focus on a particular setup, where two initially separated chaotic systems $\rm I$ and $\rm II$ are weakly coupled allowing for energy and particle exchange (see Fig.~\ref{fig:two-systems}). The expressions obtained in this setup can also be used to describe a single system, say, system ${\rm I}$, coupled to an external driving by setting $\partial S_{\rm II}/\partial E_{\rm II}=\partial S_{\rm II}/\partial N_{\rm II}=0$. We assume that the two systems are connected to each other for a short period of time $\tau$ and then detached again and allowed to equilibrate (i.e., reach a diagonal ensemble), see Fig.~\ref{fig:two-systems}. It is intuitively clear that this assumption of connecting and disconnecting the systems is not needed if the coupling between them is weak and the two systems are in an approximate stationary state with respect to the uncoupled Hamiltonian at each moment of time. This intuition can be formalized  using time-dependent perturbation theory (see Appendix~\ref{appendix:drift_diffusion_continuous}).

Repeating the same steps as in Sec.~\ref{linear_response}, one can straightforwardly generalize Eq.~\eqref{eq:raw_cyclic} to:
\begin{multline}
\frac{P(E_{\rm I}, E_{\rm II}, N_{\rm I}, N_{\rm II}, W, \delta N)}{P(E_{\rm I}+W, E_{\rm II}-W, N_{\rm I}+\delta N, N_{\rm II}-\delta N, -W, -\delta N)} \\=
\mathrm e^{S_{\rm I} (E_{\rm I}+W, N_{\rm I}+\delta N)+S_{\rm II} (E_{\rm II}-W, N_{\rm II}-\delta N)-S_{\rm I} (E_{\rm I}, N_{\rm I})-S_{\rm II} (E_{\rm II}, N_{\rm II})},\qquad\qquad
\label{eq:generalized-croocks}
\end{multline}
where 
\begin{multline*}
P(E_{\rm I}, E_{\rm II}, N_{\rm I}, N_{\rm II}, W, \delta N)\\
\equiv P(E_{\rm I}\to E_{\rm I}+W, E_{\rm II}\to E_{\rm II}-W, N_{\rm I}\to N_{\rm I}+\delta N, N_{\rm II}\to N_{\rm II}-\delta N).\qquad
\end{multline*}

If the energy and particle changes are small, one can expand the entropy and the probability distribution $P$ in Taylor series. If $W$ and $\delta N$ are small and non-extensive, then only the leading derivatives of the entropy need to be kept and we find a Crooks-type relation for two conserved quantities:
\be
P(E_{\rm I}, E_{\rm II}, N_{\rm I}, N_{\rm II}, W, \delta N)\mathrm e^{-\Delta \beta W-\Delta \kappa \delta N}\approx {P(E_{\rm I}, E_{\rm II}, N_{\rm I}, N_{\rm II}, -W, -\delta N)}
\label{eq:fundamental},
\ee
where $\Delta \beta=\beta_{\rm I}-\beta_{\rm II}$, and $\Delta \kappa=\kappa_{\rm I}-\kappa_{\rm II}$, with
\be
\kappa_{i}={\partial S_{i}\over \partial N_{i}}=-\beta_{i}\mu_{i}, \quad i={\rm I, II}
\ee
As in Sec.~\ref{sec:fluct_gibbs}, this relation is exact if the two systems $\rm I$ and $\rm II$ are described by grand canonical distributions. It is also asymptotically exact to order $1/N$ (with $N$ of the order of the number of particles in each system) irrespective of the distribution, if the two systems are extensive while the energy and particle exchanges are not (which is, e.g., the typical setup if two large macroscopic systems are connected through a surface).  For the specific case of effusion of an ideal gas between two reservoirs kept at different temperatures and chemical potentials, Eq.~\eqref{eq:fundamental} was derived microscopically in Ref.~\cite{cleuren_VandenBroeck_06}. Following Ref.~\cite{andrieux_gaspard_07}, we can use the fluctuation relation~\eqref{eq:fundamental} to derive the symmetry property of the cumulant generating function for $W$ and $\delta N$, see Eq.~\eqref{eq:symmetry_G}. Namely, multiplying both sides of Eq.~\eqref{eq:fundamental} by $\mathrm e^{\zeta W+ \delta N}$
\be
P(E_{\rm I}, E_{\rm II}, N_{\rm I}, N_{\rm II}, W, \delta N)\mathrm e^{-(\Delta \beta-\zeta) W-(\Delta \kappa-\eta) \delta N}= {P(E_{\rm I}, E_{\rm II}, N_{\rm I}, N_{\rm II}, -W, -\delta N)} \mathrm e^{\zeta W+ \delta N}
\ee
for arbitrary $\zeta$ and $\eta$, and integrating over $W$ and $\delta N$, we find
\be
G(\Delta\beta-\zeta,\Delta\kappa-\eta)=G(\zeta,\eta)
\label{eq:condition1}
\ee
together with the normalization condition $G(0,0)=1$. For the particular choices of $\zeta$ and $\eta$ the symmetry relation~\eqref{eq:condition1} is equivalent to two different Jarzynski-type relations:
\be
\langle \exp[-\Delta \beta W-\Delta \kappa\delta N]\rangle=1,\quad \langle\exp[-\Delta\beta W]\rangle=\langle \exp[-\Delta \kappa \delta N]\rangle, 
\label{eq:jarzynski_two}
\ee
where the left relation holds for $\zeta=\eta=0$ and the right for $\zeta=0$, $\eta=\Delta\kappa$. As in Sec.~\ref{linear_response} the angular brackets imply averaging over $W$ and $\delta N$ starting at  $E_{\rm I},E_{\rm II}$ and $N_{\rm I},N_{\rm II}$.

\subsection{Linear Response and Onsager Relations\label{onsager}}

Continuing with the setup of the previous section, one can perform the cumulant expansion of Eq.~\eqref{eq:jarzynski_two} up to the second order, or, alternatively, perform a Taylor expansion of the cumulant generating function and using relation~\eqref{eq:condition1}. One obtains:
\beq
&&-\Delta \beta \langle W\rangle -\Delta\kappa \langle \delta N\rangle+{\Delta \beta^2\over 2}\langle W^2\rangle_c+{\Delta \kappa^2\over 2}
\langle \delta N^2\rangle_c+\Delta \beta \Delta \kappa \langle W \delta N\rangle_c=0, \label{eq:JAR1} \\
&& -\Delta \beta \langle W\rangle +{\Delta \beta^2\over 2}\langle W^2\rangle_c = -\Delta \kappa \langle \delta N\rangle +{\Delta \kappa^2\over 2}
\langle \delta N^2\rangle_c.\label{eq:JAR2}
\eeq
First, we move all the terms to the left-hand side (LHS) of Eq.~\eqref{eq:JAR2}, then by adding and subtracting the resulting expression with Eq.~\eqref{eq:JAR1}, we obtain the two completely symmetric relations:
\be
\langle W\rangle = {\Delta \beta\over 2} \langle W^{2}\rangle_{c} + {\Delta \kappa\over 2} \langle  W\delta N\rangle_{c},\qquad
\langle\delta N\rangle = {\Delta \kappa\over 2} \langle\delta N^{2}\rangle_{c} + {\Delta \beta\over 2} \langle W\delta N\rangle_{c}.
\label{eq:result11}
\ee
If the coupling between the systems is weak, then either from Fermi's golden rule or using the same arguments as presented above Eq.~\eqref{eq:master_FP}, we conclude that $\langle W\rangle$ and $\langle \delta N\rangle$ (as well as other cumulants such as $\langle W^2\rangle_c$) are linear functions of the coupling time $\delta t$. Then one can define the energy and particle currents $J_E=\langle W\rangle/\delta t$ and $J_N=\langle \delta N \rangle/\delta t$; as well as $D_{WW}=\langle W^2\rangle_c/\delta t$,  $D_{NN}^2=\langle \delta N^2\rangle_c/\delta t$, and $D_{WN}=\langle W\delta N\rangle_c/\delta t$. Equations~\eqref{eq:result11} can then be rewritten in the matrix form~\cite{gaspard_andrieux_11}:
\be
\left(
\begin{array}{c}
J_E\\
J_N
\end {array}
\right)=\frac{1}{2}
\left(
\begin{array}{cc}
D_{WW} & D_{WN}\\
D_{WN} &  D_{NN}
\end {array}
\right)
\left(
\begin{array}{c}
\Delta \beta\\
\Delta \kappa
\end {array}
\right)
\label{eq:onsager}
\ee
These equations are analogous to Eq.~\eqref{eq:fdis} and are known as the Onsager relations. On the LHS one has the energy and particle currents, while on the RHS one has the symmetric fluctuation matrix multiplied by the thermodynamic biases ($\Delta \beta$ and $\Delta \kappa$) that drive the energy and particle currents. Note that we obtained these relations as a cumulant expansion, not as a gradient expansion. This means that they remain valid for arbitrarily large $\Delta\beta$ and $\Delta \kappa$ as long as the distribution $P(W,\delta N)$ is approximately Gaussian. For this reason, the diffusion matrix in Eq.~\eqref{eq:onsager} does not have to be the one in equilibrium, that is, does not have to correspond to the one for $\Delta\beta=\Delta\kappa=0$. When the temperature and the chemical potential gradients are small, contributions from higher order cumulants are suppressed because of the higher powers of $\Delta \beta$ and $\Delta \kappa$. As a result, the usual Onsager relations apply. 

This derivation of the Onsager relations applies to large systems. As with the Einstein relation~\eqref{eq:fdis2}, there is a correction that can be important for small systems. The derivation of this correction is analogous to the derivation of Eq.~\eqref{eq:fdis2}. The starting point is now Eq.~\eqref{eq:generalized-croocks}. One needs to expand the entropy and the probability distribution as a function of $E_{\rm I,II}\pm W$ and $N_{\rm I,II}\pm \delta N$ to second order in $W$ and $\delta N$, and then to carry out a cumulant expansion with these additional corrections. We leave the details of the derivation to Appendix~\ref{Appendix_FLUCTUATIONS} and show only the final result, which is the natural extension of Eq.~\eqref{eq:fdis2}:
\begin{eqnarray}
\langle W\rangle &=& {\Delta \beta\over 2} \langle W^{2}\rangle_{c} + {\Delta \kappa\over 2} \langle  W\delta N\rangle_{c}+{1\over 2}\partial_E \langle W^2\rangle_c +{1\over 2}\partial_N \langle W \delta N\rangle_c,\nonumber \\
\langle\delta N\rangle &=& {\Delta \kappa\over 2} \langle\delta N^{2}\rangle_{c} + {\Delta \beta\over 2} \langle W\delta N\rangle_{c}+{1\over 2}\partial_N \langle \delta N^2\rangle_c+{1\over 2}\partial_E \langle W\delta N\rangle_c.
\label{eq:result1}
\end{eqnarray}
The terms with derivatives are clearly subextensive and not important for large systems, but they can play an important role in small or mesoscopic systems. It is interesting to note that the Onsager relations can still be written in the conventional form~\eqref{eq:onsager} if one redefines the energy and particle currents as
\be
J_E={1\over \tau}\langle W\rangle-{1\over 2\tau }\partial_E \langle W^2\rangle_c- {1\over 2\tau}\partial_N \langle W \delta N\rangle_c,\ \ \ J_N={1\over \tau}\langle \delta N\rangle-{1\over 2\tau }\partial_N \langle \delta N^2\rangle_c- {1\over 2\tau}\partial_E \langle W \delta N\rangle_c \;.
\label{eq:renormalized_currents}
\ee

Let us comment that these results immediately generalize to more ($M\geq 3$) conserved quantities. For example, in the Onsager relation~\eqref{eq:onsager} one will need to use $M$-component vectors for the currents and the gradients and a symmetric $M\times M$ diffusion matrix. Similarly, one can generalize corrections to the currents~\eqref{eq:renormalized_currents} writing them using an $M$-component gradient form.

\subsection{Nonlinear Response Coefficients}

Along the lines of Refs.~\cite{andrieux_gaspard_07,gaspard_andrieux_11}, one can go beyond the Onsager relation and use the symmetry relation of the generating function~\eqref{eq:condition1} to constrain nonlinear response coefficients. It is convenient to work using a vector notation for $M$-conserved quantities $\delta \vec N$, where $\delta N_1$ stands for the energy, and $\delta N_2,\dots\,, \delta N_M$ for other conserved quantities. Similarly, let us denote by $\Delta \vec\kappa$ the gradients of affinities (or thermodynamic forces) $\kappa_\alpha =\partial S/\partial N_\alpha$ (such that $\kappa_1=\beta$) and $\Delta \kappa_\alpha$ is the difference in affinities between the systems $\rm I$ and $\rm II$. Ignoring subextensive corrections, the symmetry relation~\eqref{eq:condition1} reads
\be
G(\Delta \vec \kappa-\vec \zeta)=G(\vec \zeta),
\label{eq:symmetry_gen_function}
\ee
where we recall that
\be\left< \exp\left[-\sum_\alpha \zeta_\alpha \delta N_\alpha\right]\right>=\exp[G(\vec \zeta)].
\ee
Differentiating this equality with respect to $\zeta_\alpha$ and using that $G(0)=1$, we find
\be
J_\alpha\equiv \langle \delta N_\alpha\rangle=-{\partial G(\vec\zeta)\over \partial \zeta_\alpha}\biggr|_{\vec \zeta=0}.
\ee
To simplify the notation, we set $\tau =1$ so that $\langle \delta N_\alpha\rangle=J_\alpha$ is the current of the $\alpha$-th conserved quantity. The symmetry relation~\eqref{eq:symmetry_gen_function} thus implies that
\be
J_\alpha=-{\partial G(\vec \zeta)\over \partial\zeta_\alpha}\biggr|_{\vec \zeta=0}=-{\partial G(\Delta\vec \kappa-\vec \zeta)\over \partial\zeta_\alpha}\biggr|_{\vec \zeta=0}={\partial G(\vec \zeta)\over \partial \zeta_\alpha}\biggr|_{\vec\zeta=\Delta\vec \kappa}.
\label{eq:current:cumulant}
\ee
Let us write explicitly the cumulant expansion of the generating function $G(\vec\zeta)$:
\be
G(\vec\zeta)=-\sum_\alpha \zeta_\alpha J_\alpha+{1\over 2}\sum_{\alpha\beta} D_{\alpha\beta} \zeta_{\alpha}\zeta_\beta-{1\over 3!}\sum_{\alpha\beta \gamma} M_{\alpha\beta\gamma} \zeta_\alpha\zeta_\beta \zeta_\gamma+\dots,
\ee
where $D_{\alpha\beta}=\langle \delta N_\alpha \delta N_\beta\rangle_c$ is the covariance matrix (second-order joint cumulant matrix), $M_{\alpha\beta\gamma}$ is the third-order cumulant tensor, and so on. By substituting this expansion into Eq.~\eqref{eq:current:cumulant}, we find
\be
2J_\alpha=\sum_\beta D_{\alpha\beta} \Delta\kappa_\beta-{1\over 2}\sum_{\beta\gamma}M_{\alpha\beta\gamma}\Delta\kappa_\beta\Delta\kappa_\gamma+\dots
\ee
Using this expansion and the symmetry relations of joint cumulant tensors, such as $D_{\alpha\beta}=D_{\beta\alpha}$ and $M_{\alpha\beta\gamma}=M_{\beta\alpha\gamma}$, one can extend the Onsager reciprocity relations to higher order cumulants. As in the previous section, these results extend to externally driven systems by substituting $\Delta\vec \kappa$ by $\vec \kappa_{\rm I}$. Note that because we carry out a cumulant expansion, and {\em not} a gradient expansion, all cumulant tensors, such as $D_{\alpha,\beta}$ and $M_{\alpha\beta\gamma}$, are functions of $\Delta\vec \kappa$, that is, they are evaluated away from the global equilibrium corresponding to $\Delta\vec\kappa=0$. If one further re-expands these tensors in gradients $\Delta\vec \kappa$, one obtains a more standard gradient expansion around equilibrium~\cite{andrieux_gaspard_07}.

\subsection{ETH and the Fluctuation-Dissipation Relation for a Single Eigenstate}\label{sec:fluctuation_dissipation}

In Sec.~\ref{sec:eth_def}, we introduced the ETH ansatz for the matrix elements of physical operators in the eigenstates of a quantum chaotic Hamiltonian~\eqref{eq:ETH}. This ansatz contains two smooth functions of the mean energy $\bar E=(E_n+E_m)/2$, and the energy difference $\omega=E_m-E_n$. The first function, $O(\bar E)$, is nothing but the microcanonical average of the observable $\hat O$. The second function, $f_O(\bar E,\omega)$, contains information about the off-diagonal matrix elements of the operator $\hat O$. Let us elaborate on the second function here and show its relation to nonequal-time correlation functions of the observable $\hat O$ (see also Ref.~\cite{khatami_pupillo_13}). In parallel, we will be able to prove the fluctuation-dissipation relation for individual eigenstates.

We begin by using the ETH ansatz~\eqref{eq:ETH} to analyze the quantum fluctuations of an observable $\hat O$ in the eigenstate $|n\rangle$:
\be
\delta O_n^2=\langle n|\hat O^2|n\rangle-\langle n |\hat O|n\rangle^2=
\sum_{m\neq n} |O_{nm}|^2= \sum_{m\neq n} 
\mathrm e^{-S(E_n+\omega/2)}|f_O(E_n+\omega/2,\omega)|^2 |R_{nm}|^2,
\ee
where we used that $\bar E=(E_n+E_m)/2=E_n+\omega/2$. For concreteness, we consider the most general case in which the matrix elements of observables are complex.

Because of the ETH requirement that the function $f_O$ is smooth, the fluctuations of $|R_{nm}|^2$ average out in the sum and one can replace the summation over states $m$ by an integration over $\omega$: $\sum_{m}\to \int d\omega\, \Omega(E_n+\omega)=\int d\omega \exp[S(E_n+\omega)]$.  To shorten the notation, we will drop the subindex  $n$ in $E_n$ and, unless  otherwise specified, will identify $E$ with the energy of the energy eigenstate $|n\rangle$. Then
\be
\delta O_n^2=\sum_{m\neq n} \mathrm e^{-S(E+\omega/2)} |f_O(E+\omega/2,\omega)|^2=
\int_{-\infty}^{\infty} d\omega\, \mathrm e^{S(E+\omega)-S(E+\omega/2)} |f_O(E+\omega/2,\omega)|^2.
\label{eq:O_alpha_2}
\ee
We are interested in expectation values of few-body (usually local) operators (such as the magnetization in a given region of space) or sums of those operators (such as the total magnetization). This kind of operators generally connect states that differ by non-extensive energies, implying that the function $f_O(E+\omega/2,\omega)$ rapidly decreases with the second argument $\omega$ \cite{srednicki_99,khatami_pupillo_13} (see Figs.~\ref{fig:khatami_offdiag} and \ref{fig:FW}). On the other hand, the entropy $S(E+\omega)$ and the function $f_O$ as a function of the first argument can only change if $\omega$ changes by an extensive amount. This means that one can expand these functions in Taylor series around $\omega=0$:
\begin{align*}
S(E+\omega)-S(E+\omega/2)= {\beta\omega\over 2}+{\partial \beta\over \partial E}{3\,\omega^2\over 8}+\dots,\\
f_O(E+\omega/2,\omega)= f_O(E,\omega)+ {\partial f_O(E,\omega)\over\partial E} {\omega\over 2}+\dots.
\end{align*}
By substituting this expansion in Eq.~\eqref{eq:O_alpha_2} and keeping terms up to the linear order in $\omega$, one obtains
\be
\delta O_n^2\approx \int_{-\infty}^\infty d\omega\, \mathrm e^{\beta\omega/2} \left[|f_O(E,\omega)|^2+ {\partial |f_O(E,\omega)|^2\over\partial E} {\omega\over 2}\right].
\ee
The function $f_O(E,\omega)$ therefore determines the quantum fluctuations of the operator $\hat O$ in eigenstates of the Hamiltonian and, as a result, in the associated microcanonical ensembles.

In passing, we note that the result above shows that fluctuations of $\hat O$ within each eigenstate are slow functions of the energy. Combining this observation with an earlier discussion of fluctuations of observables in the diagonal ensemble (see Sec.~\ref{sec:eth_thermalization}), we can rewrite Eq.~\eqref{eq:fluct_O_DE} as
\be
\overline{\delta O^2}\approx \delta O_n^2+\left({\partial \bar O\over \partial E}\right)^2 \delta E^2,
\ee
where $n$ is the eigenstate corresponding to the mean energy: $E_n=\langle E\rangle$. We see that fluctuations of any observable in a diagonal ensemble have essentially two independent contributions, the first coming from fluctuations within each eigenstate and the second from the energy fluctuations. For extensive observables, these two contributions are of the same order but, for intensive observables confined to a finite subsystem, the second contribution becomes subleading and all fluctuations are essentially coming from $\delta O^2_n$. This is nothing but a manifestation of equivalence of ensembles applied to the diagonal ensemble.

The previous derivation immediately extends to more general nonequal-time correlation functions:
\be
C_O(t)\equiv \langle n | \hat O(t) \hat O(0)|n \rangle_c
\equiv \langle n | \hat O(t) \hat O(0)|n \rangle 
-\langle n |\hat O(t)|n\rangle\langle n |\hat O(0)|n\rangle,
\ee
where $\hat O(t)=\mathrm e^{i \hat {H} t}\hat O\mathrm e^{-i \hat {H} t}$ is the operator in the Heisenberg picture. Repeating the same steps as before, one finds
\be
C_O(t)\approx \int_{-\infty}^\infty d\omega\, \mathrm e^{\beta\omega/2-i\omega t} 
\left[|f_O(E,\omega)|^2+ {\partial |f_O(E,\omega)|^2\over\partial E} {\omega\over 2}\right].
\label{eq:G_O}
\ee
It is convenient to define the spectral density of the operator $\hat{O}$ as the Fourier transform of $C_O(t)$
\be
\mathcal S_O(E,\omega)=\int_{-\infty}^\infty dt\, \mathrm\, e^{i\omega t} C_O(t).
\ee
Substituting here the expression for $C_O(t)$, one obtains
\be
|f_O(E,\omega)|^2+{\omega\over 2} {\partial |f_O(E,\omega)|^2\over\partial E} 
= { \mathrm e^{-\beta\omega/2} \over 2\pi}
\mathcal S_O(E,\omega).
\label{eq:f_E_w1}
\ee
Noting that $|f_O(E,\omega)|^2$  is an even function of $\omega$ and $\omega |f_O(E,\omega)|^2$ is an odd function, changing $\omega\rightarrow-\omega$ in Eq.~\eqref{eq:f_E_w1}, and adding and subtracting the two resulting equations, one finds that
\beq
|f_O(E,\omega)|^2&=&{1\over 4\pi} \left[\mathrm e^{-\beta\omega/2} \mathcal S_O(E,\omega)
+\mathrm e^{\beta\omega/2} \mathcal S_O(E,-\omega)\right]\nonumber\\
&=&{1\over 4\pi} \left[\cosh(\beta\omega/2) \mathcal S^+_O(E,\omega)- \sinh(\beta\omega/2) \mathcal S^-_O(E,\omega)\right],\nonumber\\
{\partial |f_O(E,\omega)|^2\over\partial E}& =& {1\over 2\pi\omega} 
\left[\cosh(\beta\omega/2) \mathcal S^-_O(E,\omega)-\sinh(\beta\omega/2) \mathcal S^+_O(E,\omega)\right],
\label{eq:f_E_w2}
\eeq
where
\beq
\mathcal S^+_O(E,\omega)&=&\mathcal S_O(E,\omega)+\mathcal S_O(E,-\omega)=\int_{-\infty}^\infty dt 
\mathrm e^{i\omega t}\langle n|\{\hat O(t),\hat O(0)\}|n\rangle_c,\nonumber\\
\mathcal S^-_O(E,\omega)&=&\mathcal S_O(E,\omega)-\mathcal S_O(E,-\omega)=\int_{-\infty}^\infty dt 
\mathrm e^{i\omega t}\langle n| [\hat O(t),\hat O(0)]|n\rangle_c,
\eeq
are the Fourier transforms of the symmetric and antisymmetric correlation functions, $\{\cdot,\cdot\}$ stands for the anti-commutator and $[\cdot,\cdot]$ stands for the commutator of two operators. The  symmetric correlation function appears in the quantum fluctuations of physical observables and the antisymmetric correlation function appears in Kubo's linear response. We thus see that the absolute value of the function $f_O(E,\omega)$ and its derivative with respect to $E$ are determined by the Fourier transforms of the symmetric and antisymmetric nonequal-time correlation functions of the operator $\hat O$ taken in the many-body eigenstate state $|n\rangle$, or, equivalently, in the microcanonical ensemble consisting of a single eigenstate with the energy $E=E_n$. The phase of this function is not uniquely defined because, in Eq.~(\ref{eq:ETH}), the random function $R_{nm}$ is defined up to a random phase. One can also invert the relations~\eqref{eq:f_E_w2} and obtain
\beq
\mathcal S^+_O(E,\omega)&=& 4\pi\left[\cosh(\beta\omega/2) |f_O(E,\omega)|^2+{\omega\over 2}\sinh(\beta\omega/2){\partial |f_O(E,\omega)|^2\over\partial E}\right],\nonumber\\
\mathcal S^-_O(E,\omega)&=& 4\pi\left[\sinh(\beta\omega/2) |f_O(E,\omega)|^2+{\omega\over 2}\cosh(\beta\omega/2){\partial |f_O(E,\omega)|^2\over\partial E}\right].
\label{eq:s_pm_f}
\eeq

Let us recall that $\mathcal S(E,\omega)$ also determines Kubo's linear response susceptibility (see, e.g., Ref.~\cite{mahan_2000}):
\be
\chi_O(\omega)=i\int_0^\infty dt \mathrm e^{i\omega t}\langle [\hat O(t),\hat O(0)]\rangle=i \int_0^\infty dt \mathrm e^{i\omega t} [C_O(t)-C_O(-t)],
\ee
where we used that, for any stationary distribution (including an energy eigenstate), $\langle \hat O(0) \hat O(t)\rangle_c=\langle \hat O(-t) \hat O(0)\rangle_c=C_O(-t)$. Using Eq.~\eqref{eq:G_O}, and that
\be
\int_0^\infty dt \mathrm e^{i \nu t}=\pi \delta(\nu)+i {\cal P} \left({1\over \nu}\right),
\ee
where ${\cal P}(1/\nu)$ stands for the principal value, we find
\begin{align*}
\chi_O(E,\omega) & = 2\pi i \left[ \sinh(\beta\omega/2) |f_O(E,\omega)|^2 + \cosh(\beta\omega/2){\omega\over 2} {\partial |f_O(E,\omega)|^2\over\partial E}  \right] \\ 
& + {\cal P} \int_{-\infty}^{\infty} d\nu   \left[ 2 |f_O(E,\nu)|^2 \frac{\sinh(\beta\nu/2)}{\nu-\omega} + \frac{\partial |f_O(E,\nu)|^2}{\partial E} \frac{\nu \cosh(\beta\nu/2)}{\nu-\omega} \right]
\end{align*}
Hence, the imaginary part of Kubo's susceptibility is also determined by $|f_O(E,\omega)|^2$:
\be
\mathcal{I}\left[\chi_O(E,\omega)\right] = 2\pi  \left[|f_O(E, \omega)|^2 \sinh(\beta\omega/2)+{\omega\over 2}{\partial |f_O(E,\omega)|^2\over\partial E}\cosh(\beta\omega/2)\right]={1\over 2} \mathcal S_O^-(E,\omega).
\label{eq:chi_imaginary}
\ee

If $\hat{O}$ is a local operator, or a sum of local operators, then the terms with derivatives of the total energy become unimportant for very large systems, and Eqs.~\eqref{eq:s_pm_f} and \eqref{eq:chi_imaginary} simplify to:
\be
\mathcal S_O^+(E,\omega)\approx 4\pi \cosh(\beta\omega/2) |f_O(E,\omega)|^2,\quad \mathcal{I}\left[\chi_O(E,\omega)\right] \approx 2\pi |f_O(E,\omega)|^2\sinh(\beta\omega/2),
\ee
which imply the famous fluctuation-dissipation relation~\cite{mahan_2000}:
\be
 \mathcal S_O^+(E,\omega)\approx  2\coth \left(\frac{\beta\omega}{2}\right) \mathcal{I}\left[\chi_O(E,\omega)\right].
\label{eq:fluct_diss_rel}
\ee
We see that, as with the fluctuation theorems, the fluctuation-dissipation relation does not rely on the assumption of a Gibbs distribution. It is satisfied for every eigenstate of a chaotic Hamiltonian (away from the edges of the spectrum, where ETH is expected to hold), and hence for any stationary ensemble with non-extensive energy fluctuations. 

For finite systems, one can calculate corrections to the fluctuation-dissipation relation. For example, combining the relations in Eq.~\eqref{eq:f_E_w2}, one finds that
\begin{multline}
\left[\sinh {\beta \omega \over 2}\left(
1-{\omega^2\over 4\sigma_c^2}\right)+{\omega\over 2}\cosh{\beta\omega\over 2}\partial_E\right] \mathcal S_O^+(E,\omega)\\
=\left[\cosh {\beta \omega \over 2}\left(1-{\omega^2\over 4\sigma_c^2}\right)+{\omega\over 2}\sinh{\beta\omega\over 2}\partial_E\right] \mathcal S_O^-(E,\omega),
\label{eq:fluct_diss_corr}
\end{multline}
which replaces the standard relation
\be
\sinh{\beta\omega\over 2} \mathcal S_O^+(E,\omega)\approx \cosh{\beta\omega\over 2} \mathcal S_O^-(E,\omega),
\ee
valid for Gibbs ensembles or for individual eigenstates in very large systems. This more general fluctuation-dissipation relation for individual eigenstates~\eqref{eq:fluct_diss_corr} still connects the noise and the dissipative response but in a more complicated way.

\subsection{ETH and Two-Observable Correlations Functions}

In the previous section we established a relation between $f_{O}(E,\omega)$ and the Fourier transform $S_O(E,\omega)$ of the nonequal-time correlation function of $\hat O$, see Eq.~\eqref{eq:f_E_w2}. Here, we discuss how the ETH ansatz for two observables $\hat O^{(1)}$ and $\hat O^{(2)}$ are related to their nonequal-time (connected) correlation function $\langle n| \hat O^{(1)}(t) \hat O^{(2)}|n\rangle_c$. Because of its experimental relevance, a case of particular interest is when $\hat O^{(1)}\equiv\hat O(x_1)$ and $\hat O^{(2)}\equiv\hat O(x_2)$. This because $\langle n| \hat O(x_1,t) \hat O(x_2,0)|n\rangle_c$ determines the $\hat O$--$\hat O$ structure factor. Another commonly encountered situation correspond to $\hat O^{(1)}$ and $\hat O^{(2)}$ representing different components of some observable, such as the magnetization, the current, and the electric polarization.

We rewrite the ETH ansatz for the two observables [see Eq.~\eqref{eq:ETH}] as
\be
 O^{(j)}_{mn}=O^{(j)}(\bar E)\delta_{mn}+\mathrm e^{-S(\bar E)/2} \Upsilon^{(j)}_{mn}(E,\omega),
\label{eq:ETH_multi}
\ee
where $j=1,2$, and $\Upsilon^{(j)}_{mn}(E,\omega)\equiv f_{O^{(j)}}(E,\omega)\,R^{O^{(j)}}_{mn}$. This allows us to write
\be
\langle n| \hat O_1(t) \hat O_2(0)|n\rangle_c=\int d\omega\, \mathrm e^{\beta\omega/2-i\omega t} K_{12}(E+\omega/2, \omega),
\label{eq:corr_funct_12}
\ee
where the noise kernel
\be
K_{12}(E+\omega/2,\omega)\equiv\overline {\Upsilon^{(1)}_{nm}(E+\omega/2, \omega) \Upsilon^{(2)}_{mn}(E+\omega/2, \omega)}.
\ee
The overline indicates an average over states $|m\rangle$ within a narrow energy window, that is, at fixed $E$ and $\omega$ (or, equivalently, an average over a fictitious Random Matrix ensemble). 

It is apparent in the expressions above that the noise terms $R^{O^{(1)}}_{mn}$ and $R^{O^{(2)}}_{mn}$ must, in general, be correlated with each other, or else $\langle n| \hat O_1(t) \hat O_2(0)|n\rangle_c\equiv0$. Hence, the assumption that they are random numbers with zero mean and unit variance is oversimplifying and only applicable if we are interested in $\langle n|\hat O(t)\hat O(0)|n\rangle_c$. In order to generalize ETH to deal with nonequal-time correlations of different observables, one can still take $\Upsilon^{(j)}_{mn}$ to be Gaussian with zero mean but its noise kernel with other observables generally needs to be nonvanishing.

Inverting Eq.~\eqref{eq:corr_funct_12}, one has that
\be
K_{12}(E+\omega/2,\omega)={\mathrm e^{-\beta\omega/2}\over 2\pi}\, S_{12}(E,\omega),
\ee
where
\be
S_{12}(E,\omega)=\int_{-\infty}^\infty dt\, \mathrm e^{i\omega t} \langle n| O_1(t) O_2(0)|n\rangle_c.
\ee

For nonequal-space correlation functions in translationally invariant systems, when indexes $1$ and $2$ represent spatial coordinates $x_1$ and $x_2$ for a given observable $\hat O$, one can write
\be
K_{12}(E+\omega/2,\omega)\equiv K(E+\omega/2,\omega,x_1-x_2).
\ee
In this case it is convenient to define the spatial Fourier transform $K$ and work in the momentum space, as one does with structure factors. If the system exhibits Lorentz or Galilean invariance one can further restrict the functional form of the  noise kernel. 

We note that, as we did in Sec.~\ref{sec:fluctuation_dissipation}, one can further simplify Eq.~\eqref{eq:corr_funct_12} splitting the correlation function into its symmetric and antisymmetric parts and Taylor expanding the noise kernel $K_{1,2}(E+\omega/2,\omega)$ with respect to the first argument. However, we should stress that the extension of ETH that we have introduced in this section still needs to be verified numerically. In particular, one needs to understand the regime of validity of the Gaussian ansatz and its applicability to the study of higher order correlation functions. These important questions are left open to future investigations.

%%%%%%%%%%%%%%%%%%%%%%%%%%%%%%%%%%%%%%%%%%%%%%%%%%%%%%%%%%%%%%%%%%%%%%%%%%%%%%%%%%%%%%%%%%
\section{Application of Einstein's Relation to Continuously Driven Systems}\label{sec:sec7}
%%%%%%%%%%%%%%%%%%%%%%%%%%%%%%%%%%%%%%%%%%%%%%%%%%%%%%%%%%%%%%%%%%%%%%%%%%%%%%%%%%%%%%%%%%

\begin{figure}[!t]
\begin{center}
\includegraphics[width=7.7cm]{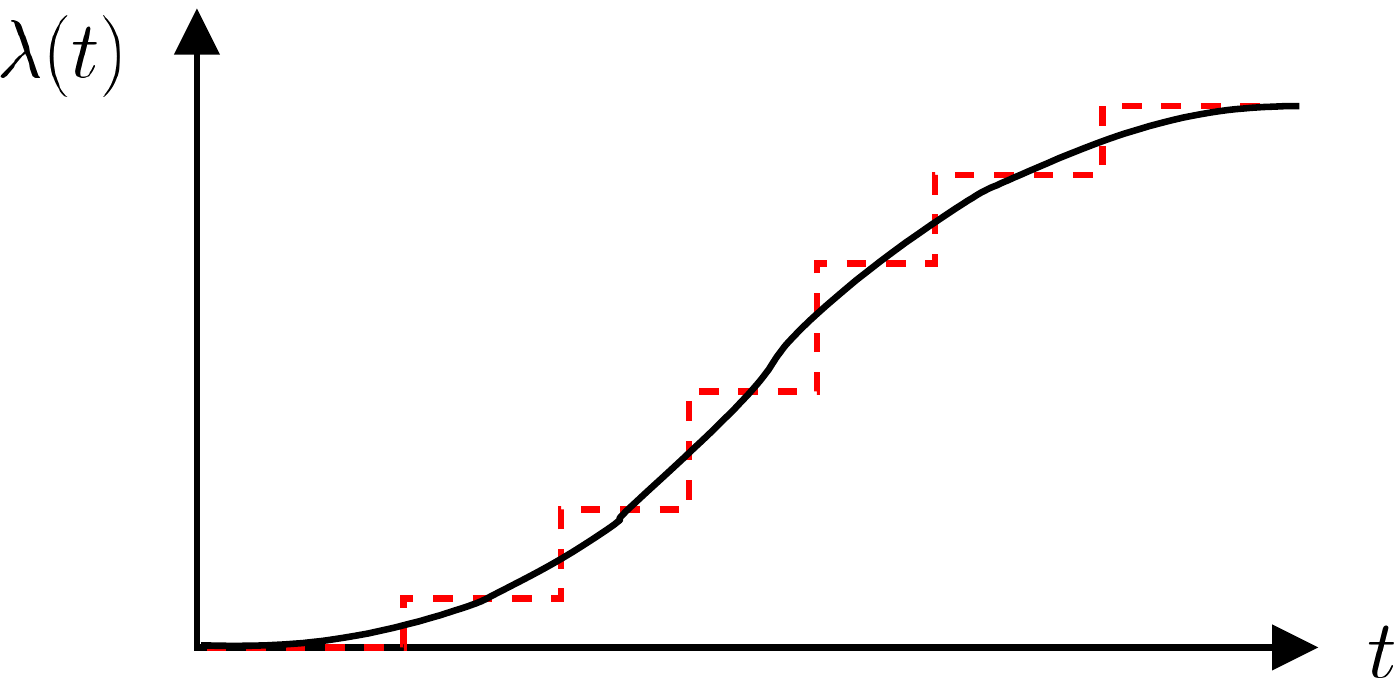}
\caption{Schematic representation of a continuous process as a series of quenches followed by relaxation to the diagonal ensemble. This approximation is justified if the relevant relaxation time in the system is short compared to the characteristic time scale associated with the change of $\lambda$.}
\label{fig:quasi-static}
\end{center}
\end{figure}

\begin{figure}[!h]
\begin{center}
\includegraphics[width=8cm]{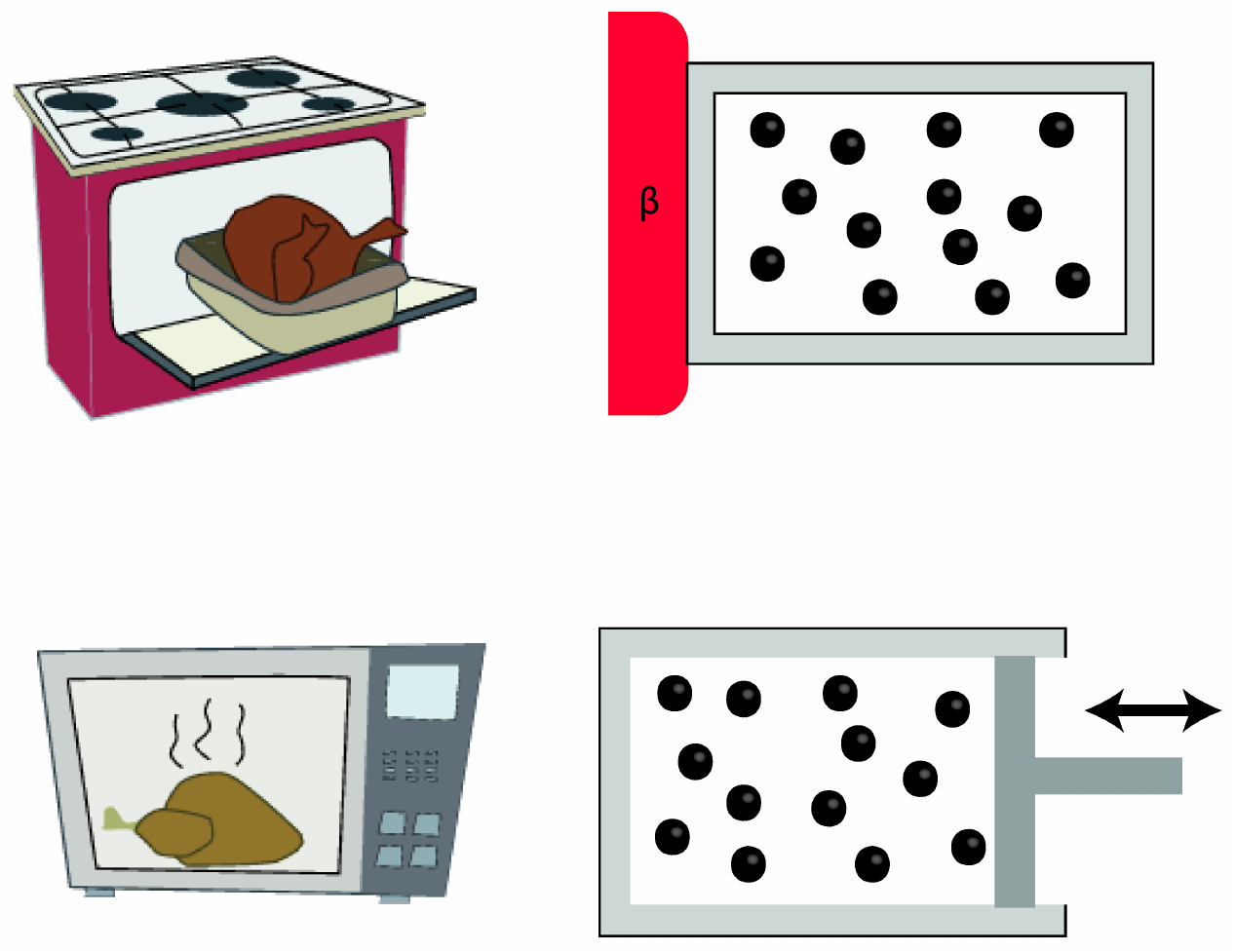}
\caption{Schematic comparison between the usual thermal heating (traditional oven, top) and an energy increase due to non-adiabatic work (microwave oven, bottom).}
\label{fig:ovens}
\end{center}
\end{figure}

In this section, we discuss several examples illustrating how one can use the Einstein relation to obtain nontrivial information about driven systems. We focus first on driven isolated systems which, in the absence of the drive, have only one conserved quantity (energy). We study the energy distribution obtained after a generic quasi-static process, where the driven system is approximately in equilibrium at each moment of time. Schematically, a quasi-static process in an isolated system can be represented as a series of small quenches and relaxation to the diagonal ensemble (see Fig.~\ref{fig:quasi-static}). As described in the previous section, the same setup applies to continuous driving protocols provided that the relevant relaxation time in the system is fast compared to time over which the energy of changes significantly.\footnote{In standard thermodynamics, by a quasi-static process one usually understands a process in which the irreversible work comes from heat exchange with a heat bath. However, such a definition is very restrictive.} As we will show, this setup, besides being common, allows one to take full advantage of the predictive power of the fluctuation theorems to derive results even if the overall energy change in the process is not small, and which, as we will see, might lead to energy distributions that are non-thermal. This setup is analogous to heating in a microwave oven. In the latter, heating occurs not due to the coupling to an external heat reservoir (like in the conventional oven) but rather due to the non-adiabatic work performed by the time-dependent electromagnetic field, see Fig.~\ref{fig:ovens}. Even though this field is periodic in time, the typical relaxation time in the system is much faster than the pulse frequency and therefore the process is quasi-static. Such a process is quasi-static but it is \textit{not} adiabatic since each electromagnetic pulse performs irreversible work $d\tilde Q$ [see Eq.~\eqref{eq:tildeQ_def}] on the system. That work accumulates and leads to heating. This heating can be described by the Fokker-Planck drift-diffusion equation, where the drift and diffusion terms are connected by Einstein's relation (see Sec.~\ref{sec:FP}).

In what follows, we apply these relations to several setups and show how they allow one to make nontrivial statements about the asymptotic energy distribution after long times. Moreover, one can even predict the existence of dynamical phase transitions. All the examples analyzed in this section are classical. The reason is that microscopic simulations of long-time dynamics in quantum chaotic systems are very difficult. It is expected, however, that the general expressions and the Fokker-Planck formalism apply equally to classical and quantum systems.

\subsection{Heating a Particle in a Fluctuating Chaotic Cavity}\label{sec:fluctpartcav}
\label{sec:cavity}

\begin{figure}[!t]
\begin{center}
\includegraphics[width=13cm]{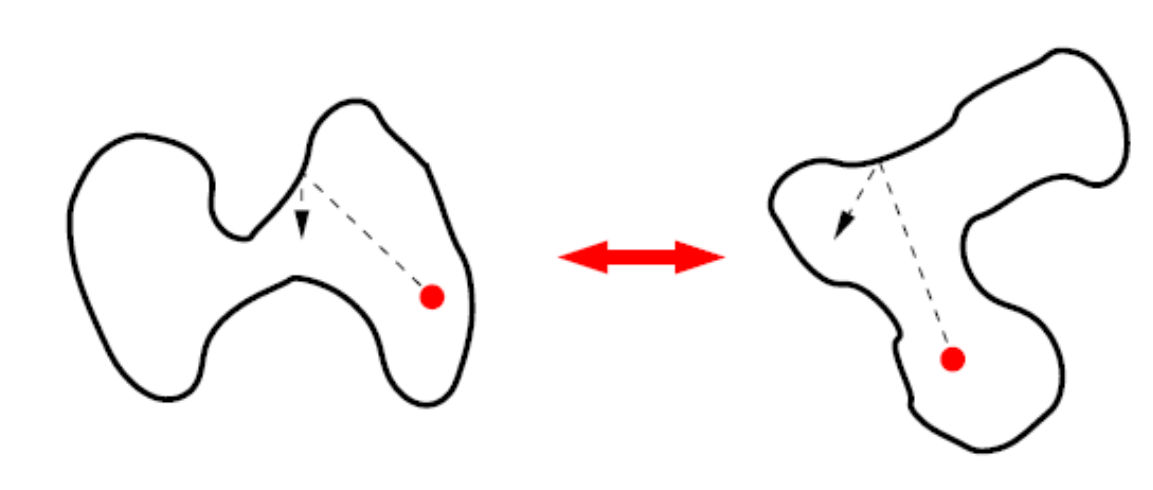}
\caption{Schematic representation of a particle moving in a two-dimensional chaotic cavity with moving boundaries. The driving protocol consists in repeatedly deforming the cavity between the two shapes shown, while keeping its volume fixed.}
\label{fig:fluctuating_cavity}
\end{center}
\end{figure}

We start by considering a very simple problem, that of a classical particle bouncing elastically in a cavity in two spatial dimensions. When the cavity is stationary, the energy of the particle is conserved. If the cavity is chaotic, in the long-time limit, the particle reaches a uniform position distribution and an isotropic momentum distribution. We consider a process in which the system is repeatedly driven by deforming the cavity. At the end of each cycle, the cavity comes back to its original shape\footnote{This requirement can be further relaxed. It is only important that the cavity comes back to the original volume, as the density of states only depends on the volume.} and the system is allowed to relax in the sense described above (see Fig.~\ref{fig:fluctuating_cavity}). In a single collision with the moving wall, the particle's kinetic energy can either increase or decrease. However, it will always increase on average and eventually the particles velocity will become much greater than the velocity of the wall, so that the work per cycle automatically becomes small. The assumption that the cavity is chaotic implies that there are no correlations between consequent collisions. If this is the case, then one can consider a continuous driving protocol instead of repeated quenches and all the results will be the same. Such a setup was analyzed by Jarzynski~\cite{jarzynski_93} followed by other works~\cite{jarzynski_swiatecki_93, blocki_brut_93, blocki_skalski_95, bunin_dalessio_11, kargovsky_anashkina_13, demers_jarzynski_15}. An interesting and nontrivial result that emerges from this analysis is a nonequilibrium exponential velocity distribution (to be contrasted with the Gaussian Maxwell distribution). Here, we analyze this problem in two ways. First, using standard kinetic considerations and then using the Einstein relation.

Let us denote the velocity of the particle as $\vec v$ and velocity of the wall as $\vec V$. Note that the wall velocity is perpendicular to the boundary and, by convention, it points outward, that is, $\vec V=V \hat{n}$, where $\hat{n}$ is the outward normal vector to the boundary. Since we assumed that the cavity deformations are volume preserving $\vec V$ averaged either over time, or over the boundary of the cavity, is zero. By elementary kinematics we know that, during a collision, the component of the relative velocity perpendicular to the boundary is reversed while the component parallel to the boundary is unchanged:
\be
\vec v \to \vec{v}\,'=\vec v -2 \left( \vec v\cdot \hat{n}- \vec V\cdot \hat{n} \right) \hat{n} = \vec v -2 \left( \vec v_\perp - \vec V \right) , 
\label{kinematic}
\ee
where, in the last equality, we have used that $\vec V=V \hat{n}$ and $\vec v_\perp \equiv v_\perp\,\hat{n}= (\vec v\cdot \hat{n})\,\hat{n}$. Note that for a collision to happen we need to have $v_\perp-V>0$ indicating that the particle and the wall are approaching each other. As a check, we verify that when the boundary is stationary, that is, $\vec V=0$, the update rule~\eqref{kinematic} simplifies to
\be
\vec v \to \vec{v}\,' =  \vec v -2 \vec v_\perp = \vec v_\parallel - \vec v_\perp\,,
\ee
where $\vec v_\parallel \equiv \vec v -\vec v_\perp$. As expected, this expression simply states that the  perpendicular component of the velocity is reversed while the parallel component is unchanged.

The energy change of the particle during a collision is
\be
\Delta E = \frac{m}{2} \left( |\vec{v}\,'|^2 - |\vec{v}|^2 \right) = 2m \left( V^2 - V v_\perp \right).
\label{eq:delta_E_cavity}
\ee
We therefore see that the sign of the energy change depends on the sign of $V^2 - V v_\perp$. Combining this result with the constraint $v_\perp-V>0$, which ensures that a collision takes place, we obtain that: i) the energy increases if the boundaries move inward, i.e., $V<0$ and ii) the energy decreases if the boundaries move outward, i.e., $V>0$. The probability of a collision per unit length, $L$, is proportional to the relative velocity between the particle and the wall $v_\perp-V$ (provided this is positive). For the collisions where the energy of the particle increases (i.e., $V<0$) the latter is 
\be
p_>={c\,  \delta t\over 8 L} (v_\perp-V) \,\,\theta\left(v_\perp-V\right)={c\,  \delta t\over 8 L} (v_\perp+|V|) \,\,\theta\left(v_\perp+|V|\right),
\ee
where $\theta$ is the Heaviside step function, $c$ is a proportionality constant (for many particles, $c$ is proportional to the particle density), and 8 was introduced for convenience. On the other hand, for collisions in which the energy of particles decreases (i.e., $V>0$) the probability of collision is 
\be
p_<={c\, \delta t\over 8 L}  (v_\perp-V)\,\, \theta\left(v_\perp-V\right)={c\,  \delta t\over 8 L} (v_\perp-|V|) \,\,\theta\left(v_\perp-|V|\right).
\ee
We are interested in the limit in which the wall moves slowly compared to the particle, i.e., $|v_\perp|\sim |v| \gg |V|$ and therefore the two-step functions above can be simplified to $\theta\left(v_\perp\right)$, indicating that $v_\perp>0$.  Rewriting expression~\eqref{eq:delta_E_cavity} for the energy increasing (decreasing) collisions as
\be
\Delta E_>=2 m \left( V^2 + |V| |v_\perp|\right),\quad \Delta E_<=2 m \left( V^2 - |V| |v_\perp|\right),
\ee
and using the expressions above for the collision probabilities, we find the average heating rate (energy drift) and the energy diffusion:
\be
\begin{split}
J_E & =\frac{1}{\delta t}\left[\Delta E_>\, p_>  + \Delta E_<\, p_< \right]= c\, m\,V^2\,\overline{|v_\perp|}\,\theta(v_\perp), \\
D_E & =\frac{1}{\delta t}\left[(\Delta E_>)^2\, p_> + (\Delta E_<)^2\, p_< \right] \approx c\, m^2\, V^2\, \overline{|v_\perp|^3}\,\theta(v_\perp),
\end{split}
\ee
where, in the second line, we have kept only the leading contribution in $|V|/v_\perp$. The overline indicates averaging over the velocity distribution. It is convenient to use polar coordinates, where $v_\perp= |\vec{v}| \cos(\phi)$ with $\phi\in[-\pi/2,\pi/2]$ to ensure $v_\perp>0$. Using that the particle energy is $E=m |\vec{v}|^2/2$, we obtain
\be
\begin{split}
J_E & = \frac{c\, m\,V^2\,}{L} \left( \frac{2 E}{m}\right)^{1/2} \frac{1}{2\pi} \int_{-\pi/2}^{\pi/2} d\phi\cos\phi, \\
D_E & = \frac{c\, m^2\,V^2\,}{L} \left( \frac{2 E}{m}\right)^{3/2} \frac{1}{2\pi} \int_{-\pi/2}^{\pi/2} d\phi \cos\phi^3.
\end{split}
\label{flat-measure}
\ee
Computing the integral over the angle $\phi$, we obtain
\be
J_E = C \sqrt{E},\quad   D_E = C \frac{4}{3} E^{3/2},
\label{AB_cavity}
\ee
where $C= c\,V^2\, \sqrt{2 m}/(L \pi)$ is a constant with dimensions of $\sqrt{\text{energy}}/\text{time}$. Note that when computing the averages over $v_\perp$ in Eq.~\eqref{flat-measure}, we have used a uniform measure, that is, we took all angles $\phi$ to be equally probable. This is justified because the cavity is assumed to be chaotic. We note that the same analysis can be carried out in an arbitrary spatial dimension $d$. The only difference arises when computing the corresponding angular integrals. A straightforward generalization of the analysis above leads to:
\be
J_E = C_d \sqrt{E},\quad   D_E = C_d \frac{4}{d+1} E^{3/2},
\label{AB_cavity_d}
\ee
where $C_d$ is an overall constant that depends on the dimensionality $d$.

These coefficients satisfy the Einstein relation~\eqref{eq:AB1}:
\be
2J_E=\beta(E) D_E+\partial_E D_E.
\ee
To see this, we note that the single-particle density of states in $d$ dimensions is $\Omega(E)\propto E^{(d-2)/2}$ and therefore 
\be
\beta(E)=\partial_E \ln[\Omega(E)]=\frac{d-2}{2 E}. 
\ee
Combining this with Eq.~\eqref{AB_cavity_d}, it is easy to see that Einstein's relation is indeed satisfied. In fact, there was no need to carry out these relatively elaborate calculations. It was sufficient to note that $J_E$ in all dimensions must be proportional to $\sqrt{E}$, which, for example, follows from the fact that the average number of collisions is proportional to $|v_\perp|\sim \sqrt{E}$. Then, the Einstein relation immediately fixes the energy diffusion $D_E$ with respect to the energy drift $J_E$.

With relation~\eqref{AB_cavity_d} in hand, we can rewrite the Fokker-Planck equation describing the heating process in an arbitrary spatial dimension $d$ as
\be
\partial_t P(E,t)=-C_d \partial_E [\sqrt{E} P(E,t)]+{2\over d+1}C_d \partial_{EE} [E^{3/2} P(E,t)].
\ee
As a preliminary step, we define a new variable $t'=t\,C_d$ and write $\partial_t=C_d\, \partial_{t'}$ so that the constant $C_d$ disappears from the Fokker-Plank equation for $P(E,t')$:
\be
\partial_{t'} P(E,t')=- \partial_E [\sqrt{E} P(E,t')]+{2\over d+1} \partial_{EE} [E^{3/2} P(E,t')]
\label{eq:P_E_tau}
\ee
We observe that, by a scaling analysis of this equation, $t'\sim \sqrt{E}$. Therefore, we can use the following scaling ansatz 
\be
P(E,t')= \frac{1}{2 \alpha t' \sqrt{E}}\,\, f\left(\frac{\sqrt{E}}{\alpha t'}\right) \equiv \frac{1}{2 t' \alpha\sqrt{E}}\,\, f(\xi).
\label{eq:ansatz}
\ee
where $\alpha$ is a constant of order one, which will be fixed later, and the prefactor $1/(2 \alpha t' \sqrt{E})$ has been chosen so that the normalization condition becomes
\be
1=\int_0^\infty dE\, P(E,t')= \int_0^\infty \frac{dE}{2 \alpha t' \sqrt{E}}\, f\left(\frac{\sqrt{E}}{\alpha t'}\right)=\int_0^\infty d\xi\, f(\xi).
\label{eq:norm}
\ee
By substituting ansatz~\eqref{eq:ansatz} into the Fokker-Planck equation~\eqref{eq:P_E_tau}, we find:
\be
2\alpha (1+d) f(\xi) + \left[ 2-d+2\alpha (1+d) \xi \right] f'(\xi) + \xi f''(\xi)=0.
\ee
It is therefore convenient to choose $\alpha^{-1}=2 (1+d)$ so that the equation assumes a particularly simple form:
\be
f(\xi) + (2-d+\xi) f'(\xi) + \xi f''(\xi)=0 \quad \Rightarrow \quad f(\xi)= \frac{1}{(d-1)!}\,\, \xi^{d-1}\,\, e^{-\xi},
\ee
where the normalization constant was fixed using Eq.~\eqref{eq:norm}.

Defining $\tau= \alpha t' = t C_d/[2(d+1)]$, we obtain the final result for the asymptotic energy distribution $P(E,\tau)$ (see also Refs.~\cite{kargovsky_anashkina_13, demers_jarzynski_15}): 
\be
P(E,\tau) = \frac{1}{2(d-1)!}  {E^{\frac{d-2}{2}}\over \tau^d}\, e^{-\sqrt{E}/\tau}.
\label{eq:P_tau}
\ee
This distribution is \textit{universal} in the sense that it does not depend on any details of the driving protocol. It is interesting to compare this result with the equilibrium canonical distribution at temperature $\beta$:
\be
P_c(E,\beta)  = \frac{1}{\Gamma[d/2]} \beta^{d/2} E^{\frac{d-2}{2}}\, e^{-\beta E}.
\label{eq:P_gibbs}
\ee
Clearly, these two distributions are different but share some properties. In particular, they both decay in energy faster than any power law so that all energy moments are well defined. However, $P(E,\tau)$ decay in energy is slower than the $P_c(E,\beta)$ one, that is, the former is ``wider". To quantify this, we compute the first and second moments of the energy with respect both to $P(E,\tau)$ and $P_c(E,\beta)$:
\be
\begin{split}
\langle E \rangle_\tau &= \int_0^\infty dE\, E P(E,\tau) = d(d+1)\tau^2, \quad \langle E^2\rangle_\tau = d(d+1)(d+2)(d+3) \tau^4 \\ 
\langle E \rangle_c &= \int_0^\infty dE\, E P_c(E,\beta) = \frac{d}{2\beta}, \quad\quad\quad\,\,\quad \langle E^2\rangle_c = \frac{d(2+d)}{4\beta^2} ,
\end{split}
\label{eq:moments}
\ee
and define the relative energy width as a figure of merit to compare the width of the two distributions:
\be
\frac{\sigma^2_\tau}{\langle E \rangle^2_\tau}\equiv\frac{\langle E^2\rangle_\tau}{\langle E\rangle^2_\tau}-1 =\frac{6+4d}{d(d+1)},
\quad \frac{\sigma^2_c}{\langle E \rangle^2_c}\equiv\frac{\langle E^2\rangle_c}{\langle E\rangle^2_c}-1 =\frac{2}{d}.
\ee
Clearly, the nonequilibrium distribution $P(E,\tau)$ is wider than $P_c(E,\beta)$ in any spatial dimension. In particular, as $d$ increases from $d=1$ to $d=\infty$, the ratio of the relative energy widths changes from $2.5$ to $2$. In the analysis above, the dimensionality $d$ only enters through the density of states. If we deal with a gas of weakly interacting particles in three dimensions, and the relaxation time of the gas is fast compared to the characteristic rate of energy change due to the cavity's motion, then the same analysis can be applied. The only difference is that $d\to 3N$, where $N$ is the number of particles. Therefore, the result that the asymptotic width of the energy distribution of a driven gas is twice as large as the width of the Gibbs distribution applies to any weakly interacting many-particle gas in any spatial dimension. The gas asymptotically approaches a universal distribution (at least has universal energy fluctuations), but it is not the canonical distribution. In the examples that follow, we will show that a generalization of this result applies to arbitrary driven interacting systems.

Since the distribution $P(E,\tau)$ is a nonequilibrium distribution, it should also have lower entropy than the equivalent Gibbs distribution. The two entropies are
\be\label{eq:entleadt}
\begin{split}
S_\tau &=-\int_0^\infty dE P(E,\tau) \ln\left[\frac{P(E,\tau)}{\Omega(E)}\right]= d (1+\ln\tau ) + \ln\left( 2\, \Gamma[d]\right), \\
S_c &=-\int_0^\infty dE P_c(E,\beta) \ln\left[\frac{P_c(E,\beta)}{\Omega(E)}\right]= \frac{d}{2}(1-\ln\beta) + \ln\left( \Gamma[d/2]\right). \\
\end{split}
\ee
In order to compare these two entropies, we evaluate $S_c$ at $\beta^{-1}=2(d+1)\tau^2$ (so that the distributions $P(E,\tau)$ and $P_c(E,\beta)$ have identical average energy) and compute $S_c-S_\tau$:
\be
S_c-S_\tau=\frac{d}{2} \ln\left[ \frac{2(d+1)}{e}\right] + \ln\left( \frac{\Gamma[d/2]}{2\Gamma[d]} \right)\;.
\ee
This function is positive and increases monotonically from $0.07$ for $d=1$ to $(1-\ln 2)/2\approx 0.15$ for $d\to\infty$. Note that, for large $d$, the first term of both entropies in Eq.~\eqref{eq:entleadt} is proportional to $d$, so the relative difference between them decreases with the dimensionality.\footnote{The $\beta$ or $\tau$ independent terms in both entropies play no role in thermodynamics.} According to our previous discussion, in a weakly interacting gas one has to substitute $d\to 3N$ so that, in the thermodynamic limit ($N\to\infty$), the driven gas has a thermal entropy up to subextensive corrections. If, conversely, we are dealing with a gas of noninteracting particles (implying that the their relaxation time is longer than the heating time) the entropy difference between the driven and the thermal gas will be extensive. This happens despite the fact that the total energy distribution of the noninteracting gas is still a Gaussian due to the central limit theorem. This extensive entropy difference can be used, for example, to build more efficient heat engines and even to beat the fundamental Carnot bound in some cases (see the discussion in Ref.~\cite{mehta_polkovnikov_13}).

Let us note that the form of the distribution~\eqref{eq:P_tau} was obtained under the assumption of constant driving, i.e., $V^2=\rm {const}$. If one uses feedback control mechanisms such that velocity of the wall is tied to the velocity of the particle, i.e., $V=V(E)$, then one can change the energy dependence of $J_E\propto V^2\sqrt{E}$ and, hence, change the resulting nonequilibrium distribution. One can even induce dynamical phase transitions (see the next example).

In passing, we note that the same results for the energy distribution have been derived in a the context of the Lorentz gas~\cite{dalessio_krapivsky_11}. This gas is defined as a system of noninteracting light particles colliding with an interacting gas of heavy particles moving with an average velocity $V$. If the ratio of the masses is very large, then there is no effect of the collision on the heavy particles so the latter serve precisely the role of moving boundaries. In this case, the behavior of the light particles can be obtained exactly via the Lorentz-Boltzmann kinetic equation. It is interesting to note that the ensemble of heavy particles can be viewed as an infinite temperature heat bath. Indeed, the average energy of heavy particles $M \langle V^2\rangle /2$, which defines temperature, diverges in the limit $M\to \infty$ at fixed $\langle V^2\rangle$. Thus, according to the general discussion of Sec.~\ref{sec:sec6}, this simple example shows that an external quasi-static driving of an isolated system is equivalent to the coupling to the infinite temperature bath. 

While the single-particle example considered here is relatively simple, it teaches us several important lessons that can be extended to many-particle systems. In particular, it shows: (i) the possibility of nonequilibrium universal distributions, and (ii) that doubly stochastic evolution for quasi-static driving protocols is sufficiently constraining to predict such universality irrespective of the details of the driving protocol (in our example, the details are encoded in the overall constant $C_d$).

\subsection{Driven Harmonic System and a Phase Transition in the Distribution Function}

Next, we consider a driven single particle in a harmonic trap~\cite{bunin_dalessio_11}. This particle is weakly coupled to a finite system composed of $N$ identical particles, so that the overall system is ergodic. The details of the larger system define the density of states $\Omega(E)$, and therefore $\beta(E)$. Repeated impulses of short duration act on the particle and drive it away from equilibrium. The time scale between impulses is taken to be larger than the equilibration time of the particle. In addition, we assume that the coupling of the particle to the rest of the system is so weak that during the impulse it can be ignored. This setup can be easily generalized to impulses acting on an extensive number of particles. The energy of the particle $\varepsilon$ (we use this notation to distinguish it from the total energy of the system $E$) between cycles is given by
\be
\varepsilon=\frac{1}{2}kx^{2}+\frac{p^2}{2}.
\label{eq:varepsilon}
\ee
For simplicity, we work in one dimension. In Eq.~\eqref{eq:varepsilon}, $x$ is the coordinate of the particle, $v$ its velocity, $m=1$ its mass, and $k$ is the spring constant. Because the system is ergodic and $N$ is large, at any given energy of the system, the probability distribution  for $(x,v)$ before the impulse is Gibbs: $\rho(x,v)\propto\exp[-\beta(E)\varepsilon]$. We take the impulse magnitude to be $F(x)\delta t$, with $\delta t$ short enough so that, during the impulse, the particle's position does not change appreciably and the coupling to the rest of the system can be ignored. Following the impulse, the momentum changes according to $p\rightarrow p+F\left(x\right)\delta t$. It is straightforward to calculate both the drift $J_E$ and the diffusion constant $D_E$~\cite{bunin_dalessio_11}:
\begin{align*}
J_E & =\left\langle \left[F\left(x\right)\right]^{2}\right\rangle \frac{\delta t^{2}}{\tau}, \\
D_E & =\frac{2}{\beta(E)}\left\langle \left[F\left(x\right)\right]^{2}\right\rangle\frac{\delta t^{2}}{\tau},\;
\end{align*}
where the angular brackets denote an average over $\rho(x,v)$ and $\tau$ is the time (or average time) between impulses. As in the previous example, one could have calculated $J_E$ and deduced $D_E$ from the Einstein relation. Note that here, in contrast to the single-particle example, the $\partial_E D_E$ term in the relation $2 J_E=\beta(E) D_E+\partial_E D_E$ is a $1/N$ correction, which is negligible. Technically, this correction would amount to the fact that, at finite but large $N$, the single-particle distribution $\rho(x,v)$ slightly deviates from a Gibbs distribution.

Next, we consider the probability distribution for the energy of the system after the application of many impulses. To proceed, we have to assume a specific form for $\Omega(E)$ [or equivalently $\beta(E)$] and for $F(x)$. For simplicity, we take $\beta\left(E\right)\propto E^{-\alpha}$ and $F\left(x\right)\propto \rm {sign}(x) |x|^{r}$ so that
\be
J_E\propto \langle x^{2r} \rangle \propto E^{\alpha r} \equiv E^s,
\ee
where we used $\langle x^2 \rangle \propto \beta(E)^{-1} \propto E^{\alpha}$, with the first proportionality following from the equipartition theorem. For large $N$, due to the central limit theorem, the energy distribution is approximately Gaussian. Therefore, to characterize the distribution, it suffices to find the mean energy as a function of time and its variance as a function of the mean energy.

To find the relation between the mean energy and its variance, to leading order in $1/N$, we can multiply the Fokker-Planck equation~\eqref{eq:fokker-planck} by $E$ and $E^2$ and integrate over all energies. This yields the following differential equations describing the time evolution of $\langle E \rangle$ and $\sigma^{2}=\langle E^2 \rangle - \langle E \rangle^2$, where angular brackets stand for averaging over the probability distribution $P(E,t)$:
\begin{eqnarray*}
 &  & \partial_{t} \la E \ra =\la J_E \ra \\
 &  & \partial_{t}\sigma^{2}=\left\langle D_E\right\rangle +2\left(\left\langle J_E\, E\right\rangle -\left\langle J_E\right\rangle \langle E \rangle\right).
\end{eqnarray*}
Combining these equations yields
\begin{equation}
\frac{\partial\sigma^{2}}{\partial \langle E \rangle}=\frac{\left\langle D_E\right\rangle +2\left(\left\langle J_E\,E\right\rangle -\left\langle J_E\right\rangle \langle E \rangle\right)}{\langle J_E\rangle} \;.
\end{equation}
Moreover, if the energy distribution $P(E)$ is narrow, as is the case of large $N$, we can evaluate the averages within the saddle-point approximation. Using the Einstein relation $J_E=\beta(E) D_E/2$, we obtain:
\begin{equation}
\frac{\partial\sigma^{2}}{\partial \la E \ra} = {2\over \beta(\la E \ra)}+2\frac{\partial_{E}J_E (\la E\ra)}{J_E(\la E\ra )}\sigma^{2}(\la E\ra).
\label{intermediate}
\end{equation}
Integrating this equation between the initial energy of the system $\la E\ra_0$ and its final energy $\la E \ra$ gives
\begin{equation}
\label{widthbeta}
\sigma^{2}(\la E\ra)=\sigma^{2}_0\frac{J_E(\la E\ra)^2}{J_E(\la E\ra_0 )^2} + 2 J_E^{2}(\la E\ra) \int_{\la E\ra_0}^{\la E\ra}\frac{dE^{\prime}}{J_E(E^{\prime})^2\beta(E^{\prime})}\;
\end{equation}
where $\sigma_0$ is the initial width of the distribution and $\la E\ra_0$ is the initial mean energy.

We now use this equation in conjunction with the results obtained for the particle in a harmonic trap. In that case, the change of the mean energy of the system is given by 
\begin{equation}
\label{s}
\partial_t \la E\ra =J_E(\la E\ra)=c \la E \ra ^s \;.
\end{equation}
Note that the values of $\alpha$ (recall that $\alpha r =s$), which define the specific heat exponents, are constrained by thermodynamic reasons to $0 <\alpha\le 1$.  The lower bound is required due to the positivity of the specific heat, and the upper bound assures that the entropy [$S(E)\propto E^{1-\alpha}$] is an increasing unbounded function of the energy (the latter condition can be violated in systems with bounded energy spectrum). To prevent the energy of the system from diverging at finite time, we require $s\le1$ [as follows from integrating Eq.~\eqref{s}].

For simplicity, we assume that $\sigma_{0}\rightarrow 0$, that is, that we are starting from a very narrow microcanonical distribution. As done in Sec.~\ref{sec:fluctpartcav}, it is useful to compare the width $\sigma^{2}$ to the equilibrium canonical width $\sigma_c^{2}=-\partial_\beta {\la E\ra } \sim {\la E\ra }^{1+\alpha}/\alpha$. This comparison reveals that, as the functional form of the impulse (specifically, the value of $r$) is changed, the system displays a {\it transition} between two behaviors. To see this, note that $\sigma^2({\la E\ra })$ is controlled by the exponent $\eta=2\alpha r-\alpha-1=2s-\alpha-1$, which determines if the integral in Eq.~\eqref{widthbeta} is controlled by its lower or upper bound: (i) When $\eta < 0$, the width is \textit{Gibbs-like} with $\sigma^2/\sigma_c^2\to 2\alpha/|\eta|$, that is, the ratio $\sigma^{2}/\sigma_c^{2}$ asymptotically approaches a constant value that can be either larger or smaller than one. Smaller widths correspond to protocols with large and negative $s$. Namely, protocols where $J_E$ is a strongly decreasing function of the energy. (ii) When $\eta>0$, there is a \textit{run-away} regime. Here, the width increases with a higher power of the energy than the canonical width: $\sigma^{2}/\sigma_c^{2}\sim E^{\eta}$. The resulting distribution is significantly wider than the canonical one. Given the constraint on the value of $s$, this regime can only be reached if $\alpha<1$ (in particular, this regime is unreachable for a driven classical ideal gas). The transition between the two regimes occurs when $\eta=0$. In this case, $\sigma^{2}/\sigma_c^{2}\sim 2 \alpha \ln (\la E\ra/\la E\ra_0)$. One can show that close to this transition, when $|\eta|\ll 1$, there is a divergent time scale (in terms of the number of impulses) required to reach the asymptotic regime, see Ref.~\cite{bunin_dalessio_11} for details. Therefore, this setup realizes a dynamical transition for the asymptotic energy distribution of the system, which is qualitatively similar to a continuous phase transition. The parameter $\eta$ plays the role of the tuning parameter.

For concreteness, consider a system with $\alpha=1/2$ (such as a Fermi liquid or the one-dimensional harmonic system above). When $r=1$, we have $\eta=2\alpha r-\alpha-1=-1/2$ and the resulting distribution is {\it Gibbs-like} with $\sigma^2/\sigma_c^2=2$. When $r=3/2$, we are at the critical regime $\eta=0$. Finally, for $r=2$, one has $\eta=1/2$ leading to the {\em run-away} regime with $\sigma^2/\sigma_c^2\sim E^{1/2}$. Note that Eq.~\eqref{widthbeta} implies that the existence of these three regimes is generic. In particular, depending on the functional form of $J_E(E)$ and $\beta(E)$, the variance of the distribution can be larger or, surprisingly, smaller than the width of the equilibrium Gibbs distribution at the same mean energy. Specifically, $\sigma^{2}(E)/\sigma^{2}_c(E)$ can be made arbitrarily small by a proper choice of $J_E(E)$. Also, the existence of the dynamical phase transition described for this simple model is only tied to whether the integral in Eq.~(\ref{widthbeta}) diverges or converges at high energy. The emergence of a nontrivial universal asymptotic behavior of the energy distribution is insensitive to the details of the driving protocol, such as the driving amplitude and shape of the pulse. 

While in the examples in this section we focused on classical systems, the same conclusions apply to driven quantum systems (see Ref.~\cite{bunin_dalessio_11} for specific examples). One can also anticipate nontrivial universal non-equilibrium distributions in driven systems with more than one conserved quantity. In this case, the Onsager relations will be responsible for constraining the mean flows of the conserved quantities and the flows of their fluctuations.

\subsection{Two Equilibrating Systems}

As a final example, we follow Ref.~\cite{bunin_kafri_13} and describe the equilibration of two weakly coupled systems, as those shown in Fig.~\ref{fig:two-systems}, but with no particle exchange ($J_N=0$). When one of the systems is much bigger than the other one, the bigger system serves as a heat bath for the small system and the two systems equilibrate at the temperature of the bath, which does not change during the equilibration process. The situation changes, however, when the two systems are comparable to each other, so that both systems are affected by the heat exchange. Assuming that the energy flow between the two systems is much slower than the characteristic relaxation time of each of the systems, and that the total energy $E_{\rm tot}$ is not fluctuating, we can again use the Fokker-Planck formalism and the analysis in the previous example. The only difference is that, instead of $\beta(E)$, we need to use $\Delta \beta(E)=\beta_{\rm I} (E_{\rm I})-\beta_{\rm II}(E_{\rm II})$, where $\beta_{\rm I}$ and $\beta_{\rm II}$ are the temperatures of systems ${\rm I}$ and ${\rm II}$, respectively, and $E_{\rm I}$ and $E_{\rm II}$ are their respective energies. The latter satisfy the constraint $E_{\rm I}+E_{\rm II}=E_{\rm tot}=\text{const}$. Then, instead of Eq.~\eqref{widthbeta}, we find the following expression for the energy fluctuations in system ${\rm I}$:
\begin{equation}
\sigma_{\rm I}^{2}(\la E_{\rm I}\ra)  = \sigma_{{\rm I}, 0}^{2}\frac{J_E(\la E_{\rm I}\ra)^2  }{J_E(\la E_{\rm I}\ra_0)^2}+ 2J_E(\la E_{\rm I}\ra)^2  \int_{\la E_{\rm I}\ra_0}^{\la E_{\rm I}\ra}\frac{1}{J_E( E^\prime)^2  \left[\beta_{\rm I}(E')-\beta_{\rm II}(E_{\rm tot}-E')\right]}dE^{\prime}\;,
\label{eq:equilibrium_approach}
\end{equation}
where $J_E$ is the rate of the heat flow into system ${\rm I}$. This equation describes the evolution of the width of the energy distribution of system ${\rm I}$. As the system equilibrates, one expects that
\be
J_E=C(E-E_{\rm I}^{\rm eq}),
\ee
where $E_{\rm I}^{\rm eq}$ is the equilibrium steady-state mean energy of system I, for a given total energy $E_{\rm tot}$. Likewise
\be
\beta_{\rm I}(E')-\beta_{\rm II}(E_\text{tot}-E')\approx -\left({1\over \sigma_{{\rm I}, c}^2}+{1\over \sigma_{{\rm II}, c}^2}\right) (E'-E_{\rm I}^{\rm eq}),
\ee
where $\sigma_{{\rm I},c}^{2}=-\partial E_{\rm I}/\partial \beta_{\rm I}$ is the variance of the energy distribution in the canonical ensemble of system I with mean energy $E_{\rm I}^{\rm eq}$ and, similarly, $\sigma_{{\rm II},c}^{2}$ is defined for system II. The two expressions above ensure that, in the steady state, when $E_{\rm I}=E_{\rm I}^{\rm eq}$, the (average) heat flux is zero and the temperatures of the two subsystems are identical. By substituting this expansion in Eq.~\eqref{eq:equilibrium_approach}, we find the asymptotic result for the energy fluctuations in system I after equilibration:
\be
\sigma_{\rm I}^{2}=\sigma_{\rm II}^{2}\approx { \sigma_{{\rm II}, c}^2  \sigma_{{\rm I}, c}^2\over  \sigma_{{\rm II}, c}^2+ \sigma_{{\rm I}, c}^2}.
\ee

If one of the systems, say system I, is much smaller than the other, then this result simply implies that, after equilibration, the energy fluctuations in both systems are given by the canonical energy fluctuations of the \textit{smallest} system. If the two systems are identical, then the energy fluctuations in either one of the systems are equal to one half of the canonical ones. 

\begin{figure}[!t]
\begin{center}
\includegraphics[height=1.9709in,width=2.8781in]{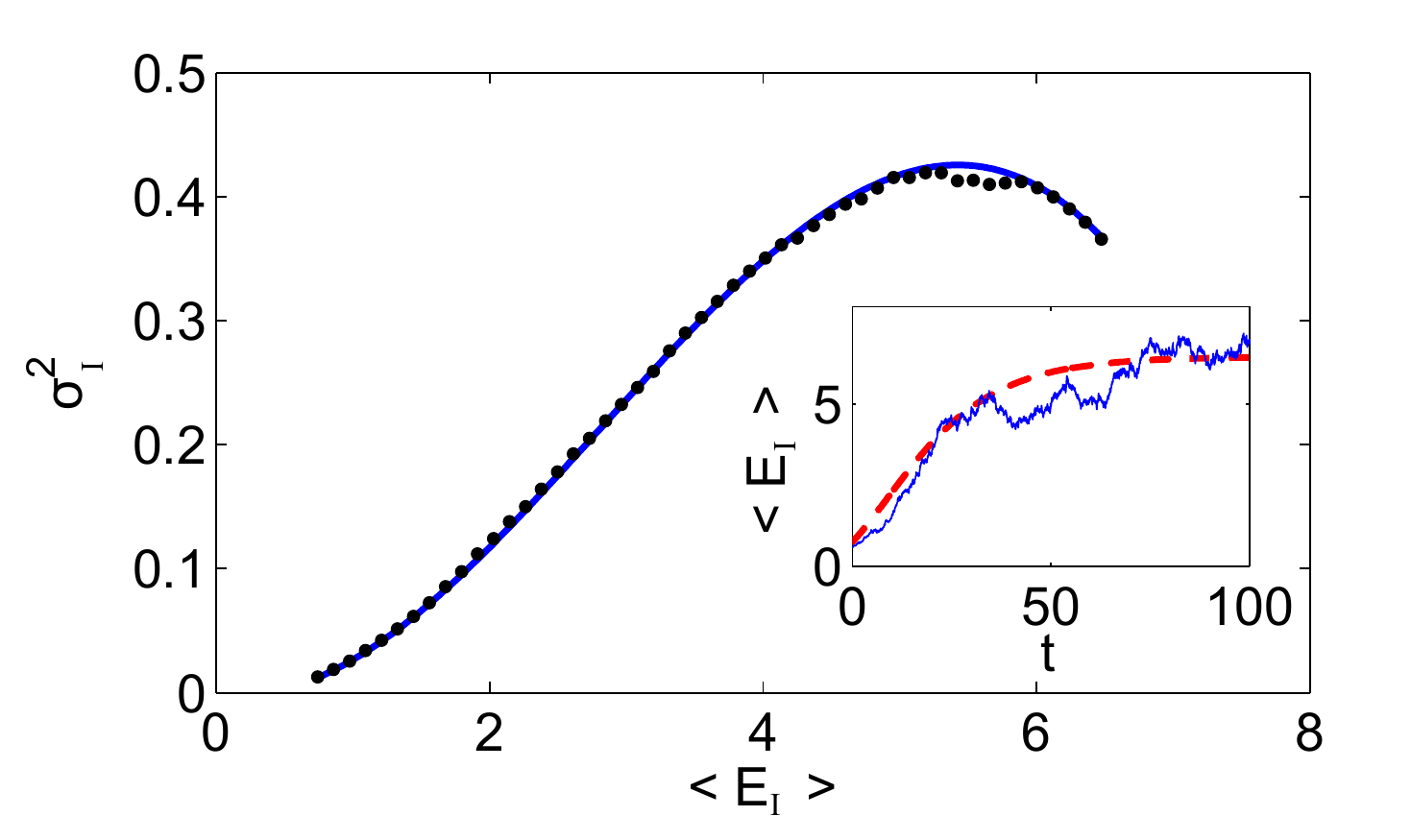}
\caption{Equilibration of a system consisting of 50 particles with two different masses confined in a box. We plot results for $\la E_{\rm I}\ra$ vs $\sigma^{2}$ (dots), and compared them to the theoretical prediction in Eq.~\eqref{eq:equilibrium_approach} (solid line). Inset: Energy $E_{\rm I}(t)$ in a single run (solid line), and the average energy $\la E_{\rm I}\ra(t)$ (dashed line). The particle numbers used in the simulations are $N_{\rm I}=30$ and $N_{\rm II}=20$. The initial velocities are sampled from a Maxwell-Boltzmann distribution with very different initial temperatures: $\beta_{\rm I}=60$ and $\beta_{\rm II}=3$. The total energy constraint is enforced by a (small) rescaling of the velocities of the $m_\text{II}$-particles. The masses are chosen to be $m_{\rm I}=10^{-4}$ and $m_\text{II}=1$ (in arbitrary units). The box is a unit cube with reflecting boundaries, and the added volume of the particles is taken to occupy a 5\% of the volume of the box.}
\label{fig:mdequil}
\end{center}
\end{figure}

Equation~\eqref{eq:equilibrium_approach} is, however, more general and can be used to study the full evolution of the distribution as the systems equilibrate (and not just the approach to the asymptotic result). Again, following Ref.~\cite{bunin_kafri_13}, let us consider a specific example of a gas of hard spheres in a box. They are simulated using an event-driven molecular dynamics~\cite{Allen_89}. The gas has $N_{\rm I}$ particles of mass $m_{\rm I}$ and $N_{\rm II}$ particles of mass $m_{\rm II}$, all of equal size. These groups of particles represent the two systems ${\rm I}$ and ${\rm II}$. This setup is similar to the Lorentz gas analyzed in Sec.~\ref{sec:cavity}, which in turn is identical to a single particle in a chaotic cavity. The difference is that here the two masses are both finite while, in the Lorentz gas analyzed earlier, one type of atoms was infinitely heavier than the other. It is straightforward to check studying the collision between two particles that, if the two masses are very different, the energy transfer in each collision is small. In this case, a significant energy transfer occurs only over many collisions. Numerical simulations were repeated for many runs to evaluate the width of the distribution as a function of the average energy. In addition, during the evolution, the energy transfer between the two systems was evaluated and Eq.~\eqref{eq:equilibrium_approach} was used to compute the width of the distribution.  The results, shown in Fig. \ref{fig:mdequil}, indeed confirm the predictions of the Fokker-Planck derivation based on the Einstein relation for the open systems~\eqref{eq:AB2}.

%%%%%%%%%%%%%%%%%%%%%%%%%%%%%%%%%%%%%%%%%%%%%%%%%%%%%%%%%%%%%%%%%%%%%%%%%%%%%%%%%%%%%%%%%%
\section{Integrable Models and the Generalized Gibbs Ensemble (GGE)}\label{sec:sec8}
%%%%%%%%%%%%%%%%%%%%%%%%%%%%%%%%%%%%%%%%%%%%%%%%%%%%%%%%%%%%%%%%%%%%%%%%%%%%%%%%%%%%%%%%%%

One of the focuses of this review has been understanding what happens in isolated nonintegrable quantum systems that are taken far from equilibrium by means of a sudden quench. We have discussed the relaxation dynamics of physical observables and their properties after relaxation. We explained that quantum chaos, through eigenstate thermalization, is the reason behind thermalization in those systems. In this section, we briefly discuss what happens in integrable systems. Such systems do not exhibit eigenstate thermalization. 

We note that the very definition of quantum integrability is a topic of debate (see, e.g., Ref.~\cite{sutherland_04,caux_mossel_11,yuzbashyan_shastry_13}), but we will not touch upon that here. A rather standard definition of quantum integrability, based on the existence of an extensive number of local operators (or, more precisely, operators that are extensive sums of local operators) $\hat I_k$ that commute with the Hamiltonian and with each other, will be sufficient for the discussion here. The requirement that the conserved quantities are local/extensive is essential and it excludes the projection operators to the eigenstates of the Hamiltonian. In fact, for any quantum Hamiltonian, integrable or not, the projection operators to its eigenstates commute with the Hamiltonian and with each other, that is, they are conserved, but they are neither local nor extensive. As we have argued for quantum chaotic systems, the existence of those conserved quantities does not preclude the thermalization of physical observables. Exactly the same can be said about higher moments of the Hamiltonian, which are separately conserved but play no role both in equilibrium thermodynamics and in the thermalization of chaotic systems~(see, e.g., the discussion in Sec.~\ref{ss:quencheth}).

\subsection{Constrained Equilibrium: the GGE}

The main difference between chaotic and integrable systems becomes apparent already at the single-particle level, see Fig.~\ref{fig:billiards}. In the classical chaotic billiard any trajectory, after some time, uniformly fills the available phase space at a given energy. As a result, the long-time average and the microcanonical ensemble average of an observable agree with each other. In contrast, in the integrable cavity the particle's motion is constrained by other conserved quantities and the time average and the microcanonical ensemble average need not agree. Nevertheless, the particle might still uniformly fill the available phase space. This means that the long-time average could still be described by an ensemble average, but it needs to be a generalized microcanonical ensemble that accounts for all conserved quantities in the system \cite{yuzbashyan_15}.

In the quantum language, this amounts to saying that, in order to describe time averages of observables in integrable systems, one needs a constrained ensemble which is built using eigenstates of the Hamiltonian involved in the dynamics (they are selected by the initial state). As in classical systems, in which conserved quantities preclude the exploration of all phase space, the failure of integrable quantum systems to exhibit eigenstate thermalization can be traced back to the fact that they have an extensive number of nontrivial (local/extensive) conserved quantities $\{\hat{I}_k\}$, as opposed to the ${\cal O}(1)$ number of extensive conserved quantities in nonintegrable systems (energy, momentum, etc). Despite the existence of the conserved quantities $\{\hat{I}_k\}$, and because of dephasing (like in nonintegrable systems), observables in integrable systems are expected to relax to stationary values and remain close to those values at most later times.
 
Remarkably, in Ref.~\cite{rigol_dunjko_07}, the previous statements were shown to hold for an integrable model of hard-core bosons. Instead of a generalized microcanonical ensemble, a generalized grand canonical one was introduced in that work. It is now known as the GGE, whose density matrix:
\begin{equation}\label{eq:GGE}
 \hat{\rho}_\text{GGE}=\frac{\exp(-\sum_k \lambda_k \hat{I}_k)}
 {\text{Tr}[\exp(-\sum_k \lambda_k \hat{I}_k)]},
\end{equation}
was obtained by maximizing the entropy \cite{jaynes_57a,jaynes_57b} under the constraints imposed by conserved quantities that make the system integrable. The values of the Lagrange multipliers were determined by requiring that, for all $k$'s, $\text{Tr}[\hat{\rho}_\text{GGE}\hat{I}_k]$ equals the expectation value of $\hat{I}_k$ in the initial state. It is a priori not obvious that the exponential form in Eq.~\eqref{eq:GGE} is warranted. For extensive integrals of motion, one can justify the exponential form in the same way as it is done in traditional statistical mechanics, namely, noting that: (i) because of the equivalence of ensembles for subsystems, which is a direct consequence of the extensivity of the conserved quantities, the precise form of the distribution is not essential, and (ii) the exponential distribution leads to statistical independence of subsystems, which is naturally expected after relaxation in a system governed by a local Hamiltonian.

The fact that observables in integrable systems do not, in general, relax to the same values seen in thermal equilibrium, and that the GGE describes few-body observables after relaxation, has been verified in a large number of studies of integrable models. These can be either solved numerically for much larger system sizes than those accessible to full exact diagonalization or analytically solved in the thermodynamic limit \cite{rigol_dunjko_07, cazalilla_06, rigol_muramatsu_06, kollar_eckstein_08, barthel_schollwock_08, iucci_cazalilla_09, iucci_cazalilla_10, cassidy_clark_11, calabrese_essler_11, calabrese_essler_12b, cazalilla_iucci_12, caux_konik_12, gramsch_rigol_12, essler_evangelisti_12, collura_sotiriadis_13a, collura_sotiriadis_13b, caux_essler_13, mussardo_13, fagotti_13, fagotti_essler_13a, fagotti_collura_14, pozsgay_14a, wright_rigol_14, vidmar_rigol_16}. We should emphasize, however, that special initial states may still lead to nearly thermal expectation values of observables in integrable systems after relaxation \cite{rigol_muramatsu_06, cramer_dawson_08, rigol_fitzpatrick_11, rigol_srednicki_12, he_rigol_12, he_rigol_13, chung_iucci_12, torres_santos_13}. Hence, finding nearly thermal results for some initial states does not automatically mean that a system is nonintegrable. In what follows we discuss several examples for which the GGE can be used. These will highlight the reasons for its applicability.

\subsubsection{Noninteracting Spinless Fermions}\label{sec:nisf}

Noninteracting systems are possibly the simplest class of integrable systems. Let us discuss one particular noninteracting system that clarifies some important features of the GGE introduced above. We focus on noninteracting spinless fermions in a one-dimensional lattice (relevant to the hard-core boson system discussed in Sec.~\ref{sec:HCBs})
\begin{eqnarray}
 {\hat H}_\text{SF} = - J \sum_{j=1}^{L-1} \left( {\hat f}_j^{\dagger} {\hat f}^{}_{j+1} + \text{H.c.} \right)
            + \sum_{j=1}^L u_j {\hat n}_j^f ,
\label{eq:nisf}
\end{eqnarray}
where $ {\hat f}_j^{\dagger}$ (${\hat f}_j^{}$) is a fermionic creation (annihilation) operator at site $j$, ${\hat n}_j^f={\hat f}_j^{\dagger}{\hat f}_j^{}$ is the site $j$ occupation operator, $J$ is the hopping parameter, and $u_j$ are arbitrary site potentials.

The single-particle Hamiltonian \eqref{eq:nisf} can be straightforwardly diagonalized: ${\hat H}_\text{SF}\hat{\gamma}^{\dagger}_k|0\rangle=\varepsilon_k\hat{\gamma}^{\dagger}_k|0\rangle$, where $\varepsilon_k$ are the single-particle eigenenergies, $|k\rangle\equiv\hat{\gamma}^{\dagger}_k|0\rangle$ are the single-particle eigenstates, and $k=1,2,\ldots,L$. The occupations of the single-particle eigenstates $\hat{\eta}^{}_k=\hat{\gamma}^{\dagger}_k\hat{\gamma}^{}_k$ immediately form a set of $L$ nontrivial \textit{nonlocal} conserved quantities for a system consisting of many noninteracting spinless fermions. These conserved quantities {\it are not} extensive. However, carrying out the GGE analysis for this set of conserved quantities, imposing that $\text{Tr}[\hat{\rho}_\text{GGE}\hat{\eta}_k]=\langle\psi_I|\hat{\eta}_k|\psi_I\rangle\equiv\eta^I_k$, one finds that the Lagrange multipliers are given by the expression \cite{rigol_dunjko_07}
\begin{equation}
\lambda_k=\ln\left[ \frac{1-\eta_k^I} {\eta_k^I}\right],
\end{equation}
that is, the Lagrange multipliers are smooth functions of $\eta_k^I$. 

\begin{figure}[!t]
\begin{center}
\includegraphics[width=0.413\textwidth]{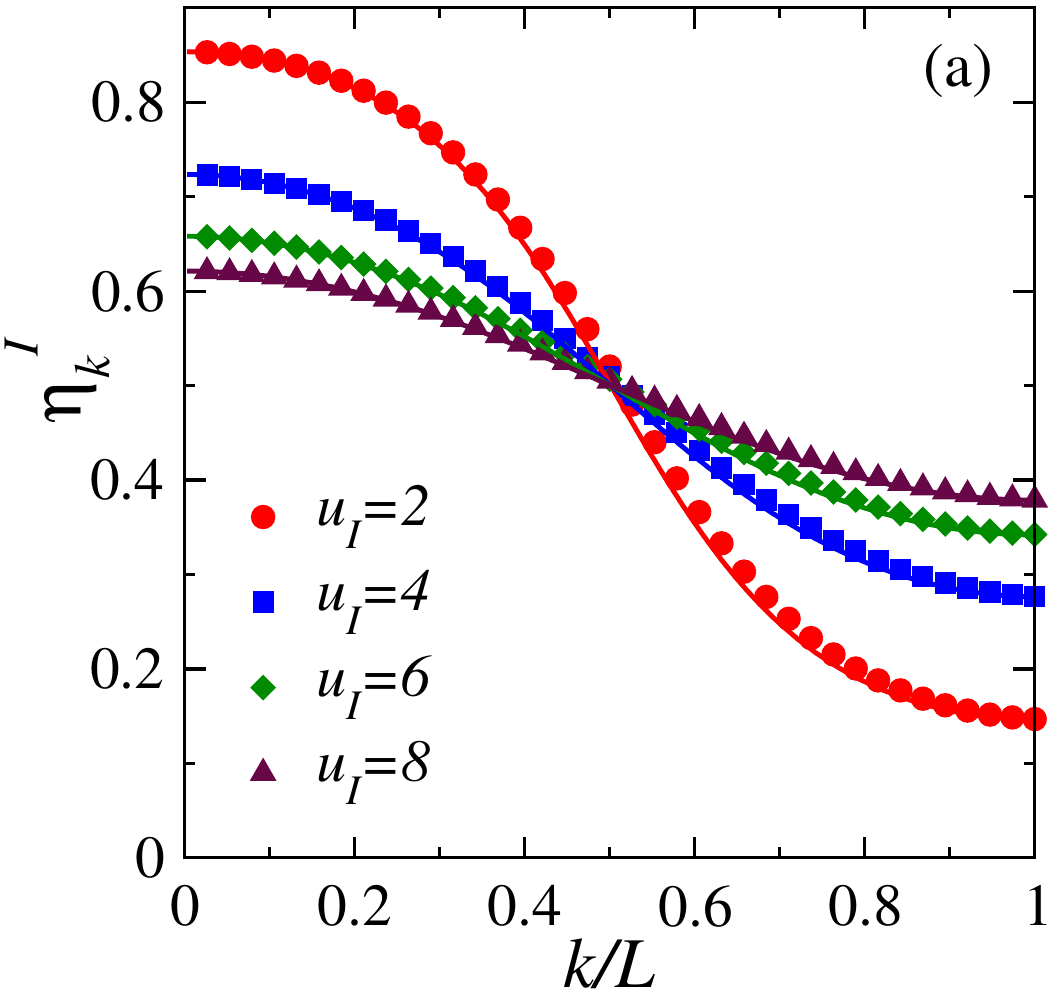}\hspace{0.7cm}
\includegraphics[width=0.4\textwidth]{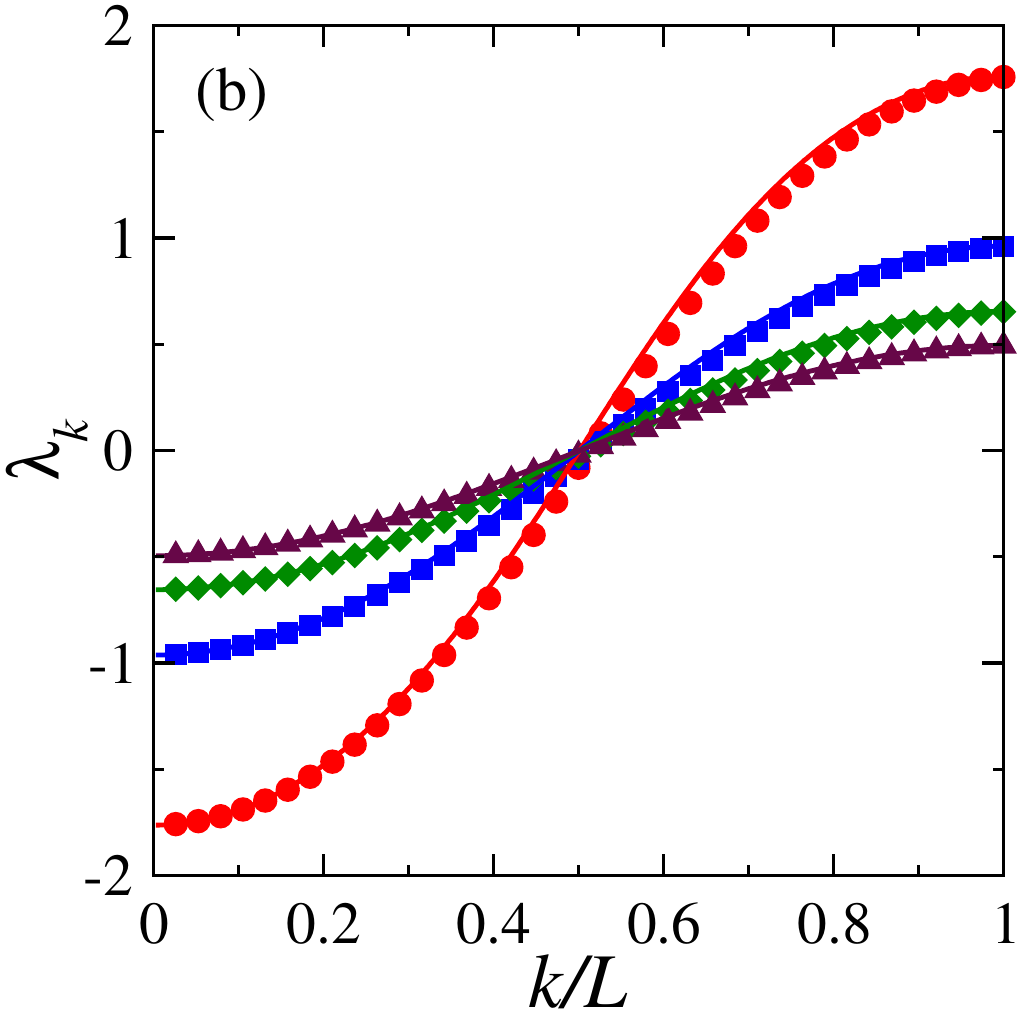}
\end{center}
\vspace{-0.15cm}
\caption{(a) Expectation value of the conserved quantities in quenches from $u_I\neq0$ (as indicated in the figure) to $u=0$. (b) The corresponding Lagrange multipliers.  The conserved quantities are ordered such that $\varepsilon_k$ increases with increasing $k$. The results denoted by symbols (lines) correspond to systems with $L=38$ (380) sites at half-filling ($N=L/2$). Adapted from Ref.~\cite{rigol_fitzpatrick_11}.}
\label{fig:nisf}
\end{figure}

In Fig.~\ref{fig:nisf}, we show results for $\eta_k^I$ and $\lambda_k$ after a quench in which a superlattice potential [$u_j=u(-1)^j$ in Eq.~\eqref{eq:nisf}] is turned off. In this quench the initial state is taken to be the ground state for $u_I\neq0$ and the time evolution is carried out under a final Hamiltonian with $u=0$. An important feature, made apparent by the results in Fig.~\ref{fig:nisf}, is that increasing the system size by a factor of 10 leads to essentially the same curve for $\eta_k^I$ vs $k$ but with 10 times the number of data points. This smooth dependence of $\lambda$ on $k'=k/L$ allows one, for sufficiently large system sizes, to define extensive integrals of motion by taking the sum $\sum_{k'\in [k''-\delta k''/2,k''+\delta k''/2]} \lambda_{k'}\hat{\eta}_{k'}$ as being equal to $\lambda_{k''}\hat{\eta}_{k''}'$ where now $\hat{\eta}_{k''}'=\sum_{k'\in [k''-\delta k''/2,k''+\delta k''/2]} \hat{\eta}_{k'}$ is extensive. Hence, the mode occupations $\hat{\eta}_k$ can be thought of as being extensive in a coarse-grained sense and the justification of the GGE exponential form [see Eq.~\eqref{eq:GGE}] presented above remains valid~\cite{cassidy_clark_11,rigol_fitzpatrick_11, vidmar_rigol_16}. The formal equivalence between the GGE constructed using occupation modes (as done here) and using extensive conserved quantities was established in Ref.~\cite{fagotti_essler_13a}. It extends beyond noninteracting systems to integrable models that may or may not be mappable to noninteracting ones.

Having justified the applicability of the GGE to the occupation modes of the single-particle Hamiltonian of noninteracting many-particle systems, one can go a step further and prove that the GGE defined this way provides exact results for the time average of all one-body observables (without finite-size errors)~\cite{ziraldo_santoro_13,he_santos_13}. The proof, following Ref.~\cite{he_santos_13}, is straightforward. Projecting the many-body time-evolving wave function $\hat{\rho}(t)=|\psi(t)\rangle\langle \psi(t)|$ onto the one-body sector, and using the fact that all eigenstates of the many-body Hamiltonian are (antisymmetrized) direct products of the single-particle states $|k\rangle$, the time evolution of the one-body density matrix can be written as
\begin{equation}\label{eq:singt}
 \hat{\rho}_\text{ob}(t)=\sum_{k,k'} c_{kk'} e^{-i(\varepsilon_k-\varepsilon_{k'})t}|k\rangle\langle k'|.
\end{equation}
In the absence of degeneracies in the {\it single-particle} spectrum, the infinite-time average of $\hat{\rho}_\text{ob}(t)$ can be written as
\begin{equation}
\overline{\hat{\rho}_\text{ob}(t)}=
\lim_{t'\rightarrow \infty} \frac{1}{t'}
\int^{t'}_0 dt\, \hat{\rho}_\text{ob}(t)
=\sum_k \eta_k^I |k\rangle\langle k|,
\end{equation}
which is, by construction, the one-body density matrix within the GGE, as $\eta^I_k=\sum_{n} |\la n | \psi_{\rm I}\ra|^2 \eta_{k,n}\equiv\text{tr}[\hat \rho_{\rm GGE}\, \hat \eta_k]$, where $\eta_{k,n}=1$ ($\eta_{k,n}=0$) if the single-particle state $|k\rangle$ is (is not) part of the particular many-body state $|n\rangle$, and $\text{Tr}[\hat \rho_{\rm GGE}\,\hat{\gamma}^{\dagger}_k\hat{\gamma}^{}_{k'}]\equiv0$ for $k\neq k'$. We should emphasize at this point that this does not mean that all one-body observables equilibrate at their GGE values (they do not, see Refs.~\cite{barthel_schollwock_08, ziraldo_santoro_13, he_santos_13, wright_rigol_14}), but simply that their time average is given by the GGE prediction.

\subsubsection{Hard-Core Bosons}\label{sec:HCBs}

We now turn our attention to hard-core bosons, described by the Hamiltonian
\begin{eqnarray}
 {\hat H}_\text{HCB} = - J \sum_{j=1}^{L-1} \left( {\hat b}_j^{\dagger} {\hat b}^{}_{j+1} + \text{H.c.} \right)
            + \sum_{j=1}^L u_j {\hat n}_j^b ,
\label{eq:nihcb}
\end{eqnarray}
where $ {\hat b}_j^{\dagger}$ (${\hat b}_j^{}$) is the hard-core boson creation (annihilation) operator at site $j$, ${\hat n}_j^b={\hat b}_j^{\dagger}{\hat b}_j^{}$ is the site $j$ occupation operator, $J$ is the hopping parameter, and $u_j$ are arbitrary site potentials. Hard-core bosons satisfy the same commutation relations as bosons but with the additional constraint that there cannot be multiple occupancy of any lattice site, i.e., $({\hat b}_j^{\dagger})^2={\hat b}_j^{2}=0$ \cite{cazalilla_citro_11}. 

The hard-core boson Hamiltonian above can be mapped onto a spin-1/2 chain through the Holstein-Primakoff transformation \cite{holstein_primakoff_40, cazalilla_citro_11} and the spin-1/2 chain onto the spinless fermion Hamiltonian in Eq.~\eqref{eq:nisf} via the Jordan-Wigner transformation \cite{jordan_wigner_28, cazalilla_citro_11}
\be\label{eq:JWT} 
\hat{b}^{\dagger}_j\rightarrow\hat{f}^{\dag}_j\prod^{j-1}_{\ell=1}e^{-i\pi \hat{f}^{\dag}_{\ell}\hat{f}^{}_{\ell}},\quad
\hat{b}^{       }_j\rightarrow\prod^{j-1}_{\ell=1} e^{i\pi \hat{f}^{\dag}_{\ell}\hat{f}^{}_{\ell}}\hat{f}_j,\quad
\hat{n}^b_j\rightarrow\hat{n}^f_j.
\ee
This means that the dynamics of the hard-core boson site occupations is identical to that of the fermions, but the momentum distribution function of hard-core bosons, which involves a Fourier transform of one-body correlations in real space, is very different from that of the fermions. The hard-core momentum distribution function can be efficiently calculated for eigenstates of the Hamiltonian \cite{rigol_muramatsu_05a}, for systems out of equilibrium \cite{rigol_muramatsu_05b}, and in the grand canonical ensemble \cite{rigol_05} using properties of Slater determinants. The conserved quantities to construct the GGE can be taken to be the same as for the noninteracting spinless fermions in Sec.~\ref{sec:nisf}, namely, single-particle mode occupations. 

\begin{figure}[!t]
\includegraphics[width=0.98\textwidth]{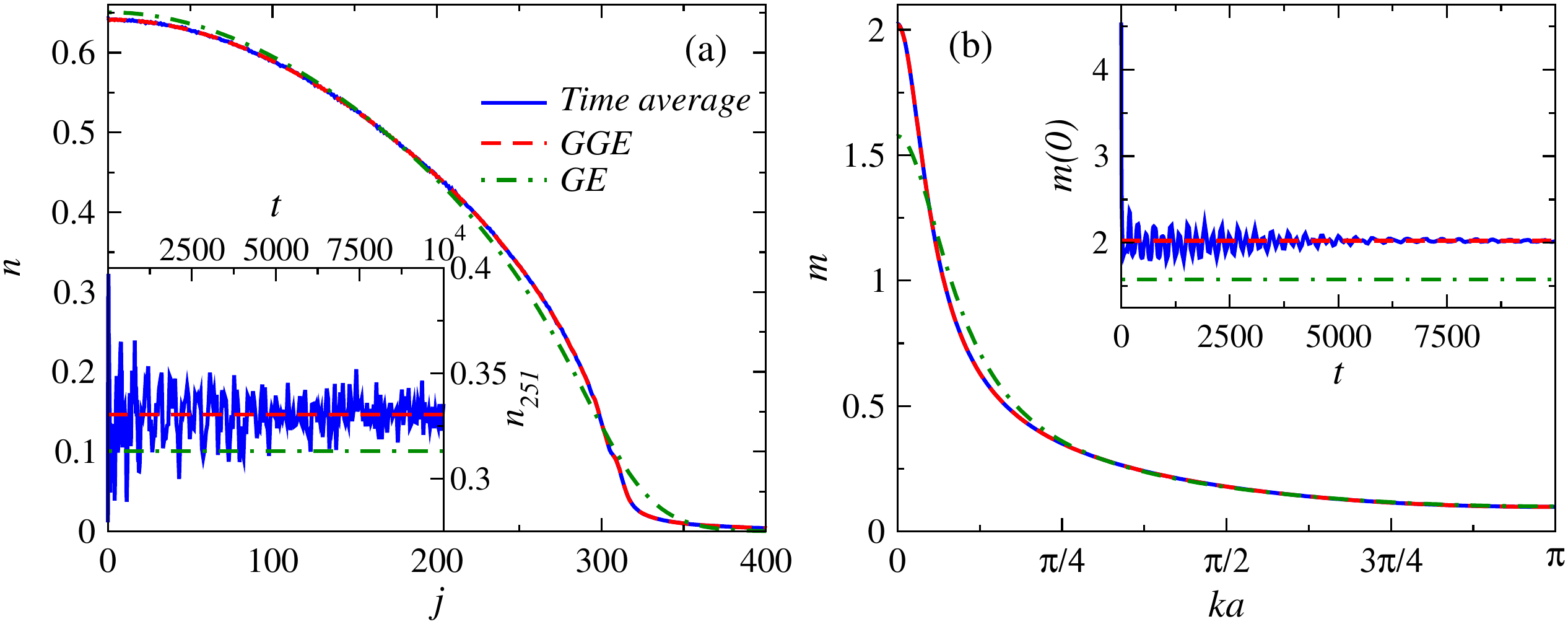}
\vspace{-0.15cm}
\caption{(Main panels) Site (a) and momentum (b) occupations in a trapped integrable system of hard-core bosons after a quench in which the initial state is the ground state of a trapped system ($v_I\neq0$) in the presence of a superlattice potential ($u_I\neq0$) and the dynamics is carried out in the presence of the same trap ($v=v_I$) but with no superlattice potential ($u=0$), see text. Results are presented for the time average of the occupations after relaxation, as well as for the GGE and the grand canonical ensemble (GE) predictions. The insets show the dynamics of the occupation of site 251 (a) and of the zero momentum occupation (b). The horizontal lines correspond to the results in the GGE and grand canonical ensemble, as depicted in the main panels. The system has 900 sites and 299 hard-core bosons. Times are reported in units of $\hbar/J$, and $k$ is reported in units of $1/a$, where $a$ is the lattice spacing. See also Ref.~\cite{rigol_muramatsu_06}.}
\label{fig:GGE}
\end{figure}

In Fig.~\ref{fig:GGE}, we show results for an integrable model of hard-core bosons in the presence of a harmonic trap after a quench in which a superlattice potential is turned off \cite{rigol_muramatsu_06} -- in this case, $u_j=v(j-L/2)^2+u(-1)^j$, where $v$ and $u$ set the strength of the harmonic trap and the superlattice potential, respectively. In the main panels, one can see the time average of the site (a) and momentum (b) occupation profiles after relaxation. They are clearly different from the predictions of a grand canonical ensemble for a system whose Hamiltonian is the one after the quench. The temperature and chemical potential of the grand canonical ensemble are fixed so that the mean energy and number of particles match those of the time-evolving state. We note that the system considered here is large enough so that the observed differences between the time-averaged profiles and the thermal predictions are not due to finite-size effects \cite{rigol_muramatsu_06}. On the other hand, the predictions of the GGE, which like the grand canonical ensemble also has energy and particle number fluctuations, are indistinguishable from the results of the time average. The insets exemplify the relaxation dynamics by depicting the time evolution of the occupation of one site [Fig.~\ref{fig:GGE}(a)] and of the zero momentum mode [Fig.~\ref{fig:GGE}(b)]. They can both be seen to relax towards, and oscillate about, the GGE prediction. The grand canonical ensemble results are clearly incompatible with the results after relaxation. The amplitude of the fluctuations about the time average reveals another qualitative difference between nonintegrable and integrable systems. In quenches involving pure states in isolated integrable systems mappable to noninteracting models (and in the absence of localization), time fluctuations only decrease as a power law of the system size \cite{cassidy_clark_11, gramsch_rigol_12, ziraldo_silva_12,campos_zanardi_13, ziraldo_santoro_13, wright_rigol_14}. This is to be contrasted to the exponential decrease of the amplitude of the time fluctuations as a function of the system size expected, and seen \cite{zangara_dente_13}, in nonintegrable systems. In Ref.~\cite{zangara_dente_13}, it was argued based on numerical experiments that the time fluctuations of observables in integrable systems that are not mappable to noninteracting ones can also decrease exponentially fast with increasing system size.

\subsection{Generalized Eigenstate Thermalization}
\label{sec:GETH}

One may ask at this point why is it that the GGE is able to describe observables after relaxation in isolated integrable systems following a quantum quench. After all, Eq.~\eqref{eq:timeevolution} is still dictating the dynamics and, once eigenstate thermalization does not occur, one might expect that the results after relaxation will depend on the \textit{exponentially} large (in the system size) number of parameters $C_n\equiv \la n | \psi_{\rm I}\ra$ that are set by the initial state while the GGE depends only on a \textit{polynomially} large number of parameters. As discussed in Ref.~\cite{cassidy_clark_11}, the validity of the GGE can be understood in terms of a generalization of eigenstate thermalization in integrable systems. Namely, if eigenstates of integrable Hamiltonians with similar distributions of conserved quantities have similar expectation values of physical observables (we call this phenomenon generalized eigenstate thermalization\footnote{What changes from nonintegrable to integrable Hamiltonians is that, in the latter, the expectation values of few-body observables in any eigenstate are determined by the values of all conserved quantities and not just the energy.}), then the GGE will describe those observables after relaxation following a quench.  This can be understood as follows. In the diagonal ensemble (after quenches to integrable systems), the fluctuations of each extensive conserved quantity are expected to be subextensive for physical initial states -- as we showed for the energy in Sec.~\ref{ss:quencheth}. This, together with the fact that the GGE is constructed to have the same expectation values of conserved quantities as the diagonal ensemble and combined with generalized eigenstate thermalization, is what leads to the agreement between the GGE and the results after relaxation.

Numerical evidence for the occurrence of generalized eigenstate thermalization was presented in Refs.~\cite{cassidy_clark_11, he_santos_13} for integrable hard-core boson systems (similar to those in Fig.~\ref{fig:GGE}), and in Ref.~\cite{vidmar_rigol_16} for the transverse field Ising model. In addition, for some observables, the occurrence of generalized eigenstate thermalization was proved analytically in Ref.~\cite{vidmar_rigol_16}. In what follows, we review results for the transverse field Ising model.

The relation between the transverse field Ising model and the hard-core boson model in Eq.~\eqref{eq:nihcb} can be understood as follows. The hard-core boson chain can be mapped onto a spin-1/2 chain via:
\be
\hat S^{z}_j=\hat b_j^\dagger \hat b^{}_j-\frac12,\quad \hat S^+_j=\hat b^\dagger_j,\quad \hat S^-_j=\hat b_j.
\ee
After these substitutions, using that $\hat S_j^+=\hat S_j^x+i \hat S_j^y$ and $\hat S_j^-=\hat S_j^x-i \hat S_j^y$, the Hamiltonian~(\ref{eq:nihcb}) reads
\be
 {\hat H}_{XX} = - 2J \sum_{j=1}^{L-1} \left( \hat S^x_j \hat S^x_{j+1}+\hat S^y_j \hat S^y_{j+1}\right) +\sum_{j=1}^L u_j \left({\hat S}^z_j+\frac{1}{2}\right) ,
\label{eq:nihcb_spin}
\ee
This Hamiltonian is known as the isotropic $XY$ model (or the $XX$ model) in a transverse field. Its anisotropic version, for periodic boundary conditions in the presence of a uniform field of strength $h$, can be written as 
\be
 {\hat H}_{XY} = - J\frac{(1+\gamma)}{2} \sum_{j=1}^{L} \hat \sigma^x_j \hat \sigma^x_{j+1}-J\frac{(1-\gamma)}{2}\sum_{j=1}^{L} \hat \sigma^y_j \hat \sigma^y_{j+1}+h \sum_{j=1}^L {\hat \sigma^z}_j ,
\label{eq:nihcb_spin_anis}
\ee
where we used that $\hat{S}^\alpha_j=\hat\sigma^\alpha_j/2$, and $\gamma$ is the anisotropy parameter. In the hard-core boson language, $\gamma\neq0$ leads to non-number-conserving terms  of the form $(\hat b_j^\dagger \hat b_{j+1}^\dagger+\text{H.c.})$. In the extreme anisotropic limit $\gamma=1$, the model in Eq.~\eqref{eq:nihcb_spin_anis} is known as the transverse field Ising model~\cite{sachdev_11}.

\begin{figure}[!t]
\includegraphics[width=1\textwidth]{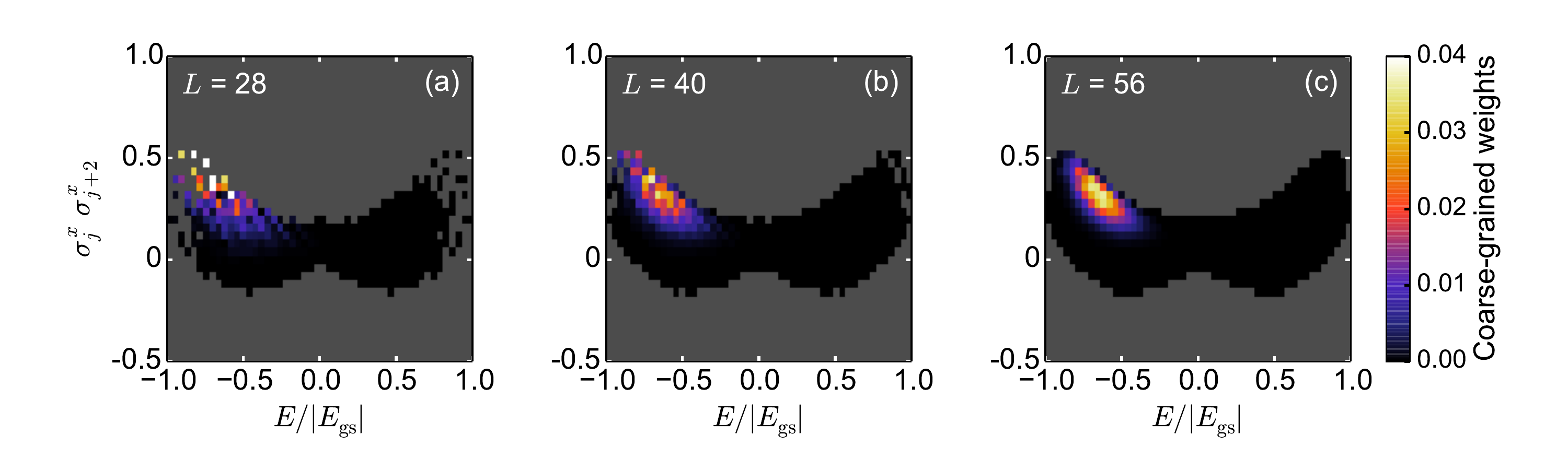}
\includegraphics[width=1\textwidth]{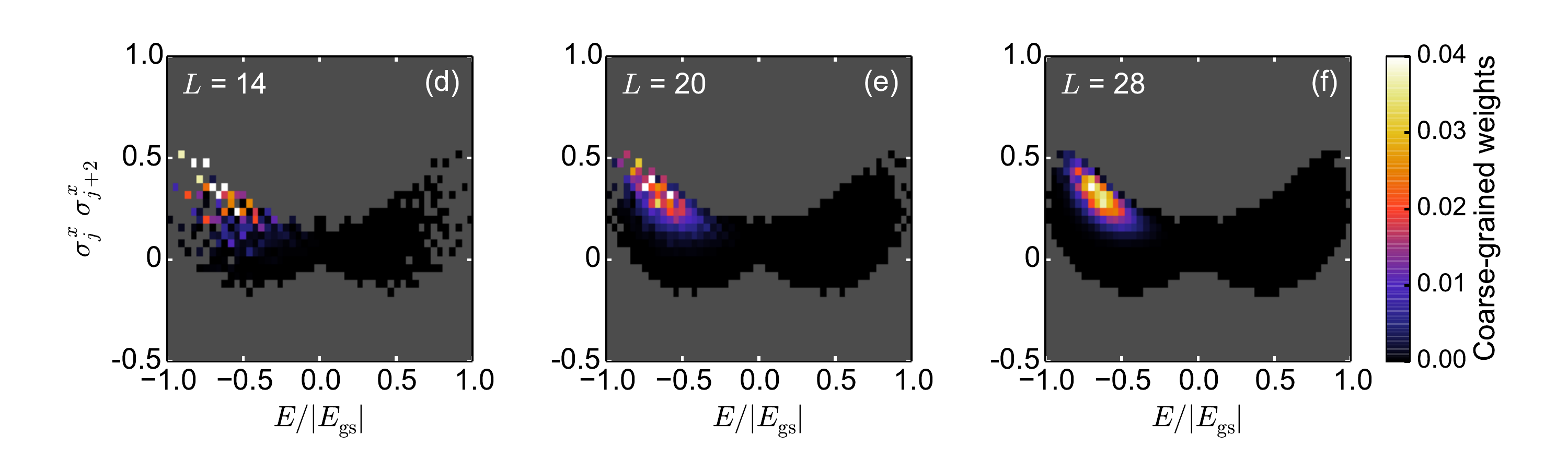}
\caption{Density plots of the weights of the eigenstates of the transverse field Ising Hamiltonian [Eq.~\eqref{eq:nihcb_spin_anis} with $\gamma=1$ and $J=1$] as a function of their eigenenergies and of the eigenstate expectation values of the next-nearest neighbor correlation function $\hat \sigma^x_j\hat \sigma^x_{j+2}$. Panels (a)-(c) depict the weights in the diagonal ensemble and panels (d)-(f) depict the weights in the GGE, in each case for three different system sizes. The initial state is the ground state for $h_I=0.1$, and after the quench $h=1.5$. Black pixels mark the presence of eigenstates (with vanishing weight), while gray pixels signal their absence. Colored pixels show the nonvanishing weights in the diagonal ensemble [panels (a)-(c)] and in the GGE [panels (d)-(f)]. These results were provided by Lev Vidmar (see also Ref.~\cite{vidmar_rigol_16}).}
\label{fig_sysyr1_density_080}
\end{figure}

In Fig.~\ref{fig_sysyr1_density_080}, we show results for the weights of the many-body eigenstates of the transverse field Ising model (color coded in the scale on the right) as a function of the energy of the eigenstates and of the eigenstate expectation values of $\hat \sigma^x_j\hat \sigma^x_{j+2}$. In the top panels, we show weights in the diagonal ensemble and in the bottom panels we show weights in the GGE, in each case for three different system sizes. The black regions mark the existence of eigenstates with the corresponding eigenenergies and eigenstate expectation values, but with vanishing weight in the ensembles. The fact that those black regions do not narrow with increasing system size (they can be seen to slightly widen) is to be contrasted to the results in Fig.~\ref{fig:nk_ETH} for nonintegrable systems. The contrast makes apparent that in the transverse field Ising model eigenstate thermalization does not occur.

More remarkably, Fig.~\ref{fig_sysyr1_density_080} shows that the eigenstates of the Hamiltonian with a significant weight in the diagonal ensemble and the GGE are located in approximately the same small region in the plane defined by the eigenstate energies and expectation values of $\hat \sigma^x_j\hat \sigma^x_{j+2}$  (see Ref.~\cite{vidmar_rigol_16} for similar results for other observables). In Ref.~\cite{vidmar_rigol_16}, it was shown that, in both ensembles, the width of the energy distribution and of the distribution of expectation values of $\hat \sigma^x_j\hat \sigma^x_{j+2}$ vanishes as $1/\sqrt{L}$ with increasing system size. If one adds to this finding the fact that, in both ensembles, the mean value of the energy is the same by construction and the expectation value of $\hat \sigma^x_j\hat \sigma^x_{j+2}$ is found to agree, as in previous examples, one concludes that the eigenstates of the final Hamiltonian that determine the results in the diagonal ensemble and in the GGE in the thermodynamic limit are located at the same point in the aforementioned plane. Generalized eigenstate thermalization is reflected by the fact that the width of the distribution of eigenstate expectation values vanishes with increasing system size. These results make apparent that the exact distribution of weights in the diagonal ensemble and the GGE is irrelevant, the overwhelming majority of the states they sample have the same expectation values of the observable. This is why the GGE can predict the expectation values of observables in integrable systems after relaxation following a quench, even though the number of parameters required to construct the GGE increases polynomially with the system size while for the diagonal ensemble it increases exponentially with the system size. In this spirit, in Ref.~\cite{caux_essler_13} it was argued that a single representative state is sufficient to describe the relaxed state of integrable systems after a quench in the thermodynamic limit. This statement is indeed very reminiscent of ETH for nonintegrable systems.

\subsubsection{Truncated GGE for the Transverse Field Ising Model}\label{sec:GGEtrex}       

As for hard-core bosons, for the transverse field Ising model in Fig.~\ref{fig_sysyr1_density_080} the GGE was constructed using occupations of single-particle fermionic quasiparticles (Bogoliubov fermions) \cite{vidmar_rigol_16}. Alternatively, one can construct a different equivalent set of integrals of motion, which are explicitly local and extensive. Following Ref.~\cite{fagotti_essler_13a}, these integrals of motion can be ordered according to their locality and come in pairs $\hat I_k^{+}$ and $\hat I_{k}^-$, such that $I_k^{+,-}$ contains sums of products of up to $k+2$ neighboring spin operators \cite{fagotti_essler_13a}. They can be written as
\begin{align}
&\hat I_0^+=\hat H=-J \sum_{j}\hat{{\cal S}}^{xx}_{j,j+1} +h\sum_{j} {\hat \sigma^z}_j ,\nonumber\\
&\hat I_1^{+}=-J\sum_j (\hat{{\cal S}}^{xx}_{j,j+2}-\hat \sigma^z_j)-h \sum_j (\hat{{\cal S}}^{xx}_{j,j+1}+\hat{{\cal S}}^{yy}_{j,j+1}) ,\nonumber\\
&\hat I_{n\geq2}^{+}=-J\sum_j (\hat{{\cal S}}^{xx}_{j,j+n+1}+\hat{{\cal S}}^{yy}_{j,j+n-1})-h \sum_j (\hat{{\cal S}}^{xx}_{j,j+n}+\hat{{\cal S}}^{yy}_{j,j+n}) ,\nonumber\\
&\hat I_n^-=-J \sum_{j} (\hat{{\cal S}}^{xy}_{j,j+n+1}-\hat{{\cal S}}^{yx}_{j,j+n+1}),
\end{align}
where $\hat{{\cal S}}^{\alpha\beta}_{j,j+l}=\sigma^\alpha_j[\sigma^z_{j+1}\ldots\sigma^z_{j+l-1}]\sigma^\beta_{j+l}$.

\begin{figure}[!t]
\begin{center}
 \includegraphics[width=0.66\textwidth]{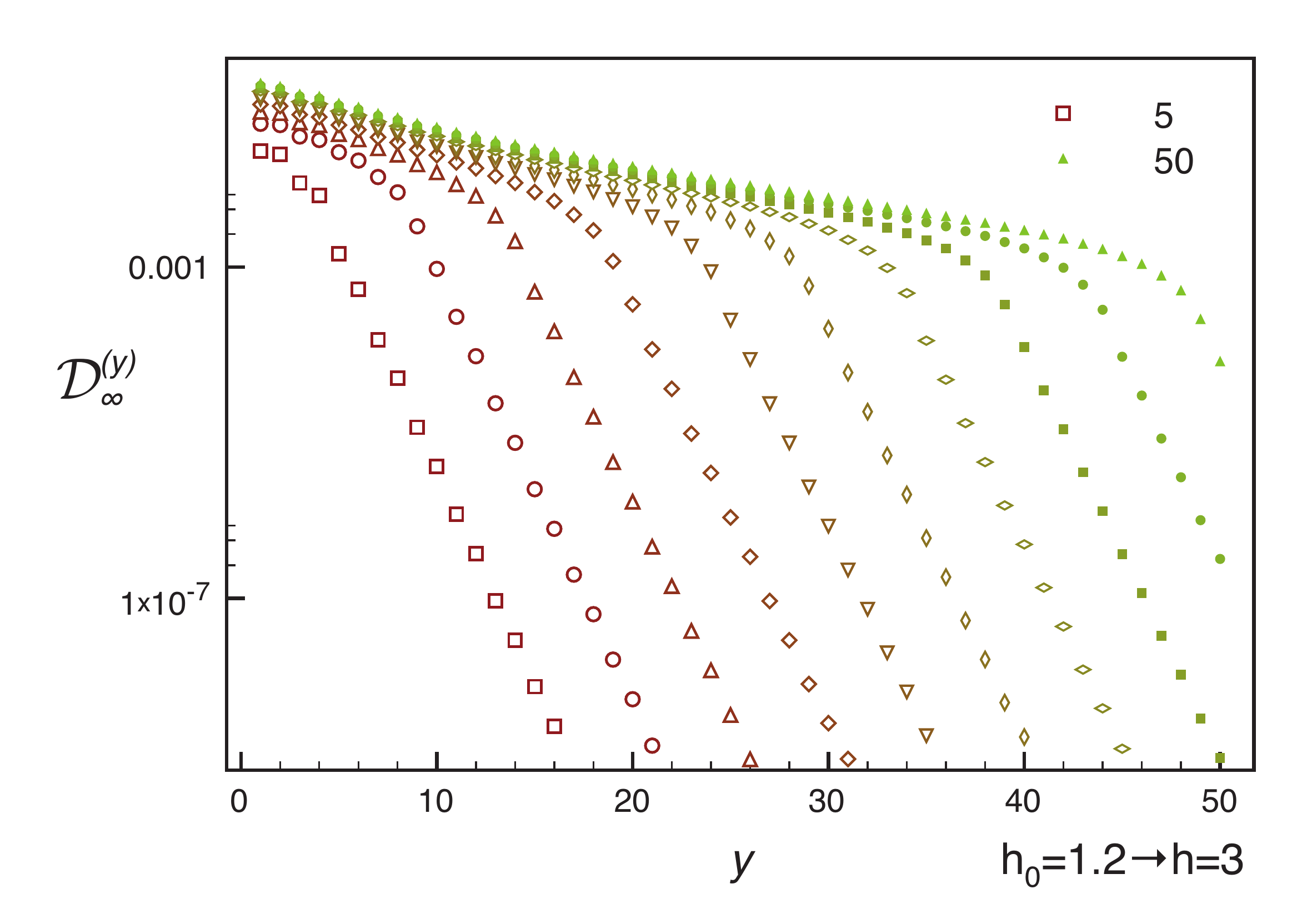}
\end{center}
\vspace{-0.5cm}
\caption{Norm distance $\mathcal D^{(y)}_\infty$, in the thermodynamic limit, between reduced density matrices in the GGE and in the truncated GGE obtained by imposing local conservation laws with support in at most $y+1$ consecutive sites, for a particular quench in the transverse Ising model. The subsystem size ranges from $l = 5$ to $l = 50$. The color and size of the symbols in the figure change gradually as a function of the subsystem size $l$. For $y > l$, the norm distance decays exponentially in $y$, with an $l$-independent decay constant. From Ref.~\cite{fagotti_essler_13a}.}
\label{fig:fagotti_gge}
\end{figure}

An important question one might ask is, given a desired accuracy for some observable, how many conserved quantities are needed for the GGE to describe the result after relaxation. On this point, the locality of the observable and of the conserved quantities included in the GGE turn out to be crucial. For the model above, Fagotti et al.~showed numerically that if one is interested in the reduced density matrix of a subsystem of size $L'$, then only the first $L'$ integrals of motion: $I_k^{+,-}$ with $k\lesssim L'$, that is, the integrals of motion that can ``fit'' on the subsystem, are important \cite{fagotti_essler_13a}. All other ``less local'' integrals of motion have an exponentially small effect on the subsystem (see Fig.~\ref{fig:fagotti_gge}). This is expected to be generic in integrable models, whether they are mappable or not to noninteracting ones.

\subsection{Quenches in the XXZ model}
\label{sec:XXZ}

An area of much current interest within the far from equilibrium dynamics of integrable systems is that of quenches in models that are not mappable to noninteracting ones. One model in this class, which is particularly important due to its relevance to experiments with ultracold bosons in one-dimensional geometries, is the Lieb-Liniger model. Studies of quenches within this model, in which repulsive interactions were suddenly turned on, revealed that the expectation values of conserved quantities diverge. As a result, a straightforward implementation of the GGE is not possible \cite{kormos_shashi_13,nardis_wouters_14}. A lattice regularization for this problem was discussed in Refs.~\cite{kormos_shashi_13,pozsgay_14a} (generalized eigenstate thermalization was argued to occur in this regularized model \cite{pozsgay_14a}). Despite progress in constructing GGEs for field theories \cite{essler_mussardo_14}, an explicit construction of the GGE for the Lieb-Liniger is still lacking.

\begin{figure}[!t]
\begin{center}
\includegraphics[width=0.8\columnwidth]{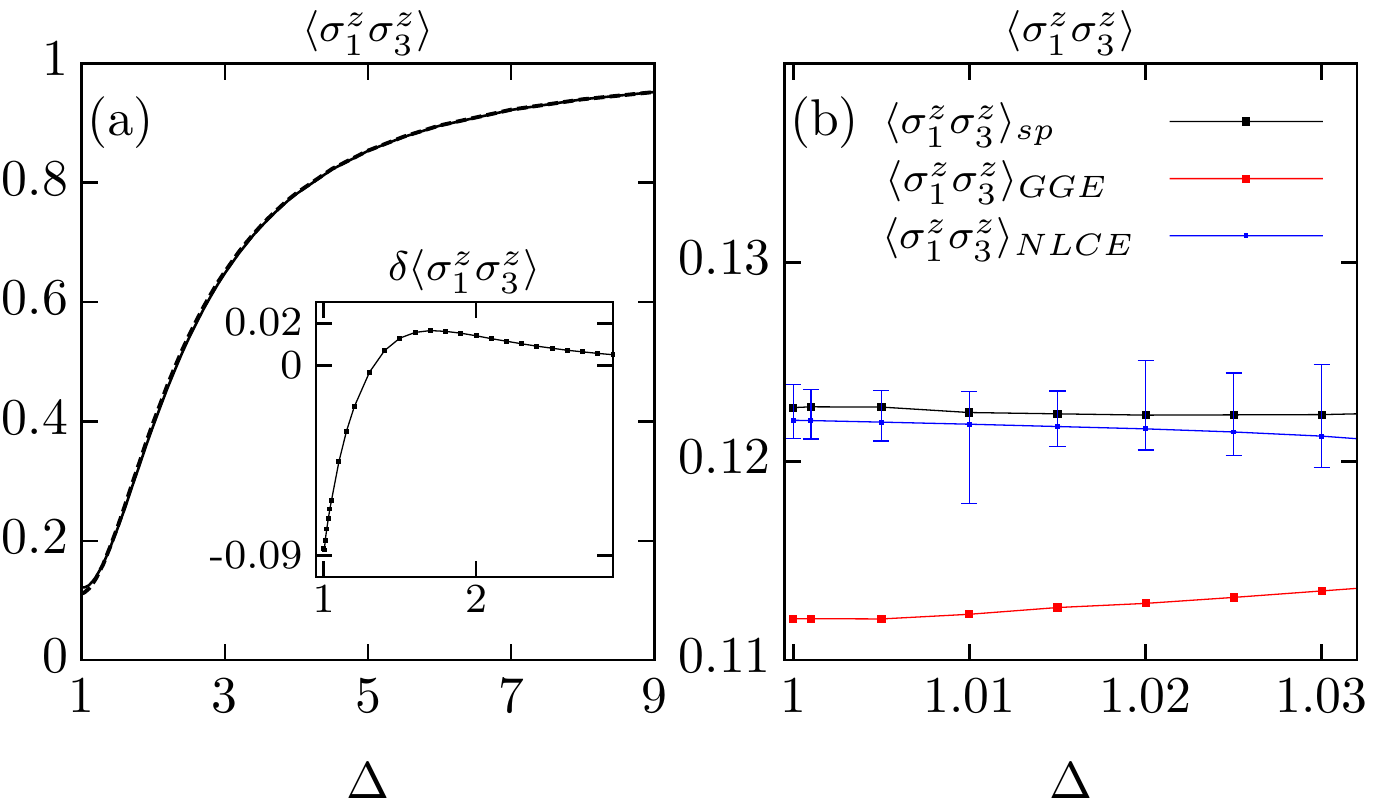}
\caption{\label{fig:nnn_correlator} Quenches in the $XXZ$ chain in the thermodynamic limit. (a) Next-nearest neighbor correlations $\langle \hat \sigma_1^z\hat \sigma_3^z\rangle$ following a quench starting from a N{\'e}el state to a finite value of the anisotropy parameter $\Delta$. Results are reported for the expected steady state obtained using Bethe ansatz (solid line) and for the GGE prediction (dashed line). (b) Comparison between the steady state $\langle \hat \sigma_1^z\hat \sigma_3^z\rangle_\text{sp}$, the GGE $\langle \hat \sigma_1^z\hat \sigma_3^z\rangle_\text{GGE}$, and the NLCE for the diagonal ensemble $\langle \hat \sigma_1^z\hat \sigma_3^z\rangle_\text{NLCE}$ results close to the isotropic point. The GGE results are seen to depart from the others. Error bars in the NLCE data display an interval of confidence. Inset in (a): Relative difference between the steady-state Bethe ansatz result and the GGE, $\delta \langle \hat \sigma_1^z\hat \sigma_3^z\rangle = \left( \langle \hat \sigma_1^z\hat \sigma_3^z\rangle_\text{GGE} - \langle \hat \sigma_1^z\hat \sigma_3^z\rangle_\text{sp} \right)/\left| \langle \hat \sigma_1^z\hat \sigma_3^z\rangle_\text{sp} \right|$. From Ref.~\cite{wouters_denardis_14}.}
\end{center}
\end{figure}

Another model that has attracted much recent interest, and which is the focus of this subsection, is the $XXZ$ model
\be
 {\hat H}_{XXZ} = - J \sum_{j=1}^{L} \left(\hat \sigma^x_j \hat \sigma^x_{j+1}+\hat \sigma^y_j \hat \sigma^y_{j+1} + \Delta\, \hat \sigma^z_j \hat \sigma^z_{j+1}\right).
\label{eq:xxz}
\ee
This model is, up to a possible boundary term, mappable onto the models in Eqs.~\eqref{eq:fermionHam} and \eqref{eq:HCBHam} when $J'=V'=0$. Studies of quenches in the $XXZ$ model revealed that the GGE constructed using all known local conserved quantities at the time failed to describe few-body observables after relaxation \cite{wouters_denardis_14, pozsgay_mestyan14, mierzejewski_prelovssek_14, goldstein_andrei_14}. In Fig.~\ref{fig:nnn_correlator}(a), we show results for the next-nearest neighbor correlation $\langle \hat \sigma_1^z\hat \sigma_{3}^z\rangle$ in this model after a quench from an initial N{\'e}el state to a finite value of the anisotropy parameter $\Delta$ [equivalent to $V/(2J)$ in Eq.~\eqref{eq:HCBHam} for $J'=V'=0$]. The results expected for that correlation in the steady state, which were obtained in the thermodynamic limit using Bethe ansatz, ($\langle \hat \sigma_1^z\hat \sigma_3^z\rangle_\text{sp}$, continuous line) are almost indistinguishable from the GGE results ($\langle \hat \sigma_1^z\hat \sigma_3^z\rangle_\text{GGE}$, dashed line). However, plotting the relative difference $\delta \langle \hat \sigma_1^z\hat \sigma_3^z\rangle = \left( \langle \hat \sigma_1^z\hat \sigma_3^z\rangle_\text{GGE} - \langle \hat \sigma_1^z\hat \sigma_3^z\rangle_\text{sp} \right)/\left| \langle \hat \sigma_1^z\hat \sigma_3^z\rangle_\text{sp} \right|$, see the inset in Fig.~\ref{fig:nnn_correlator}(a), reveals that they are not identical. The differences are largest close to the isotropic Heisenberg point. Near that point, calculations using numerical linked cluster expansions (NLCEs) for the diagonal ensemble \cite{rigol_14a} after the same quench \cite{rigol_14b} agree with the steady-state predictions obtained using Bethe ansatz, see Fig.~\ref{fig:nnn_correlator}(b). The discrepancy between the GGE results and the others suggested that, given the set of conserved quantities selected, generalized eigenstate thermalization did not occur in this model (in Ref.~\cite{pozsgay_14b}, it was argued that it fails for integrable models that support bound states). Hence, other local conserved quantities (not known at the time) were expected to also be important. The extra conserved quantities needed were recently found \cite{ilievski_medenjak_15, ilievski_denardis_15} and the GGE constructed using them has been shown to describe the steady state of observables after relaxation following the quench for the $XXZ$ model described above \cite{ilievski_denardis_15}.

It is important to emphasize at this point that we expect generalized eigenstate thermalization to be a generic phenomenon in integrable systems (as eigenstate thermalization is in nonintegrable systems) and that, as a result, GGEs allow one to describe observables in integrable systems after relaxation. However, in contrast to nonintegrable systems in which the conserved quantities are trivial to find and, as a result, traditional statistical mechanics can be used almost as a black box, the same is not true in integrable systems. For the latter, a careful analysis needs to be done (specially for models that are not mappable to noninteracting ones) in order to identify the appropriate conserved quantities needed to construct the GGE.

In all quenches discussed so far for integrable systems, mappable or not mappable to noninteracting ones, the initial states were taken to be eigenstates (mostly ground states) of an integrable model. One may wonder whether the lack of thermalization we have seen in those quenches is a result of the special initial states selected. In order to address this question, NLCEs were used in Ref.~\cite{rigol_16} to study the diagonal ensemble results (in the thermodynamic limit) for observables in quenches to the $J'=V'=0$ hard-core boson model in Eq.~\eqref{eq:HCBHam}, which is the $XXZ$ model \eqref{eq:xxz} written in the hard-core boson language. The initial states for those quenches were taken to be thermal equilibrium states of Hamiltonian~\eqref{eq:HCBHam} for $\Lambda\equiv J'_I=V'_I\neq0$, that is, thermal equilibrium states of a nonintegrable model. Those are the kind of initial states that one expects to have usually in experiments.

\begin{figure}[!t]
 \includegraphics[width=0.49\linewidth]{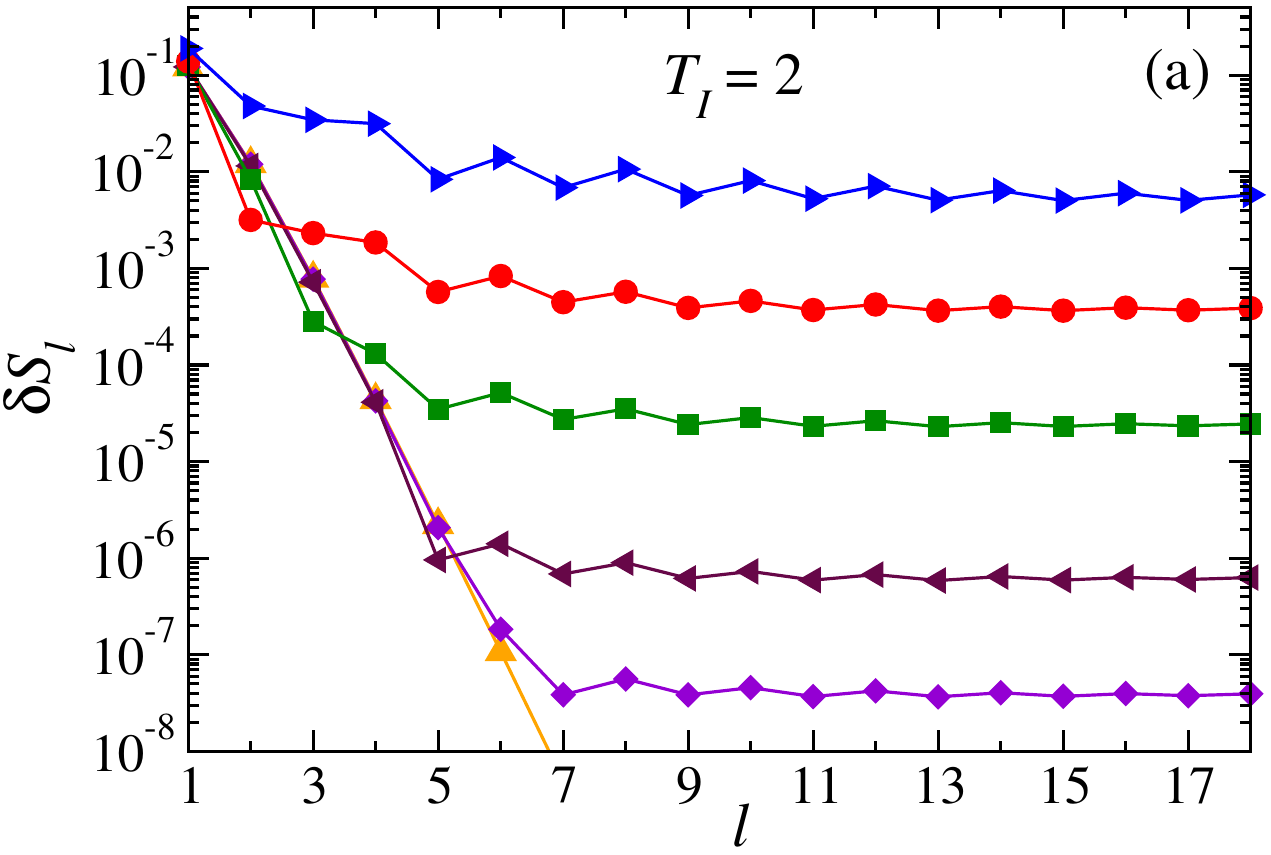} 
 \includegraphics[width=0.49\linewidth]{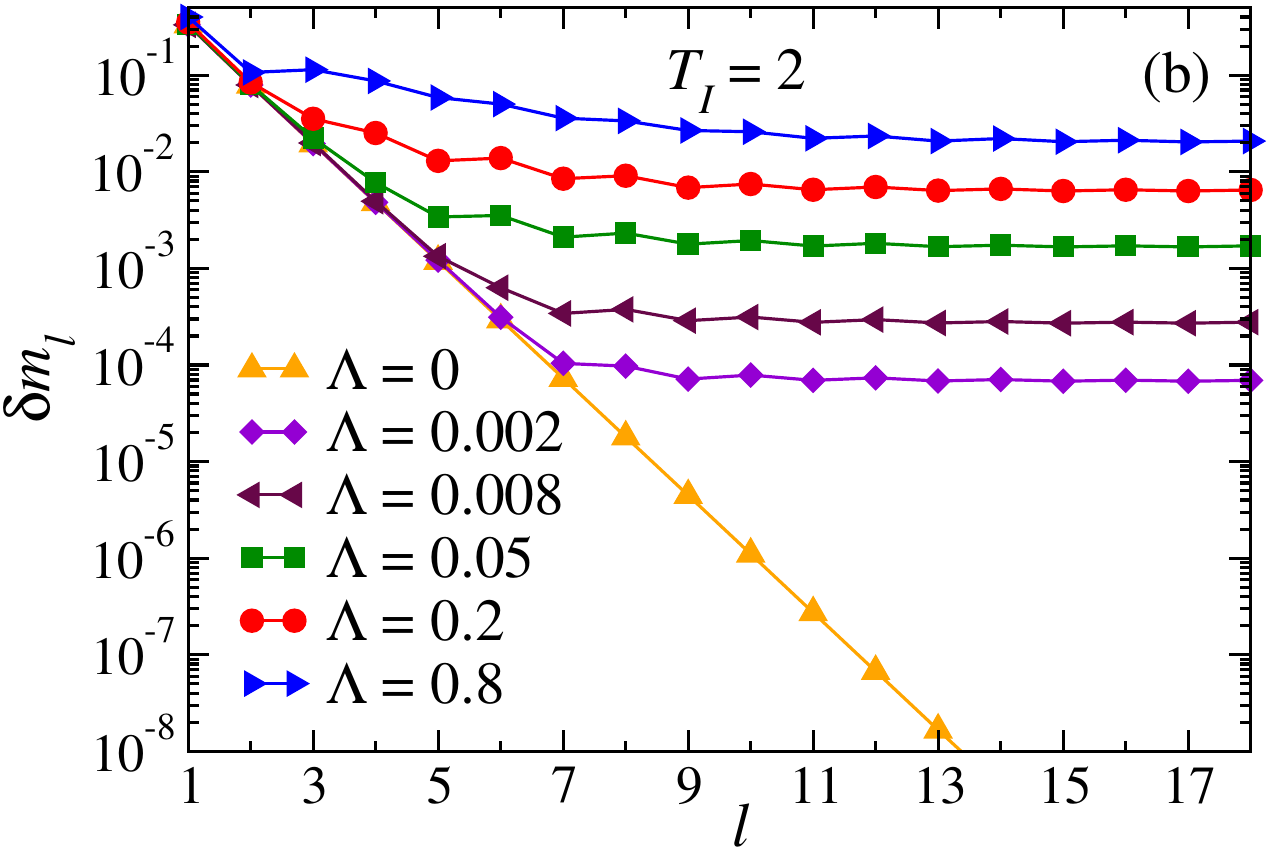}
 \caption{NLCE results after quenches in which the initial state is a thermal equilibrium state of the nonintegrable Hamiltonian \eqref{eq:HCBHam} with $\Lambda\equiv J'_I=V'_I\neq0$ and the system is quenched to $J'=V'=0$, that is, to the integrable $XXZ$ model written in the hard-core boson language. $J=V=1$ remains unchanged during the quench. (a) Relative entropy differences $\delta S_l$ [see Eq.~\eqref{eq:NLCES}] and (b) relative momentum distribution differences $\delta m_{l}$ [see Eq.~\eqref{eq:NLCEM}] vs the order $l$ of the NLCE for the diagonal ensemble. Results are shown for $T_I=2$ and six values of $\Lambda$. $\delta S_l$ and $\delta m_{l}$ for $\Lambda=0$, that is, in the absence of a quench, decrease exponentially fast with the order $l$ of the NLCE. All other differences saturate to a nonvanishing value reflecting lack of thermalization. Similar results were obtained in Ref.~\cite{rigol_16} for other initial temperatures and observables. Adapted from Ref.~\cite{rigol_16}.}
 \label{fig:NLCEs}
\end{figure}

In Fig.~\ref{fig:NLCEs}, we show results for the relative entropy differences
\be\label{eq:NLCES}
 \delta S_l=\frac{S^\text{GE}_{18}-S^\text{DE}_l}
 {S^\text{GE}_{18}},
\ee
between the grand canonical ensemble (GE) and the diagonal ensemble (DE) predictions, and the relative momentum distribution differences
\be\label{eq:NLCEM}
 \delta m_l=\frac{\sum_k\left|{m_k}^\text{DE}_{l}-{m_k}^\text{GE}_{18}\right|}
 {\sum_k {m_k}^\text{GE}_{18}},
\ee
also between the GE and the DE predictions, plotted as a function of the order $l$ of the NLCE for the diagonal ensemble. The initial states were taken to have a temperature $T_I=2J$ (the results for other initial temperatures are qualitatively similar \cite{rigol_16}). After the quench, the temperature and chemical potential in the grand canonical ensemble were fixed so that the mean energy and number of bosons per site agree (up to machine precision) with those in the diagonal ensemble. The NLCE was carried out up to order $l=18$. $S^\text{GE}_{18}$ and ${m_k}^\text{GE}_{18}$ were checked to be converged to the thermodynamic limit result up to machine precision (see Ref.~\cite{rigol_16} for details).

The results for $\delta S_l$ and $\delta m_l$ in Fig.~\ref{fig:NLCEs} are qualitatively similar. For $\Lambda=0$, that is, in the absence of a quench ($J'_I=V'_I=J'=V'=0$), one can see that $\delta S_l$ and $\delta m_l$ vanish exponentially fast with increasing $l$, i.e., the convergence of the NLCE expansion to the thermodynamic limit result is exponential in $l$. However, as soon as $\Lambda\neq0$, that is, as soon as there is a quench, the differences saturate to a nonzero value.\footnote{The quantitative difference between the values at which each relative difference saturates is related to the fact that $\delta S_{l\rightarrow\infty}\propto\Lambda^2$ while $\delta m_{l\rightarrow\infty}\propto\Lambda$. This is something that was argued analytically and demonstrated numerically for $\delta S_{l=18}$ and $\delta m_{l=18}$ in Ref.~\cite{rigol_16}, but that it is not important for the discussion here.} This means that the entropy and the momentum distribution function in the diagonal ensemble are different from their grand canonical counterpart in the {\it thermodynamic limit}. This is to be contrasted with the opposite quench, from integrable to nonintegrable points, for which the numerical results are consistent with vanishing $\delta S_{l\rightarrow\infty}$ and $\delta m_{l\rightarrow\infty}$ \cite{rigol_16}. For quenches to the integrable point, the fact that $\delta S_{l\rightarrow\infty}\neq0$ [as suggested by Fig.~\ref{fig:NLCEs}(a)] means that the energy distribution is not a smooth Gaussian function (or else $\delta S_{l\rightarrow\infty}=0$, see Sec.~\ref{sec:fundamental_relation}). Hence, the sparseness of the energy density appears to be a generic feature in physically relevant quenches and not a consequence of specially fine-tuned initial states. This is why, after the quench, observables such as the momentum distribution function do not thermalize [$\delta m_{l\rightarrow\infty}\neq0$ as suggested by Fig.~\ref{fig:NLCEs}(b)]. The latter phenomenon also appears to be generic. These results highlight how careful one needs to be when using typicality arguments \cite{tasaki_98, goldstein_lebowitz_06, popescu_06}. Those arguments might lead one to conclude that a fine-tuning of the initial state is needed for integrable systems not to thermalize after a quench, while this appears not to be the case in physically relevant situations.

\subsection{Relaxation of Weakly Non-Integrable Systems: Prethermalization and Quantum Kinetic Equations}
\label{sec:kinetic_equations}

Integrable systems are unlikely to be found in nature. Nevertheless there are many examples of models which are nearly integrable, where the integrability breaking terms are irrelevant for relatively long times. As early as in 1834, Scott Russel observed the soliton created by a boat in a narrow canal near Edinburgh~\cite{russel_1844}. In our language, the solitary wave is an example of a macroscopic non-thermalizing perturbation. It was not until 30 years later that it was realized that this phenomenon can be attributed to the integrability of the Korteweg-de Vries (KdV) equation, which approximately describes water waves in narrow one-dimensional channels~\cite{kdv_1895}. Since the KdV equation only provides an approximate description of the problem, one can expect that after long times the soliton will decay and the system will thermalize. Similarly, in a recent experiment with ultracold atoms~\cite{kinoshita_wenger_06}, the lack of thermalization of the one-dimensional bosonic gas was attributed to the integrability of the Lieb-Liniger model that quite accurately describes those systems \cite{cazalilla_citro_11}. Like the KdV equation, the Lieb-Liniger model provides only an approximate description of the experimental system and there are various integrability breaking corrections that need to be taken into account at long times (see, e.g., Ref.~\cite{mazets_11}). 

For nearly integrable systems, one can naturally expect a relatively fast relaxation to an approximate steady state determined by the integrable model, and then a much slower relaxation to the true thermal equilibrium. Such a scenario is now known under the name of prethermalization. This term was introduced by Berges et al.~in the context of cosmology~\cite{berges_borsanyi_04}, though the ideas of multi-time thermalization are much older. Recently, several different prethermalization scenarios have been explored both theoretically and experimentally. Just to name a few: relaxation of weakly interacting fermions after an interaction quench~\cite{moeckel_kehrein_08, moeckel_kehrein_09, kollar_eckstein_08, eckstein_kollar_09, eckstein_hackl_10, kollar_wolf_11},  prethermalization plateaus in various one-dimensional nonintegrable systems~\cite{kitagawa_imambekov_11, gring_kuhnert_12, mitra_13, kaminishi_mori_14, tavora_rosch_14, essler_kehrein_14, fagotti_collura_15}, prethermalization in interacting spin systems~\cite{babadi_demler_15}, two-dimensional superfluids with slow vortices and other topological defects~\cite{mathey_polkovnikov_10, barnett_polkovnikov_11, stamper-kurn_ueda_13}, prethermalization after turning on a long-range interaction in a spinless Fermi gas in two dimensions \cite{nessi_iucci_14}, the emergence of nonthermal fixed points, and, in particular, the emergence of turbulence~\cite{gurarie_95, scheppach_berges_09, nowak_schole_12}. On the latter, it is actually interesting to note that a GGE based on momentum occupation numbers (for the limit of weakly interacting particles) can be used to explain Kolmogorov's law~\cite{gurarie_95}.

In few-particle classical systems, the KAM theorem ensures that chaotic motion does not appear immediately after one breaks integrability. Instead, one can have coexistence of regions of chaotic and regular motion. As the strength of integrability breaking perturbation increases, chaotic regions spread and eventually occupy all available phase space (see, e.g., Fig.~\ref{fig:kicked_rotor}). It is not known whether a similar scenario is realized in many-particle systems. If situations like that exist, so that an extensive number of integrals of motion can survive small integrability breaking perturbations then, instead of prethermalization, one can anticipate relaxation to a GGE defined with respect to deformed integrals of motion. Such deformations have been discussed in the literature for transitionally invariant integrable systems with small integrability breaking perturbations~\cite{shimshoni_andrei_03, boulat_mehta_07, essler_kehrein_14}, and, in the context of many-body localization, for weakly interacting disordered systems~\cite{serbyn_papic_13a, huse_oganesyan_14}. At the moment, it is unclear whether in non-disordered extended systems (either classical or quantum) in the thermodynamic limit there can be a finite threshold for ergodicity breaking. Thus, it is not clear whether the emerged deformed GGEs can only describe transient states (although potentially long lived) or can represent true steady states. The former scenario is probably more generic but we are not aware of any strong evidence for it.

Arguably, the most successful approach for describing relaxation of weakly interacting (i.e., weakly nonintegrable) systems to equilibrium is the kinetic theory~(see, e.g., Ref.~\cite{lifshitz_pitaevskii_06}). Recently, Stark and Kollar~\cite{stark_kollar_13} derived kinetic equations using time-dependent perturbation theory applied to the GGE. These equations describe the relaxation from a prethermalized GGE to a thermal state. Below, we closely follow that work, extending it to arbitrary integrable systems. 

Let us assume that we have an integrable system described by the Hamiltonian $\hat{H}_0$ and a weak integrability breaking perturbation $\hat{H}'$
\be
\hat{H}=\hat{H}_0 + \hat{H}',
\ee
The Hamiltonian $\hat{H}_0$ commutes with a set of mutually commuting linearly independent integrals of motion $\hat{I}_k$, i.e., $[\hat{H}_0, \hat{I}_k]=0$. For example, in the spirit of what was discussed in Sec.~\ref{sec:nisf}, in a system of interacting fermions or bosons, these integrals of motion can be the occupations of the single-particle eigenstates.

Let us now assume that the system is prepared in some nonequilibrium initial state, for example, by a quantum quench, and we are interested in its relaxation. If $\hat{H}'$ is a weak perturbation of $\hat{H}_0$, then, at short times after a quench, the system ``does not know'' that it is nonintegrable and the effect of $\hat{H}'$ on the dynamics is small and the system relaxes to an appropriate GGE, possibly described by deformed integrals of motion of $\hat H_0$. This was found to be the case numerically in several systems  (see, e.g., Refs.~\cite{stark_kollar_13, essler_kehrein_14}.) At long times, $\hat{H}'$ is expected to lead to relaxation to thermal equilibrium. 

Since $\hat{H}'$ is assumed weak compared to $\hat{H}_0$, the dynamics generated by $\hat{H}'$ is slow compared to the dynamics generated by $\hat{H}_0$. This time scale separation translates into the fact that, at each moment of the evolution, the system is approximately stationary with respect to $\hat{H}_0$ so that it can be described by a GGE with \textit{slowly} evolving Lagrange multipliers plus a \textit{small} correction $\delta \hat{\rho}(t)$. This leads to the following ansatz for the density matrix of the system:
\begin{equation}
\begin{split}
\hat{\rho}(t) &\equiv \hat{\rho}_\text{GGE}(t) + \delta \hat{\rho}(t), \\
\hat{\rho}_\text{GGE}(t) &\equiv  \frac{\exp(-\sum_k \lambda_k(t) \hat{I}_k)}
 {\text{Tr}[\exp(-\sum_k \lambda_k(t) \hat{I}_k)]} .
 \label{eq:ansatz_rho}
 \end{split}
\end{equation}
If the system thermalizes, one expects that the Lagrange multiplier associated with the energy approaches the inverse temperature $\beta$, while all others approach zero.

Since $\hat{H}'$ is small, it is convenient to go to the interaction picture with respect to $\hat{H}_0$:
\be
\hat{\rho_I} (t)=\mathrm e^{-i \hat{H}_0 t} \hat{\rho}(t)\, \mathrm e^{i \hat{H}_0 t},\qquad \hat{H}'(t)=\mathrm e^{i \hat{H}_0 t} \hat{H}' \mathrm e^{-i \hat{H}_0 t}. 
\ee
where von Neumann's equation becomes:
\be
i \partial_t \hat{\rho}_I(t)=[\hat{H}'(t), \hat{\rho}_I(t)].
\label{eq:vnm_rho}
\ee
Our strategy will be to solve the von Neumann equation~\eqref{eq:vnm_rho} using time-dependent perturbation theory with the GGE ansatz as the initial condition. This will allow us to find small changes in the expectation values of the integrals of motion $\langle \hat{I}_k\rangle$, which in turn define the Lagrange multipliers in the GGE~\eqref{eq:ansatz_rho}. In this way, the slow evolution of $\hat{\rho}_\text{GGE}$ is determined self-consistently:
\be
d_t \langle \hat{I}_k\rangle=d_t \left({\rm Tr} \bigl[\hat{I}_k \hat{\rho}_I(t)\bigr]\right)={\rm Tr}\left(\hat{I}_k \partial_t [\hat{\rho}_I(t)]\right)=i
{\rm Tr}\left(\hat I_k[\hat \rho_I(t) , H'(t)]\right), 
\label{eq:dI_dt}
\ee
where we used Eq.~\eqref{eq:vnm_rho}, and that $\hat I_k$ is commute with $\hat H_0$ and thus remain time independent in the interaction picture. To leading order of perturbation theory in $\hat H'(t)$, we have $\hat \rho_I(t)\approx \hat \rho_{GGE}$,\footnote{We note that the GGE density matrix is not affected by the transformation to the interaction picture.} and therefore
\be
d_t\langle \hat{I}_k\rangle\approx i {\rm Tr}\left(\hat{I}_k \left[\hat{\rho}_\text{GGE}, \hat{H}'(t)\right] \right)=0.
\ee
The last equality follows from the cyclic property of trace and the fact that $\hat{\rho}_\text{GGE}$ and $\hat{I}_k$ commute. Therefore, we have to go to the next order of perturbation theory: $\hat \rho_I(t)\approx \hat \rho_{GGE}+\delta \hat \rho_I(t)$, where from Eq.~\eqref{eq:vnm_rho}
\be
\delta \hat{\rho}_I(t) \approx i \int_{t_0}^t dt' [\hat{\rho}_\text{GGE}, \hat{H}'(t')] =i \int_{0}^{t-t_0} d\tau [\hat{\rho}_\text{GGE}, \hat{H}'(t-\tau)].
\ee
Here, $t_0$ is some arbitrary time in the past. By substituting this correction to Eq.~\eqref{eq:dI_dt}, we obtain
\begin{multline}
d_t\langle \hat{I}_k\rangle  = i {\rm Tr}\left(\hat{I}_k [\delta \hat{\rho}_I(t), \hat{H}'(t)] \right)\approx - \int_{0}^{t-t_0} d\tau\, {\rm Tr}\left(\hat{I}_k \left[ \left[ \hat{\rho}_\text{GGE}, \hat{H}'(t-\tau)\right], \hat{H}'(t)\right] \right) \\
= -\int_{0}^{t-t_0} d\tau \langle [[\hat{I}_k, \hat{H}'(t-\tau)],\hat{H}'(t)]\rangle_\text{GGE}, 
\label{kin_eq_main}
\end{multline}
where, once again, we have used the cyclic property of the trace and the fact that $\hat{\rho}_\text{GGE}$ and $\hat{I}_k$ commute. To simplify this expression further, we note that the correlation functions appearing in Eq.~\eqref{kin_eq_main} depend only on time differences. Also, because the relaxation of nearly conserved integrals of motion is very slow compared to the time scales set by $\hat{H}_0$, the correlation functions appearing in the integral above decay fast so one can take the limit $t-t_0\to \infty$.  After these simplifications, one obtains the desired quantum kinetic equations for the integrals of motion
\be
d_t \langle \hat{I}_k \rangle\approx -\int_0^\infty d t \langle [[\hat{I}_k, \hat{H}' (0)],\hat{H}'(t)]\rangle_\text{GGE}.
\label{kin_general}
\ee
Both the expectation value on the LHS and RHS of the equation above can be written in terms of the Lagrange multipliers specifying $\hat{\rho}_\text{GGE}$ [see Eq.~\eqref{eq:ansatz_rho}]. Solving these equations, it is possible to determine the evolution of the Lagrange multipliers and therefore the slow relaxation of the GGE to the thermal equilibrium state. Being a set of coupled \textit{scalar} equations, Eq.~\eqref{kin_general} is much simpler than the original von Neumann equation.

It is instructive to rewrite the kinetic equations using the Lehman representation in the basis of $\hat{H}_0$. Using the identity 
\be
\int_0^\infty dt \mathrm e^{i (\epsilon_n-\epsilon_m) t}=\pi \delta(\epsilon_n-\epsilon_m)+{\cal P}{i\over \epsilon_n-\epsilon_m},
\ee
and, for simplicity, assuming that both the Hamiltonian and the integrals of motion are real, we rewrite Eq.~\eqref{kin_general} as
\be
d_t \langle \hat{I}_k\rangle =2\pi \sum_{nm} (\rho^{GGE}_{nn}-\rho^{GGE}_{mm}) \langle n| \hat{I}_k|n\rangle |\langle n|\hat{H}'|m\rangle|^2\delta (\epsilon^0_n-\epsilon_m^0).
\label{eq:kin_gen_lehmann}
\ee
where $\epsilon^0_n$ is the eigenvalue of $\hat{H}_0$ corresponding to eigenstate $|n\ra$, i.e., $\hat{H}_0|n\ra = \epsilon^0_n |n\ra$. In this form, it becomes clear that the thermal distribution (where $\rho_{nn}$ is only a function of energy) is a stationary solution of these kinetic equations, i.e., $d_t \langle \hat{I}_k\rangle=0$ for any $\hat{I}_k$. Also, the delta function of $(\epsilon^0_n-\epsilon_m^0)$ ensures that $d_t \langle \hat{H}_0\rangle=0$. So relaxation to thermal equilibrium occurs in the presence of energy conservation. Both properties are, of course, expected from general considerations.

Let us now apply the kinetic equation~\eqref{kin_eq_main} to a common setup dealing with a gas of weakly interacting particles, bosons or fermions. For simplicity, we assume that they are spinless. Also, to shorten notations, we will use a scalar notation for the momentum modes, keeping in mind that this can be a vector index. Then the Hamiltonian reads
\be
\hat{H}_0=\sum_k \epsilon_k \hat{c}_k^\dagger \hat{c}^{}_k
\ee
For the integrability breaking term, we take the usual (normal ordered) density-density interactions
\be
\hat{H}'=\sum_{ij} V(i,j) \hat{c}_i^\dagger \hat{c}_j^\dagger \hat{c}^{}_j \hat{c}^{}_i=\sum_{k_1,k_2 ,k_3, k_4} \hat{c}_{k_1}^\dagger \hat{c}_{k_2}^\dagger V_{k_1, k_2, k_3, k_4}  \hat{c}^{}_{k_3} \hat{c}^{}_{k_4}.
\ee
For translationally invariant interactions, $V_{k_1,k_2,k_3,k_4}$ is nonzero only when $k_1+k_2=k_3+k_4$, and it depends only on the transferred momentum $q=k_1-k_3$. But, since our formalism applies even if translational invariance is broken, we will keep the interaction matrix element in the most general form. The obvious integrals of motion are the momentum occupation numbers $ \hat{n}_k=\hat{c}_k^\dagger \hat{c}^{}_k$. Let us first compute the commutator
\be
[\hat n_{k'}, \hat{H}']=2\sum_{k_2,k_3,k_4} [\hat{c}_{k'}^\dagger \hat{c}_{k_2}^\dagger V_{k',k_2,k_3,k_4} \hat{c}^{}_{k_3} \hat{c}^{}_{k_4}- \hat{c}_{k_2}^\dagger \hat{c}_{k_3}^\dagger V_{k_2,k_3,k',k_4} \hat{c}^{}_{k'} \hat{c}^{}_{k_4}],
\label{nk_hint_commutator}
\ee
where we used the invariance of the interaction matrix element with respect to permutation of $k_1$ with $k_2$ and $k_3$ with $k_4$. Plugging this into Eq.~\eqref{kin_general} and using Wick's theorem, which works for any GGE with quadratic integrals of motion, we find
\be
\dot n_{k'} \approx 16\pi \sum_{k_2, k_3, k_4} (\tilde n_{k'} \tilde n_{k_2} n_{k_3} n_{k_4}-n_{k'} n_{k_2} \tilde n_{k_3} \tilde n_{k_4}) |V_{k',k_2,k_3,k_4}|^2\delta(\epsilon_{k'}+\epsilon_{k_2}-\epsilon_{k_3}-\epsilon_{k_4}),
\label{eq:intermediateki}
\ee
where $\tilde n_k=1\pm n_k$ with a plus sign referring to bosons and a minus sign referring to fermions and $n_k = \la \hat{n}_k \ra$. Classical kinetic equations are obtained by taking the limit $n_k\ll 1$ and effectively replacing $\tilde n_k$ by unity. Solving these kinetic equations can be tedious, but it is numerically feasible for very large systems. Let us check that the thermal distribution is a fixed point of these kinetic equations. For example, for fermions, the equilibrium distribution reads
\be
n_{k}={1\over 1+\exp[\beta (\epsilon_k-\mu)]}
\ee
then
\begin{multline}
\tilde n_{k'} \tilde n_{k_2} n_{k_3} n_{k_4}-n_{k'} n_{k_2} \tilde n_{k_3} \tilde n_{k_4}=(1-n_{k'}-n_{k_2})n_{k_3} n_{k_4}-n_{k'}n_{k_2}(1-n_{k_3}-n_{k_4}) \\=\left[e^{\beta(\epsilon_{k'}+\epsilon_{k_2}-2\mu)}-e^{\beta(\epsilon_{k_3}+\epsilon_{k_4}-2\mu)}\right]n_{k'}n_{k_2} n_{k_3} n_{k_4}=0,
\end{multline}
where the last equality relies on the total energy conservation. With more effort, one can show that the equilibrium fixed distribution is the attractor of the kinetic equations. 

This example connects the ideas of GGE as a generic stationary state of integrable systems, prethermalized states as slowly evolving GGE states, and the kinetic theory as the perturbation theory in time describing the final evolution to thermal equilibrium. As previously mentioned, more work is needed to understand the generality of this approach and its applicability to systems other than those for which one can take the occupations of the single-particle eigenstates to be the nearly conserved quantities.

\section*{Acknowledgments}
This work was supported by the Army Research Office: Grant No.~W911NF1410540 (L.D., A.P, and M.R.), the U.S.-Israel Binational Science Foundation: Grant No.~2010318 (Y.K. and A.P.), the Israel Science Foundation: Grant No.~1156/13 (Y.K.), the National Science Foundation: Grants No.~DMR-1506340 (A.P.) and PHY-1318303 (M.R.), the Air Force Office of Scientific Research: Grant No.~FA9550-13-1-0039 (A.P.), and the Office of Naval Research: Grant No.~N000141410540 (M.R.). The computations were performed in the Institute for CyberScience at Penn State. We acknowledge discussions with Marin Bukov, Guy Bunin, Fabian Essler, Tarun Grover, Wen Wei Ho, Christopher Jarzynski, Ehsan Khatami, Michael Kolodrubetz, Albion Lawrence, Maxim Olshanii, Daniel Podolsky, Peter Reimann, Lea Santos, Mark Srednicki, Hal Tasaki, Lev Vidmar, Emil Yuzbashyan, and thank an anonymous referee for many useful comments.
  
%%%%%%%%%%%%%%%%%%%%%%%%%%%%%%%%%%%%%%%%%%%%%%%%%%%%%%%%%%%%%%%%%%%%%%%%%%%%%%%%%%%%%%%%%%
%%%%%%%%%%%%%%%%%%%%%%%%%%%%%%%%%%%%%%%%%%%%%%%%%%%%%%%%%%%%%%%%%%%%%%%%%%%%%%%%%%%%%%%%%%  
%%% appendices
%%%%%%%%%%%%%%%%%%%%%%%%%%%%%%%%%%%%%%%%%%%%%%%%%%%%%%%%%%%%%%%%%%%%%%%%%%%%%%%%%%%%%%%%%%
%%%%%%%%%%%%%%%%%%%%%%%%%%%%%%%%%%%%%%%%%%%%%%%%%%%%%%%%%%%%%%%%%%%%%%%%%%%%%%%%%%%%%%%%%%
\begin{appendices}
%%%%%%%%%%%%%%%%%%%%%%%%%%%%%%%%%%%%%%%%%%%%%%%%%%%%%%%%%%%%%%%%%%%%%%%%%%%%%%%%%%%%%%%%%%
%%%%%%%%%%%%%%%%%%%%%%%%%%%%%%%%%%%%%%%%%%%%%%%%%%%%%%%%%%%%%%%%%%%%%%%%%%%%%%%%%%%%%%%%%%

%%%%%%%%%%%%%%%%%%%%%%%%%%%%%%%%%%%%%%%%%%%%%%%%%%%%%%%%%%%%%%%%%%%%%%%%%%%%%%%%%%%%%%%%%%
\section{The Kicked Rotor}\label{app:kicked}
%%%%%%%%%%%%%%%%%%%%%%%%%%%%%%%%%%%%%%%%%%%%%%%%%%%%%%%%%%%%%%%%%%%%%%%%%%%%%%%%%%%%%%%%%%

In this appendix, we discuss how chaos emerge in the simplest setup, namely, a driven single-particle system in one dimension. In the presence of driving there is no energy conservation and, as a result, the system is not integrable. The Hamiltonian of the \textit{classical} kicked rotor reads
\be
H(p,x,t)={p^2\over 2}-K\cos(x)\sum_{n=-\infty}^{\infty}\delta(t-nT).
\ee
If one thinks of $p$ as angular momentum and $x$ as the canonically conjugate angle, this Hamiltonian describes a freely rotating particle that is periodically kicked at times $t=nT$. We choose the minus sign in front of $K$ so that the time-averaged Hamiltonian reduces to a conventional pendulum with an energy minimum at $x=0$ (there is no loss of generality as this sign can always be changed by $x\to x+\pi$). For simplicity, in what follows we refer to $p$ and $x$ as momentum and position, respectively.

The equations of motion for the kicked rotor are
\be
{dx\over dt}=\left\{ x, H \right\} = {\partial H\over \partial p}=p,\quad 
{dp\over dt}=\left\{ p, H \right\} = -{\partial H\over \partial x}=-K\sin(x)\sum_{n=-\infty}^{\infty}\delta(t-nT).
\ee
These equations can be easily integrated between kicks. Let us denote by $p_n$ and $x_n$ the momentum and the position of the particle, respectively, just before the $n$-th kick, i.e., at time $t=nT-\epsilon$, where $\epsilon\to 0$. Then the equations of motion result in the following recursion relations
\be
x_{n+1}=x_n+T\, p_{n+1}, \qquad
p_{n+1}=p_n-K\sin(x_{n}).
\label{chirikov_map}
\ee
These equations provide a discrete map (known as the Chirikov standard map) that allows one to uniquely determine the position and the momentum of the particle. If $nT<t<(n+1)T$, then $p(t)=p_{n+1}$ and $x(t)=x_{n}+p_{n+1} (t\,{\rm mod}\, T)$. Note that one can confine momentum to any periodic interval of length $2\pi/T$. Indeed, from Eq.~(\ref{chirikov_map}), it is obvious that shift of the momentum by $2\pi/T$ and the coordinate by $2\pi$ leaves the map invariant. Let us analyze the time evolution that follows from this map. The dynamics is determined by the kick strength $K$, the period $T$, and the initial values of $p$ and $x$. The map depends only on the product $KT$ so we can set $T=1$ and analyze the map as a function of $K$ keeping in mind that $K\ll1$ is equivalent to the short period limit. 

If $K\ll 1$ and $p_0\ll 1$, from Eqs.~\eqref{chirikov_map}, one can see that both $p$ and $x$ change very little during each period. Hence, instead of solving discrete equations, one can take the continuum limit and write
\be
{\partial x\over \partial n}\approx p,\quad 
{\partial p\over \partial n}=-K\sin(x)\;\rightarrow {\partial^2 x\over \partial n^2}\approx -K\sin(x).
\ee
This equation describes the motion of a pendulum in a cosine potential, which is regular. Depending on the initial conditions there are two types of trajectories, corresponding to oscillations ($p\ll K$) and full rotations ($p\gg K$). A careful analysis shows that one does not need to assume that $p$ is initially small, the only crucial assumption is that $K\ll 1$. 

Next, one needs to check the stability of the obtained trajectories. It might happen that, if one includes corrections to the continuum approximation, chaos occurs. However, as proved by Kolmogorov, Arnold, and Moser (KAM) \cite{kolmogorov_54,arnold_63,moser_62}, this is not the case. As mentioned in the Introduction, the KAM theorem states that regular motion is stable against small perturbations. For the kicked rotor problem, one can check the stability of the solution above perturbatively. In particular, Eqs.~\eqref{chirikov_map} can be written as
\be
x_{n+1}-2x_n+x_{n-1}=-K\sin(x_n).
\ee
If $K$ is small, we can assume that $x$ is a nearly continuous variable. By expanding in Taylor series one gets
\be \label{eq:pekr}
{d^2x\over dn^2}+{1\over 12} {d^4x\over dn^4}\approx -K\sin(x).
\ee
From the unperturbed solution, we see that (at least in the localized regime) the natural frequency of oscillations is $\sqrt{K}$. This means that, in Eq.~\eqref{eq:pekr}, the term with the fourth derivative is proportional to $K^2$, that is, it is small when $K\ll 1$. 

When $K$ is large, the continuum approximation for the map fails. $p$ and $x$ ``jump'' from kick to kick. Since both are determined modulo $2\pi$, one may guess that the motion is chaotic. A rigorous analytical proof that this is the case does not exist. Hence, we discuss indications for the occurrence of chaos for large values of $K$ by analyzing the stability of the fixed points of the map:
\be
x_{n+1}=x_n+p_{n+1}=x_n, \quad p_{n+1}=p_n-K\sin(x_n)=p_{n}.
\ee
There are only two possible solutions: $p_n=0, x_n=0$ and $p_n=0,x_n=\pi$. Now, let us perturb the trajectories and see whether they remain stable. The linearized equations read
\be
\delta x_{n+1}-2\delta x_n + \delta x_{n-1}=-K\cos(x_n)\delta x_n=\mp K \delta x_n,
\label{linearized_kicked_rotor}
\ee
where the minus and plus signs refer to the fixed points $x=0$ and $x=\pi$, respectively. In Eq.~\eqref{linearized_kicked_rotor}, one might recognize the equation of motion of coupled harmonic oscillators, where $\pm K$ plays the role of the frequency squared. For a harmonic chain, it is standard to introduce normal Fourier modes, i.e, $\lambda=\exp[i q]$. Here, we need to be careful because the frequency, $\sqrt{\pm K}$, can be imaginary. Because this is a translationally invariant system, we seek the solution of the form $\delta x_{n+1}=\lambda \delta x_n=\lambda^n \delta x_0$. Using our ansatz for the solution, Eq.~\eqref{linearized_kicked_rotor} reduces to a simple quadratic equation
\be
\lambda^2-(2\mp K)\lambda+1=0,
\ee
which has two solutions
\be
\lambda_{1,2}=1\mp {K\over 2}\pm\sqrt{\mp K+{K^2\over 4}}.
\ee
Let us analyze these solutions separately for the two fixed points. For $x=0$, corresponding to the ``$-$'' sign, we have two solutions
\be
\lambda_{1,2}=1-{K\over 2}\pm \sqrt{{K^2\over 4}-K}.
\ee
For $0<K<4$, the expression in the square root is negative leading to an imaginary contribution to $\lambda$. In the same range of $K$, the absolute value of the real part of $\lambda$ is smaller than one. This means that the solution is stable. Indeed, if one introduces a small deviation to the stable position then, as the discrete time $n$ increases, that deviation does not grow. Moreover, in this range, we can check that
\be
|\lambda^2|=(1-K/2)^2+K-K^2/4=1
\ee
implying that $\lambda=\exp[iq]$, as for a harmonic chain. This means that any small deviation leads to oscillations around the fixed point. 

If $K>4$, the outcome of introducing a small deviation is completely different. This is because now there are two real solutions for $\lambda$. The solution with the negative sign,
\be
\lambda_2=1-{K\over 2}-\sqrt{{K^2\over 4}-K},
\ee
satisfies $|\lambda_2|>1$. This means that any small deviation from the fixed point grows exponentially in time without bound, at least in the linearized regime. This exponential growth does not prove that the motion is chaotic, but is a strong indicator of it. The exponent characterizing the rate of growth, $\ln(\lambda)$, is called the Lyapunov exponent. In chaotic systems with many degrees of freedom, there are many Lyapunov exponents. Typically, the largest one determines the rate of divergence of nearby phase-space trajectories.

The analysis of the other fixed point, with $x=\pi$, is even simpler
\be
\lambda_{1,2}=1+ {K\over 2}\pm\sqrt{K+{K^2\over 4}}.
\ee
Clearly, for any positive $K$, there are two real solutions with one larger than one, that is, this point is always unstable. This is not surprising since this fixed point corresponds to the situation where a mass sits at a potential maximum. It is interesting that if, instead of $\delta$ kicks, one applies a fast periodic drive to the pendulum: $K=K_0+a\sin(\nu t)$, one can stabilize $x=\pi$ to be an equilibrium position. This is known as the Kapitza effect (or Kapitza pendulum), and can be seen to occur in many physical systems \cite{kapitza_51,broer_hoveijn_04, bukov_dalessio_14}. 

\begin{figure}[!t]
\includegraphics[width=14.7cm]{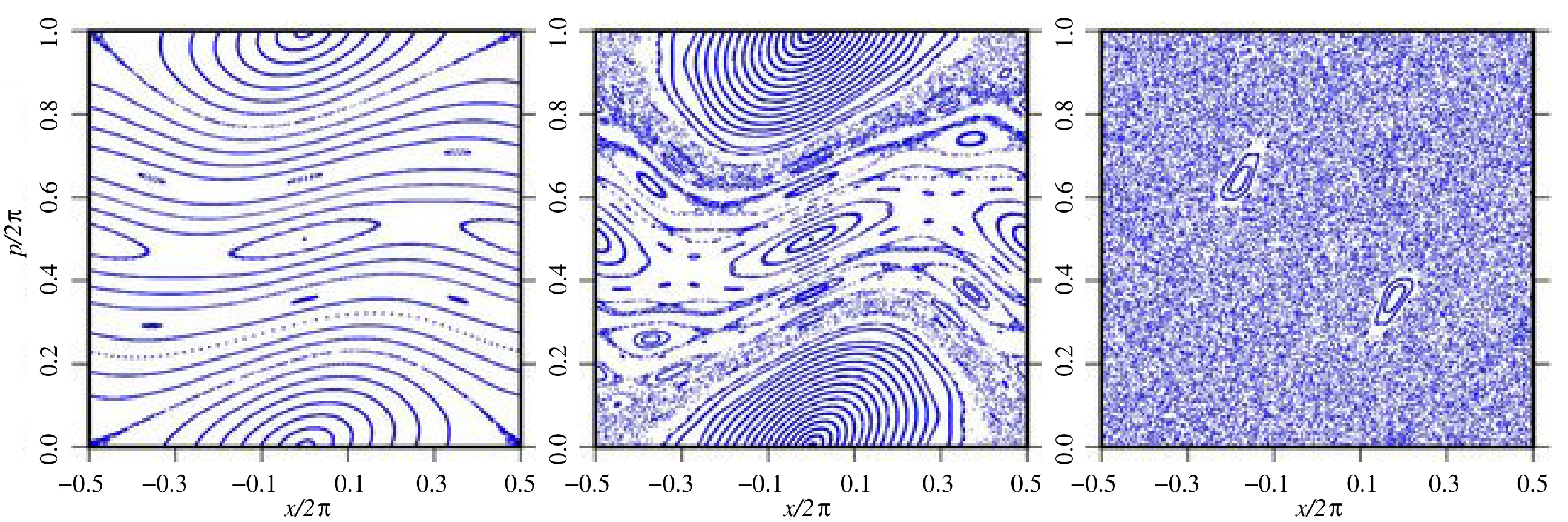}
\caption{Phase-space portrait (Poincare cross-section) of the kicked rotor for different values of the parameter $K$. From left to right, $K=0.5$, 0.971635, and 5. Images taken from scholarpedia~\cite{chirikov_shepelyansky_08}.}
\label{fig:kicked_rotor}
\end{figure}

In Fig.~\ref{fig:kicked_rotor}, we show phase-space portraits of the kicked rotor for different values of $K$. For small values of $K$ (left panel), the motion is essentially regular everywhere except in the vicinity of the unstable fixed point $x=\pi$, $p=0$. As $K$ increases, chaotic regions gradually cover a larger and larger fraction of phase space.  At $K=K_c$ (center panel), with $0.971635\lesssim K_c<63/64$ \cite{greene_79,mackay_83,mackay_percival_85}, there is a percolation transition. Isolated chaotic regions for $K<K_c$ percolate through phase space for $K>K_c$. This implies that the system can increase its energy without bound. For  sufficiently large values of $K$, chaotic regions cover phase space almost entirely (right panel).

The Chirikov standard map can be quantized. A discussion of the quantum map can be found in Ref.~\cite{chirikov_shepelyansky_08}, and references therein.

%%%%%%%%%%%%%%%%%%%%%%%%%%%%%%%%%%%%%%%%%%%%%%%%%%%%%%%%%%%%%%%%%%%%%%%%%%%%%%%%%%%%%%%%%%
\section{Zeros of the Riemann Zeta Function}\label{app:riemann}
%%%%%%%%%%%%%%%%%%%%%%%%%%%%%%%%%%%%%%%%%%%%%%%%%%%%%%%%%%%%%%%%%%%%%%%%%%%%%%%%%%%%%%%%%%

Here, we discuss a remarkable manifestation of RMT, which highlights a connection between GUE statistics and prime numbers. There is no clear understanding of the origin of this connection and, at the moment, it represents one of the biggest mysteries associated with prime numbers. The Riemann zeta function $\zeta(s)$ is formally defined (for $\Re(s)>1$) as
\be
\zeta(s)=\sum_{n\geq 1} {1\over n^s} \;.
\ee
For other values of $s$, it is defined by an appropriate analytic continuation through the integral:
\be
 \zeta(s) = \frac{1}{\Gamma(s)} \int_{0}^{\infty} \frac{x ^ {s-1}}{e ^ x - 1} \mathrm{d}x 
\ee
As proved by Euler in 1859, the Riemann zeta function is related to prime numbers (again for $\Re(s)>1$):
\be
\zeta(s)=\prod_{p=\rm prime} {1\over 1-p^{-s}}.
\label{zeta_s}
\ee
The proof of this result is simple and elegant. Notice that we can construct a function $I_2(s)$
\be
\quad I_2(s)= \zeta(s)-{1\over 2^s}\zeta(s)=1+{1\over 3^s}+{1\over 5^s}+\dots\ ,
\ee
which is a sum that lacks terms that are inverse powers of integer multiples of 2. Similarly, one can construct $I_3(s)$
\be
I_3(s)=I_2(s)-{1\over 3^s} I_2(s)=\zeta(s)\left(1-{1\over 2^s}\right)\left(1-{1\over 3^s}\right)=
1+{1\over 5^s}+{1\over 7^s}+{1\over 11^s}+\dots,
\ee
which is a sum that lacks terms that are inverse powers of integer multiples of 2 and 3. Continuing this exercise, and using the fundamental theorem of arithmetic, i.e., that any number has a unique decomposition into prime numbers, ones proves that, as $n\to \infty$, $I_n(s)\to 1$, and hence proves Eq.~\eqref{zeta_s}. 

Equation~\eqref{zeta_s} allows one to map $\zeta(s)$ onto the partition function of a noninteracting harmonic chain in which the frequencies of the normal modes are related to the prime numbers. The partition function for a single oscillator is
\be
z_p(\beta)=\sum_n \exp[-\beta \omega_p n]={1\over 1-\exp[-\beta \omega_p]}.
\ee
If we associate prime numbers with normal modes, $\omega_p=\ln(p)$, and require that 
$\beta=s$, then
\be
Z(\beta)=\prod_p z_p(\beta)=\zeta(\beta).
\ee
The (complex) zeros of the zeta function are thus the (complex) zeros of the partition function of this model. The zeros of the partition function are known as Fisher zeros~\cite{fisher_67}, which are closely related to Yang-Lee zeros~\cite{yang_lee_52}. Condensation of these zeros near the real temperature axis is an indicator of a phase transition~\cite{fisher_67}. Recently, it was conjectured that concentration of Fisher zeros at complex infinity is related to the ergodicity of the system~\cite{heyl_vojta_13}. Physically, the Fisher zeros correspond to the zeros of the Fourier transform of the energy distribution function $P(E)$ (closely connected to the Loshmidt echo and zeros of the work distribution for the quench problems~\cite{heyl_polkovnikov_13}). Indeed
\be
P(E)={1\over Z(\beta)}\sum_n \exp[-\beta E_n]\delta (E_n-E)
\ee
where
\be
Z(\beta)=\sum_n \exp[-\beta E_n]
\ee
is the partition function. Hence
\be
\widetilde W(\tau)\equiv \int_{-\infty}^\infty dE\, P(E) \exp[i E\tau]={1\over Z(\beta)} \sum_n
\exp[-(\beta-i\tau) E_n]={Z(\beta-i\tau)\over Z(\beta)}.
\ee
Thus, in the physics language, the complex zeros of the partition function $Z(\beta-i\tau)$ correspond to the zeros of the Fourier transform of the energy distribution for a system of phonons with normal modes given by the natural logarithm of the prime numbers. 

\begin{figure}[!t]
\begin{center}
 \includegraphics[width=0.6\textwidth,angle=-90]{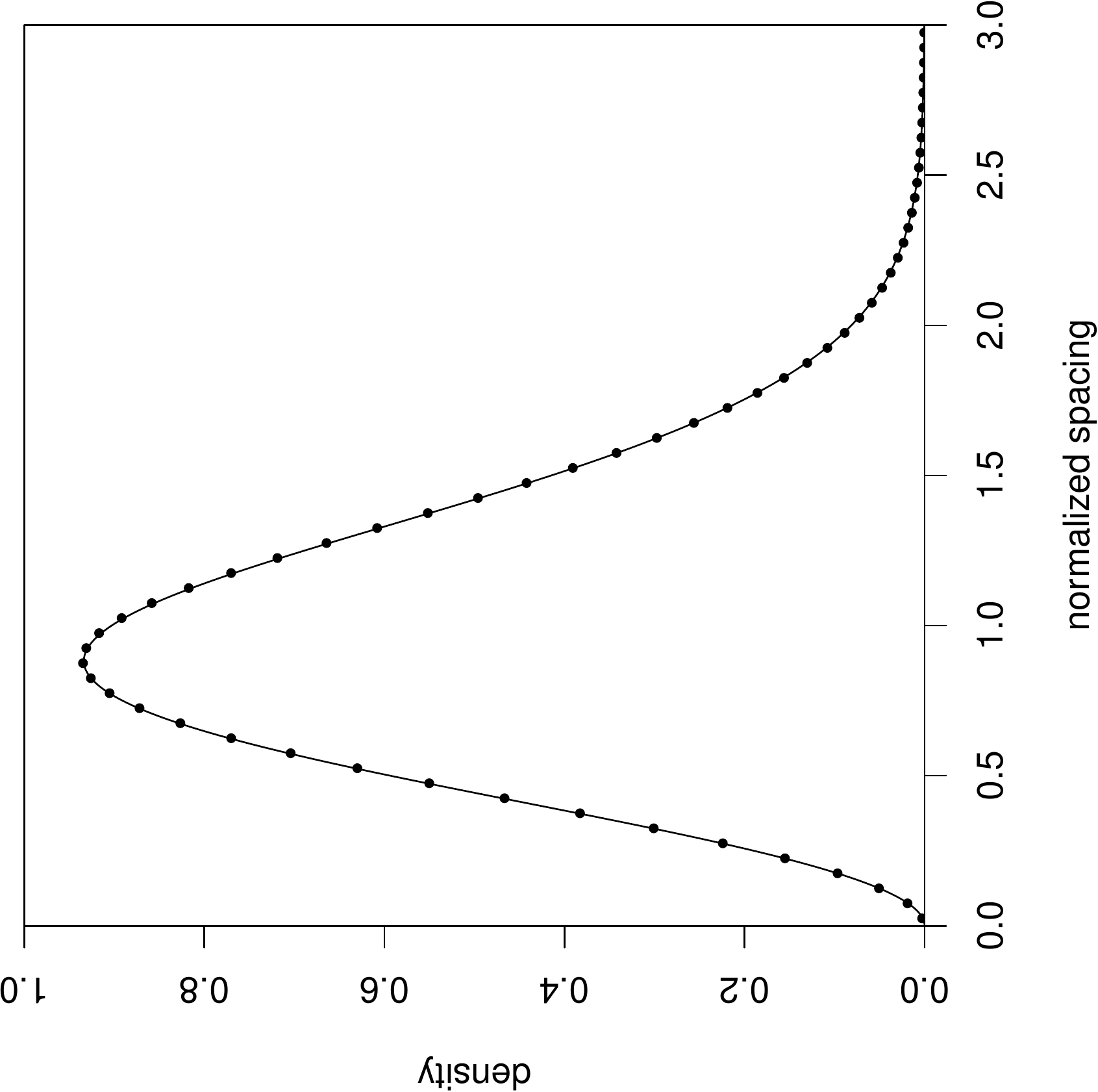}
\end{center}
\vspace{-0.1cm}
\caption{Distribution of the spacings between approximately one billion zeros of the Riemann zeta function near zero number $10^23 + 17,368,588,794$, and the statistics of level spacings in the Gaussian unitary ensemble (continuous line). These results were provided by Andrew Odlyzko (see also Ref. [316]).}
\label{fig:riemann_zeta}
\end{figure}

Riemann's zeta function has many fascinating properties. One of them is that the nontrivial zeros of $\zeta(s)$, i.e., zeros that are nonnegative integers, lie on the line $\Re(s)=1/2$. This conjecture is called the Riemann hypothesis and it remains one of the greatest unsolved problems in mathematics. By now, it has been checked for the first $10^{22}$ zeros \cite{odlyzko_99}. Remarkably, the distribution of the normalized spacings of the zeros of Riemann's zeta function is the same as that of level spacings in the GUE (see Fig.~\ref{fig:riemann_zeta}). This agreement hints, as mentioned before, a deep connection between prime numbers and random numbers.

%%%%%%%%%%%%%%%%%%%%%%%%%%%%%%%%%%%%%%%%%%%%%%%%%%%%%%%%%%%%%%%%%%%%%%%%%%%%%%%%%%%%%%%%%%
\section{The Infinite Temperature State as an Attractor\label{app:inftematt}}
%%%%%%%%%%%%%%%%%%%%%%%%%%%%%%%%%%%%%%%%%%%%%%%%%%%%%%%%%%%%%%%%%%%%%%%%%%%%%%%%%%%%%%%%%%

To prove that the infinite temperature distribution is an attractor, let us recall that any stochastic matrix $\boldsymbol{M}$ or, equivalently, Markov matrix has one eigenvalue $\lambda_0=1$, while all the other eigenvalues have absolute value less or equal than one, i.e., $|\lambda_{\alpha>0}|\le 1$ \cite{VanKampen}. The eigenvalue $\lambda_0=1$ clearly corresponds to the steady state of the system while the others denote processes where the probability distribution decays to the steady state. The \textit{left} and \textit{right} eigenvectors corresponding to the eigenvalue $\lambda_0=1$ satisfy the relation
\be
L_0 \boldsymbol{M} = L_0,\quad \boldsymbol{M} R_0 =R_0.
\ee 
Note that, by construction, the right eigenvector $R_0$ is the steady-state probability distribution of the system so that its elements should, by normalization, sum to $1$. By the conservation of probability, i.e., $\sum_m M_{n\to m}=1$, and by direct substitution, it is easy to see that the \textit{left} eigenvector is given by the constant vector $L_0=(1,1,\cdots,1)$. Note that with this choice $L_0 \cdot R_0=1$. Of course, the \textit{right} eigenvector $R_0$ depends on the details of the Markov matrix $\boldsymbol{M}$ and in general has a nontrivial structure. 

To see this, let us decompose the vector of initial probabilities $\rho^{(0)}_{nn}$ in terms of the \textit{right} eigenvectors as\footnote{In this discussion, we ignore cases in which the Markov matrix cannot by diagonalized. The proof can be extended to these cases as well.} 
\be
\rho^{(0)}=\sum_{\alpha=0}^D c_\alpha R_\alpha = R_0 + \sum_{\alpha>0}^D c_\alpha R_\alpha
\ee
where we have used the fact that the coefficients $c_\alpha$ are determined by projection on the \textit{left} eigenvectors $c_\alpha=\sum_n \rho^{(0)}_{nn} (L_\alpha)_n$ and therefore $c_0=1$. Plugging the expression above into the master equation~\eqref{master2}, we obtain
\be
\rho^{(N)}=\sum_{\alpha=0}^D c_\alpha \left(\lambda_\alpha\right)^N R_\alpha  \approx R_0 \label{eq:inf_T}
\ee
where we have used that $c_0=1$, $\lambda_0=1$, and assumed that $|\lambda_{\alpha>0}|<1$ so that  $\lambda_{\alpha>0}^N\approx 0$. This equation shows that $\rho^{(N)}$ approaches the stationary state $R_0$ exponentially fast in the number of processes $N$. The only exception to this result is when the doubly stochastic matrix admits other eigenvalues with absolute value one. This situation, however, is not generic. It corresponds to systems that are not ergodic so that some portions of the configuration space cannot be accessed from others, see Appendix~\ref{appendix:birkhoff}.

Doubly stochastic matrices $\boldsymbol{p}$ are a special subgroup of Markov matrices which, besides satisfying the conservation of probability $\sum_m p_{n\to m}=1$, satisfy the additional constraint $\sum_n p_{n\to m}=1$. This additional property allows one to prove that the \textit{right} eigenvector corresponding to the $\lambda_0=1$ eigenvalue has the specific form $R^\text{ds}_0=(1/D,\cdots,1/D)^T$, where $D$ is the dimension of the matrix $\boldsymbol{p}$. Therefore, for a doubly stochastic matrix, the stationary state is the ``infinite temperature state". To this end, we simply check explicitly that $\boldsymbol{p}\,R^\text{ds}_0 =R^\text{ds}_0$:
\be
\boldsymbol{p}\,\, R^\text{ds}_0 = \boldsymbol{p} \left(\begin{array}{c}
1/D\\
\vdots\\
1/D
\end{array}\right) = \left(\begin{array}{c}
D^{-1} \sum_m p_{1\rightarrow m}\\
D^{-1} \sum_m p_{2\rightarrow m} \\
\vdots
\end{array}\right) = \left(\begin{array}{c}
D^{-1}\\
D^{-1} \\
\vdots
\end{array}\right) =R^\text{ds}_0,
\ee
where the third equality follows directly from the doubly stochastic condition~\eqref{eq:double_stochastic}. Next, we prove that $|\lambda_\alpha| \le 1$. Let us assume that there is an eigenvalue $\lambda_\alpha$ larger than one. Then, the corresponding eigenvector $R_\alpha$ grows exponentially under the repeated action of $\boldsymbol{p}$, that is, $\boldsymbol{p}^N R_\alpha = \lambda_\alpha^N R_\alpha$. This implies that there are entries of $\boldsymbol{p}^N$ that are larger than one. However, $\boldsymbol{p}^N$ is a product of doubly stochastic matrices so it is itself a doubly stochastic matrix (see Sec.~\ref{sec:prop}). Therefore, all its entries need to be smaller than one, see Eq.~\eqref{eq:bound}. As a result, we conclude that there cannot be any eigenvalue larger than one.

%%%%%%%%%%%%%%%%%%%%%%%%%%%%%%%%%%%%%%%%%%%%%%%%%%%%%%%%%%%%%%%%%%%%%%%%%%%%%%%%%%%%%%%%%%
\section{Birkhoff's Theorem and Doubly Stochastic Evolution\label{appendix:birkhoff}}
%%%%%%%%%%%%%%%%%%%%%%%%%%%%%%%%%%%%%%%%%%%%%%%%%%%%%%%%%%%%%%%%%%%%%%%%%%%%%%%%%%%%%%%%%%

Birkhoff's theorem~\cite{marshall_olkin_79} states that any doubly stochastic matrix is given by the convex sum of permutation matrices $\boldsymbol{\Pi}_\alpha$:
\be
{\bf p} = \sum_\alpha k_\alpha \boldsymbol{\Pi}_\alpha,\quad \sum_\alpha k_\alpha =1, \quad 0 \le k_\alpha\le 1. \label{birkhoff}
\ee
We can then rewrite the doubly stochastic master equation~\eqref{eq:master} as
\be
\rho^{(1)}= {\bf p}\,\rho^{(0)} = \sum_\alpha k_\alpha \boldsymbol{\Pi}_\alpha \rho^{(0)}
\ee
When only one permutation matrix contributes, the master equation simply describes a perfect transfer of population between two states. For example, for a three level system in which only one permutation matrix contributes, say, between states one and two, the permutation matrix is given by
\be
\Pi_{1\leftrightarrow 2} = \left(\begin{array}{ccc}
0 & 1 & 0\\
1 & 0 & 0\\
0 & 0 & 1
\end{array}\right)
\ee
and the master equation above simply reduces to:
\be
\rho^{(1)}_{11}=\rho^{(0)}_{22},\quad \rho^{(1)}_{22}=\rho^{(0)}_{11}, \quad \rho^{(1)}_{33}=\rho^{(0)}_{33}
\label{simple-master}
\ee
If we apply the master equation with this transition matrix, the system will enter a cycle that will neither lead to an entropy increase nor to an energy increase.

In general, however, many permutation matrices contribute to the master equation. Using Birkhoff's theorem, it is easy to see that the most general doubly stochastic matrix for the three level system considered above is
\be
{\bf p} = \left(\begin{array}{ccccc}
\delta +\beta & & \alpha+\eta & &\gamma+\mu \\
\alpha+\mu &  &\delta +\gamma & &\beta+\eta \\
\gamma+\eta & &\beta+\mu & &\delta +\alpha
\end{array}\right), \label{eq:mostgeneral}
\ee
which describes the transfer of probabilities between states 1 and 2 with weight $\alpha$, between states 2 and 3 with weight $\beta$, between states 1 and 3 with weight $\gamma$, no transfer of probabilities with weight $\delta$, cyclic permutations $1\to 2\to 3\to 1$ with probability $\eta$, and cyclic permutations $3\to 2\to 1\to 3$ with probability $\mu$. Note that the sum of the weights is one, that is, $\alpha+\beta+\gamma+\delta+\eta+\mu=1$, so that the matrix above is doubly stochastic (i.e., the sum of each row and each column is one). Also note that, while ${\bf p}$ is not symmetric in general, it is symmetric when only pairwise permutation are present.

The most general master equation for a three level system, using ${\bf p}$ in Eq.~\eqref{eq:mostgeneral}, has the form: 
\be
\begin{split}
\rho^{(1)}_{11}&=(\delta + \beta) \rho^{(0)}_{11}+(\alpha+\eta) \rho^{(0)}_{22}+(\gamma+\mu) \rho^{(0)}_{33}\\ 
\rho^{(1)}_{22}&=(\alpha+\mu) \rho^{(0)}_{11}+(\delta +\gamma) \rho^{(0)}_{22}+(\beta+\eta) \rho^{(0)}_{33} \\
\rho^{(1)}_{33}&=(\gamma+\eta) \rho^{(0)}_{11}+(\beta+\mu) \rho^{(0)}_{22}+ (\delta +\alpha) \rho^{(0)}_{33},
\end{split}
\ee
and is already quite complicated when compared to the particular case in Eq.~\eqref{simple-master}. The complexity increases as the number of states increases. However, as we discussed in the main text, the doubly stochastic form of the transition matrix leads to several important general consequences (see also Ref.~\cite{allahverdyan_hovhannisyan_11}).

%%%%%%%%%%%%%%%%%%%%%%%%%%%%%%%%%%%%%%%%%%%%%%%%%%%%%%%%%%%%%%%%%%%%%%%%%%%%%%%%%%%%%%%%%%
\section{Proof of $\langle W\rangle\ge 0$ for Passive Density Matrices and Doubly-Stochastic Evolution\label{appendix:kelvin}}
%%%%%%%%%%%%%%%%%%%%%%%%%%%%%%%%%%%%%%%%%%%%%%%%%%%%%%%%%%%%%%%%%%%%%%%%%%%%%%%%%%%%%%%%%%

In this appendix, we prove that $\langle W\rangle\ge 0$ if the initial density matrix is passive and the evolution is doubly stochastic (see also Ref.~\cite{thirring_book}). 

Let us arrange the energy levels in order of increasing energies and, hence, by passivity, decreasing occupation probabilities, that is, $E_1\leq E_2\leq\dots E_D$ and $\rho^{(0)}_{11}\geq \rho^{(0)}_{22} \geq\dots \rho^{(0)}_{DD}$. We also assume that the Hilbert space size $D$ is finite (one can always take the limit $D\to\infty$ at the end). The general expression for the average work is
\be
\begin{split}
\langle W\rangle &=\sum_{m} \rho_{mm}^{(1)} E_m - \sum_n \rho_{nn}^{(0)} E_n \\
 &= \sum_{m,n} \rho_{nn}^{(0)} p_{n\rightarrow m} E_m - \sum_n \rho_{nn}^{(0)} E_n = \sum_{n} \rho_{nn}^{(0)} \left[ \sum_m p_{n\rightarrow m} E_m - E_n \right]
\end{split}
\label{step1}
\ee
where the sums over $n$ and $m$ go from $1$ to $D$. Next, we define
\be
\Delta_{n}^k\equiv\rho_{nn}^{(0)}-\rho_{D-k,D-k}^{(0)}. 
\label{eq:def_delta}
\ee 
Clearly, from the passivity condition~\eqref{eq:passive}, $\Delta_{n}^0=\rho_{nn}^{(0)}-\rho_{DD}^{(0)}\ge0$ . 
Then, we rewrite Eq.~\eqref{step1} as:
\be
\langle W\rangle=\sum_{n=1}^D \Delta_{n}^0 \left[ \sum_{m=1}^D p_{n\rightarrow m} E_m - E_n \right] + K^0,
\label{step2}
\ee
where, using the doubly stochastic condition for transition rates, that is, $\sum_{n=1}^D p_{n\to m}=1$, one can show that 
\be
K^0= \rho_{DD}^{(0)} \sum_{n=1}^D  \left[ \sum_{m=1}^D p_{n\rightarrow m} E_m - E_n \right] = 0.
\ee
Finally, noting that $\Delta_{D}^0=0$, we rewrite~\eqref{step2} as
\be
\langle W\rangle=\sum_{n=1}^{D-1} \Delta_{n}^0 \left[ \sum_{m=1}^D p_{n\rightarrow m} E_m - E_n \right].
\label{step3}
\ee

Next, we write $\Delta_n^0$ in terms of $\Delta_n^1$, that is, $\Delta_n^0 = \Delta_n^1 + \rho_{D-1,D-1}^{(0)} - \rho_{D,D}^{(0)}$, and plug it in the equation above to obtain
\be
\langle W\rangle=\sum_{n=1}^{D-1} \Delta_{n}^1 \left[ \sum_{m=1}^{D} p_{n\rightarrow m} E_m - E_n \right] + K^1,
\label{step4}
\ee
where 
\be
K^1=\left(\rho_{D-1,D-1}^{(0)}-\rho_{D,D}^{(0)}\right) \sum_{n=1}^{D-1} \left[ \sum_{m=1}^{D} p_{n\rightarrow m} E_m - E_n \right] \ge 0.
\ee
To see why $K^1\ge 0$ note that: (i) the passivity condition, Eq.~\eqref{eq:passive}, implies $\rho_{D-1,D-1}^{(0)}\ge\rho_{D,D}^{(0)}$, and that (ii) the remaining sum can be rewritten as
\be
\begin{split}
&\sum_{n=1}^{D-1} \left[ \sum_{m=1}^{D} p_{n\rightarrow m} E_m - E_n \right] = \sum_{m,n=1}^D p_{n\rightarrow m} E_m - \sum_{m=1}^D p_{D\rightarrow m} E_m - \sum_{n=1}^{D-1} E_n \\
& = \sum_{m=1}^D E_m - \sum_{n=1}^{D-1} E_n - \sum_{m=1}^D p_{D\rightarrow m} E_m  
= E_D - \sum_{m=1}^D p_{D\rightarrow m} E_m \ge E_D - \sum_{m=1}^D p_{D\rightarrow m} E_D = 0,
\end{split}
\ee
where we have used the doubly stochastic~\eqref{eq:double_stochastic} condition multiple times. Finally, noting that $\Delta_{D-1}^1=0$, we rewrite Eq.~\eqref{step4} as
\be
\langle W\rangle\ge \sum_{n=1}^{D-2} \Delta_{n}^1 \left[ \sum_{m=1}^{D} p_{n\rightarrow m} E_m - E_n \right]. 
\label{step5}
\ee
Equation~\eqref{step5} is similar to Eq.~\eqref{step3} except that the external sum in Eq.~\eqref{step5} extends only up to $D-2$ and not to $D-1$ as in Eq.~\eqref{step3}.

The proof continues iteratively. For example, in the next iteration, we write  $\Delta_n^1$ in terms of $\Delta_n^2$, that is, $\Delta_{n}^1=\Delta_{n}^2+\rho_{D-2,D-2}^{(0)}-\rho_{D-1,D-1}^{(0)}$ to obtain
\be
\langle W\rangle\ge \sum_{n=1}^{D-2} \Delta_{n}^2 \left[ \sum_{m=1}^{D} p_{n\rightarrow m} E_m - E_n \right] + K^2 
\ee
where 
\be
K^2=(\rho_{D-2,D-2}^{(0)}-\rho_{D-1,D-1}^{(0)}) \sum_{n=1}^{D-2} \left[ \sum_{m=1}^{D} p_{n\rightarrow m} E_m - E_n \right]
\ee
The prefactor $(\rho_{D-2,D-2}^{(0)}-\rho_{D-1,D-1}^{(0)})$ is positive by the passivity condition, Eq.~\eqref{eq:passive}. Moreover, using similar steps as above, we rewrite the sum as
\be
\begin{split}
\sum_{n=1}^{D-2} \left[ \sum_{m=1}^{D} p_{n\rightarrow m} E_m - E_n \right] &= E_D + E_{D-1} - \sum_{m=1}^D  p_{(D-1)\rightarrow m} E_m - \sum_{m=1}^D  p_{D\rightarrow m} E_m \\
&= E_D + E_{D-1} - 2 E_{D-1} + \sum_{m=1}^D  \left[p_{(D-1)\rightarrow m} +p_{D\rightarrow m} \right] \delta E_m \\
& \ge  E_D + E_{D-1} - 2 E_{D-1} + \left[ p_{(D-1)\rightarrow D} +p_{D\rightarrow D} \right] \delta E_D \\
& \ge E_D + E_{D-1} - 2 E_{D-1} + (E_{D-1}-E_D) = 0,
\end{split}
\ee
where we have defined $\delta E_m \equiv E_{D-1}-E_m$. In the third line, we have used that $\delta E_m\ge 0$ for any $m\le D-1$ and, in the fourth line, we have used that $p_{(D-1)\rightarrow D} +p_{D\rightarrow D}\le 1$, which is guaranteed by the doubly stochastic condition $\sum_m p_{m\rightarrow D}=1$. We therefore conclude that $K_2\ge0$. Finally, noting that $\Delta^2_{D-2}=0$ we arrive at:
\be
\langle W\rangle\ge \sum_{n=1}^{D-3} \Delta_{n}^2 \left[ \sum_{m=1}^{D} p_{n\rightarrow m} E_m - E_n \right] 
\label{step6}
\ee
where now the external sum extends only up to $D-3$. 

Clearly, comparing Eqs.~\eqref{step3}, \eqref{step5} and \eqref{step6}, we see that at each iteration the upper limit of the external sum decreases by one and the index $k$ in $\Delta_n^k$ increases by one. Continuing this iterative process, eventually, the external sum will include only one element proportional to $\Delta_{1}^{D-1}$, which is zero by definition [see Eq.~\eqref{eq:def_delta}]. Therefore one can conclude that $\langle W\rangle\ge0$.

%%%%%%%%%%%%%%%%%%%%%%%%%%%%%%%%%%%%%%%%%%%%%%%%%%%%%%%%%%%%%%%%%%%%%%%%%%%%%%%%%%%%%%%%%%
\section{Derivation of the Drift Diffusion Relation for Continuous Processes}\label{appendix:drift_diffusion_continuous}
%%%%%%%%%%%%%%%%%%%%%%%%%%%%%%%%%%%%%%%%%%%%%%%%%%%%%%%%%%%%%%%%%%%%%%%%%%%%%%%%%%%%%%%%%%

We already showed how one can derive the drift diffusion relation~\eqref{eq:fdis1} by means of a cumulant expansion of the Evans-Searles fluctuation relation~\eqref{eq:raw_result}. Here, we show how the same result can be derived directly from the ETH ansatz \eqref{eq:ETH} applied to continuous driving protocols. In particular, let us focus on a setup in which the external parameter $\lambda$, conjugate to the observable $\hat O=-\partial_\lambda \hat H$, changes in time at a constant rate. For example, this parameter can be the position of a macroscopic object moving in some media (quantum or classical).

Within leading order in adiabatic perturbation theory (see Ref.~\cite{dalessio_polkovnikov_14} for further details), the energy dissipation in the system is given by (we have set $\hbar=1$ so that the energy has dimension of $\text{time}^{-1}$):
\be
{d\tilde Q\over dt}\approx \dot \lambda^2 \sum_{n,m} {\rho_n-\rho_m\over E_m-E_n}\langle n| \hat O|m\rangle\langle m|\hat O|n\rangle\delta(E_n-E_m),
\label{eq:dQ/dt}
\ee
where $\rho_n$ and $\rho_m$ are the stationary probabilities to occupy the many-body eigenstates $|n\rangle$ and $|m\rangle$ corresponding to the energies $E_n$ and $E_m$, respectively. All matrix elements and energies here correspond to the instantaneous value of $\lambda$.  For the Gibbs distribution, $\rho_n\propto \exp[-\beta E_n]$, it is easy to see that
\be
 {\rho_n-\rho_m\over E_m-E_n}\delta (E_n-E_m)=\beta \rho_n \delta(E_n-E_m)
\ee
and Eq.~\eqref{eq:dQ/dt} reduces to the standard expression for the energy dissipation (see, e.g., Ref.~\cite{sivak_crooks_12}). As in many places in this review, let us focus instead on the dissipation from a single many-body energy eigenstate, $\rho_n=\delta_{n,n_0}$. If the relation holds for any eigenstate, it holds for any stationary distribution with subextensive energy fluctuations. For a single eigenstate, Eq.~\eqref{eq:dQ/dt} becomes
\be
{d\tilde Q\over dt}= 2\dot\lambda^2 \sum_{m\neq n_0} {1\over E_m-E_{n_0}}\langle n_0| \hat O|m\rangle\langle m|\hat O|n_0\rangle\delta(E_{n_0}-E_m),
\label{eq:dQ/dt1}
\ee
Let us now use the ETH ansatz~\eqref{eq:ETH} and, as usual, replace the summation over the eigenstates by an integration over $\omega=E_m-E_{n_0}\equiv E_m-E$: $\sum_m \to\int d\omega\exp[S(E+\omega)]$ (for simplicity, we drop the index $n_0$ in the energy). Then
\be
{d\tilde Q\over dt}= 2\dot\lambda^2 {\cal P}\int\, d\omega\,\frac{\mathrm e^{S(E+\omega)-S(E+\omega/2)}}{\omega}  |f_O(E+\omega/2,\omega)|^2 \delta(\omega),
\ee
where ${\cal P}\int$ stands for the principal value of the integral. Noting that
\beq
\mathrm e^{S(E+\omega)-S(E+\omega/2)}&\approx& \mathrm e^{\beta \omega/2}=1+{\beta\omega\over 2}+\dots,\nonumber\\ 
|f_O(E+\omega/2,\omega)|^2&=& |f_O(E,\omega)|^2+{\omega\over 2}\partial_E |f_O(E,\omega)|^2 +\dots,
\eeq
and using the fact that $|f_O(E,\omega)|^2$ is an even function of $\omega$, we find
\be
{d\tilde Q\over dt}\equiv J_E= \dot\lambda^2 [\beta |f_O(E,0)|^2+\partial_E |f_O(E,0)|^2].
\label{eq:JE_cont}
\ee
We note that, formally, the function $|f_O(E,0)|^2$ diverges in the thermodynamic limit (see Fig.~\ref{fig:FW}). However, physically, this divergence is cutoff by the inverse relaxation time in the system, which defines the broadening of the $\delta$-function in Eq.~\eqref{eq:dQ/dt}.

Similarly, from adiabatic perturbation theory, one can show that 
\be
{d (\delta E^2)\over dt}\equiv D_E=2\dot \lambda^2 \sum_{m\neq n_0} \langle n_0| \hat O|m\rangle\langle m|\hat O|n_0\rangle\delta(E_{n_0}-E_m)=2\dot\lambda^2 |f_O(E,0)|^2.
\label{eq:DE_cont}
\ee
Comparing Eqs.~\eqref{eq:JE_cont} and \eqref{eq:DE_cont}, we recover the desired drift-diffusion relation~\eqref{eq:fdis1}.

%%%%%%%%%%%%%%%%%%%%%%%%%%%%%%%%%%%%%%%%%%%%%%%%%%%%%%%%%%%%%%%%%%%%%%%%%%%%%%%%%%%%%%%%%%
\section{Derivation of Onsager Relations \label{Appendix_FLUCTUATIONS}} 
%%%%%%%%%%%%%%%%%%%%%%%%%%%%%%%%%%%%%%%%%%%%%%%%%%%%%%%%%%%%%%%%%%%%%%%%%%%%%%%%%%%%%%%%%%

In this appendix, we derive Eqs.~\eqref{eq:fdis2} and~\eqref{eq:result1}. Since Eq.~\eqref{eq:result1} is more general, we show its derivation first and then obtain Eq.~\eqref{eq:fdis2} as a special case.  First, in the Crooks relation~\eqref{eq:result1}, we expand the entropy and the probability distribution as a function of $E_{\rm I,II}\pm W$ and $N_{\rm I,II}\pm \delta N$ to second order in $W$ and $\delta N$ to obtain
\begin{align}
& P(E_{\rm I}, E_{\rm II}, N_{\rm I}, N_{\rm II}, W,\delta N) \exp \left[  -\Delta \beta W - \Delta \kappa \delta N  -  \frac{W^2}{2} \partial_{E}\Delta\beta - \frac{\delta N^2}{2}\partial_N \Delta \kappa\right.\\
 & \quad \quad -  \left. \frac{W \delta N}{2} (\partial_N \Delta \beta+\partial_E \delta \kappa) \right] =  \exp\left[  W  \partial_{E} + \delta N  \partial_{N} \right] P(E_{\rm I}, E_{\rm II}, N_{\rm I}, N_{\rm II}, -W,-\delta N), \nonumber 
\end{align}
where $\Delta \beta = \beta_{\rm I} - \beta_{\rm II}$,  $\Delta \kappa = \kappa_{\rm I} - \kappa_{\rm II}$, $\beta_{\rm I}=\partial_{\rm E_{I}} S_{\rm I}$, $\kappa_{\rm I}=\partial_{\rm \delta N_{I}} S_{\rm I}$, and similarly for II (${\rm I}\to{\rm II}$). The partial derivatives $\partial_E$ and $\partial_N$ are understood as derivatives with respect to energy and particle exchange: $\partial_E f(E_{\rm I}, E_{\rm II})\equiv \partial_{E_{\rm I}} f(E_{\rm I}, E_{\rm II})-\partial_{E_{\rm II}} f(E_{\rm I}, E_{\rm II})$. Next, one needs to integrate over $W$ and $\delta N$ and perform the cumulant expansion to second order. Following the discussion after Eq.~\eqref{eq:cumfirsteq}, we keep only the terms linear in the cumulants to obtain
\be
\begin{split}
&-\Delta \beta \la W\ra - \Delta \kappa \la \delta N \ra - \frac{\la W^2\ra_c }{2}  \partial_{E}\Delta\beta - \frac{\la \delta N^2\ra_c}{2}  \partial_{N}\Delta \kappa  -  \frac{\la W \delta N \ra_c}{2} \left( \partial_{N}\Delta \beta + \partial_{E} \Delta\kappa \right) \\ 
& + \frac{(\Delta \beta)^2}{2} \la W^2\ra_c + \frac{(\Delta \kappa)^2}{2} \la \delta N^2 \ra_c + \Delta \beta \Delta \kappa \la W \delta N \ra_c \\
&=  - \partial_E \la W \ra - \partial_N \la \delta N \ra + \frac{1}{2} \partial^2_{EE} \la W^2\ra_c +  \frac{1}{2} \partial^2_{NN} \la \delta N^2\ra_c +  \partial^2_{EN} \la W \delta N\ra_c \,.
\label{intermediate_Croock} 
\end{split}
\ee
Using the following identities:
\begin{align*}
\la W^2 \ra_c \partial_{E}\Delta\beta  =& \partial_E \left(\Delta \beta \la W^2 \ra_c \right) - \Delta \beta \partial_E  \la W^2 \ra_c, \\
\la \delta N^2 \ra_c \partial_{N}\Delta\kappa =& \partial_N \left(\Delta \kappa \la \delta N^2 \ra_c \right) - \Delta \kappa \partial_N  \la \delta N^2 \ra_c, \\
\la W \delta N \ra_c \left( \partial_{N} \Delta \beta + \partial_{E}\Delta \kappa \right)  =& \partial_N (\Delta \beta \la W \delta N \ra_c) - \Delta \beta \partial_N \la W \delta N \ra_c \\
& + \partial_E (\Delta \kappa \la W \delta N \ra_c) - \Delta \kappa \partial_E \la W \delta N \ra_c,
\end{align*}
we can rewrite Eq.~\eqref{intermediate_Croock} as
\begin{align}
& \left( \Delta \beta - \partial_E \right) \left[ - \la W \ra + \frac{\Delta \beta}{2} \la W^2\ra_c + \frac{\Delta \kappa}{2} \la W \delta N \ra_c + \frac{1}{2} \partial_E \la W^2\ra_c +  \frac{1}{2} \partial_N \la W \delta N \ra_c \right] +\\
& \left( \Delta \kappa - \partial_N \right) \left[ - \la \delta N \ra + \frac{\Delta \kappa}{2} \la \delta N^2 \ra_c + \frac{\Delta \beta}{2} \la W \delta N \ra_c  + \frac{1}{2} \partial_N \la \delta N^2\ra_c + \frac{1}{2} \partial_E \la W \delta N \ra_c \right] =0\,.  \nonumber 
\end{align}
Since this relation holds for any value of $\Delta \beta$ and $\Delta \kappa$, each term in the square brackets must be zero, leading to Eq.~\eqref{eq:result1}. By assuming that the systems do not exchange particles, that is, that $\delta N=0$, we recover Eq.~\eqref{eq:fdis2}.

\end{appendices}

\bibliographystyle{tADP}
\bibliography{./Review.bib}

\begin{thebibliography}{100}
\newcommand{\noopsort}[1]{}
\newcommand{\printfirst}[2]{#1}
\newcommand{\singleletter}[1]{#1}
\newcommand{\switchargs}[2]{#2#1}
\providecommand{\url}[1]{\normalfont{#1}}
\providecommand{\urlprefix}{Available at }

\bibitem{ma_book_85}
S.K. Ma, \emph{Statistical mechanics}, World Scientific, Singapore,
  Philadelphia, 1985.

\bibitem{feynman_book_98}
R.P. Feynman, \emph{Statistical Mechanics: A Set of Lectures}, Westview Press;
  2 edition, 1998.

\bibitem{boltzmann_96}
L.~Boltzmann, Ann. Phys. (Leipzig) 57 (1896), p. 773.

\bibitem{boltzmann_97}
L.~Boltzmann, Ann. Phys. (Leipzig) 60 (1897), p. 392.

\bibitem{gibbs_book_65}
J.W. Gibbs, \emph{Elementary Principles in Statistical Mechanics}, Dover, New
  York, 1960.

\bibitem{penrose_book_65}
O.~Penrose, \emph{Foundations of Statistical Mechanics}, Pergamon, Oxford,
  1970.

\bibitem{lebowitz_penrose_73}
J.L. Lebowitz and O.~Penrose, Physics Today 2 (1973), p.~23.

\bibitem{sinai_63}
Y.G. Sinai, Sov. Math Dokl. 4 (1963), pp. 1818--1822.

\bibitem{sinai_70}
Y.G. Sinai, Russian Mathematical Surveys 25 (1970), pp. 137--191.

\bibitem{bunimovich_79}
L.A. Bunimovich, Commun. Math. Phys. 65 (1979), p. 295.

\bibitem{simanyi_04}
N.~Simanyi, Ann. Henri Poincar\'e 5 (2004), p. 203.

\bibitem{poincare_90}
H.~Poincar\'e, Acta Math. 13 (1890), p.~1.

\bibitem{faddeev_tachtajan_07}
L.D. Faddeev and L.A. Takhtajan, \emph{Hamiltonian Methods in the Theory of
  Solitons (Classics in Mathematics)}, Springer, 2007.

\bibitem{fermi_pasta_55}
E.~Fermi, J.~Pasta, and S.~Ulam, \emph{Studies of nonlinear problems}, Tech.
  {R}ep.,  1955.

\bibitem{dauxois_08}
T.~Dauxois, Physics Today 61 (2008), pp. 55--57.

\bibitem{berman_izrailev_2005}
G.P. Berman and F.M. Izrailev, Chaos 15 (2005), p. 015104.

\bibitem{kozik_svistunov_09}
E.~Kozik and B.~Svistunov, J. of Low Temp. Phys. 156 (2009), p. 215.

\bibitem{bouchad_92}
J.P. Bouchaud, J. Phys. I France 2 (1992), p. 1705.

\bibitem{levesque_verlet_93}
D.~Levesque and L.~Verlet, Journal of Statistical Physics 72, pp. 519--537.

\bibitem{jarzynski_97}
C.~Jarzynski, Phys. Rev. Lett. 78 (1997), p. 2690.

\bibitem{crooks_99}
G.E. Crooks, Phys. Rev. E 60 (1999), p. 2721.

\bibitem{collin_05}
D.~Collin, F.~Ritort, C.~Jarzynski, S.B. Smith, I.~{Tinoco Jr}, and
  C.~Bustamante, Nature 437 (2005), p. 231.

\bibitem{kurchan_00}
J.~Kurchan, arXiv:cond-mat/0007360  (2000).

\bibitem{tasaki_00}
H.~Tasaki, arXiv:cond-mat/0009244  (2000).

\bibitem{campisi_hanggi_11}
M.~Campisi, P.~H\"anggi, and P.~Talkner, Rev. Mod. Phys. 83 (2011), p. 771.

\bibitem{deutsch_91}
J.M. Deutsch, Phys. Rev. A 43 (1991), pp. 2046--2049.

\bibitem{srednicki_94}
M.~Srednicki, Phys. Rev. E 50 (1994), pp. 888--901.

\bibitem{rigol_dunjko_08}
M.~Rigol, V.~Dunjko, and M.~Olshanii, Nature 452 (2008), p. 854.

\bibitem{srednicki_99}
M.~Srednicki, J. Phys. A 32 (1999), p. 1163.

\bibitem{vonneumann_29}
J.~von Neumann, Zeitschrift f\"ur Physik 57 (1929), pp. 30--70, translated to
  English in European Phys. J. H {\bf 35}, 201 (2010).

\bibitem{wigner_55}
E.~Wigner, Ann. of Math. 62 (1955), pp. 548--564.

\bibitem{bloch_dalibard_review_08}
I.~Bloch, J.~Dalibard, and W.~Zwerger, Rev. Mod. Phys. 80 (2008), pp. 885--964.

\bibitem{cazalilla_citro_11}
M.A. Cazalilla, R.~Citro, T.~Giamarchi, E.~Orignac, and M.~Rigol, Rev. Mod.
  Phys. 83 (2011), pp. 1405--1466.

\bibitem{greiner_mandel_02}
M.~Greiner, O.~Mandel, T.W. H\"ansch, and I.~Bloch, Nature 419 (2002), pp.
  51--54.

\bibitem{will_best_10}
S.~Will, T.~Best, U.~Schneider, L.~Hackerm\"uller, D.S. L\"uhmann, and
  I.~Bloch, Nature 465 (2010), pp. 197--201.

\bibitem{will_best_11}
S.~Will, T.~Best, S.~Braun, U.~Schneider, and I.~Bloch, Phys. Rev. Lett. 106
  (2011), p. 115305.

\bibitem{will_iyer_15}
S.~Will, D.~Iyer, and M.~Rigol, Nat. Commun. 6 (2015), p. 6009.

\bibitem{kinoshita_wenger_06}
T.~Kinoshita, T.~Wenger, and D.S. Weiss, Nature 440 (2006), p. 900.

\bibitem{gring_kuhnert_12}
M.~Gring, M.~Kuhnert, T.~Langen, T.~Kitagawa, B.~Rauer, M.~Schreitl, I.~Mazets,
  D.A. Smith, E.~Demler, and J.~Schmiedmayer, Science 337 (2012), pp.
  1318--1322.

\bibitem{langen_erne_15}
T.~Langen, S.~Erne, R.~Geiger, B.~Rauer, T.~Schweigler, M.~Kuhnert,
  W.~Rohringer, I.E. Mazets, T.~Gasenzer, and J.~Schmiedmayer, Science 348
  (2015), p. 207.

\bibitem{trotzky_chen_12}
S.~Trotzky, Y.A. Chen, A.~Flesch, I.P. McCulloch, U.~Schollw\"ock, J.~Eisert,
  and I.~Bloch, Nature Phys. 8 (2012), p. 325.

\bibitem{winkler_thalhammer_06}
K.~Winkler, G.~Thalhammer, F.~Lang, R.~Grimm, J.H. Denschlag, A.J. Daley,
  A.~Kantian, H.P. B\"uchler, and P.~Zoller, Nature 441 (2006), p. 853.

\bibitem{xia_zundel_15}
L.~Xia, L.~Zundel, J.~Carrasquilla, J.M. Wilson, M.~Rigol, and D.S. Weiss,
  Nature Phys. 11 (2015), p. 316.

\bibitem{childress_dutt_06}
L.~Childress, M.V.G. Dutt, J.M. Taylor, A.S. Zibrov, F.~Jelezko, J.~Wrachtrup,
  P.R. Hemmer, and M.D. Lukin, Science 314 (2006), p. 281.

\bibitem{wall_brida_11}
S.~Wall, D.~Brida, S.R. Clark, H.P. Ehrke, D.~Jaksch, A.~Ardavan, S.~Bonora,
  H.~Uemura, Y.~Takahashi, T.~Hasegawa, H.~Okamoto, G.~Cerullo, and
  A.~Cavalleri, Nature Physics 7 (2011), p. 114.

\bibitem{basov_averitt_11}
D.N. Basov, R.D. Averitt, D.~van~der Marel, M.~Dressel, and K.~Haule, Rev. Mod.
  Phys. 83 (2011), p. 471.

\bibitem{berges_boguslavski_14}
J.~Berges, K.~Boguslavski, S.~Schlichting, and R.~Venugopalan, Phys. Rev. D 89
  (2014), p. 074011.

\bibitem{dziarmaga_10}
J.~Dziarmaga, Adv. Phys. 59 (2010), pp. 1063--1189.

\bibitem{polkovnikov_sengupta_11}
A.~Polkovnikov, K.~Sengupta, A.~Silva, and M.~Vengalattore, Rev. Mod. Phys. 83
  (2011), pp. 863--883.

\bibitem{yukalov_11}
V.I. Yukalov, Laser Phys. Lett. 8 (2011), pp. 485--507.

\bibitem{nandkishore_huse_14}
R.~Nandkishore and D.A. Huse, Annual Review of Condensed Matter Physics 6
  (2015), pp. 15--38.

\bibitem{eisert_friesdorf_15}
J.~Eisert, M.~Friesdorf, and C.~Gogolin, Nature Physics 11 (2015), p. 124.

\bibitem{gogolin_eisert_15}
C.~Gogolin and J.~Eisert, Reports on Progress in Physics 79 (2016), p. 056001.

\bibitem{cazalilla_rigol_10}
M.A. Cazalilla and M.~Rigol, New J. Phys. 12 (2010), p. 055006.

\bibitem{daley_rigol_14}
A.J. Daley, M.~Rigol, and D.S. Weiss, New J. Phys. 16 (2014), p. 095006.

\bibitem{lichtenberg_liebermann_book_85}
A.J. Lichtenberg and M.A. Liebermann, \emph{Regular and chaotic dynamics (2nd
  ed)}, Springer, Berlin, New York, 1992.

\bibitem{chaosbook}
P.~Cvitanovic, R.~Artuso, R.~Mainieri, G.~Tanner, and G.~Vattay, \emph{Chaos:
  Classical and Quantum, \url{ChaosBook.org}}, Niels Bohr Institute,
  Copenhagen, 2012.

\bibitem{arnold_89}
V.I. Arnold, \emph{Mathematical Methods of Classical Mechanics},
  Springer-Verlag, New York, 1989.

\bibitem{jose_saletan_book_98}
J.V. Jos{\' e} and E.J. Saletan, \emph{Classical Dynamics: A Contemporary
  Approach}, Cambridge Univ. Press, Cambridge, 1998.

\bibitem{stockmann_10}
H.J. St\"ockmann, Scholarpedia 5 (2010), p. 10243.

\bibitem{landau_lifshitz_1_76}
L.D. Landau and E.M. Lifshitz, \emph{Mechanics, Third Edition (Volume 1)},
  Butterworth-Heinemann, 1976.

\bibitem{kolmogorov_54}
A.N. Kolmogorov, Dokl. Akad. Nauk. SSR 98 (1954), pp. 527--530.

\bibitem{arnold_63}
V.I. Arnold, Russian Math. Survey 18 (1963), pp. 13--40.

\bibitem{moser_62}
J.K. Moser, Nach. Akad. Wiss. G\"ottingen, Math. Phys. Kl. II 1 (1962), pp.
  1--20.

\bibitem{lichtenberg_lieberman_92}
A.J. Lichtenberg and M.A. Lieberman, \emph{Regular and Chaotic Dynamics},
  Applied Mathematical Sciences, Vol. 38, Springer-Verlag, New York, 1992.

\bibitem{broer_hoveijn_04}
H.W. Broer, I.~Hoveijn, M.~van Noort, C.~Simo, and G.~Vegter, Journal of
  Dynamics and Differential Equations 16 (2004), p. 897.

\bibitem{chirikov_79a}
B.V. Chirikov, Physics Reports 52 (1979), pp. 263--379.

\bibitem{chirikov_79b}
G.Casati, B.V. Chirikov, F.M. Izrailev, and J.~Ford, \emph{Stochastic behavior
  of a quantum pendulum under a periodic perturbation}, in \emph{Stochastic
  Behavior in Classical and Quantum Hamiltonian Systems}, G.~Casati and
  J.~Ford, eds., Lecture Notes in Physics, Vol.~93, Springer Berlin Heidelberg,
   1979, pp. 334--352.

\bibitem{santos_rigol_10a}
L.F. Santos and M.~Rigol, Phys. Rev. E 81 (2010), p. 036206.

\bibitem{santos_rigol_10b}
L.F. Santos and M.~Rigol, Phys. Rev. E 82 (2010), p. 031130.

\bibitem{guhr_muller_98}
T.~Guhr, A.~M\"uller, Groeling, and H.A. Weidenm\"uller, Physics Reports 299
  (1998), pp. 189--425.

\bibitem{alhassid_00}
Y.~Alhassid, Rev. Mod. Phys. 72 (2000), p. 895.

\bibitem{mehta_04}
M.L. Mehta, \emph{Random Matrices}, Amsterdam: Elsevier/Academic Press, 2004.

\bibitem{kravtsov_09}
V.E. Kravtsov, \emph{Random matrix theory: {Wigner-Dyson} statistics and
  beyond}, arXiv:0911.0639.

\bibitem{hillery_oconnell_84}
M.~Hillery, R.F. O'Connell, M.O. Scully, and E.P. Wigner, Physics Reports 106
  (1984), pp. 121--167.

\bibitem{polkovnikov_10}
A.~Polkovnikov, Annals of Phys. 325 (2010), p. 1790.

\bibitem{landau_lifshitz_3_81}
L.D. Landau and E.M. Lifshitz, \emph{Quantum Mechanics, Third Edition:
  Non-Relativistic Theory (Volume 3)}, Butterworth-Heinemann, 1981.

\bibitem{stone_05}
D.~Stone, Physics Today 58 (2005), p.~37.

\bibitem{gutzwiller_71}
M.C. Gutzwiller, J. Math. Phys. 12 (1971), p. 343.

\bibitem{rudnik_08}
Z.~Rudnick, Notices of the AMS 55 (2008), p.~32.

\bibitem{wigner_57}
E.~Wigner, Ann. of Math. 65 (1957), pp. 203--207.

\bibitem{wigner_58}
E.~Wigner, Ann. of Math. 67 (1958), pp. 325--326.

\bibitem{dyson_62}
F.J. Dyson, J. Math. Phys. 3 (1962), p. 140.

\bibitem{reichl_04}
L.E. Reichl, \emph{The transition to Chaos: Conservative Classical Sysstems and
  Quantum Manifestations}, Springer, New York, 2004.

\bibitem{bohigas_giannoni_84}
O.~Bohigas, M.J. Giannoni, and C.~Schmit, Phys. Rev. Lett. 52 (1984), pp. 1--4.

\bibitem{bogomolny_georgeot_92}
E.B. Bogomolny, B.~Georgeot, M.J. Giannoni, and C.~Schmit, Phys. Rev. Lett. 69
  (1992), p. 1477.

\bibitem{brody_flores_81}
T.A. Brody, J.~Flores, J.B. French, P.A. Mello, A.~Pandey, and S.S.M. Wong,
  Rev. Mod. Phys. 53 (1981), p. 385.

\bibitem{berry_tabor_77}
M.V. Berry and M.~Tabor, Proc. Roy. Soc. A 356 (1977), pp. 375--394.

\bibitem{pandey_ramaswamy_91}
A.~Pandey and R.~Ramaswamy, Phys. Rev. A 43 (1991), pp. 4237--4243.

\bibitem{berry_77}
M.V. Berry, J. Phys. A 10 (1977), p. 2083.

\bibitem{flambaum_izrailev_97a}
V.V. Flambaum and F.M. Izrailev, Phys. Rev. E 55 (1997), pp. R13--R16.

\bibitem{flambaum_izrailev_97b}
V.V. Flambaum and F.M. Izrailev, Phys. Rev. E 56 (1997), pp. 5144--5159.

\bibitem{kota_14}
V.K.B. Kota, Lecture Notes in Physics 884 (2014).

\bibitem{bochhoff_83}
K.H. B\"ochhoff, \emph{Nuclear Data for Science and Technology}, Springer,
  Brussels and Luxemburg, 1983.

\bibitem{wintgen_friedrich_87}
D.~Wintgen and H.~Friedrich, Phys. Rev. A 35 (1987), pp. 1464--1466.

\bibitem{bohigas_haq_83}
O.~Bohigas, R.U. Haq, and A.~Pandey, \emph{Fluctuation Properties of Nuclear
  Energy Levels and Widths : Comparison of Theory with Experiment}, in
  \emph{Nuclear Data for Science and Technology}, Reidel, Dordrecht (1983), p.
  809.

\bibitem{berry_robnik_84}
M.V. Berry and M.~Robnik, Journal of Physics A: Mathematical and General 17, p.
  2413.

\bibitem{rudnik_81}
M.~Robnik, J. Phys. A: Math. Gen. 14, p. 3195.

\bibitem{rigol_santos_10}
M.~Rigol and L.F. Santos, Phys. Rev. A 82 (2010), p. 011604(R).

\bibitem{kollath_roux_10}
C.~Kollath, G.~Roux, G.~Biroli, and A.~Laeuchli, J. Stat. Mech.  (2010), p.
  P08011.

\bibitem{santos_borgonovi_12a}
L.F. Santos, F.~Borgonovi, and F.M. Izrailev, Phys. Rev. Lett. 108 (2012), p.
  094102.

\bibitem{santos_borgonovi_12b}
L.F. Santos, F.~Borgonovi, and F.M. Izrailev, Phys. Rev. E 85 (2012), p.
  036209.

\bibitem{atas_bogomolny_13a}
Y.Y. Atas, E.~Bogomolny, O.~Giraud, and G.~Roux, Phys. Rev. Lett. 110 (2013),
  p. 084101.

\bibitem{atas_bogomolny_13b}
Y.Y. Atas, E.~Bogomolny, O.~Giraud, P.~Vivo, and E.~Vivo, Journal of Physics A:
  Mathematical and Theoretical 46 (2013), p. 355204.

\bibitem{modak_mukerjee_14a}
R.~Modak, S.~Mukerjee, and S.~Ramaswamy, Phys. Rev. B 90 (2014), p. 075152.

\bibitem{modak_mukerjee_14b}
R.~Modak and S.~Mukerjee, New J. Phys. 16 (2014), p. 093016.

\bibitem{kota_sahu_98}
V.K.B. Kota and R.~Sahu, Physics Letters B 429 (1998), p.~1.

\bibitem{landau_lifshitz_5_80}
L.D. Landau and E.M. Lifshitz, \emph{Statistical Physics, Third Edition, Part
  1, Volume 5}, Butterworth-Heinemann, 1980.

\bibitem{santos_polkovnikov_12}
L.F. Santos, A.~Polkovnikov, and M.~Rigol, Phys. Rev. E 86 (2012), p.
  010102(R).

\bibitem{deutsch_li_13}
J.M. Deutsch, H.~Li, and A.~Sharma, Phys. Rev. E 87 (2013), p. 042135.

\bibitem{mierzejewski_prosen_13}
M.~Mierzejewski, T.~Prosen, D.~Crivelli, and P.~Prelov\ifmmode~\check{s}\else
  \v{s}\fi{}ek, Phys. Rev. Lett. 110 (2013), p. 200602.

\bibitem{kim_huse_13}
H.~Kim and D.A. Huse, Phys. Rev. Lett. 111 (2013), p. 127205.

\bibitem{khlebnikov_kruczenski_14}
S.~Khlebnikov and M.~Kruczenski, Phys. Rev. E 90 (2014), p. 050101.

\bibitem{bardarson_pollman_12}
J.H. Bardarson, F.~Pollmann, and J.E. Moore, Phys. Rev. Lett. 109 (2012), p.
  017202.

\bibitem{vosk_altman_13}
R.~Vosk and E.~Altman, Phys. Rev. Lett. 110 (2013), p. 067204.

\bibitem{serbyn_papic_13}
M.~Serbyn, Z.~Papi\'c, and D.A. Abanin, Phys. Rev. Lett. 110 (2013), p. 260601.

\bibitem{nanduri_kim_13}
A.~Nanduri, H.~Kim, and D.A. Huse, Phys. Rev. B 90 (2013), p. 064201.

\bibitem{ponte_papic_14}
P.~Ponte, Z.~Papi\ifmmode~\acute{c}\else \'{c}\fi{}, F.~Huveneers, and D.A.
  Abanin, Phys. Rev. Lett. 114 (2015), p. 140401.

\bibitem{neil_roushan_16}
C.~Neill, P.~Roushan, M.~Fang, Y.~Chen, M.~Kolodrubetz, Z.~Chen, A.~Megrant,
  R.~Barends, B.~Campbell, B.~Chiaro, A.~Dunsworth, E.~Jeffrey, J.~Kelly,
  J.~Mutus, P.J.J. O'Malley, C.~Quintana, D.~Sank, A.~Vainsencher, J.~Wenner,
  T.C. White, A.~Polkovnikov, and J.M. Martinis, arXiv:1601.00600  (2016).

\bibitem{kaufman_tai_16}
A.M. Kaufman, M.E. Tai, A.~Lukin, M.~Rispoli, R.~Schittko, P.M. Preiss, and
  M.~Greiner, arXiv:1603.04409  (2016).

\bibitem{daley_pichler_12}
A.J. Daley, H.~Pichler, J.~Schachenmayer, and P.~Zoller, Phys. Rev. Lett. 109
  (2012), p. 020505.

\bibitem{islam_ma_15}
R.~Islam, R.~Ma, P.M. Preiss, M.E. Tai, A.~Lukin, M.~Rispoli, and M.~Greiner,
  Nature 528.

\bibitem{garrison_grover_15}
J.R. Garrison and T.~Grover, arXiv:1503.00729  (2015).

\bibitem{page_93}
D.N. Page, Phys. Rev. Lett. 71 (1993), p. 1291.

\bibitem{santos_polkovnikov_11}
L.F. Santos, A.~Polkovnikov, and M.~Rigol, Phys. Rev. Lett. 107 (2011), p.
  040601.

\bibitem{neuenhahn_marquardt_12}
C.~Neuenhahn and F.~Marquardt, Phys. Rev. E 85 (2012), p. 060101.

\bibitem{casati_chirikov_93}
G.~Casati, B.V. Chirikov, I.~Guarneri, and F.M. Izrailev, Phys. Rev. E 48
  (1993), p. R1613(R).

\bibitem{casati_chirikov_96}
G.~Casati, B.~Chirikov, I.~Guarneria, and F.~Izrailev, Physics Letters A 223
  (1996), p. 430.

\bibitem{flambaum_izrailev_00}
V.V. Flambaum and F.M. Izrailev, Phys. Rev. E 61 (2000), pp. 2539--2542.

\bibitem{rigol_16}
M.~Rigol, Phys. Rev. Lett. 116 (2016), p. 100601.

\bibitem{cassidy_clark_11}
A.C. Cassidy, C.W. Clark, and M.~Rigol, Phys. Rev. Lett. 106 (2011), p. 140405.

\bibitem{he_santos_13}
K.~He, L.F. Santos, T.M. Wright, and M.~Rigol, Phys. Rev. A 87 (2013), p.
  063637.

\bibitem{vidmar_rigol_16}
L.~Vidmar and M.~Rigol, J. Stat. Phys.  (2016), p. 064007.

\bibitem{polkovnikov_11}
A.~Polkovnikov, Annals of Phys. 326 (2011), p. 486.

\bibitem{rigol_14a}
M.~Rigol, Phys. Rev. Lett. 112 (2014), p. 170601.

\bibitem{rigol_fitzpatrick_11}
M.~Rigol and M.~Fitzpatrick, Phys. Rev. A 84 (2011), p. 033640.

\bibitem{he_rigol_12}
K.~He and M.~Rigol, Phys. Rev. A 85 (2012), p. 063609.

\bibitem{gurarie_13}
V.~Gurarie, J. Stat. Mech.  (2013), p. P02014.

\bibitem{kormos_bucciantini_14}
M.~Kormos, L.~Bucciantini, and P.~Calabrese, EPL (Europhysics Letters) 107
  (2014), p. 40002.

\bibitem{collura_kormos_14}
M.~Collura, M.~Kormos, and P.~Calabrese, J. Stat. Mech.  (2014), p. P01009.

\bibitem{goldstein_lebowitz_10}
S.~Goldstein, J.L. Lebowitz, R.~Tumulka, and N.~Zanghi, European Phys. J. H 35
  (2010), pp. 173--200.

\bibitem{tasaki_98}
H.~Tasaki, Phys. Rev. Lett. 80 (1998), pp. 1373--1376.

\bibitem{goldstein_lebowitz_06}
S.~Goldstein, J.L. Lebowitz, R.~Tumulka, and N.~Zangh\`\i{}, Phys. Rev. Lett.
  96 (2006), p. 050403.

\bibitem{popescu_06}
S.~Popescu, A.J. Short, and A.~Winter, Nature Phys. 2 (2006), p. 754.

\bibitem{rigol_srednicki_12}
M.~Rigol and M.~Srednicki, Phys. Rev. Lett. 108 (2012), p. 110601.

\bibitem{srednicki_96}
M.~Srednicki, J. Phys. A 29 (1996), p. L75.

\bibitem{jensen_shankar_85}
R.V. Jensen and R.~Shankar, Phys. Rev. Lett. 54 (1985), pp. 1879--1882.

\bibitem{feingold_peres_86}
M.~Feingold and A.~Peres, Phys. Rev. A 34 (1986), pp. 591--595.

\bibitem{wilkinson_87}
M.~Wilkinson, Journal of Physics A: Mathematical and General 20 (1987), p.
  2415.

\bibitem{prosen_94}
T.~Prosen, Annals of Phys. 235 (1994), pp. 115--164.

\bibitem{eckhardt_fishman_95}
B.~Eckhardt, S.~Fishman, J.~Keating, O.~Agam, J.~Main, and K.~M\"uller, Phys.
  Rev. E 52 (1995), pp. 5893--5903.

\bibitem{hortikar_srednicki_98}
S.~Hortikar and M.~Srednicki, Phys. Rev. E 57 (1998), pp. 7313--7316.

\bibitem{akkermans_montambaux_07}
E.~Akkermans and G.~Montambaux, \emph{Mesoscopic Physics of Electrons and
  Photons}, Cambridge University Press, 2007.

\bibitem{khatami_pupillo_13}
E.~Khatami, G.~Pupillo, M.~Srednicki, and M.~Rigol, Phys. Rev. Lett. 111
  (2013), p. 050403.

\bibitem{lux_muller_14}
J.~Lux, J.~M\"uller, A.~Mitra, and A.~Rosch, Phys. Rev. A 89 (2014), p. 053608.

\bibitem{reimann_15}
P.~Reimann, Phys. Rev. Lett. 115 (2015), p. 010403.

\bibitem{rigol_09a}
M.~Rigol, Phys. Rev. Lett. 103 (2009), p. 100403.

\bibitem{rigol_09b}
M.~Rigol, Phys. Rev. A 80 (2009), p. 053607.

\bibitem{steinigeweg_herbrych_13}
R.~Steinigeweg, J.~Herbrych, and P.~Prelov\ifmmode~\check{s}\else \v{s}\fi{}ek,
  Phys. Rev. E 87 (2013), p. 012118.

\bibitem{beugeling_moessner_14}
W.~Beugeling, R.~Moessner, and M.~Haque, Phys. Rev. E 89 (2014), p. 042112.

\bibitem{kim_14}
H.~Kim, T.N. Ikeda, and D.A. Huse, Phys. Rev. E 90 (2014), p. 052105.

\bibitem{steinigeweg_khodja_14}
R.~Steinigeweg, A.~Khodja, H.~Niemeyer, C.~Gogolin, and J.~Gemmer, Phys. Rev.
  Lett. 112 (2014), p. 130403.

\bibitem{khodja_steinigeweg_15}
A.~Khodja, R.~Steinigeweg, and J.~Gemmer, Phys. Rev. E 91 (2015), p. 012120.

\bibitem{beugeling_moessner_15}
W.~Beugeling, R.~Moessner, and M.~Haque, Phys. Rev. E 91 (2015), p. 012144.

\bibitem{khatami_rigol_12}
E.~Khatami, M.~Rigol, A.~Rela\~no, and A.M. Garcia-Garcia, Phys. Rev. E 85
  (2012), p. 050102(R).

\bibitem{genway_ho_12}
S.~Genway, A.F. Ho, and D.K.K. Lee, Phys. Rev. A 86 (2012), p. 023609.

\bibitem{biroli_kollath_10}
G.~Biroli, C.~Kollath, and A.M. L\"auchli, Phys. Rev. Lett. 105 (2010), p.
  250401.

\bibitem{roux_10}
G.~Roux, Phys. Rev. A 81 (2010), p. 053604.

\bibitem{sorg_vidmar_14}
S.~Sorg, L.~Vidmar, L.~Pollet, and F.~Heidrich-Meisner, Phys. Rev. A 90 (2014),
  p. 033606.

\bibitem{mondaini_fratus_16}
R.~Mondaini, K.R. Fratus, M.~Srednicki, and M.~Rigol, Phys. Rev. E 93 (2016),
  p. 032104.

\bibitem{ikeda_watanabe_13}
T.N. Ikeda, Y.~Watanabe, and M.~Ueda, Phys. Rev. E 87 (2013), p. 012125.

\bibitem{alba_15}
V.~Alba, Phys. Rev. B 91 (2015), p. 155123.

\bibitem{mukerjee_oganesyan_06}
S.~Mukerjee, V.~Oganesyan, and D.~Huse, Phys. Rev. B 73 (2006), p. 035113.

\bibitem{ott_mirandes_04}
H.~Ott, E.~de~Mirandes, F.~Ferlaino, G.~Roati, G.~Modugno, and M.~Inguscio,
  Phys. Rev. Lett. 92 (2004), p. 160601.

\bibitem{fertig_ohara_05}
C.D. Fertig, K.M. O'Hara, J.H. Huckans, S.L. Rolston, W.D. Phillips, and J.V.
  Porto, Phys. Rev. Lett. 94 (2005), p. 120403.

\bibitem{strohmaier_takasu_07}
N.~Strohmaier, Y.~Takasu, K.~G\"unter, R.~J\"ordens, M.~K\"ohl, H.~Moritz, and
  T.~Esslinger, Phys. Rev. Lett. 99 (2007), p. 220601.

\bibitem{schneider_hacke_12}
U.~Schneider, L.~Hackerm\"uller, J.P. Ronzheimer, S.~Will, S.~Braun, T.~Best,
  I.~Bloch, E.~Demler, S.~Mandt, D.~Rasch, and A.~Rosch, Nature Phys 8 (2012),
  pp. 213--218.

\bibitem{ronzheimer_schreiber_13}
J.P. Ronzheimer, M.~Schreiber, S.~Braun, S.S. Hodgman, S.~Langer, I.P.
  McCulloch, F.~Heidrich-Meisner, I.~Bloch, and U.~Schneider, Phys. Rev. Lett.
  110 (2013), p. 205301.

\bibitem{kollath_lauchli_07}
C.~Kollath, A.M. L\"auchli, and E.~Altman, Phys. Rev. Lett. 98 (2007), p.
  180601.

\bibitem{manmana_wessel_07}
S.R. Manmana, S.~Wessel, R.M. Noack, and A.~Muramatsu, Phys. Rev. Lett. 98
  (2007), p. 210405.

\bibitem{cramer_flesch_08a}
M.~Cramer, A.~Flesch, I.P. McCulloch, U.~Schollw\"ock, and J.~Eisert, Phys.
  Rev. Lett. 101 (2008), p. 063001.

\bibitem{cramer_flesch_08b}
A.~Flesch, M.~Cramer, I.P. McCulloch, U.~Schollw\"ock, and J.~Eisert, Phys.
  Rev. A 78 (2008), p. 033608.

\bibitem{eckstein_kollar_09}
M.~Eckstein, M.~Kollar, and P.~Werner, Phys. Rev. Lett. 103 (2009), p. 056403.

\bibitem{banuls_cirac_11}
M.C. Ba\~nuls, J.I. Cirac, and M.B. Hastings, Phys. Rev. Lett. 106 (2011), p.
  050405.

\bibitem{zangara_dente_13}
P.R. Zangara, A.D. Dente, E.J. Torres-Herrera, H.M. Pastawski, A.~Iucci, and
  L.F. Santos, Phys. Rev. E 88 (2013), p. 032913.

\bibitem{crooks_98}
G.E. Crooks, Journal of Statistical Physics 90 (1998), pp. 1481--1487.

\bibitem{seifert_12}
U.~Seifert, Rep. Prog. Phys. 75 (2012), p. 126001.

\bibitem{thirring_book}
W.~Thirring, \emph{Quantum Mathematical Physics. Second Ed.}, Springer-Verlag.
  Berlin, Heidelberg, 2002.

\bibitem{lifshitz_pitaevskii_06}
E.M. Lifshitz and L.P. Pitaevskii, \emph{Physical Kinetics}, Landau Lifshitz
  Course of Theoretical Physics, Vol. 10, Elsevier, 2006.

\bibitem{VanKampen}
N.G.V. Kampen, \emph{Stochastic Processes in Physics and Chemistry}, Elsevier.
  Amsterdam, The Netherlands, 2003.

\bibitem{marshall_olkin_79}
A.W. Marshall and I.~Olkin, \emph{Inequalities: Theory of Majorization},
  AcademicPress, NewYork, 1979.

\bibitem{polkovnikov_08}
A.~Polkovnikov, Phys. Rev. Lett. 101 (2008), p. 220402.

\bibitem{bunin_dalessio_11}
G.~Bunin, L.~D'Alessio, Y.~Kafri, and A.~Polkovnikov, Nature Phys. 7 (2011), p.
  913.

\bibitem{allahverdyan_hovhannisyan_11}
A.E. Allahverdyan and K.V. Hovhannisyan, EPL (Europhysics Letters) 95 (2011),
  p. 60004.

\bibitem{ikeda_sukumichi_15}
T.N. Ikeda, N.~Sakumichi, A.~Polkovnikov, and M.~Ueda, Annals of Physics 354
  (2015), p. 338.

\bibitem{ji_fine_2011}
K.~Ji and B.V. Fine, Phys. Rev. Lett. 107 (2011), p. 050401.

\bibitem{bukov_dalessio_14}
M.~Bukov, L.~D'Alessio, and A.~Polkovnikov, Advances in Physics 64 (2015), pp.
  139--226.

\bibitem{kullback_leibler_51}
S.~Kullback and R.A. Leibler, Annals of Mathematical Statistics 22 (1951),
  p.~79.

\bibitem{lenard_78}
A.~Lenard, Journal of Statistical Physics 19 (1978), pp. 575--586.

\bibitem{balian_06}
R.~Balian, \emph{From microphysics to macrophysics. 2nd Ed.}, Springer, 2006.

\bibitem{dalessio_polkovnikov_14}
L.~D'Alessio and A.~Polkovnikov, Annals of Physics 345 (2014), p. 141.

\bibitem{kolodrubetz_clark_12}
M.~Kolodrubetz, B.K. Clark, and D.A. Huse, Phys. Rev. Lett. 109 (2012), p.
  015701.

\bibitem{bonnes_essler_14}
L.~Bonnes, F.H. Essler, and A.M. L\"auchli, Phys. Rev. Lett. 113 (2014), p.
  187203.

\bibitem{allahverdyan_nieuwenhuizen_05}
A.E. Allahverdyan and T.M. Nieuwenhuizen, Phys. Rev. E 71 (2005), p. 046107.

\bibitem{bochkov_kuzovlev_77}
G.N. Bochkov and Y.E. Kuzovlev, JETP 45 (1977), p. 125.

\bibitem{bochkov_kuzovlev_81}
G.N. Bochkov and Y.E. Kuzovlev, Physica A 106 (1981), p. 443.

\bibitem{jarzynski_11}
C.~Jarzynski, Annual Review Condensed Matter Physics 2 (2011), pp. 329--351.

\bibitem{andrieux_gaspard_07}
D.~Andrieux and P.~Gaspard, J. Stat. Mech.  (2007), p. P02006.

\bibitem{evans_searles_94}
D.J. Evans and D.J. Searles, Phys. Rev. E 50 (1994), p. 1645.

\bibitem{pradhan_kafri_levine_08}
P.~Pradhan, Y.~Kafri, and D.~Levine, Phys. Rev. E 77 (2008), p. 041129.

\bibitem{Krapivsky_2010}
P.L. Krapivsky, S.~Redner, and E.~Ben-Naim, \emph{A Kinetic View of Statistical
  Physics}, Cambridge University Press, 2010.

\bibitem{jarzynski_92}
C.~Jarzynski, Phys. Rev. A 46 (1992), p. 7498.

\bibitem{jarzynski_93}
C.~Jarzynski, Phys. Rev. E 48 (1993), p. 4340.

\bibitem{bunin_kafri_13}
G.~Bunin and Y.~Kafri, Journal of Physics A: Mathematical and Theoretical 46
  (2013), p. 095002.

\bibitem{ott_79}
E.~Ott, Phys. Rev. Lett. 42 (1979), p. 1628.

\bibitem{cohen_00}
D.~Cohen, Ann. Phys. (NY) 283 (2000), p. 175.

\bibitem{cleuren_VandenBroeck_06}
B.~Cleuren, C.V. den Broeck, and R.~Kawai, Phys. Rev. E 74 (2006), p. 021117.

\bibitem{gaspard_andrieux_11}
P.~Gaspard and D.~Andrieux, J. Stat. Mech.  (2011), p. P03024.

\bibitem{mahan_2000}
G.D. Mahan, \emph{Many-Particle Physics (Physics of Solids and Liquids), Third
  edition}, Springer, 2000.

\bibitem{jarzynski_swiatecki_93}
C.~Jarzynski and W.J. \'Swiatecki, Nucl. Phys. A 552 (1993), p.~1.

\bibitem{blocki_brut_93}
J.~Blocki, F.~Brut, and W.J. Swiatecki, Nuclear Physics A 554 (1993), pp. 107
  -- 117.

\bibitem{blocki_skalski_95}
J.~Blocki, J.~Skalski, and W.J. Swiatecki, Nuclear Physics A 594 (1995), pp.
  137 -- 155.

\bibitem{kargovsky_anashkina_13}
A.V. Kargovsky, E.I. Anashkina, O.A. Chichigina, and A.K. Krasnova, Phys. Rev.
  E 87 (2013), p. 042133.

\bibitem{demers_jarzynski_15}
J.~Demers and C.~Jarzynski, Phys. Rev. E 92 (2015), p. 042911.

\bibitem{mehta_polkovnikov_13}
P.~Mehta and A.~Polkovnikov, Annals of Phys. 332 (2013), p. 110.

\bibitem{dalessio_krapivsky_11}
L.~D'Alessio and P.L. Krapivsky, Phys. Rev. E 83 (2011), p. 011107.

\bibitem{Allen_89}
M.P. Allen and D.J. Tildesley, \emph{Computer simulation of liquids}, Oxford
  University Press, 1989.

\bibitem{sutherland_04}
B.~Sutherland, \emph{Beautiful Models}, World Scientific, Singapore, 2004.

\bibitem{caux_mossel_11}
J.S. Caux and J.~Mossel, J. Stat. Mech.  (2011), p. P02023.

\bibitem{yuzbashyan_shastry_13}
E.A. Yuzbashyan and B.S. Shastry, J. Stat. Phys. 150 (2013), p. 704.

\bibitem{yuzbashyan_15}
E.A. Yuzbashyan, Annals of Physics 367 (2016), pp. 288 -- 296.

\bibitem{rigol_dunjko_07}
M.~Rigol, V.~Dunjko, V.~Yurovsky, and M.~Olshanii, Phys. Rev. Lett. 98 (2007),
  p. 050405.

\bibitem{jaynes_57a}
E.T. Jaynes, Phys. Rev. 106 (1957), pp. 620--630.

\bibitem{jaynes_57b}
E.T. Jaynes, Phys. Rev. 108 (1957), pp. 171--190.

\bibitem{cazalilla_06}
M.A. Cazalilla, Phys. Rev. Lett. 97 (2006), p. 156403.

\bibitem{rigol_muramatsu_06}
M.~Rigol, A.~Muramatsu, and M.~Olshanii, Phys. Rev. A 74 (2006), p. 053616.

\bibitem{kollar_eckstein_08}
M.~Kollar and M.~Eckstein, Phys. Rev. A 78 (2008), p. 013626.

\bibitem{barthel_schollwock_08}
T.~Barthel and U.~Schollw\"ock, Phys. Rev. Lett. 100 (2008), p. 100601.

\bibitem{iucci_cazalilla_09}
A.~Iucci and M.A. Cazalilla, Phys. Rev. A 80 (2009), p. 063619.

\bibitem{iucci_cazalilla_10}
A.~Iucci and M.A. Cazalilla, New J. Phys. 12 (2010), p. 055019.

\bibitem{calabrese_essler_11}
P.~Calabrese, F.H.L. Essler, and M.~Fagotti, Phys. Rev. Lett. 106 (2011), p.
  227203.

\bibitem{calabrese_essler_12b}
P.~Calabrese, F.H.L. Essler, and M.~Fagotti, J. Stat. Mech.  (2012), p. P07022.

\bibitem{cazalilla_iucci_12}
M.A. Cazalilla, A.~Iucci, and M.C. Chung, Phys. Rev. E 85 (2012), p. 011133.

\bibitem{caux_konik_12}
J.S. Caux and R.M. Konik, Phys. Rev. Lett. 109 (2012), p. 175301.

\bibitem{gramsch_rigol_12}
C.~Gramsch and M.~Rigol, Phys. Rev. A 86 (2012), p. 053615.

\bibitem{essler_evangelisti_12}
F.H.L. Essler, S.~Evangelisti, and M.~Fagotti, Phys. Rev. Lett. 109 (2012), p.
  247206.

\bibitem{collura_sotiriadis_13a}
M.~Collura, S.~Sotiriadis, and P.~Calabrese, Phys. Rev. Lett. 110 (2013), p.
  245301.

\bibitem{collura_sotiriadis_13b}
M.~Collura, S.~Sotiriadis, and P.~Calabrese, J. Stat. Mech.  (2013), p. P09025.

\bibitem{caux_essler_13}
J.S. Caux and F.H.L. Essler, Phys. Rev. Lett. 110 (2013), p. 257203.

\bibitem{mussardo_13}
G.~Mussardo, Phys. Rev. Lett. 111 (2013), p. 100401.

\bibitem{fagotti_13}
M.~Fagotti, Phys. Rev. B 87 (2013), p. 165106.

\bibitem{fagotti_essler_13a}
M.~Fagotti and F.H.L. Essler, Phys. Rev. B 87 (2013), p. 245107.

\bibitem{fagotti_collura_14}
M.~Fagotti, M.~Collura, F.H.L. Essler, and P.~Calabrese, Phys. Rev. B 89
  (2014), p. 125101.

\bibitem{pozsgay_14a}
B.~Pozsgay, J. Stat. Mech.  (2014), p. P10045.

\bibitem{wright_rigol_14}
T.M. Wright, M.~Rigol, M.J. Davis, and K.V. Kheruntsyan, Phys. Rev. Lett. 113
  (2014), p. 050601.

\bibitem{cramer_dawson_08}
M.~Cramer, C.M. Dawson, J.~Eisert, and T.J. Osborne, Phys. Rev. Lett. 100
  (2008), p. 030602.

\bibitem{he_rigol_13}
K.~He and M.~Rigol, Phys. Rev. A 87 (2013), p. 043615.

\bibitem{chung_iucci_12}
M.C. Chung, A.~Iucci, and M.A. Cazalilla, New J. Phys. 14 (2012), p. 075013.

\bibitem{torres_santos_13}
E.J. Torres-Herrera and L.F. Santos, Phys. Rev. E 88 (2013), p. 042121.

\bibitem{ziraldo_santoro_13}
S.~Ziraldo and G.E. Santoro, Phys. Rev. B 87 (2013), p. 064201.

\bibitem{holstein_primakoff_40}
T.~Holstein and H.~Primakoff, Phys. Rev. 58 (1940), pp. 1098--1113.

\bibitem{jordan_wigner_28}
P.~Jordan and E.~Wigner, Z. Phys. 47 (1928), p. 631.

\bibitem{rigol_muramatsu_05a}
M.~Rigol and A.~Muramatsu, Phys. Rev. A 72 (2005), p. 013604.

\bibitem{rigol_muramatsu_05b}
M.~Rigol and A.~Muramatsu, Mod. Phys. Lett. 19 (2005), p. 861.

\bibitem{rigol_05}
M.~Rigol, Phys. Rev. A 72 (2005), p. 063607.

\bibitem{ziraldo_silva_12}
S.~Ziraldo, A.~Silva, and G.E. Santoro, Phys. Rev. Lett. 109 (2012), p. 247205.

\bibitem{campos_zanardi_13}
L.~Campos~Venuti and P.~Zanardi, Phys. Rev. E 87 (2013), p. 012106.

\bibitem{sachdev_11}
S.~Sachdev, \emph{Quantum Phase Transitions. 2nd Ed.}, Cambridge University
  Press, 2011.

\bibitem{kormos_shashi_13}
M.~Kormos, A.~Shashi, Y.Z. Chou, J.S. Caux, and A.~Imambekov, Phys. Rev. B 88
  (2013), p. 205131.

\bibitem{nardis_wouters_14}
J.~De~Nardis, B.~Wouters, M.~Brockmann, and J.S. Caux, Phys. Rev. A 89 (2014),
  p. 033601.

\bibitem{essler_mussardo_14}
F.H.L. Essler, G.~Mussardo, and M.~Panfil, Phys. Rev. A 91 (2015), p. 051602.

\bibitem{wouters_denardis_14}
B.~Wouters, J.~De~Nardis, M.~Brockmann, D.~Fioretto, M.~Rigol, and J.S. Caux,
  Phys. Rev. Lett. 113 (2014), p. 117202.

\bibitem{pozsgay_mestyan14}
B.~Pozsgay, M.~Mesty\'an, M.A. Werner, M.~Kormos, G.~Zar\'and, and G.~Tak\'acs,
  Phys. Rev. Lett. 113 (2014), p. 117203.

\bibitem{mierzejewski_prelovssek_14}
M.~Mierzejewski, P.~Prelov\ifmmode~\check{s}\else \v{s}\fi{}ek, and T.~Prosen,
  Phys. Rev. Lett. 113 (2014), p. 020602.

\bibitem{goldstein_andrei_14}
G.~Goldstein and N.~Andrei, Phys. Rev. A 90 (2014), p. 043625.

\bibitem{rigol_14b}
M.~Rigol, Phys. Rev. E 90 (2014), p. 031301(R).

\bibitem{pozsgay_14b}
B.~Pozsgay, J. Stat. Mech.  (2014), p. P09026.

\bibitem{ilievski_medenjak_15}
E.~Ilievski, M.~Medenjak, and T.~Prosen, Phys. Rev. Lett. 115 (2015), p.
  120601.

\bibitem{ilievski_denardis_15}
E.~Ilievski, J.~De~Nardis, B.~Wouters, J.S. Caux, F.H.L. Essler, and T.~Prosen,
  Phys. Rev. Lett. 115 (2015), p. 157201.

\bibitem{russel_1844}
J.S. Russel, Rep. Br. Assoc. Adv. Sci.  (1844), pp. 311--390.

\bibitem{kdv_1895}
D.~Korteweg and G.~de~Vries, Philos. Mag. 39 (1895), pp. 422--443.

\bibitem{mazets_11}
I.~Mazets, Eur. Phys. J. D 65 (2011), p.~43.

\bibitem{berges_borsanyi_04}
J.~Berges, S.~Bors\'anyi, and C.~Wetterich, Phys. Rev. Lett. 93 (2004), p.
  142002.

\bibitem{moeckel_kehrein_08}
M.~Moeckel and S.~Kehrein, Phys. Rev. Lett. 100 (2008), p. 175702.

\bibitem{moeckel_kehrein_09}
M.~Moeckel and S.~Kehrein, Ann. Phys. 324 (2009), p. 2146.

\bibitem{eckstein_hackl_10}
M.~Eckstein, A.~Hackl, S.~Kehrein, M.~Kollar, M.~Moeckel, P.~Werner, and F.A.
  Wolf, Eur. Phys. J. Special Topics 180 (2010), p. 217.

\bibitem{kollar_wolf_11}
M.~Kollar, F.A. Wolf, and M.~Eckstein, Phys. Rev. B 84 (2011), p. 054304.

\bibitem{kitagawa_imambekov_11}
T.~Kitagawa, A.~Imambekov, J.~Schmiedmayer, and E.~Demler, New J. Phys. 13
  (2011), p. 073018.

\bibitem{mitra_13}
A.~Mitra, Phys. Rev. B 87 (2013), p. 205109.

\bibitem{kaminishi_mori_14}
E.~Kaminishi, T.~Mori, T.N. Ikeda, and M.~Ueda, Nat. Phys. 11.

\bibitem{tavora_rosch_14}
M.~Tavora, A.~Rosch, and A.~Mitra, Phys. Rev. Lett. 113 (2014), p. 010601.

\bibitem{essler_kehrein_14}
F.H.L. Essler, S.~Kehrein, S.R. Manmana, and N.J. Robinson, Phys. Rev. B 89
  (2014), p. 165104.

\bibitem{fagotti_collura_15}
M.~Fagotti and M.~Collura, arXiv:1507.02678  (2015).

\bibitem{babadi_demler_15}
M.~Babadi, E.~Demler, and M.~Knap, Phys. Rev. X 5 (2015), p. 041005.

\bibitem{mathey_polkovnikov_10}
L.~Mathey and A.~Polkovnikov, Phys. Rev. A 81 (2010), p. 033605.

\bibitem{barnett_polkovnikov_11}
R.~Barnett, A.~Polkovnikov, and M.~Vengalattore, Phys. Rev. A 84 (2011), p.
  023606.

\bibitem{stamper-kurn_ueda_13}
D.M. Stamper-Kurn and M.~Ueda, Rev. Mod. Phys. 85 (2013), p. 1191.

\bibitem{nessi_iucci_14}
N.~Nessi, A.~Iucci, and M.A. Cazalilla, Phys. Rev. Lett. 113 (2014), p. 210402.

\bibitem{gurarie_95}
V.~Gurarie, Nucl.Phys. B 441 (1995), p. 569.

\bibitem{scheppach_berges_09}
C.~Scheppach, J.~Berges, and T.~Gasenzer, Phys.Rev.A 81 (2010), p. 033611.

\bibitem{nowak_schole_12}
B.~Nowak, J.~Schole, D.~Sexty, and T.~Gasenzer, Phys. Rev. A 85 (2012), p.
  043627.

\bibitem{shimshoni_andrei_03}
E.~Shimshoni, N.~Andrei, and A.~Rosch, Phys. Rev. B 68 (2003), p. 104401.

\bibitem{boulat_mehta_07}
E.~Boulat, P.~Mehta, N.~Andrei, E.~Shimshoni, and A.~Rosch, Phys. Rev. B 76
  (2007), p. 214411.

\bibitem{serbyn_papic_13a}
M.~Serbyn, Z.~Papi\'c, and D.A. Abanin, Phys. Rev. Lett. 111 (2013), p. 127201.

\bibitem{huse_oganesyan_14}
D.~Huse and V.~Oganesyan, Phys. Rev. B 90 (2014), p. 174202.

\bibitem{stark_kollar_13}
M.~Stark and M.~Kollar, arXiv:1308.1610  (2013).

\bibitem{kapitza_51}
P.L. Kapitza, Soviet Phys. JETP 21 (1951), p. 588.

\bibitem{chirikov_shepelyansky_08}
B.~Chirikov and D.~Shepelyansky, Scholarpedia 3 (2008), p. 3550.

\bibitem{greene_79}
J.M. Greene, J. Math. Phys. 20 (1979), p. 1183.

\bibitem{mackay_83}
R.S. MacKay, Physica D 7 (1983), p. 283.

\bibitem{mackay_percival_85}
R.S. MacKay and I.~Percival, Comm. Math. Phys. 94 (1985), p. 469.

\bibitem{fisher_67}
M.E. Fisher, \emph{The Nature of Critical Points}, in \emph{Boulder Lectures in
  Theoretical Physics}, Vol.~7, University of Colorado, Boulder (1965).

\bibitem{yang_lee_52}
C.N. Yang and T.D. Lee, Phys. Rev. 87 (1952), p. 404.

\bibitem{heyl_vojta_13}
M.~Heyl and M.~Vojta, Phys. Rev. B 92 (2015), p. 104401.

\bibitem{heyl_polkovnikov_13}
M.~Heyl, A.~Polkovnikov, and S.~Kehrein, Phys. Rev. Lett. 110 (2013), p.
  135704.

\bibitem{odlyzko_99}
A.M. Odlyzko, \emph{The {$10^{22}$}-nd zero of the {R}iemann zeta function}, in
  \emph{Dynamical, spectral, and arithmetic zeta functions ({S}an {A}ntonio,
  {TX}, 1999)}, Contemp. Math., Vol. 290, Amer. Math. Soc., Providence, RI,
  2001, pp. 139--144.

\bibitem{sivak_crooks_12}
D.A. Sivak and G.E. Crooks, Phys. Rev. Lett. 108 (2012), p. 190602.

\end{thebibliography}

\end{document}